\documentclass[12pt,preprint]{aastex}
\usepackage{ulem}
\usepackage{color}

\usepackage{mathtools}

\newcommand{\eb}{{\boldsymbol{e}}}
\newcommand{\rb}{{\boldsymbol{r}}}
\newcommand{\Sb}{{\bf{S}}}
\newcommand{\Ub}{{\bf{U}}}
\newcommand{\Ubt}{\widetilde{\Ub}}
\newcommand{\Vb}{{\bf{V}}}
\newcommand{\xb}{{\boldsymbol{x}}}

\newcommand{\kappab}{{\boldsymbol{\kappa}}}

\newcommand{\Hnuc}{H_{\rm nuc}}
\newcommand{\omegadot}{\dot\omega}
\newcommand{\gammaonebar}{\overline{\Gamma_1}}

\renewcommand{\d}{{\mathrm{d}}}

\newcommand{\gtaprx}{\lower .1ex\hbox{\rlap{\raise .6ex\hbox{\hskip .3ex
        {\ifmmode{\scriptscriptstyle >}\else
                {$\scriptscriptstyle >$}\fi}}}
        \kern -.4ex{\ifmmode{\scriptscriptstyle \sim}\else
                {$\scriptscriptstyle\sim$}\fi}}}
\newcommand{\ltaprx}{\lower .1ex\hbox{\rlap{\raise .6ex\hbox{\hskip .3ex
        {\ifmmode{\scriptscriptstyle <}\else
                {$\scriptscriptstyle <$}\fi}}}
        \kern -.4ex{\ifmmode{\scriptscriptstyle \sim}\else
                {$\scriptscriptstyle\sim$}\fi}}}

\begin{document}

\title{High-Resolution Simulations of Convection Preceding Ignition in Type Ia Supernovae
       Using Adaptive Mesh Refinement}
\shorttitle{High-Resolution SNe Ia Convection}
\shortauthors{Nonaka et al.}

\author{A.~Nonaka\altaffilmark{1},
        A.~J.~Aspden\altaffilmark{1,2},
        M.~Zingale\altaffilmark{3},
        A.~S.~Almgren\altaffilmark{1},
        J.~B.~Bell\altaffilmark{1},
        S.~E.~Woosley\altaffilmark{4}}

\altaffiltext{1}{Center for Computational Sciences and Engineering,
                 Lawrence Berkeley National Laboratory,
                 Berkeley, CA 94720, USA}

\altaffiltext{2}{School of Engineering,
                 University of Portsmouth,
                 Portsmouth, Hants, PO1 3DJ, UK}

\altaffiltext{3}{Department of Physics \& Astronomy,
                 Stony Brook University,
		 Stony Brook, NY 11794-3800, USA}

\altaffiltext{4}{Department of Astronomy \& Astrophysics,
                 The University of California, Santa Cruz,
                 Santa Cruz, CA 95064, USA}

\begin{abstract}
We extend our previous three-dimensional, full-star simulations of 
the final hours of convection preceding ignition in Type Ia 
supernovae to higher resolution
using the adaptive mesh refinement capability of our low Mach number code,
MAESTRO.  We report the statistics of the ignition of the first flame
at an effective 4.34~km resolution,
and general flow field properties at an effective 2.17~km resolution.
We find that off-center ignition is likely, with radius of 50~km most 
favored and a likely range of 40 to 75~km.
This is consistent with our previous coarser (8.68~km resolution) simulations, 
implying that we have achieved sufficient resolution in our determination 
of likely ignition radii.  The dynamics of the last few hot spots
preceding ignition suggest that a multiple ignition scenario is not likely.
With improved resolution, we can more clearly see 
the general flow pattern in the convective region, characterized by a 
strong outward plume with a lower speed recirculation.
We show that the convective core is turbulent with a Kolmogorov spectrum and
has a lower turbulent intensity and larger integral length scale than previously
thought (on the order of 16~km~s$^{-1}$ and 200~km, respectively), and we discuss 
the potential consequences for the first flames.
\end{abstract}

\section{Introduction}
For the Chandrasekhar mass white dwarf
(single-degenerate) progenitor model of Type Ia supernovae (SNe Ia), the
location of the first flames greatly affects the outcome of the
explosion (see for example
\citealt{niemeyer:1996,plewa:2004,livne:2005,garciasenz:2005}).  The
convective state leading up to ignition is highly nonlinear and the
ignition results from a hot temperature perturbation near the center
of the white dwarf.  Once the temperature exceeds $\sim$$8\times
10^8$~K, a hot spot burns faster than it can cool via expansion
\citep{nomoto1984iii}, igniting a flame.  In earlier studies on white dwarf
convection in SNe Ia
(\citealt{paper4}, henceforth Z09; \citealt{wdconvect}, henceforth Z11), 
we performed three-dimensional, full-star
simulations of the final $\sim$3 hours of convection in a
white dwarf leading up to the ignition of the first flames.  We
followed the nonlinear rise in the temperature approaching ignition
and showed that the ignition is likely to take place off-center 
(50~km most favored, with a likely range of 40 to 75~km, and an
outer limit of 100~km) in an outward flowing parcel of fluid.  Our 
results differed from the two-dimensional 
wedge simulation of \cite{hoflichstein:2002} which argued that the ignition 
is closer to the center ($\sim 30$~km), and is driven by inflow compression.

It is important to understand how robust our results for the likely ignition 
radius are to resolution.  With the recently implemented
adaptive mesh refinement (AMR) capability in our low Mach number code,
MAESTRO (\citealt{multilevel}, henceforth N10), 
we are now able to study the final minutes
of convection up to ignition at unprecedented resolution.
We are also interested in the likelihood of multiple ignition points.
Detailed visualizations of the evolution of the last few hot
spots preceding ignition will be used to examine this scenario.
Previous studies with
an anelastic code showed that a dipole flow dominates the flow
(\citealt{kuhlen-ignition:2005}, also seen in the follow-up studies
shown in \citealt{woosley-scidac2007,ma:2011}).  Our results for
non-rotating white dwarfs also saw this feature.  Here we examine
this structure at higher resolution.  

Higher resolution is also important for resolving the 
turbulence and capturing the turbulent cascade.
Simulations have shown that the flame needs to accelerate considerably
beyond its laminar value for the resulting energetics to match
observations.  The primary mechanism for this acceleration is thought
to be instabilities and the interaction with turbulence
\citep{mullerarnett1986,livne1993,khokhlov:1995,niemeyerhillebrandt1995,niemeyerwoosley1997}.
A popular view is that the flame interacts with turbulence 
generated by the flame itself via instabilities.  The vast majority of simulations
to date have only considered this flame-generated turbulence during
the explosion phase.
\cite{aspden-bubble} suggested that turbulent entrainment was the dominant
mechanism for enhancing the burning rate, and that the local flame speed, 
whether laminar or turbulent, was largely unimportant.  As the
flame encounters lower densities and broadens, the turbulence may be
able to directly affect the flame structure (at this point, the flame
is said to be in the ``distributed burning regime'').  The altering of
the flame by turbulence has been the subject of many studies, both
semi-analytic and one-dimensional calculations with a model for
turbulence \citep{lisewski:2000,pan2008,Woosley:2009} and multi-dimensional
numerical
simulations \citep{roepke2004,Aspden:2008,Aspden:2010,Aspden:2011}.  If the
conditions are right, the flame may transition to a detonation in this
regime \citep{khokhlov:1997,niemeyerwoosley1997,Woosley:2009,Woosley:2011}.

What are not well known are the characteristics of the turbulence that
already exists at ignition from the centuries-long convective period.
The very first flame(s) that ignite will form flame ``bubbles'' that
buoyantly rise away from the center as they burn outward.  These
bubbles will deform due to shear instabilities and interact with the
pre-existing turbulence and wrinkle \citep{garciasenzwoosley1995,
bychkovliberman,iapichino,zingaledursi,iapichino:2010,aspden-bubble}.
If the turbulence is strong enough, it could potentially disrupt the
flames or even quench them.  Additionally, the initial convective
velocity field has been shown to introduce large asymmetries in the
burning \citep{livne:2005}.

In this paper, we expand upon our previous studies of the final 
hours of convection leading up to the ignition of the first flames in 
Type Ia supernovae.  In Z09, we used 13.2~km resolution; in Z11, we
used 8.68~km resolution.  Here, we use the AMR capability of MAESTRO to 
compute ignition statistics at 4.34~km resolution, and general flow field 
properties at 2.17~km resolution.  

This paper is organized as follows.  In Section \ref{sec:Numerical Methodology}
we give an overview of the MAESTRO algorithm, including our latest improvements
for both regridding and adding an additional level of refinement to a simulation 
in progress.  In Section \ref{Sec:Results}, we describe our new high-resolution
simulations.  We examine the ignition statistics and compare them to our previous 
results in order to show that we have achieved sufficient resolution in our 
determination of likely ignition radii.  We determine the likelihood of multiple
ignition points by examining the dynamics of the last few hot spots leading
up to ignition.  We provide visualizations of
the convective flow field to gain a better understanding of the flow structure.
We include a detailed analysis of the turbulent nature of the flow field,
and discuss the implications for the first flames.  Finally, in Section 
\ref{Sec:Conclusions and Discussion}, we summarize and conclude.

\section{Numerical Methodology}\label{sec:Numerical Methodology}
MAESTRO is a finite-volume, AMR hydrodynamics code for low Mach number
astrophysical flows.  In our low Mach number formulation, sound waves
have been analytically removed, allowing for a time step 
based on the fluid velocity CFL constraint rather than the sound speed
CFL constraint, while retaining compressibility effects due to
background stratification, reaction heating, and compositional
changes.  The algorithm is described in full detail in
N10.  We note that the low Mach number equations do not enforce that 
the Mach number remain small; rather, if the
dynamics of the flow are such the Mach number does remain small, then these equations are
valid approximations for the evolution of the flow.  Thus, MAESTRO is not suitable
for post-ignition calculations, where we expect the Mach number to quickly approach
or exceed unity.  Also, our low Mach number approach assumes that the background
state is spherical; thus, any deformation due to rotation is not accounted for
in the background state.

We now summarize the algorithm, and then
describe the new procedures for dynamically changing the grids as well
as adding an additional level of refinement to a simulation in
progress.  For the simulations in this paper, we begin with data
from Z11, in which we computed the last $\sim$3 hours of
convection in a non-rotating white dwarf up to the point of ignition
using 8.68~km resolution ($576^3$ computational cells; the problem domain is
5000 km on a side) and no AMR.  We expand upon this simulation by adding a
level of refinement a few minutes before ignition and examining the
ignition statistics for a 4.34~km ($1152^3$ effective grid cells)
resolution simulation.  Next, we will add an additional level of AMR
to examine the turbulent flow field in a 2.17~km ($2304^3$ effective grid
cells) resolution simulation.  Computer allocation limits prevent
us from running 2.17~km resolution simulations to ignition, even with
the efficiency gains provided by AMR.

\subsection{MAESTRO Overview}
MAESTRO is based on the BoxLib software framework~\citep{rendleman-hyper}, which provides
infrastructure for block-structured AMR applications, and includes
linear solvers that scale to 100,000 cores on the current generation of 
supercomputers (see \citealt{scidac-scaling} for details).  We use a finite-volume
approach, where each computational cell stores the average value of a state variable
in that cell.  The domain is decomposed into a nested
hierarchy of logically rectangular Cartesian grids with computational cell width
$\Delta x^{\ell}$ in each direction (the grids at the coarsest
level are associated with level index $\ell = 1$, the first level of refinement with 
$\ell = 2$, etc.), and a refinement ratio of two in each spatial direction.
We solve a system of coupled PDEs containing advection and reaction terms
constrained by an equation of state that takes the form of a divergence
constraint on the velocity field.  This constraint is enforced using a projection
method, which requires linear solvers to solve a variable-coefficient Poisson
equation.

\begin{figure}
\begin{center}
   \begin{minipage}[t]{1.25in}
      \vspace{0.25in}
      \includegraphics[width=1.25in]{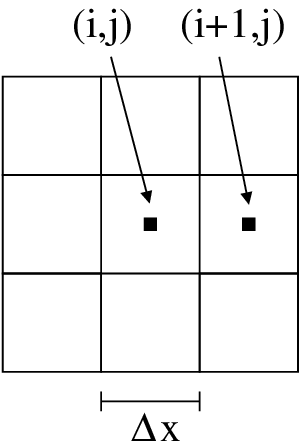}
   \end{minipage}
   \hfill
   \begin{minipage}[t]{2.25in}
      \vspace{0.75in}
      \includegraphics[width=2.25in]{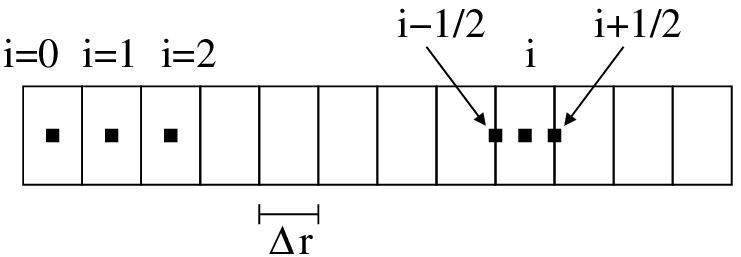}
   \end{minipage}
   \hfill
   \begin{minipage}[t]{2.5in}
      \vspace{0in}
      \includegraphics[width=2.0in]{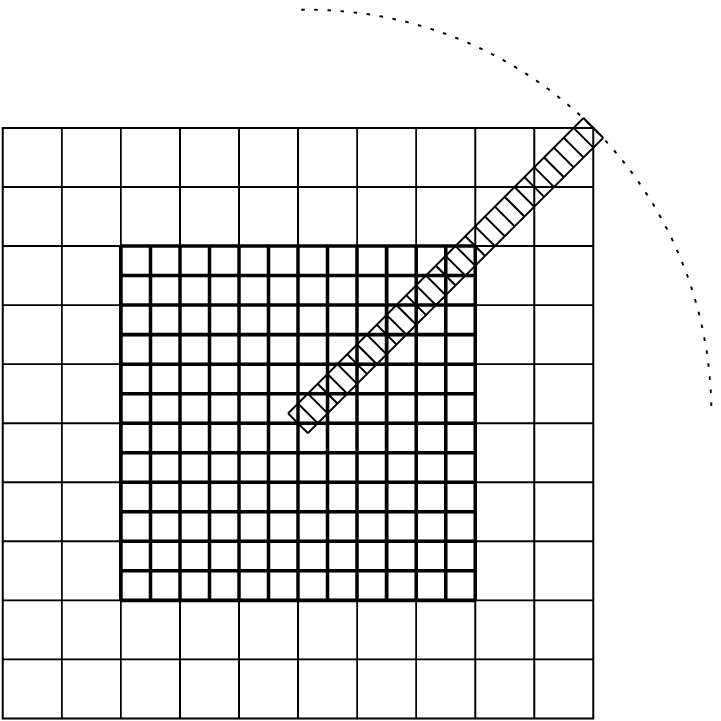}
   \end{minipage}
\caption{\label{fig:grid} (Left) For data on the Cartesian grid (shown here in two 
dimensions), we use a cell-centered convention to denote the average value over the 
computational cell.  (Center) The base state variables live on a one-dimensional 
radial array, and can live at cell centers or edges.  
(Right) A graphical depiction of how the base state and Cartesian grid are related.  
Note that there is no direct alignment between the radial cell centers and the
Cartesian grid cell centers.}
\end{center}
\end{figure}

One feature that makes MAESTRO different from standard AMR hydrodynamics codes is the
presence of base state variables, which are functions of radius and
time, $(r,t)$, as opposed to Cartesian grid quantities which are functions of all
spatial dimensions and time, $(\xb,t)$.  We represent base state
variables using a one-dimensional, time-dependent array.  The base
state array has a constant grid spacing, $\Delta r = \Delta
x^{\ell_{\rm max}}/5$, where $\ell_{\rm max}$ is the finest level in
the simulation, and due to the spherical nature of our problem, does
not directly align with the Cartesian grid.  Figure \ref{fig:grid} shows
a depiction of the Cartesian grid, one-dimensional radial array, and a
graphical representation of how they relate to each other.  Some base
state variables are cell-centered, and others are defined on edges.  Each
of the base state variables is computed directly from other base
state variables and/or Cartesian grid variables.  The base
state density obeys an evolution equation within each time step
(described below).
We require frequent mapping from the base state to the
Cartesian grid, and vice versa.  In N10, we describe how
we interpolate a base state quantity onto the Cartesian grid, as well
as a ``lateral average'' procedure that determines the average value
of a Cartesian grid quantity at a particular radius and maps that
value onto the radial array.

In the following overview, all variables are assumed to live on the Cartesian grid, 
unless noted otherwise.  MAESTRO solves the equations of reacting flow constrained 
by an equation of state in the form of a divergence constraint.  The 
species are evolved according to
\begin{equation}
\frac{\partial(\rho X_k)}{\partial t} = -\nabla\cdot(\rho X_k\Ub) + \rho\omegadot_k,\label{eq:species}
\end{equation}
where $\rho$ is the density, $X_k$ is the mass fraction of species $k$,
$\Ub$ is the velocity field, and $\omegadot_k$ is the creation rate of
species $k$ due to reactions.  We note that the density can be determined at 
any time using
\begin{equation}
\rho = \sum_k (\rho X_k),\label{eq:density}
\end{equation}
and thus density does not have to be explicitly evolved in time.  

We define a base state density, $\rho_0(r,t)$, that represents the average
value of density at a particular radius.  The base state density has its own
evolution equation, as described below.  The base state (thermodynamic) pressure,
$p_0(r,t)$, is computed using the condition of hydrostatic equilibrium, 
\begin{equation}
\nabla p_0 \equiv \frac{\partial p_0}{\partial r} = -\rho_0 g,\label{eq:HSE}
\end{equation}
where the gravity, $g(r,t)$ is computed by integrating $\rho_0$ assuming 
piecewise-constant profiles of $\rho_0$ within each radial cell.

In general, given $\rho$ and $X_k$, we could 
derive the temperature from the specific enthalpy, $h$, evolved as
\begin{equation}
\frac{\partial(\rho h)}{\partial t} = -\nabla\cdot(\rho h\Ub) + \frac{Dp_0}{Dt} + \rho\Hnuc,
\end{equation}
where $\Hnuc$ is
the energy generation rate from reactions.  In practice, we adopt the
prescription used in Z09, Z11 and derive the temperature from
$\rho, X_k,$ and $p_0,$ effectively decoupling the enthalpy from the problem.  In the future,
we will seek ways to evolve the enthalpy in a manner that minimizes the drift
from the equation of state. 

The velocity field is evolved according to
\begin{equation}
\frac{\partial\Ub}{\partial t} = -\Ub\cdot\nabla\Ub  - \frac{1}{\rho}\nabla\pi - \frac{\rho-\rho_0}{\rho} g\eb_r,\label{eq:momentum}
\end{equation}
where $\pi$ is the perturbational pressure, i.e., the local deviation of the total
pressure from $p_0$, and $\eb_r$ is the unit vector 
in the outward radial direction.  The evolution of the thermodynamic variables 
($\rho, X_k$, and $p_0$) are constrained by the equation of state,
which we represent as a divergence constraint on the velocity field,
\begin{equation}
\nabla\cdot(\beta_0\Ub) = \beta_0\left(S - \frac{1}{\gammaonebar p_0}\frac{\partial p_0}{\partial t}\right).\label{eq:U divergence}
\end{equation}
Here, $\beta_0(r,t)$ is a base state quantity that captures the expansion/contraction 
of a fluid parcel as it changes altitude, and $S$ is a local source term that captures 
the compressibility effects due to reactions and compositional changes.  The quantity 
$\gammaonebar(r,t)$ is a base state variable representing the average at constant
radius of $\Gamma_1 = \partial \log p / \partial \log \rho |_s$, where
$s$ is the entropy.  A full derivation of this constraint, the form
of $\beta_0$ and $S$, and the numerical projection can be found in 
\cite{ABRZ:I,ABRZ:II,ABNZ:III}.

The evolution equation for $\rho_0$ is
\begin{equation}
\frac{\partial\rho_0}{\partial t} = -\frac{\partial\left(\rho_0w_0\right)}{\partial r} - \frac{\partial\eta_\rho}{\partial r},\label{eq:rho0}
\end{equation}
where $w_0(r,t)$ is the base state expansion velocity, which accounts for the expansion 
of the atmosphere due to large-scale heating.  We compute this term by integrating a 
one-dimensional version of the divergence constraint (Eq. [\ref{eq:U divergence}]).  
The quantity $\eta_\rho(r,t)$ is 
a base state quantity that accounts for changes in background stratification
due to large-scale convection (see \citealt{ABNZ:III} and N10).

The velocity field can be decomposed into the base state velocity and a local 
velocity, $\Ubt(\xb,t)$, that governs the local dynamics,
\begin{equation}
\Ub(\xb,t) = w_0(r,t)\eb_r + \Ubt(\xb,t).
\end{equation}
We follow the approach in N10, where we compute the evolution of these terms 
separately, and thus evolve $\Ubt$ subject to a perturbational form of 
Equations (\ref{eq:momentum}) and (\ref{eq:U divergence}).

We note that the base state quantities $\rho_0, p_0, \beta_0, \gammaonebar$, and 
$\eta_\rho$ are stored on radial cell-centers, whereas $w_0$ is stored at radial edges.

To summarize, we advance Equations (\ref{eq:species}), (\ref{eq:momentum}),
and (\ref{eq:rho0}), subject to Equations (\ref{eq:density}), (\ref{eq:HSE}) and 
(\ref{eq:U divergence}).  We use a second-order predictor-corrector approach in which
we discretize the advection terms using an unsplit Godunov method, compute
the effect of reactions on a cell-by-cell basis using the VODE stiff ODE package, and
couple these processes using Strang splitting.
We enforce the divergence constraint on velocity using a projection method, which
uses multigrid to solve a variable-coefficient Poisson equation 
for the update for the perturbational pressure, $\pi$.

\subsection{Regridding and Adding a Level of Refinement}
Regridding is the process of redefining the AMR grid structure based on
user-specified refinement criteria.  The regridding algorithm
also uses interpolation stencils to initialize data on newly created refined
grids from underlying coarse data.
Here we have improved the regridding algorithm described in
N10 and have also implemented a new algorithm for adding 
an additional level of refinement to a simulation in progress.  Our 
approach to AMR uses a nested hierarchy of logically rectangular grids 
with successively finer grids at higher levels.  This is based on the 
strategy introduced for gas dynamics by \citet{berger-colella}, extended 
to the incompressible Navier-Stokes equations by \citet{AlmBelColHowWel98}, 
and extended to low Mach number reacting flows by \citet{pember-flame} 
and \citet{DayBell:2000}.  We refer the reader to these works for more 
details.  The complication in applying these methods to MAESTRO is the 
presence of the time-dependent base state variables.  We refer the 
reader to N10 for the MAESTRO-specific implementation 
including details on creating and managing the grid hierarchy, 
communication between levels, and the implementation of AMR with 
time-dependent base state variables.

We note an error in the Cartesian grid regridding procedure as 
described in Section 5.1 of N10.  For problems with three or
more total AMR levels, we require that each grid at level $\ell+1$ 
be a distance of at least four (not two as previously reported) 
level $\ell$ cells from the boundary between level $\ell$ and level 
$\ell-1$ grids; this allows us to always fill ``ghost cells'' at 
level $\ell+1$ from the level $\ell$ data (or the physical boundary 
conditions, if appropriate).

The major change regarding the regridding of the Cartesian grid data
is in the way we interpolate coarse data to fill newly created fine
grids.  Our piecewise-linear interpolation algorithm applied
to $\rho X_k$ causes an artificial buoyancy term to appear in the 
momentum equation, leading to spurious
velocities emanating from the coarse-fine interface.  The basic idea
of the improved algorithm is to interpolate $\rho'$ and $X_k$ 
separately, rather than $\rho X_k$, to initialize data on 
the newly created refined grids.

The variables on the Cartesian grid that we need to interpolate are $\Ub$, $\rho$, 
$\rho X_k$, $\nabla\pi$, and $S$.  The base state does not change structure,
but we still need to recompute $\rho_0$, $p_0$, $\beta_0$, and 
$\gammaonebar$ to be consistent with the data on the Cartesian grid.
We keep the original values of $w_0$ and $\eta_\rho$.
Here are the steps for regridding:
\begin{enumerate}

\item Starting with level 1 and user-defined refinement criteria, tag all level 1
Cartesian cells that satisfy the criteria for refinement.  Create the level 2 grids, 
and initialize the level 2 data by copying from the previous grid structure where possible.
Otherwise, use piecewise linear interpolation from underlying coarse cells to initialize any 
newly created refined regions, including ghost cells.  Continue to add additional levels 
of refinement in this way until all data on the grids at level $\ell_{\rm max}$ are filled in.
There is a slight modification to the interpolation procedure for $\rho X_k,$
where we first interpolate $\rho' = \rho - \rho_0$ and $X_k = (\rho X_k)/\rho$
to newly refined regions and then construct $\rho X_k = (\rho' + \rho_0)X_k$.

\item Recompute $\rho_0$ by calling the lateral average routine, then use Equation 
(\ref{eq:HSE}) to compute $p_0$.

\item Recompute $T$ and $\Gamma_1$ on each Cartesian cell using the equation of 
state.  Recompute $\gammaonebar$  by calling the lateral average routine.
Then, recompute $\beta_0$ as described in N10.

\item Compute a new appropriate $\Delta t$.

\end{enumerate}

The procedure for adding an additional level of refinement to a simulation in progress
is largely based on the standard regridding procedure, except that now the base
state array will have twice as many cells since $\Delta r$ is based on the resolution
of the finest Cartesian grid, i.e., $\Delta r = \Delta x^{\ell_{\rm max}}/5$.
To add one additional level of refinement to a simulation in progress:
\begin{enumerate}

\item Perform Step 1 in the regridding procedure defined above, except allow 
for an additional level of refinement.

\item Define a new base state array with twice the resolution, i.e.,
$\Delta r = \Delta x^{\ell_{\rm max}}/5$.  Call the lateral average routine to 
compute $\rho_0$ and use Equation (\ref{eq:HSE}) to recompute $p_0$.

\item Recompute $T$ and $\Gamma_1$ on each Cartesian cell using the equation of 
state.  Call the lateral average routine to compute $\gammaonebar$.  Then,
recompute $\beta_0$ as described in N10.

\item The base state variable $w_0$ is edge-centered.  We compute $w_0$ on the finer
base state array using direct injection from the previous coarser base state array on 
aligning edges, and piecewise-linear interpolation on non-aligning edges.

\item The base state variable $\eta_{\rho}$ is cell-centered.  We interpolate 
$\eta_\rho$ onto the finer base state array using piecewise-linear interpolation 
from the previous coarser base state array.\footnote{
In practice, we store $\eta_\rho$ on radial 
cell centers and edges as separate arrays.  We interpolate the radial 
cell-centered and edge-based arrays onto the finer base state arrays 
separately, rather than simply interpolate the radial cell-centered array 
onto the finer base state array, and then arithmetically average to get the 
radial edge-centered array.  In the future, we will run in the latter 
mode for simplicity, noting that the effects of this change are very minor, and
that both methods are second-order.}

\item Compute a new appropriate $\Delta t$.

\end{enumerate}

\section{Results}\label{Sec:Results}
We now focus on one particular simulation performed in Z11, in
which we computed the last $\sim$3 hours of convection
up to the point of ignition for a non-rotating white dwarf using 8.68~km resolution 
($576^3$ computational cells) and no AMR.  As before, we define ignition as the time when 
the maximum temperature exceeds $8\times 10^8$~K.  Here is a summary of our results
from that simulation:

\begin{itemize}

\item The plots of peak temperature, peak radial velocity, and peak Mach number as a 
function of time each exhibited a gradual, non-linear rise until the peak temperature exceeded
$\sim$$7\times 10^8$~K.  Then, the rise in each field became much steeper, 
with ignition following shortly afterwards.

\item The first cell to ignite was 25.7~km off-center, and had an outward radial velocity
of 5.1~km~s$^{-1}$.

\item For the last $\sim$3 minutes preceding ignition, the average radius of the hottest cell
was 52.3~km with a standard deviation of 25.5~km.

\item Histograms of the radius of the hottest cell during the final $\sim$3 minute preceding ignition
averaged over small time intervals indicated that

\begin{itemize}

\item The favored ignition radius was 50~km, with a likely range of 40~km to 75~km, and
an outer limit of 100~km.

\item Nearly all of the hot spots had an outward radial velocity.

\item These two results were consistent within any smaller time window 
within the final $\sim$3 minutes.

\end{itemize}

\item Visualizations of the convective flow field showed a dipole structure in the 
interior convectively unstable core, and a sharp boundary between the interior 
and the stably stratified outer portion of the star. 

\end{itemize}

In this section, we examine the robustness of the ignition results at higher 
resolution.  Then, we use new visualization techniques to follow the last
few hot spots preceding ignition to determine the likelihood of multiple
ignition points.  We also visualize the overall convective flow field
to show the detailed fine-scale structure, as well as a more coherent picture 
of the large-scale features.  Finally, we analyze the turbulent spectrum 
to quantify the extent to which we have resolved the 
turbulent cascade, and discuss the effect that turbulence could have on
the first few flames.

We note that we do not consider a high-resolution rotating white dwarf
at this time.  A $576^3$ rotating simulation developed high velocities on the
surface of the star at the poles, likely due to the deformation of the
star.  In our lower resolution rotating runs in Z11, we saw a similar feature,
but the velocities did not become large enough to restrict our time step as
they do for the higher-resolution runs.  A
potential future solution to this would be to reformulate the base
state in MAESTRO to deal with equipotentials instead of a spherical radius.

\subsection{Problem Setup}
The 8.68~km resolution simulation in Z11 followed the last
$\sim$10500~s preceding ignition.  The simulation required $\sim$6 
million CPU hours on the Jaguarpf Cray XT5 at OLCF.  Assuming perfect scaling and
no AMR overhead, a 4.34~km simulation from $t=0$ would require a
factor of $\sim$4 more computational resources (since the time step is a factor
of two smaller, and with our tagging criteria we have approximately the same
number of cells at levels 1 and 2, so there are twice as many
total grid cells).  Due to computer allocation limits,
running 4.34~km resolution from $t=0$ is infeasible, so instead we add an
additional level of refinement to an 8.68~km simulation at a time corresponding
to $\sim$250~s preceding ignition.  
The edge of the star lies 
where $\rho_0\approx 1\times 10^5$~g~cm$^{-3}$, corresponding to a radius
of $r\approx 1890$~km.  We refine all level 1 cells where 
$\rho > 5\times 10^7$~g~cm$^{-3}$ ($r\approx 1225$~km), which more 
than encompasses the convective region (the convective region boundary lies 
approximately where $\rho_0\approx 1.26\times 10^8$~g~cm$^{-3}$, with
$r\approx 1030$~km).  
This new simulation has 4.34~km resolution ($1152^3$ effective grid cells).  
We note that since this problem is 
highly non-linear, we do not expect the ignition to occur at exactly the same time.
In fact, the 4.34~km simulation takes $\sim$350~s to ignite.
Approximately 100~s into the $1152^3$ simulation, we add another level 
of refinement, tagging all level 2 cells where $\rho > 1\times 10^8$~g~cm$^{-3}$
($r\approx 1080$~km).
This second new simulation has 2.17~km resolution ($2304^3$ effective grid cells).
We run the 2.17~km simulation for $\sim$80~s, 
and not to ignition (again due to computer allocation limits).

\begin{figure}
\begin{center}
\includegraphics[width=3in]{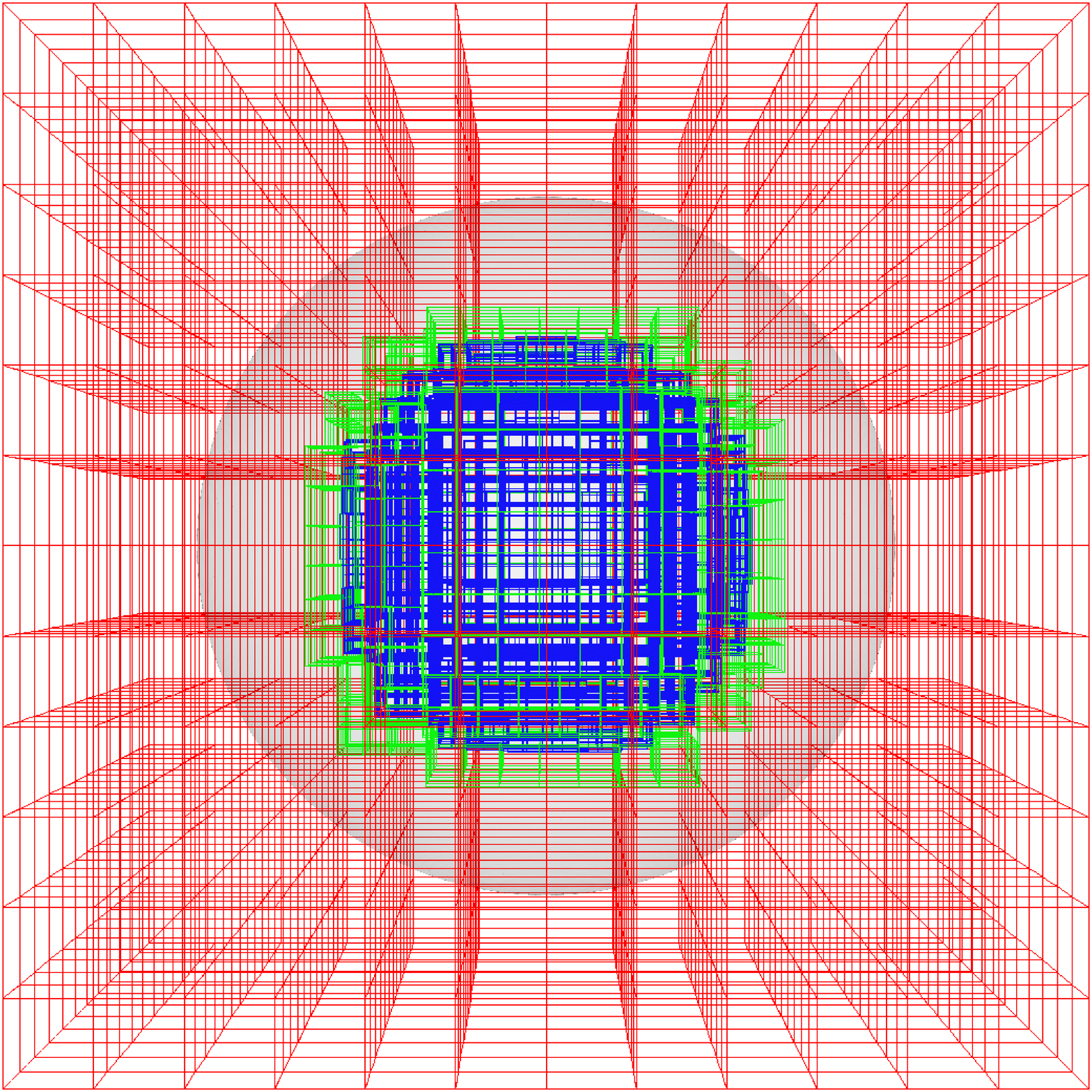}
\includegraphics[width=3in]{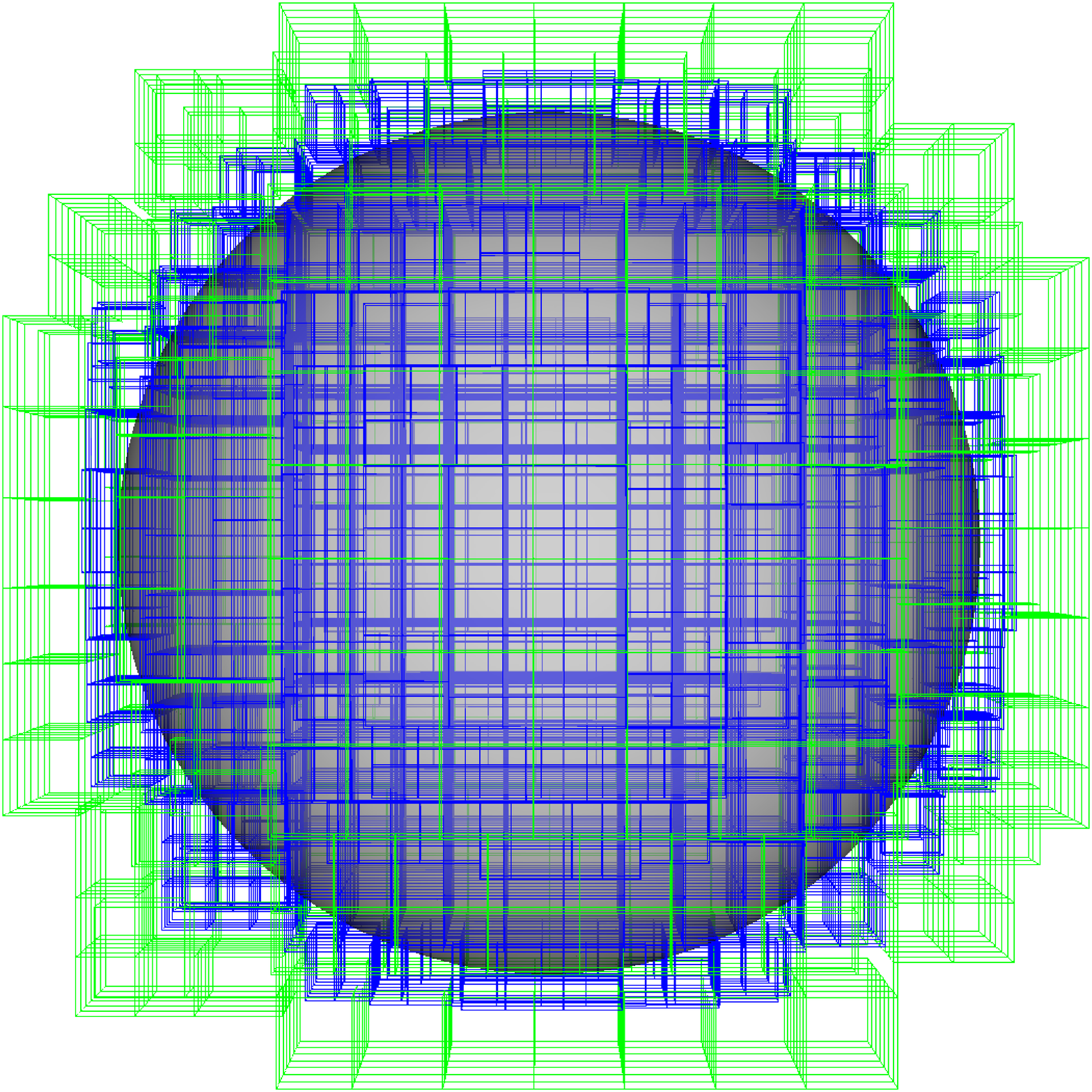}
\caption{\label{fig:wd_2304_grid} Grid structure for our three-level
simulations.  The base grid has $576^3$ grid cells (8.68~km resolution), and the refined
grids have effective $1152^3$ (4.34~km) and $2304^3$  (2.17~km) grid cells.  The red, 
green, and blue
outlines indicate boxes which can contain up to $64^3$ grid cells.  
(Left) The shaded  region indicates the edge of the star, defined by the location where 
$\rho = 10^5$~g~cm$^{-3}$ at $r\approx 1890$~km.  (Right) In this zoom-in, the shaded 
region indicates the edge of the convective region, defined by the location 
where $\rho\approx 1.26\times 10^8$~g~cm$^{-3}$ at $r\approx 1030$~km.  The finest grids 
contain the entire convective region.}
\end{center}
\end{figure}

The resulting three-level grid structure is shown in Figure \ref{fig:wd_2304_grid}.
The grids adaptively change as the simulation progresses, but since the overall 
base state density profile of the star is relatively constant (as shown in 
Z11), the grids do not change significantly over time.  Some
specific details concerning this grid structure are as follows.
\begin{itemize}

\item The red grids are at 8.68~km resolution.  There are 1728 red grids, 
      each of which has $48^3$ grid cells ($\sim$191 million grid cells total).

\item The green grids are at 4.34~km resolution.  There are
      1736 green grids of varying size, with a maximum of $48^3$ cells per grid.
      ($\sim$141 million grid cells total).

\item The blue grids are at 2.17~km resolution.  There are
      3646 blue grids of varying size, with a maximum of $64^3$ cells per grid.
      ($\sim$654 million grid cells total).

\end{itemize}
By contrast, a simulation without AMR at 2.17~km resolution would contain 
$2304^3 = 12.2$ billion grid cells, or a factor of $\sim$12 more grid cells
than the AMR simulation.

For the simulations in this paper, we use the recently implemented
hierarchical approach to parallelism described in \cite{scidac-scaling}.
We use a coarse-grain parallelization strategy to distribute grids to nodes, 
where the nodes communicate with each other using MPI.
We also use a fine-grain parallelization strategy in the physics-based modules
and the linear solvers, in which we use OpenMP to spawn a thread on
each core on a node.  Each thread operates on a portion of the associated grid.
Grids at each level of refinement are distributed independently.
This approach allows for MAESTRO (in particular the linear solvers)
to scale to $\sim$100,000 cores.  All simulations were performed
on the Jaguarpf Cray XT5 at OLCF using 1,728 MPI processes with
6 threads per MPI process (10,368 total cores).

\subsection{General Behavior}
\begin{figure}
\begin{center}
\includegraphics{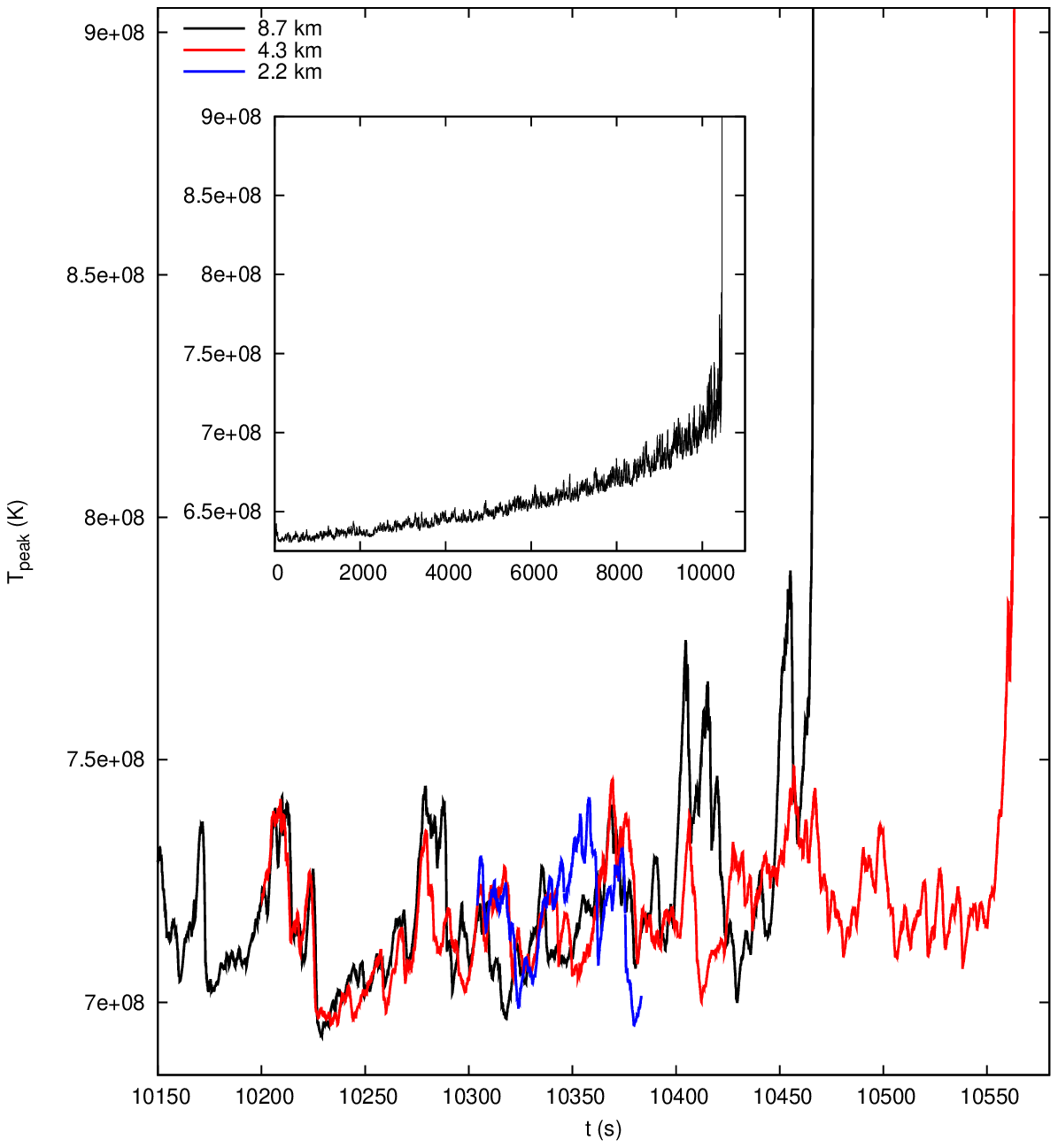}
\caption{\label{fig:temp_compare} Peak temperature leading up to
  ignition for the 8.68~km, 4.34~km, and 2.17~km simulations.
  The inset plot shows the long-time behavior of the 8.68~km simulation 
  originally presented in Z11.}
\end{center}
\end{figure}

We begin by reproducing some of the diagnostics used in Z11 
using data from the 4.34~km and 2.17~km simulations.
In Figure \ref{fig:temp_compare}, we plot the peak temperature
as a function of time for the 4.34~km and 2.17~km resolution 
simulations, and also include original 8.68~km resolution data for comparison.
First, we see that over the last few minutes, the temperature profiles
have the same general trend.  The peak temperature value steadily grows
with time, including fluctuations of several percent.  Once the star
ignites, the peak temperature rapidly increases beyond $8\times 10^8$~K.
We consider the local temperature peaks preceding ignition to be ``failed''
ignition points, i.e., hot bubbles that are not quite hot enough
to cause the temperature to run away.
The ignition time for the 8.68~km and 4.34~km simulations differ
by $\sim$100~s.  Due to the highly non-linear nature of this problem,
this result is not particularly surprising.  At the beginning of the
4.34~km simulation, we notice that the peak temperature curves track each other 
very well for the first $\sim$80~s (from time range 10200--10280~s) before the 
curves begin to show different behavior.  This is not particularly surprising
either since the 4.34~km solution begins as an interpolated imprint of the 8.68~km 
simulation.  After $\sim$80~s, we say that the peak temperature in the 4.34~km 
simulation has decorrelated from the 8.68~km simulation, and we expect the 
statistical properties of the hot spots near the center of the star to be 
consistent with an independent 4.34~km simulation initialized at time $t=0$.
We still expect 
the general convective flow field to look qualitatively similar for a longer 
period of time.  The 2.17~km simulation shows similar behavior; after initializing 
the simulation from the 4.34~km data, it takes $\sim$40 seconds for the peak
temperature curves to decorrelate.

We would like to comment on the time step used in these simulations.  Using an
advective CFL number of 0.7, the average time steps over the time range
10300--10380~s are approximately 0.027~s (for the 8.68~km simulation), 
0.016~s (4.34~km), and 0.010~s (2.17~km).  The time steps
do not quite change by a factor of two with refinement since the peak
velocities do not necessarily lie in the refined convective region.
We also want to comment on the efficiency of the low Mach number
formulation as compared to an explicit, fully compressible approach.
In \cite{scidac-restart}, we showed that the 8.68~km simulation took
a time step of a factor of $\sim$70 larger than a compressible code,
yet a low Mach number time step takes approximately 2.5 times as long 
given the same computational resources, yielding an overall efficiency 
increase of $\sim$28 over a compressible code.  This comparison is 
especially meaningful because we compared to 
the CASTRO \citep{castro} compressible code,
which is based on the same BoxLib framework as MAESTRO, and 
uses the same unsplit Godunov advection formulism, same equation of state, 
and same reaction network ODE solver.

\begin{figure}
\begin{center}
\includegraphics{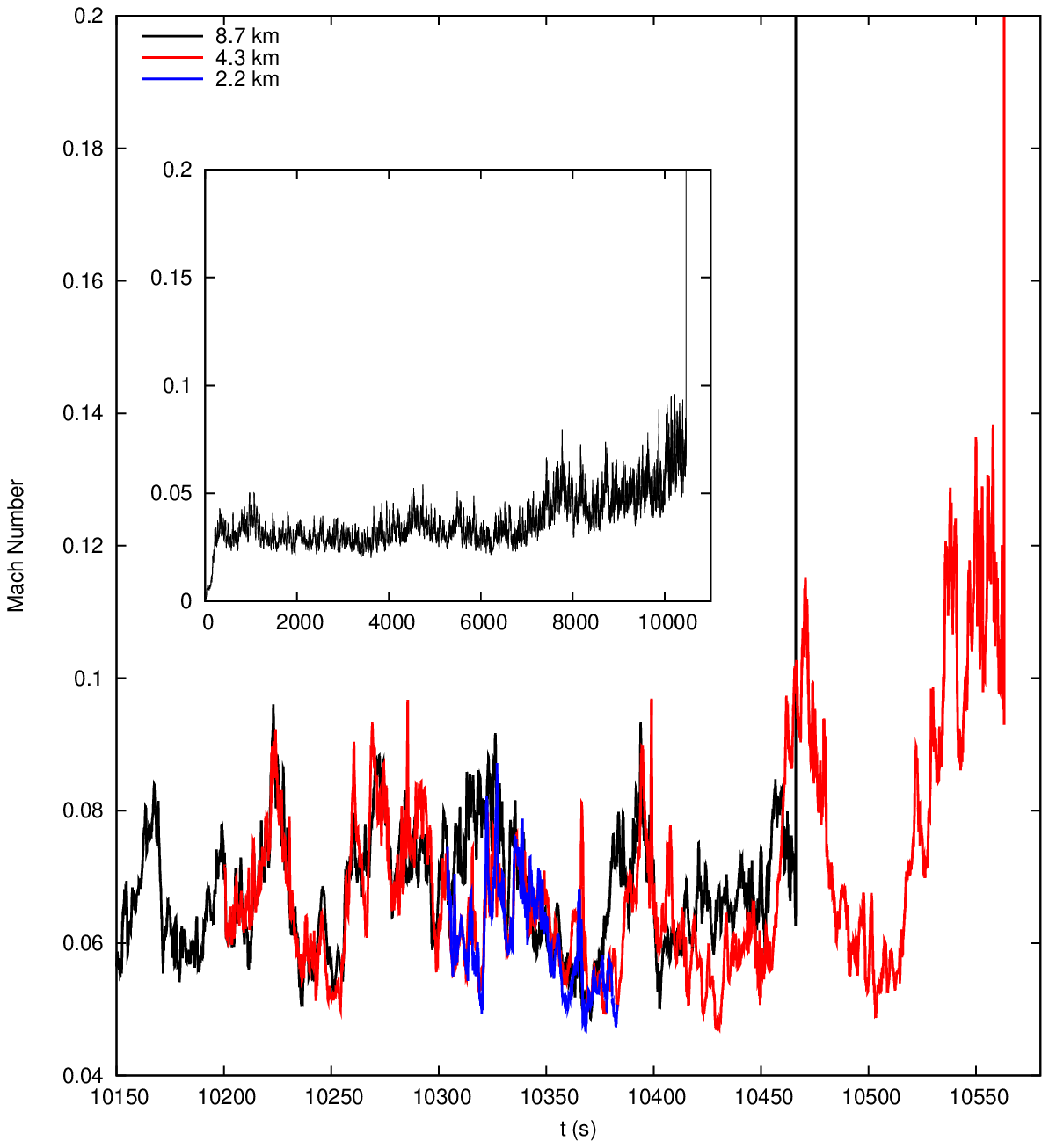}
\caption{\label{fig:mach_compare} Peak Mach number leading up to
  ignition for the 8.68~km, 4.34~km, and 2.17~km simulations.
  The inset plot shows the long-time behavior of the 8.68~km simulation
  originally presented in Z11.}
\end{center}
\end{figure}

\begin{figure}
\begin{center}
\includegraphics{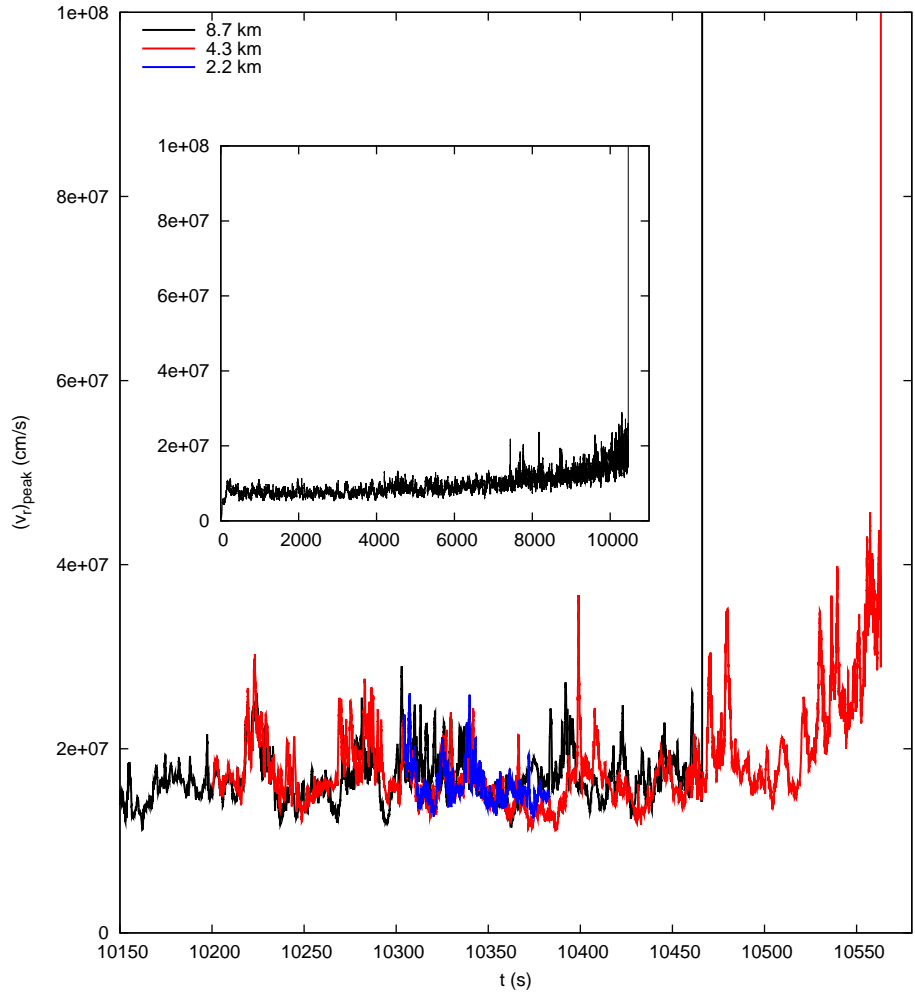}
\caption{\label{fig:radvel_compare} Peak radial velocity leading up to
  ignition for the 8.68~km, 4.34~km, and 2.17~km simulations.
  The inset plot shows the long-time behavior of the 8.68~km simulation
  originally presented in Z11.}
\end{center}
\end{figure}

Next, in Figures \ref{fig:mach_compare} and \ref{fig:radvel_compare} 
we plot the peak Mach number and peak radial velocity as a function of time.
We see the same general behavior as in the 8.68~km simulation, where the peak Mach
number and radial velocity remain relatively constant until the final seconds
preceding ignition, where the values rapidly increase.

The next quantities of interest are the radius of the first ignited cell and
its corresponding outward radial velocity.  In the 4.34~km simulation,
the radius of the ignited cell is 41.3~km, with outward radial velocity of
9.5~km~s$^{-1}$.  We compare these values to those reported in Table 1 of
Z11; the 8.68~km simulation had an ignition
cell radius of 25.7~km with outward radial velocity of 5.1~km~s$^{-1}$.

We would also like to examine the ignition radius and radial velocity if we were to define
ignition as $1.1\times 10^9$~K rather than $8\times 10^8$~K.  However, by advancing the
solution using our advective CFL condition, the simulation quickly becomes unphysical.
Specifically, if we continue to let the white dwarf evolve past $8\times 10^8$~K,
over the next $\sim$0.5~s ($\sim$60 time steps), the peak temperature steadily climbs to 
$\sim$$9\times 10^8$~K while the peak Mach number holds steady at $\sim$0.1.  Then, over the next few
time steps, the temperature unphysically spikes to $\sim$$8\times 10^9$~K, with the peak Mach number
quickly climbing to well over 1000.  Our low Mach number model has obviously broken down, so these
results are not physical.  To remedy this situation, and to advance our simulation to $1.1\times 10^9$~K,
we apply a heuristic time step limiter, which attempts to reduce the time step so the peak temperature does
not increase by more than $\sim$1\% each step.  We limit the time step using
\begin{equation}
\Delta t = \min\left[\Delta t_{\rm CFL}, \frac{\Delta t_{\rm CFL}}{100}\frac{T_{\rm max}^{n-1}}{T_{\rm max}^n-T_{\rm max}^{n-1}} \right],
\end{equation}
where $\Delta t_{\rm CFL}$ is the time step computed using our standard advective CFL condition, 
$T_{\rm max}^n$ is the maximum temperature in the domain at the current 
time step, and $T_{\rm max}^{n-1}$ is the maximum temperature in the domain from 
the previous time step.  In doing this, we find that that we reach $1.1\times 10^9$~K at 0.57~s after
$8\times 10^8$~K, the ignition point has advected to a larger
radius (48.9~km), and the ignition point radial velocity has increased to $v_r = 14.0$~km~s$^{-1}$.  
These results are not surprising, given the ignition conditions at $8\times 10^8$~K reported above.

\begin{figure}
\begin{center}
\includegraphics{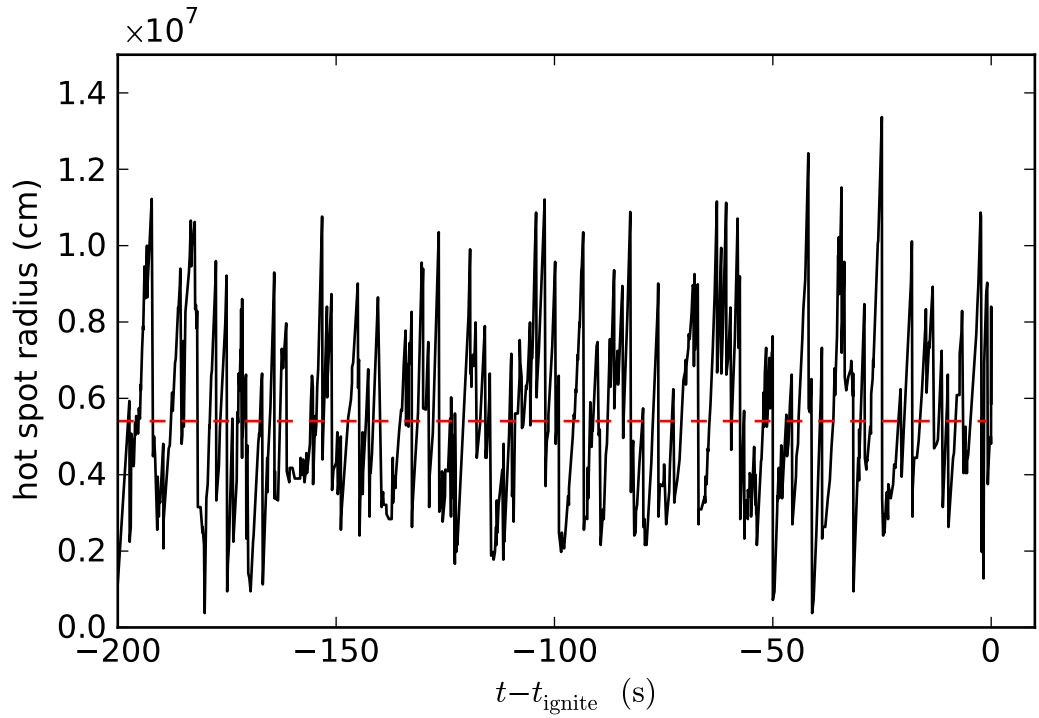}
\caption{\label{fig:hotspot_radius} Radial location of the hottest cell as a 
function of time for the 4.34~km simulation.  Only the last 200 s before 
ignition are shown. Here we see that right up to the end of the calculation the hot 
spot location changes rapidly. The horizontal horizontal dashed line 
indicates the average radial 
position of the hot spot from 200 s to 1 s before ignition.}
\end{center}
\end{figure}

In Z11 we studied the time history of the hottest cell over the last
few minutes.  We gathered statistics to help us in our determination of likely
ignition radii, and repeat the same diagnostics here.
In Figure \ref{fig:hotspot_radius}, we show the radius of the hottest cell
as a function of time for the final seconds preceding ignition for the 4.34~km simulation.
In Table 2 of Z11,
we computed the average radius of the hottest zone, and its standard deviation, for the last
200~s and 100~s preceding ignition.  For the 4.34~km simulation, over the last 200~s, the
average hot spot radius and standard deviation are 54.0~km and 22.1~km (as compared to
52.3~km and 25.5~km for the 8.68~km simulation).  Over the last 100~s, the
average hot spot radius and standard deviation are 54.7~km and 22.5~km (as compared to
54.7~km and 27.3~km for the 8.68~km simulation).  This tells us two things.  First, 
the hot spot
statistics do not seem to change much whether we consider the final 200~s or 100~s preceding
ignition.  Second, the results are very similar to the 8.68~km simulation, implying that
we have sufficient resolution in our determination of likely ignition radii.

\begin{figure}
\begin{center}
\includegraphics[width=3.0in]{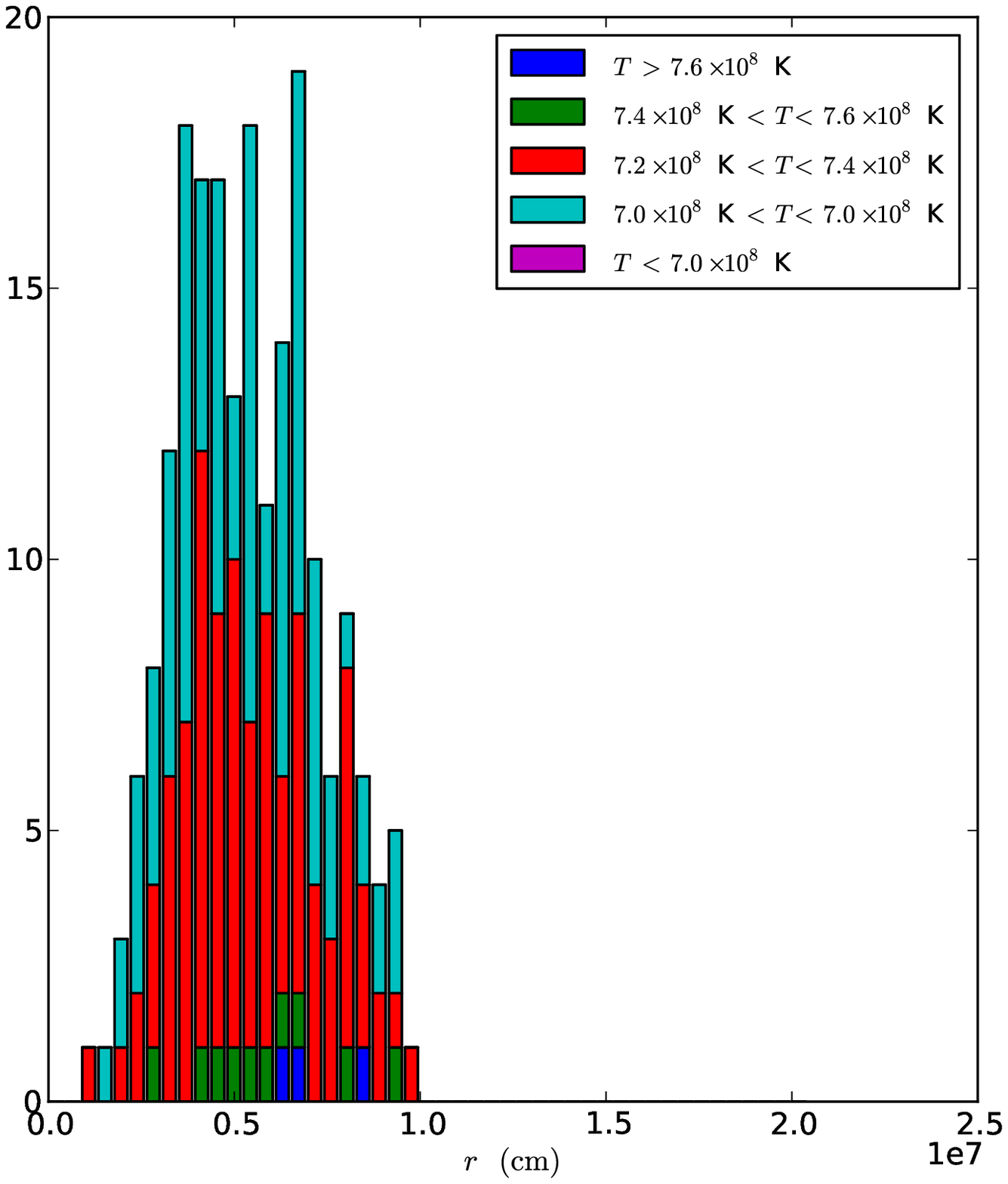}
\includegraphics[width=3.0in]{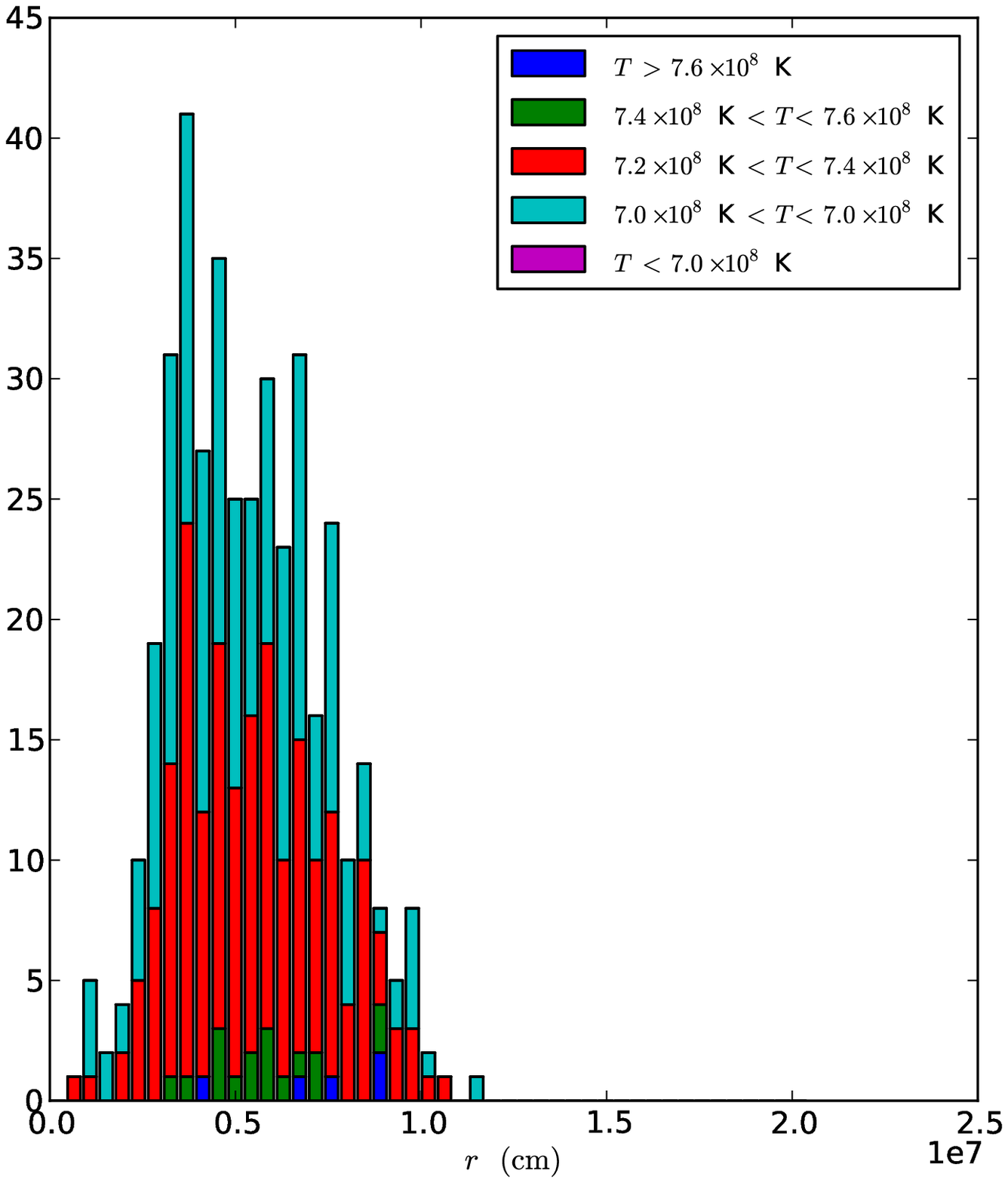}
\caption{\label{fig:T histograms} Histograms of the hottest cell, sorted by radius, 
with the colors representing the average temperature of the hottest cell over the
averaging interval for the 4.34~km simulation with
(Left) $\Delta t_{\rm hist} = 1.0~s$ and (Right) $\Delta t_{\rm hist} = 0.5~s$.}
\end{center}
\end{figure}

\begin{figure}
\begin{center}
\includegraphics[width=3.0in]{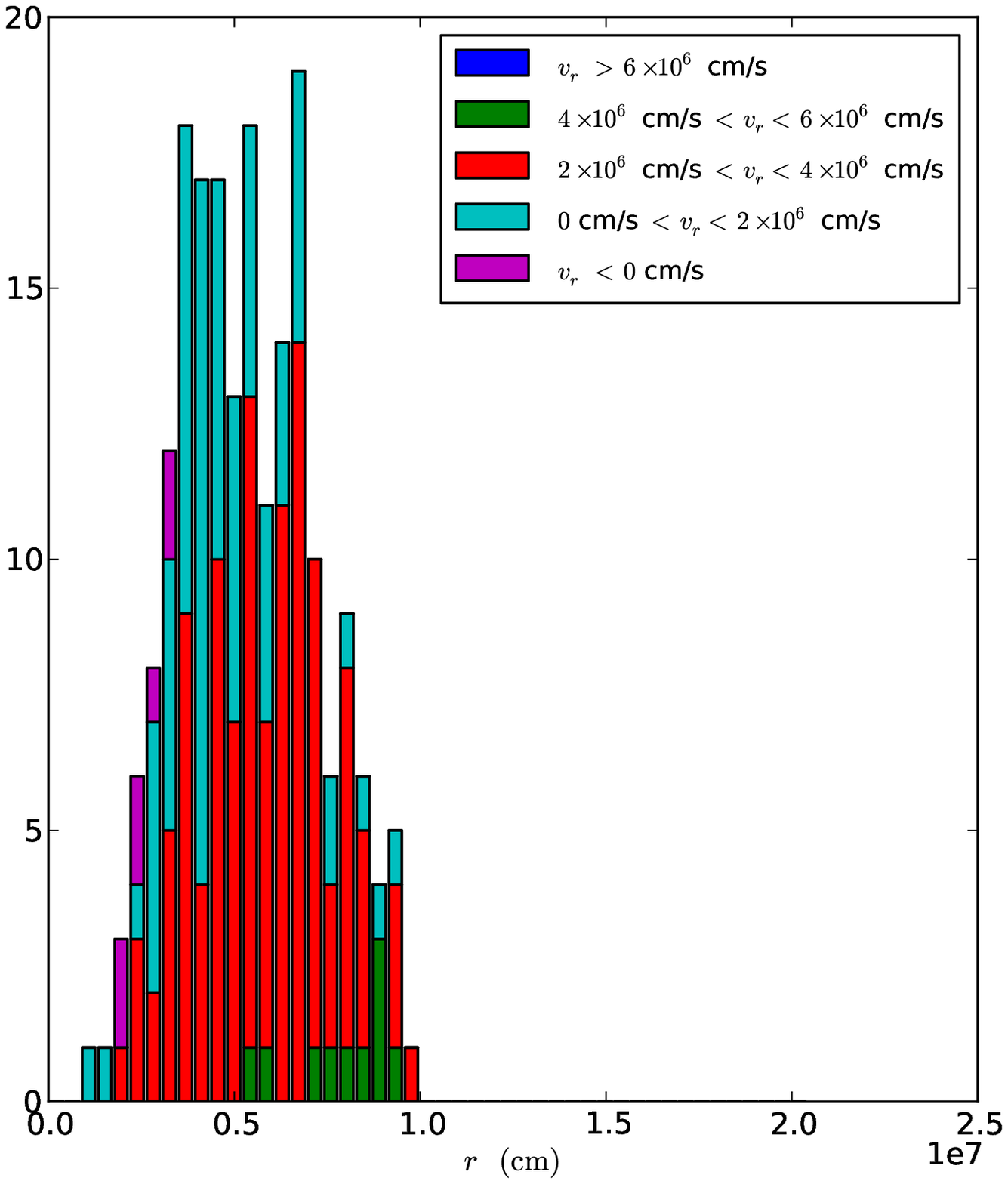}
\includegraphics[width=3.0in]{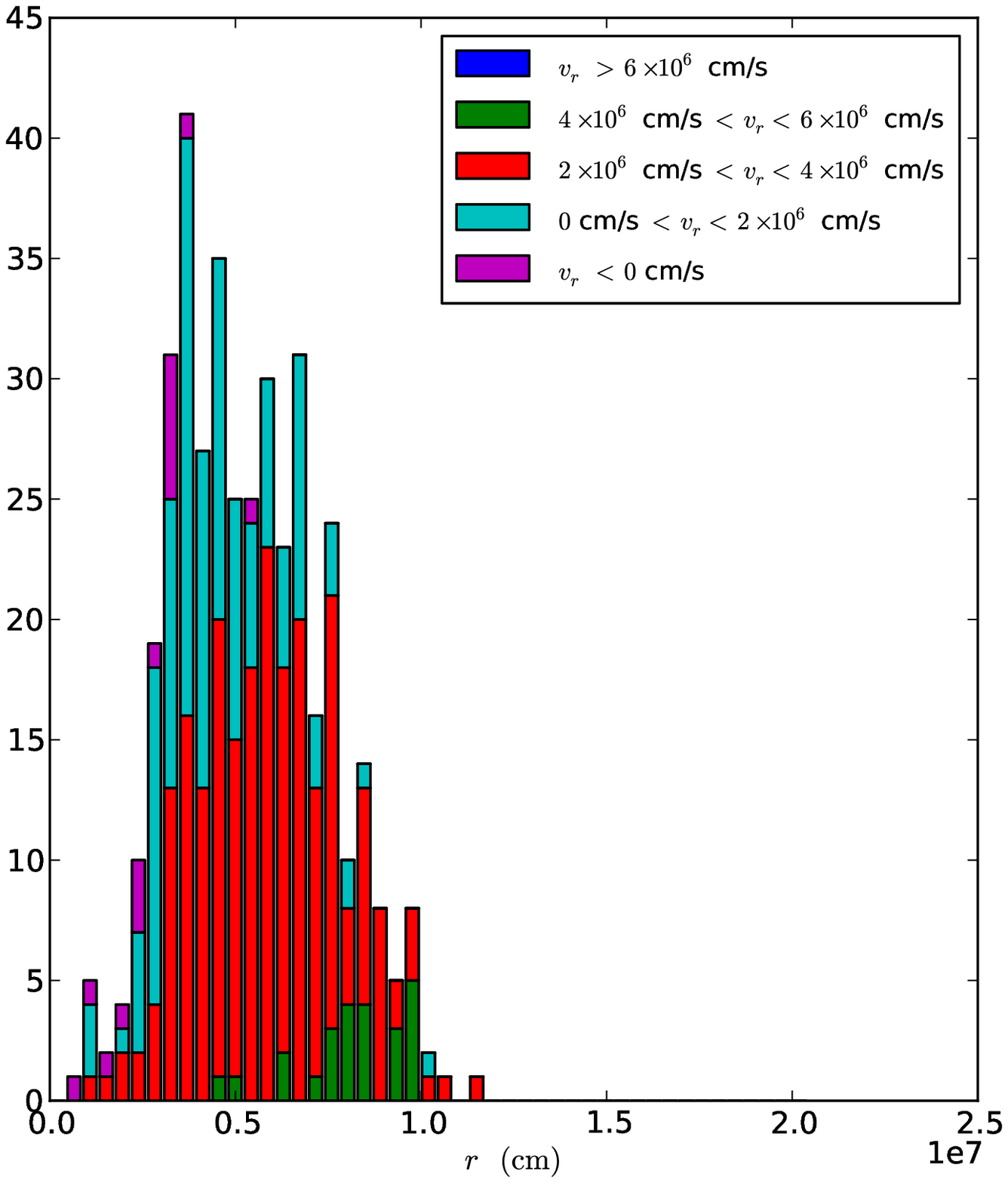}
\caption{\label{fig:v_r histograms} Histograms of the hottest cell, sorted by radius, 
with the colors representing the average radial velocity of the hottest cell over
the averaging interval for the 4.34~km simulation with
(Left) $\Delta t_{\rm hist} = 1.0~s$ and (Right) $\Delta t_{\rm hist} = 0.5~s$.}
\end{center}
\end{figure}

\begin{figure}
\begin{center}
\includegraphics[width=3.0in]{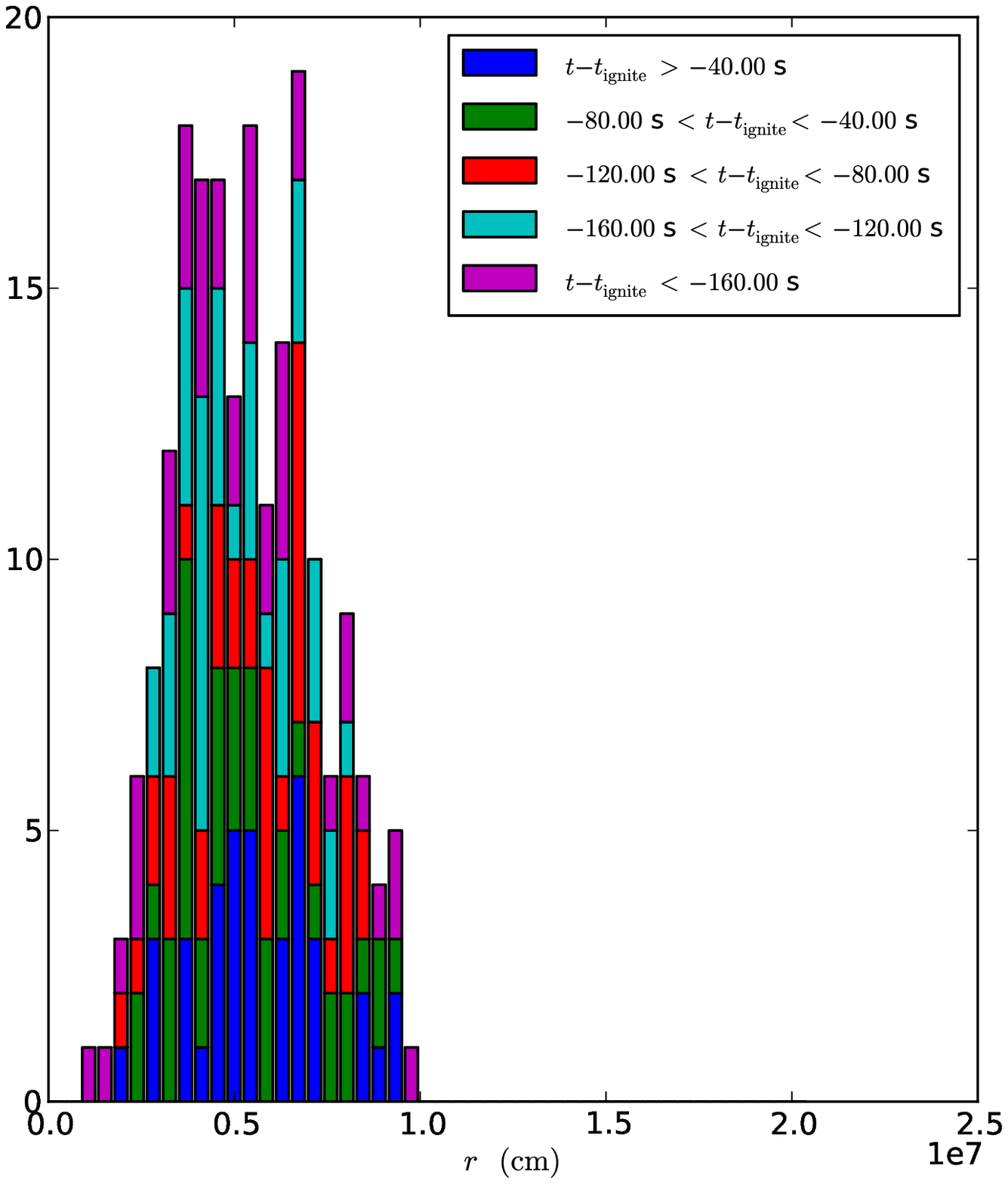}
\includegraphics[width=3.0in]{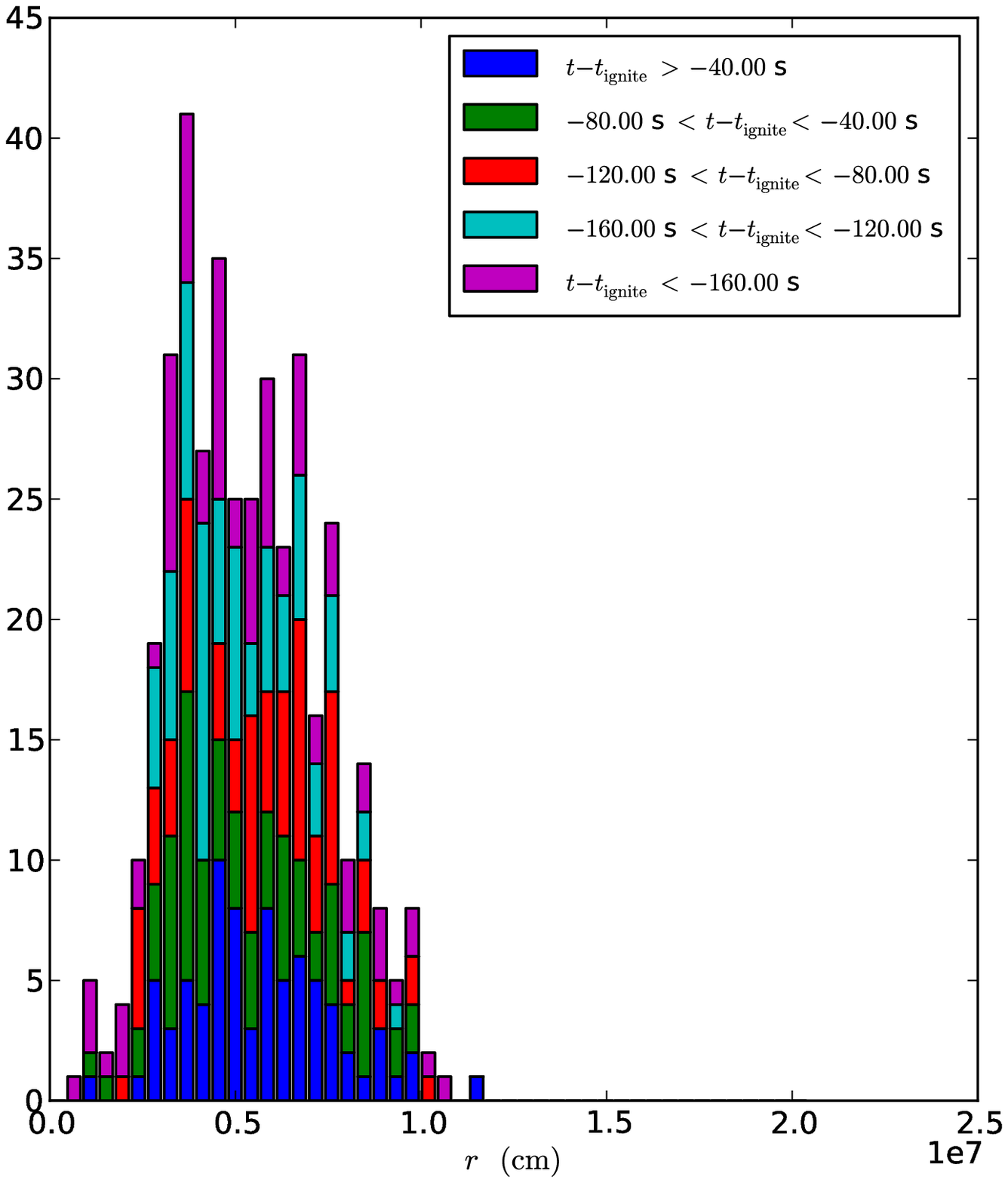}
\caption{\label{fig:t_ignitehistograms} Histograms of the hottest cell, sorted by 
radius, with the colors representing time to ignition for the 4.34~km simulation
with (Left) $\Delta t_{\rm hist} = 1.0~s$ and (Right) $\Delta t_{\rm hist} = 0.5~s$.}
\end{center}
\end{figure}

Next, as in Z11, we break the final approach to ignition into small time
intervals and look at properties of the flow within each time interval.  We consider 
the last 200~s preceding ignition, and use time intervals of 
$\Delta t_{\rm hist} = 1.0$~s and $0.5$~s.  Within each time interval, we compute the 
average radius of the hottest cell, the average temperature of the hottest cell, 
and the average radial velocity of the hottest cell.  We sort this data into histograms
to understand the statistics of the last few hot bubbles preceding ignition.
In each of the following figures, we show histograms for both 
$\Delta t_{\rm hist} = 1.0$~s and $0.5$~s.
Figure \ref{fig:T histograms} shows histograms of the hottest cell, sorted by radius, 
with the colors representing the average temperature of the hottest cell over the
averaging interval.
Figure \ref{fig:v_r histograms} shows histograms of the hottest cell, sorted by radius, 
with the colors representing the average radial velocity of the hottest cell over
the averaging interval.
Figure \ref{fig:t_ignitehistograms} shows histograms of the hottest cell, sorted by 
radius, with the colors representing time to ignition.
Overall, the results are consistent with our observations in Z11, which
we now summarize.  Some general observations:
\begin{itemize}
\item From each set of histograms, we see that hot spot is most likely to be found 
      between 40~km and 75~km off-center.  This is consistent with both 
      Figure \ref{fig:hotspot_radius} and the histograms from Z11.
      However, we do not see the slight extended tail observed in Z11,
      which indicated a slight preference for the hot spots 
      to lie at larger radii within the distribution.
\item For each set of histograms, we observe that the results are essentially the 
      same regardless of whether $\Delta t_{\rm hist} = 1.0$~s or $0.5$~s is used
      as the averaging interval.
\end{itemize}
Some figure-specific observations:
\begin{itemize}
\item In Figure \ref{fig:T histograms}, within each temperature interval, the 
      overall shape of the distribution appears roughly the same, with a peak 
      slightly greater than 50 km, indicating that hot spots of all temperatures 
      can appear at any radius in the distribution.
\item In Figure \ref{fig:v_r histograms}, nearly all of the hot spots have 
      an outward radial velocity.  Also, there is a tendency for the hot spots at 
      larger radii to be associated with larger values of $v_r$ as expected, since
      the flow will carry them away from the center.
\item In Figure \ref{fig:t_ignitehistograms}, we see a reasonably symmetric 
      distribution for all cases, indicating that the hot spot radius does not
      depend strongly on time to ignition.
\end{itemize}

\begin{figure}
\begin{center}
\includegraphics[width=3.0in]{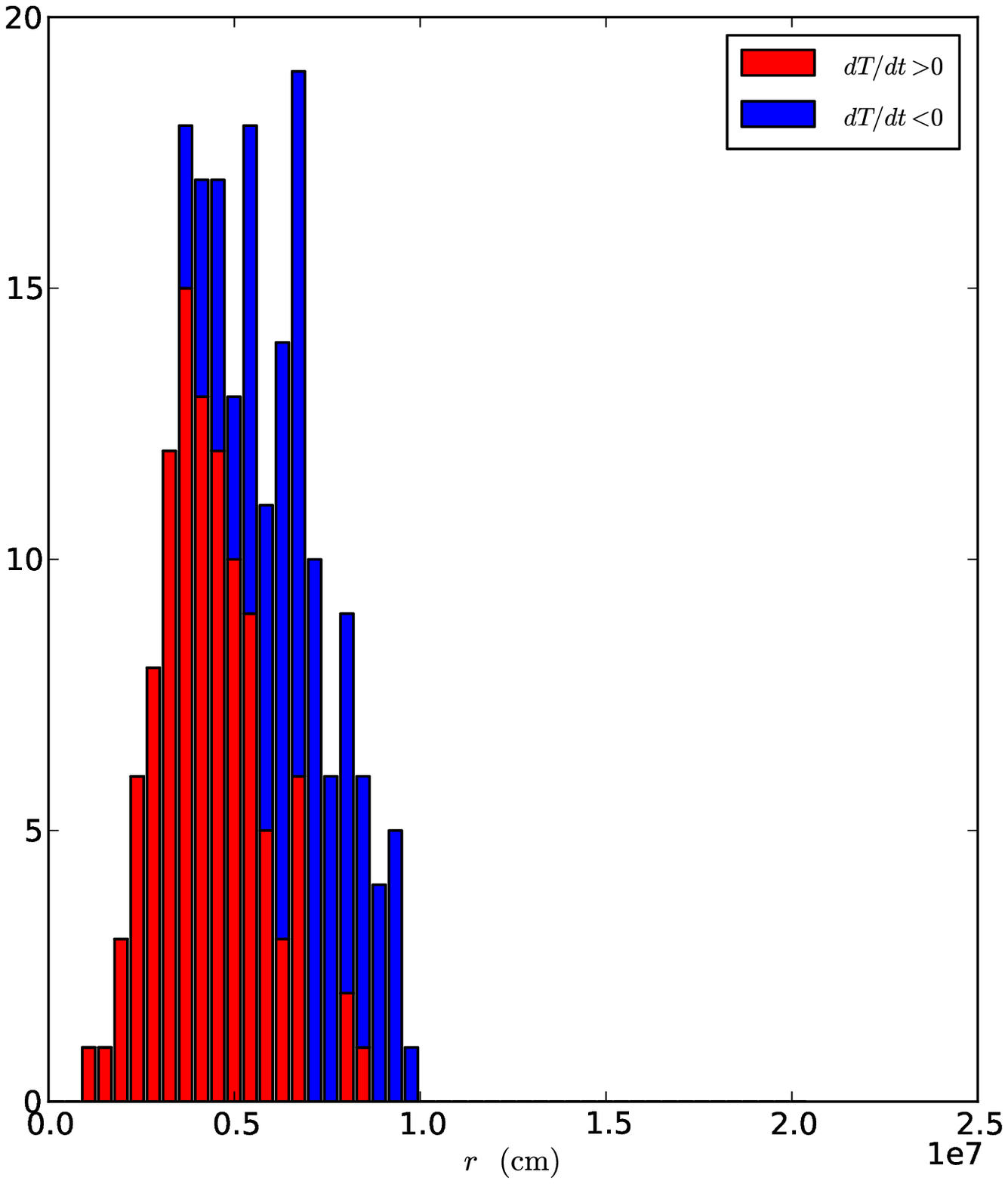}
\includegraphics[width=3.0in]{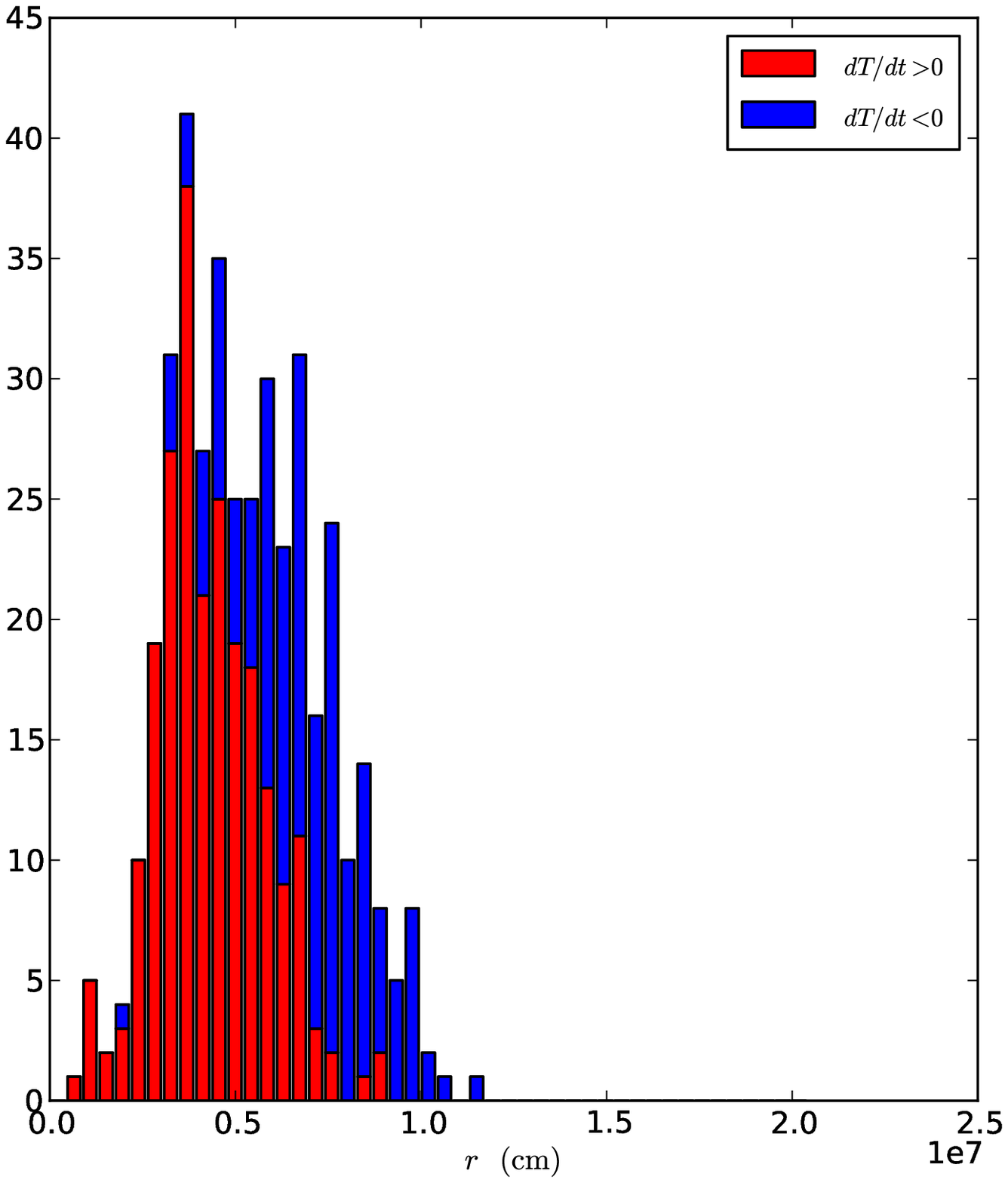}
\caption{\label{fig:dT/dt_histograms} Histograms of the hottest cell, sorted by 
radius, with the colors indicating whether the temperature of the hottest cell
is increasing or decreasing with time for the 4.34~km simulation with 
(Left) $\Delta t_{\rm hist} = 1.0~s$ and (Right) $\Delta t_{\rm hist} = 0.5~s$.}
\end{center}
\end{figure}

Next, we include a new histogram where we examine whether the hottest cell
is increasing or decreasing in temperature.  Figure \ref{fig:dT/dt_histograms}
shows histograms of the hottest cell, sorted by radius, with the colors indicating
whether the temperature of the hottest cell is increasing or decreasing with time.
We observe that when the hottest zone is within 40~km of the center it is almost always
heating up, and when the hottest zone is outside of 75~km it is almost always cooling
down.  This 40 to 75~km range is consistent with the previous histograms.  Another conclusion 
is that it seems highly unlikely that ignition will
occur outside of 75~km since hot cells beyond that radius are most likely cooling down.
This is in contrast to our result from Z11, where we claimed that
100~km was an outer limit for ignition radii.

\subsection{Hot Spot Analysis}
We are interested in the likelihood of multiple ignition points, so we
now take a closer look at the dynamics of the last few hot spots preceding 
ignition.  In the diagnostics we have presented, we have only been able to
track the hottest cell in the simulation.  We do not know, for example, if there
are other hot spots elsewhere in the star that are almost as hot as the
hottest zone.  It is possible that at the time of ignition, there are one or more
cells not directly connected to the ignition cell that have almost reached the ignition
threshold.  In a multiple ignition scenario, such cells could also ignite very shortly
after the initial ignition.  Since the white dwarf explodes within
a few seconds of ignition, a multiple ignition scenario would require another ignition
point to develop within the early phases of the explosion for it to have any
meaningful impact.  We wish to examine the temperature field very
close to the ignition time to get a feel for how likely the multiple ignition
scenario is.

\begin{figure}
\begin{center}
\includegraphics{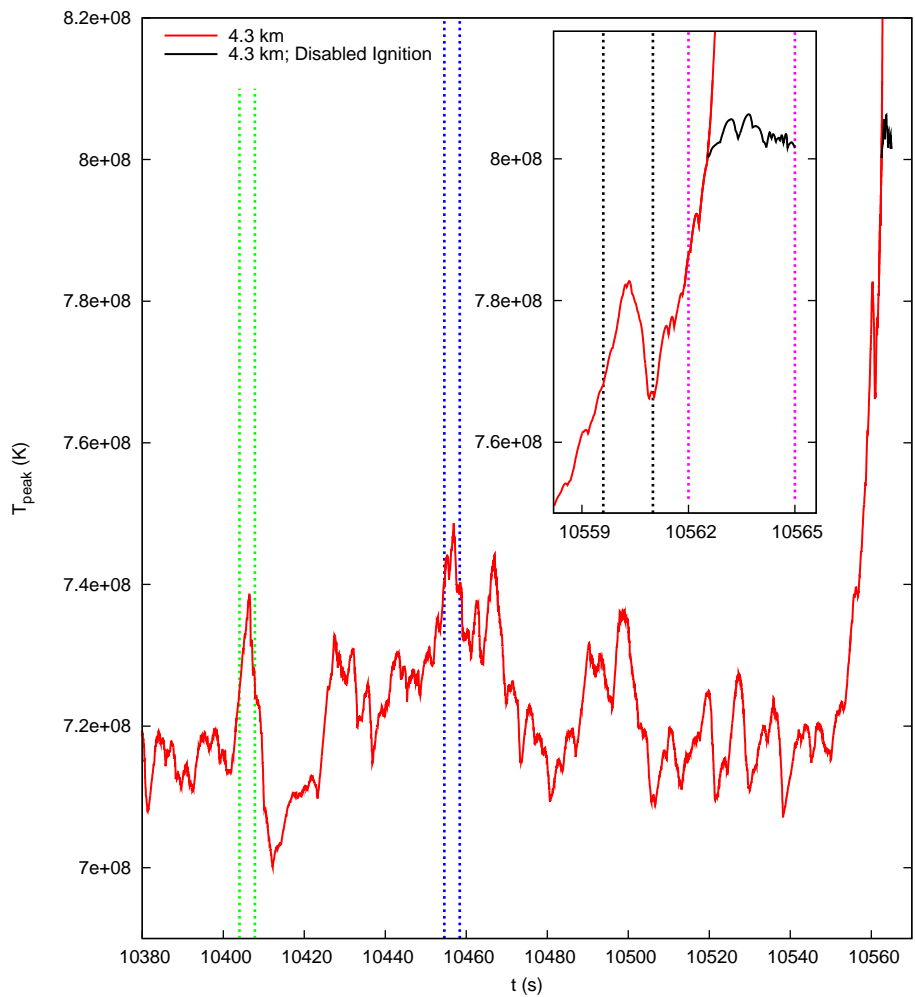}
\caption{\label{fig:temp_1152} Peak temperature during the $\sim$200~s preceding 
  ignition for the 4.34~km simulation.  The dashed vertical lines 
  indicate time ranges where we will examine whether there are multiple hot spots.  The
  inset plot is a zoom-in of the final $\sim$5~s preceding ignition.  The 
  black curve follows the maximum temperature for a simulation where we disable burning 
  in all cells with $T>8\times 10^8$~K.}
\end{center}
\end{figure}

We have previously defined a failed ignition point as a spike in
the plot of the peak temperature vs.~time that does not run away.  
We begin by examining the temperature
distribution in the star during three failed ignition points preceding ignition.
Figure \ref{fig:temp_1152} is a zoom-in of Figure \ref{fig:temp_compare} for the final
minutes preceding ignition for the 4.34~km simulation.  Three failed ignition
points preceding ignition are encapsulated within the green, blue, and black dotted 
lines.  We will examine the temperature distribution in each of these time ranges
to see if there are hot spots elsewhere in the star.

\begin{figure}
\begin{center}
\includegraphics[width=1.5in]{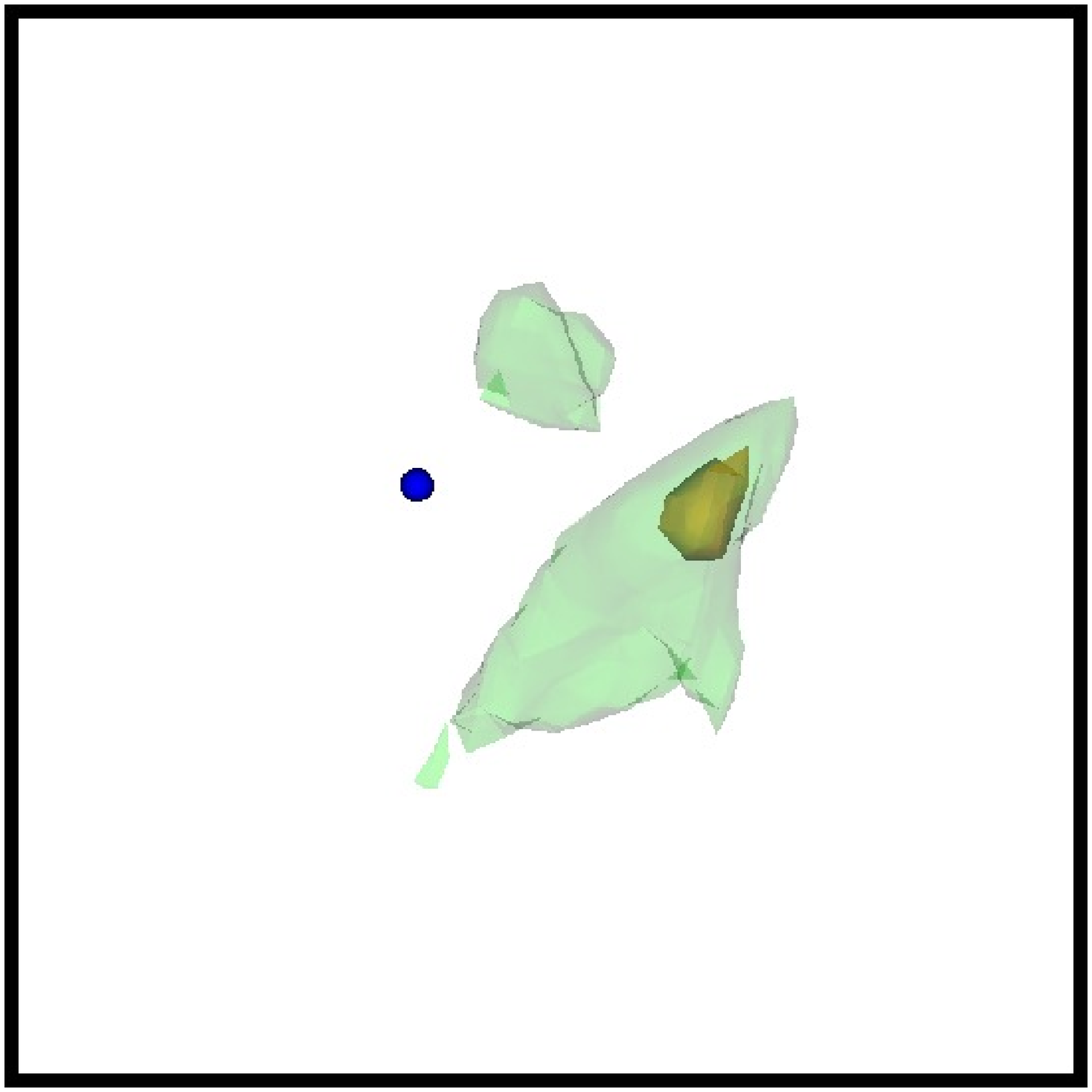}
\includegraphics[width=1.5in]{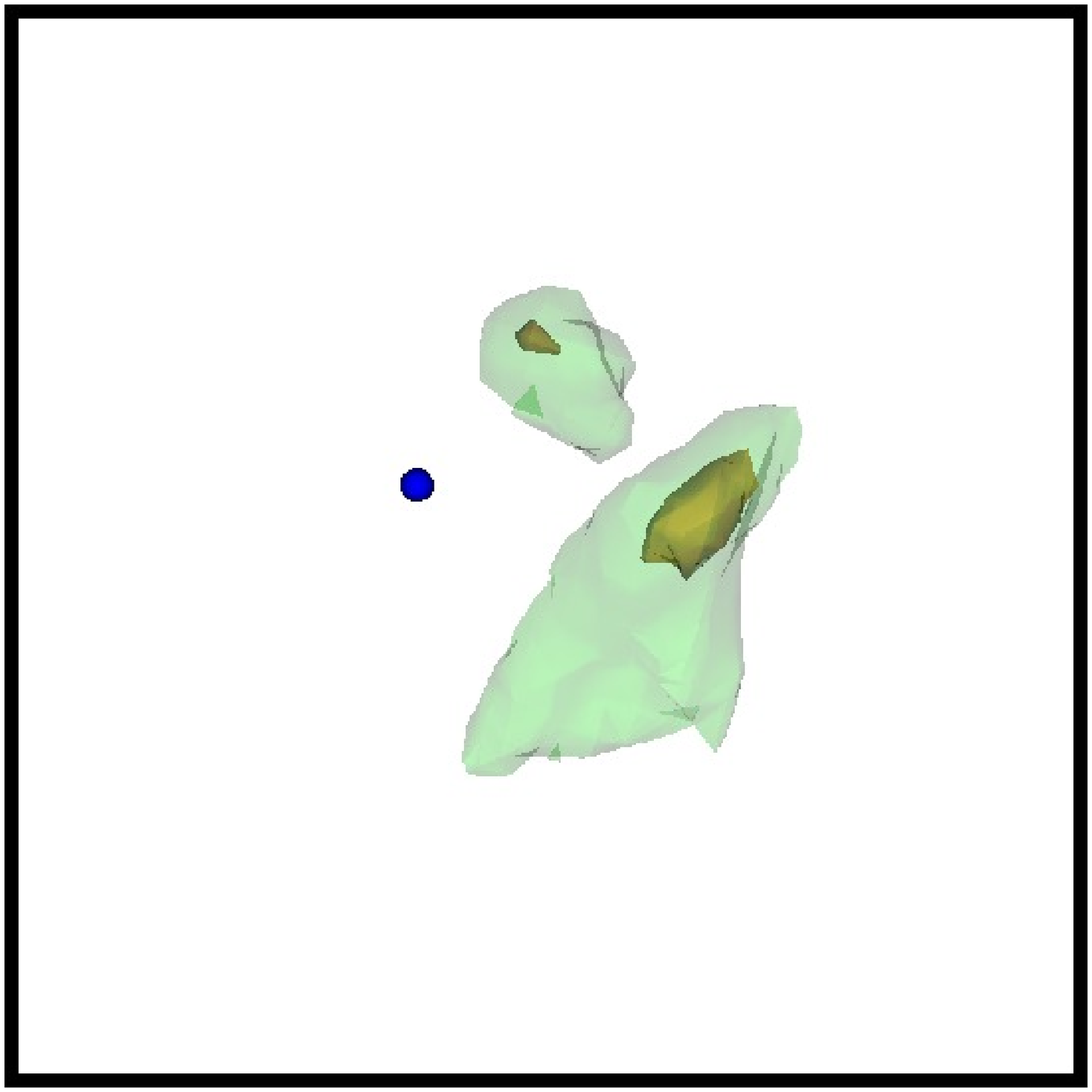}
\includegraphics[width=1.5in]{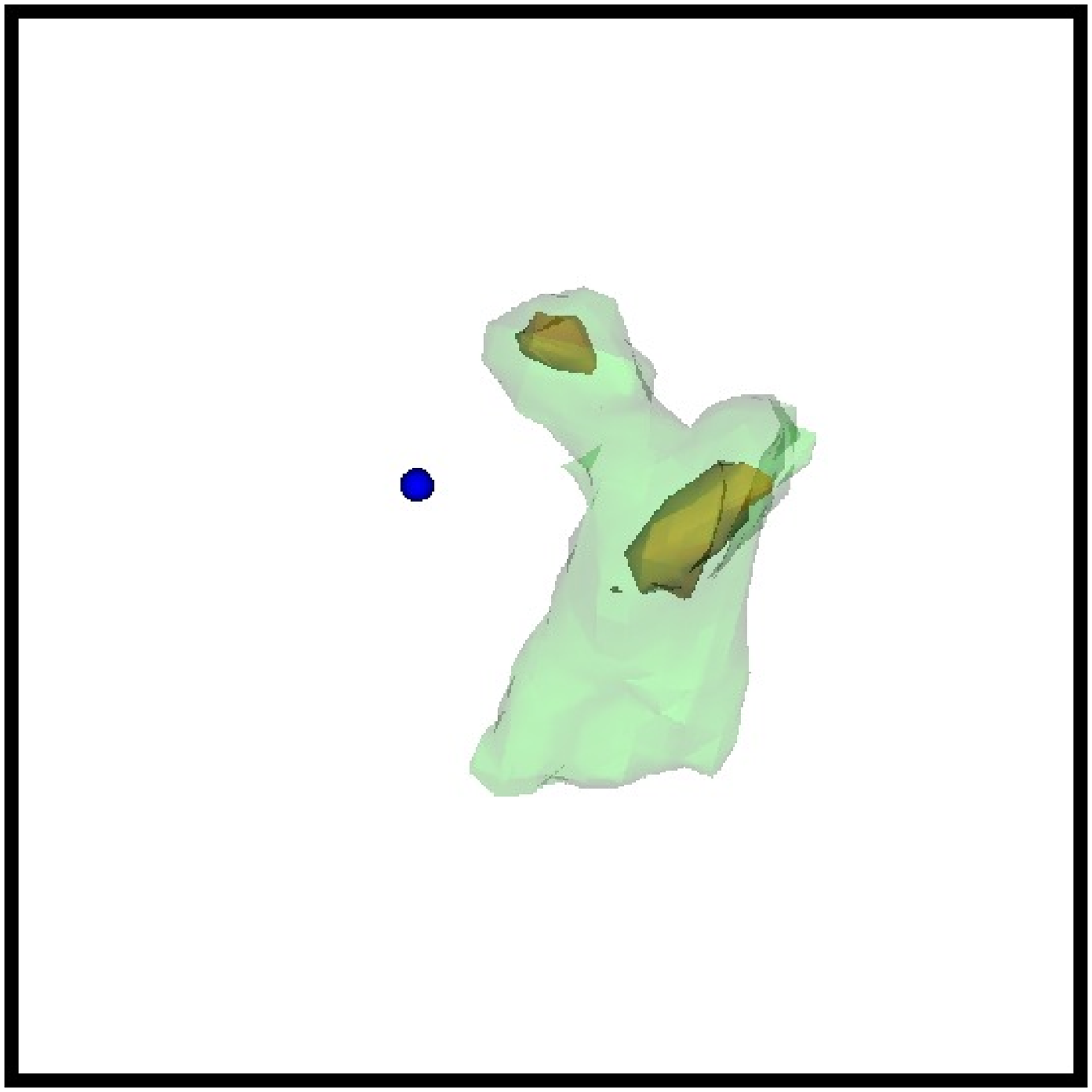}
\includegraphics[width=1.5in]{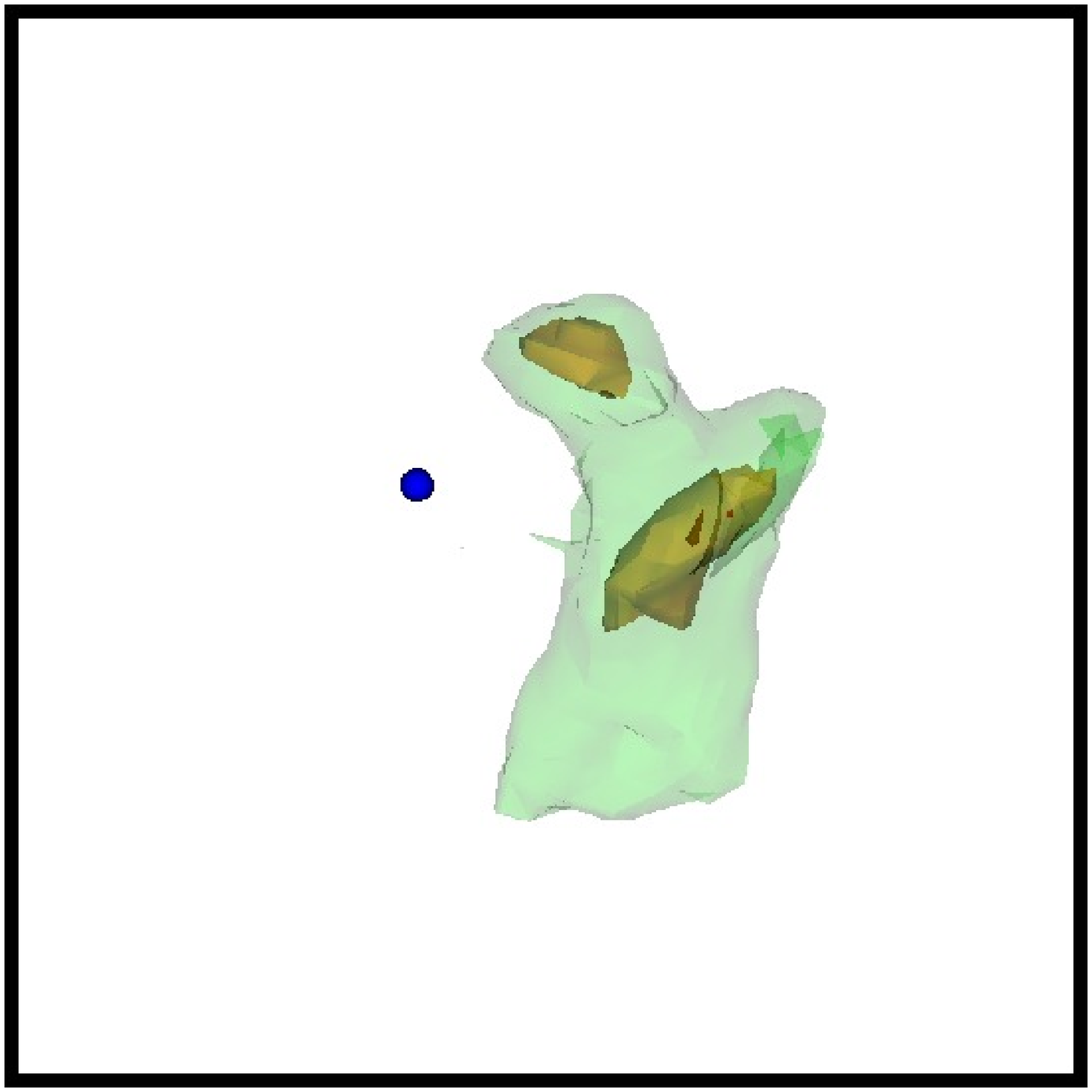}\\
\includegraphics[width=1.5in]{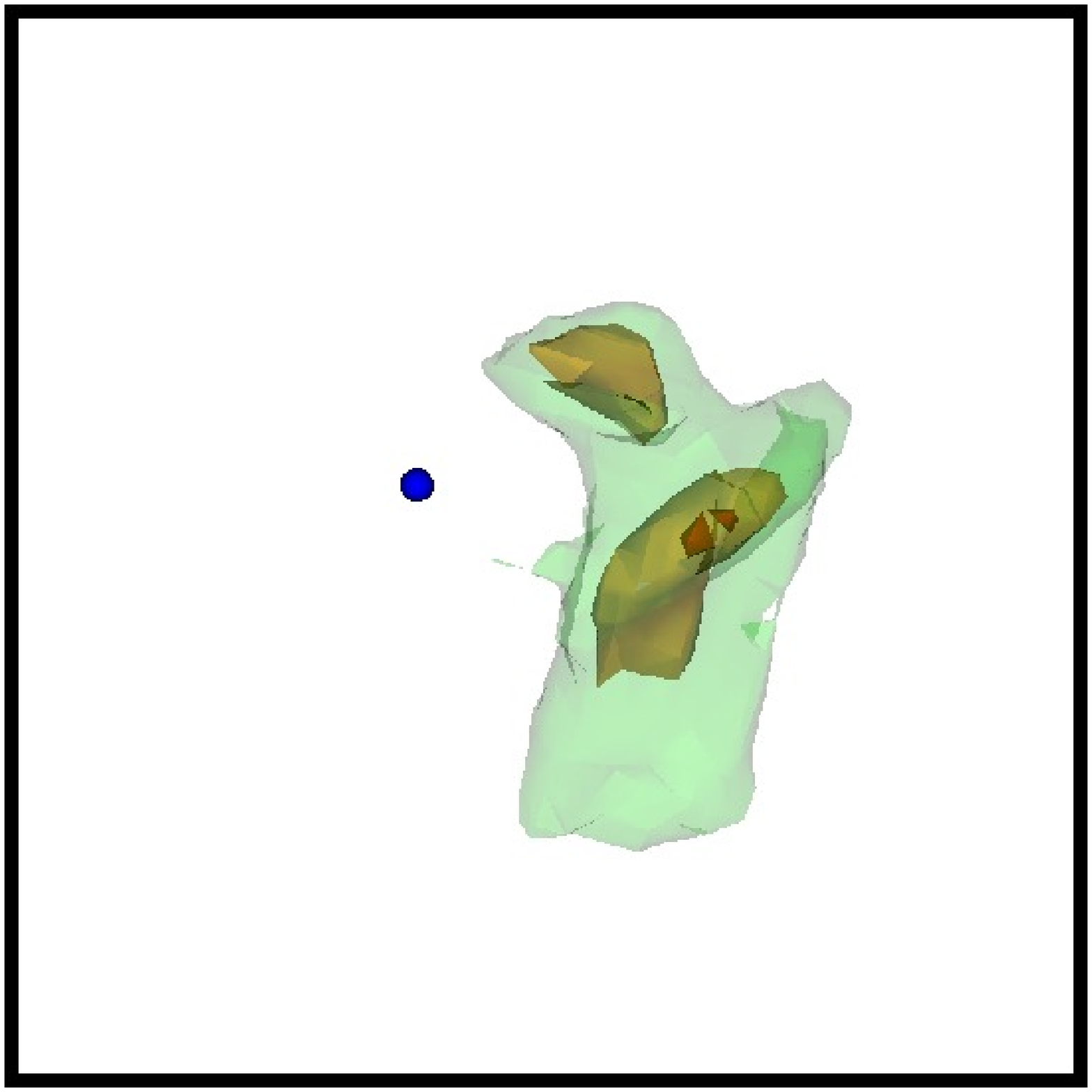}
\includegraphics[width=1.5in]{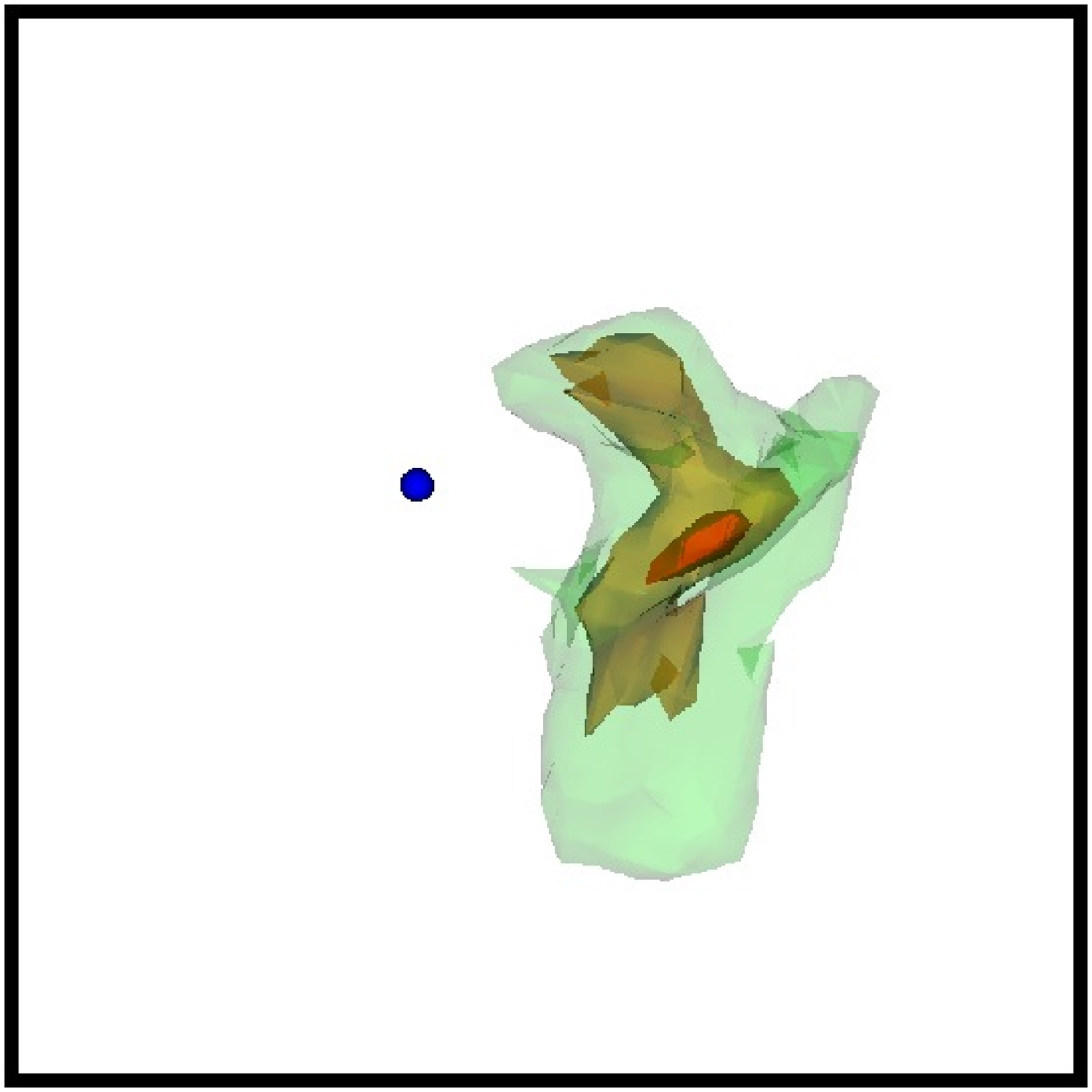}
\includegraphics[width=1.5in]{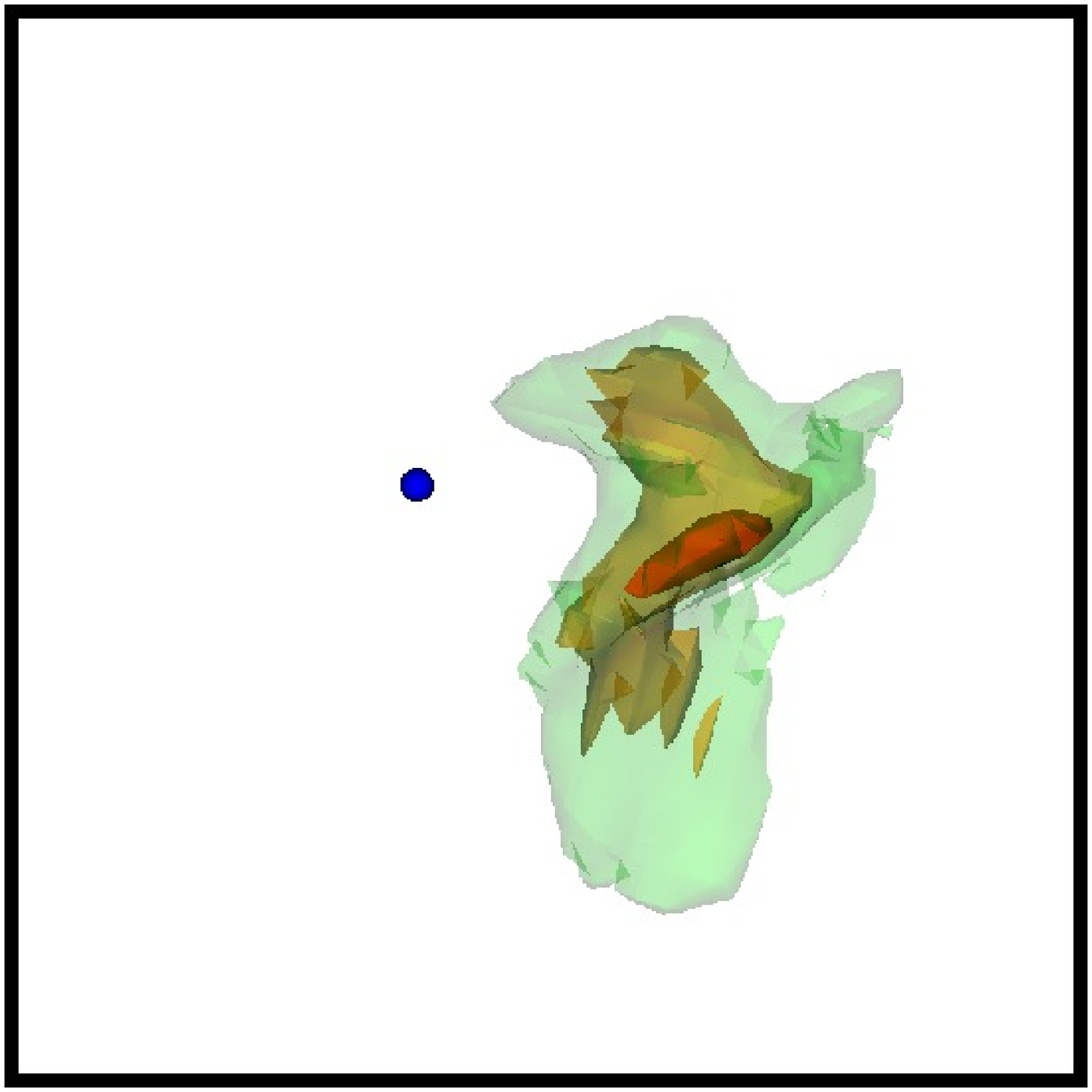}
\includegraphics[width=1.5in]{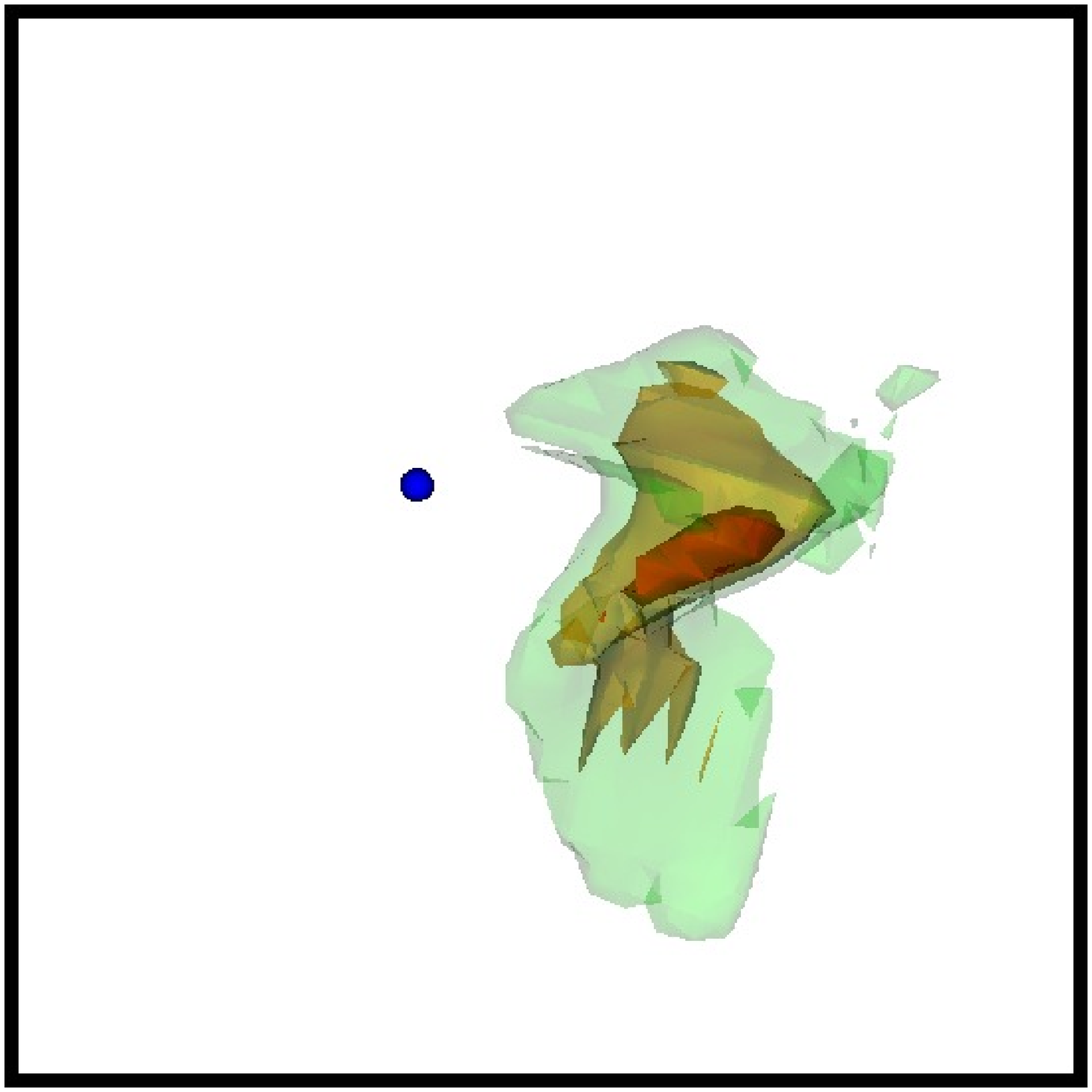}\\
\includegraphics[width=1.5in]{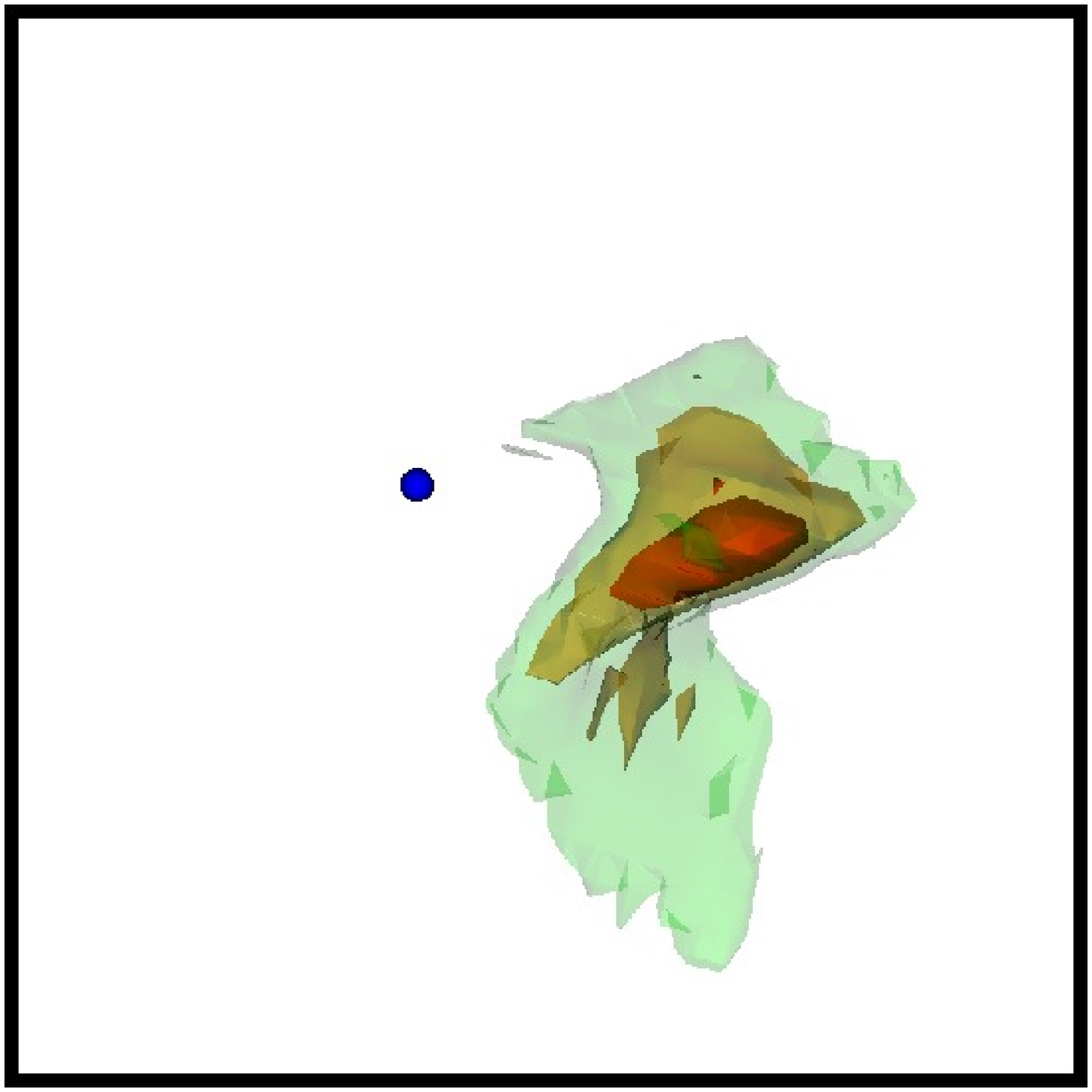}
\includegraphics[width=1.5in]{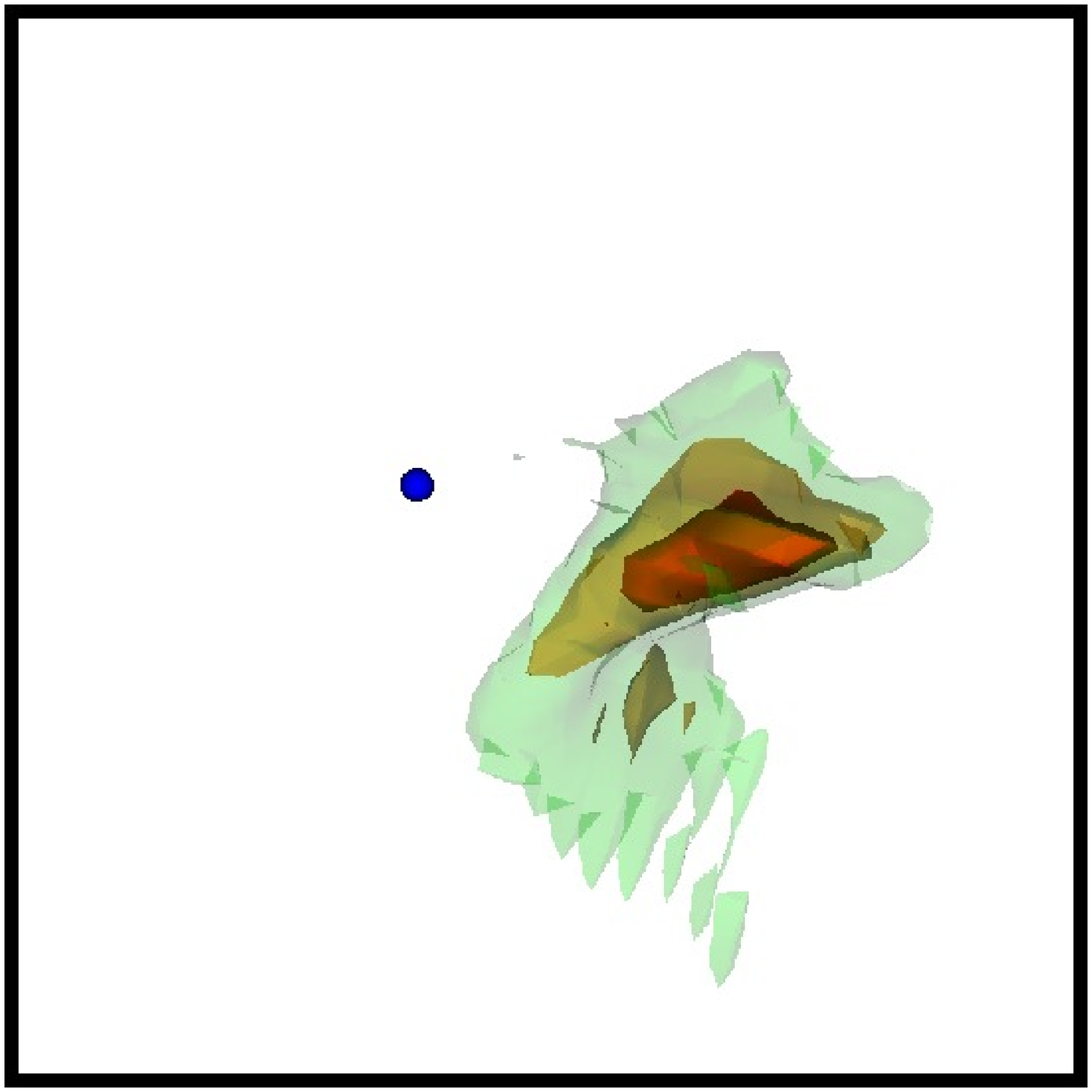}
\includegraphics[width=1.5in]{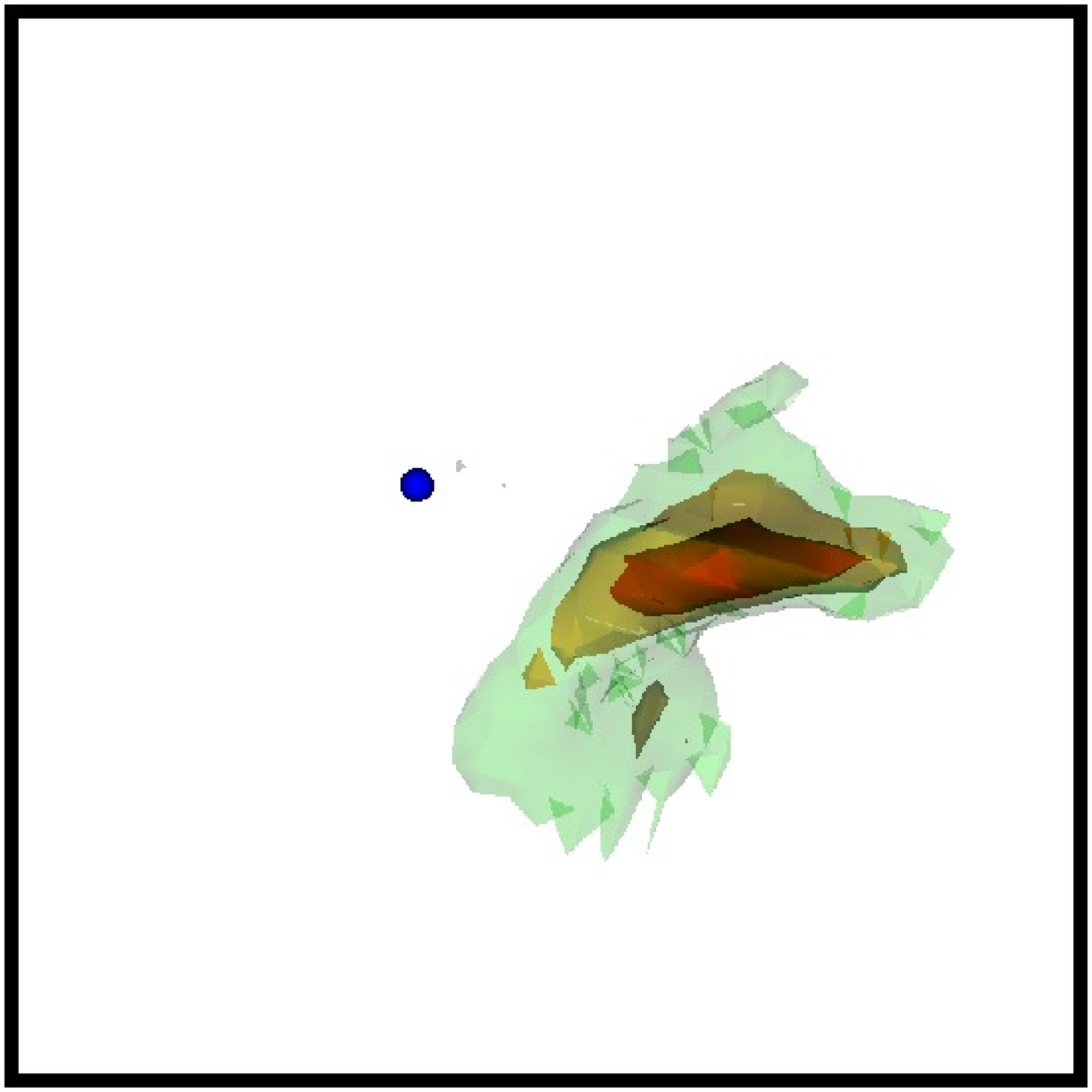}
\includegraphics[width=1.5in]{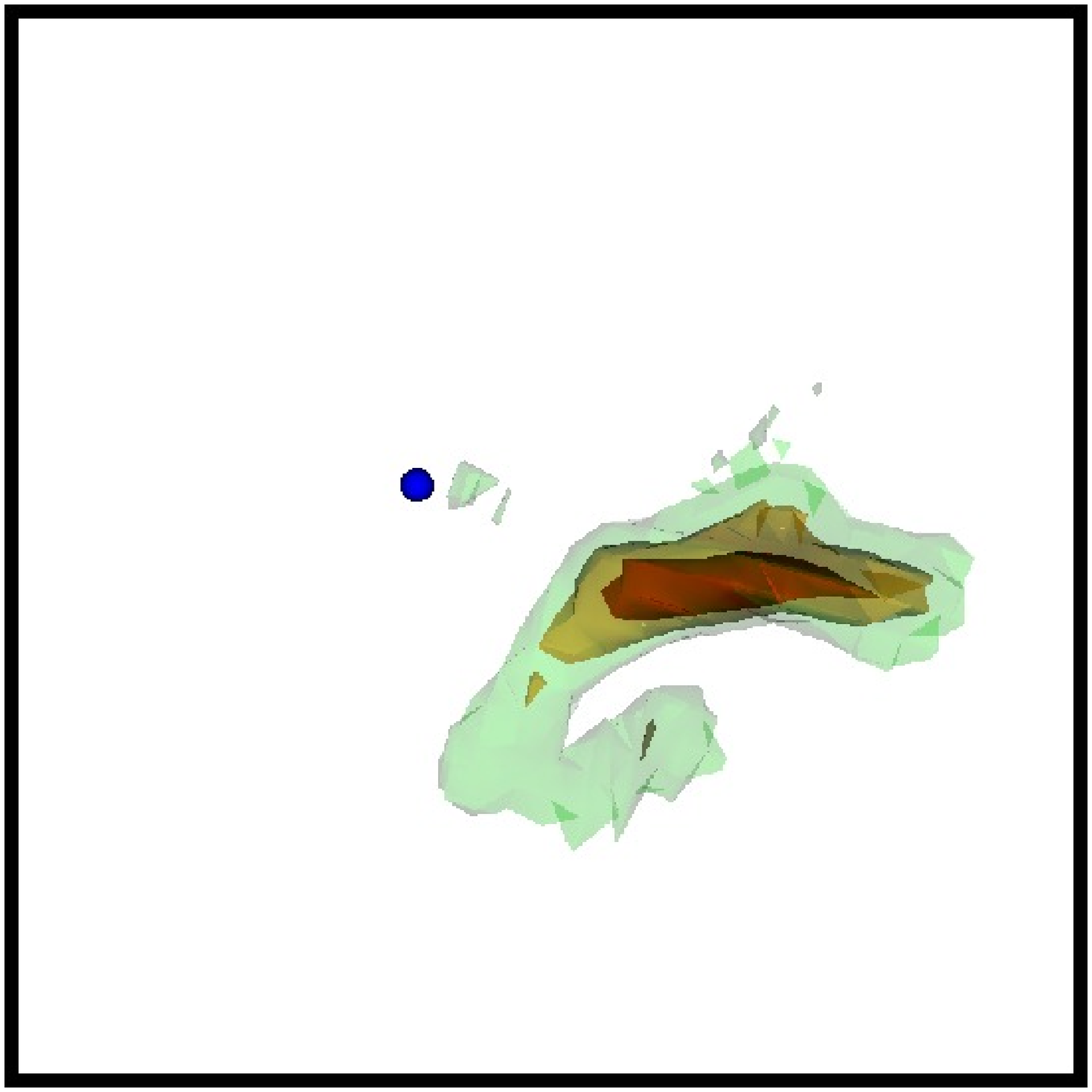}\\
\includegraphics[width=1.5in]{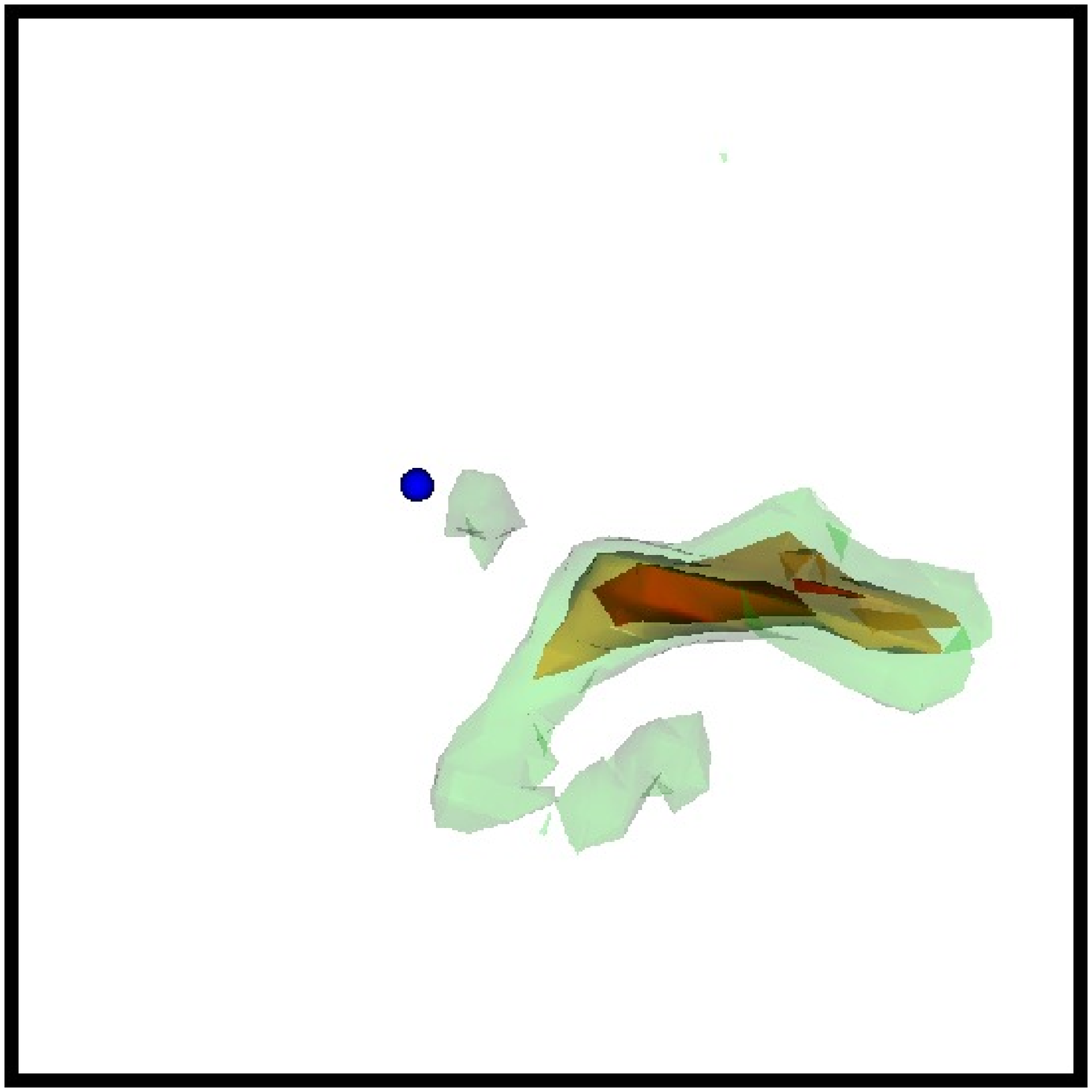}
\includegraphics[width=1.5in]{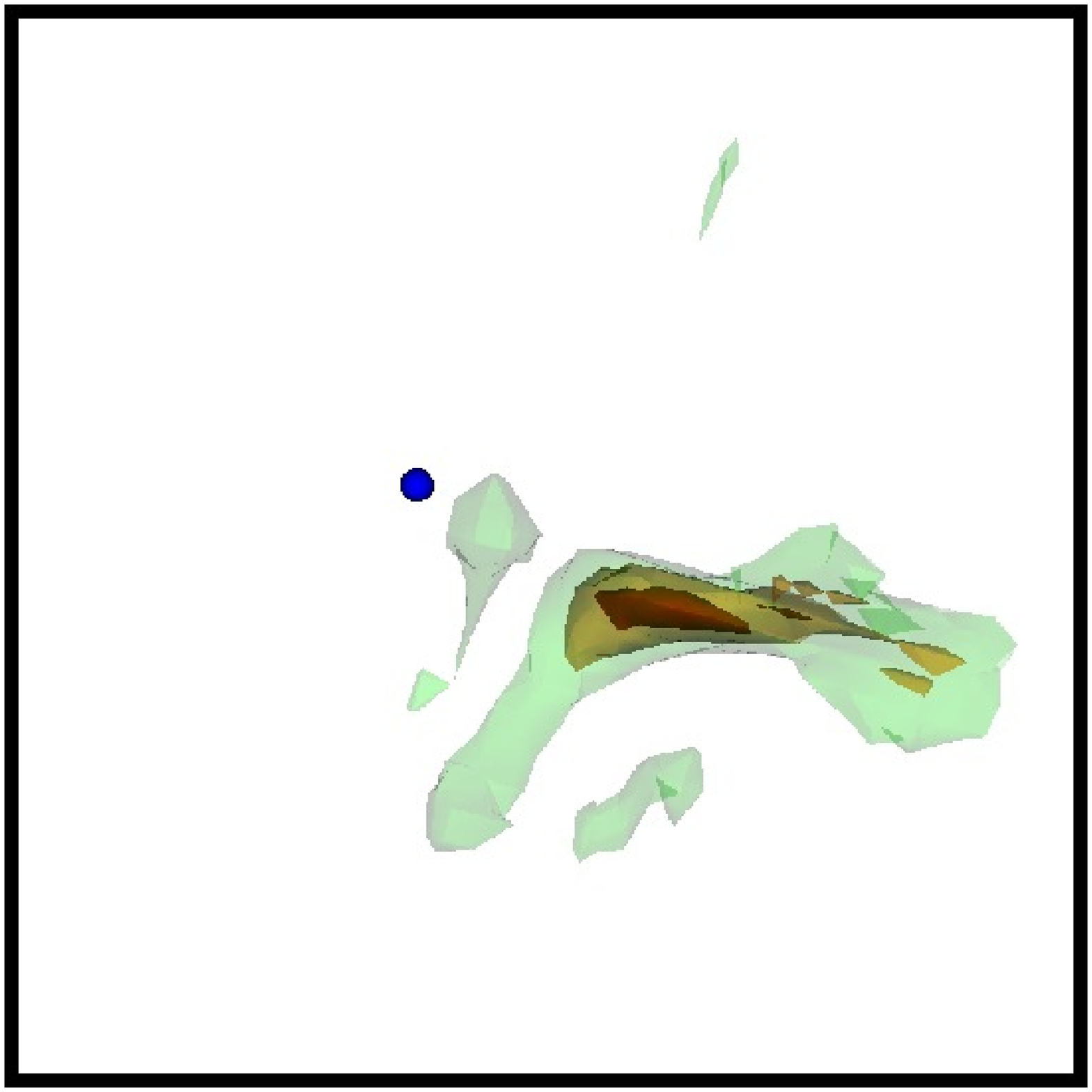}
\includegraphics[width=1.5in]{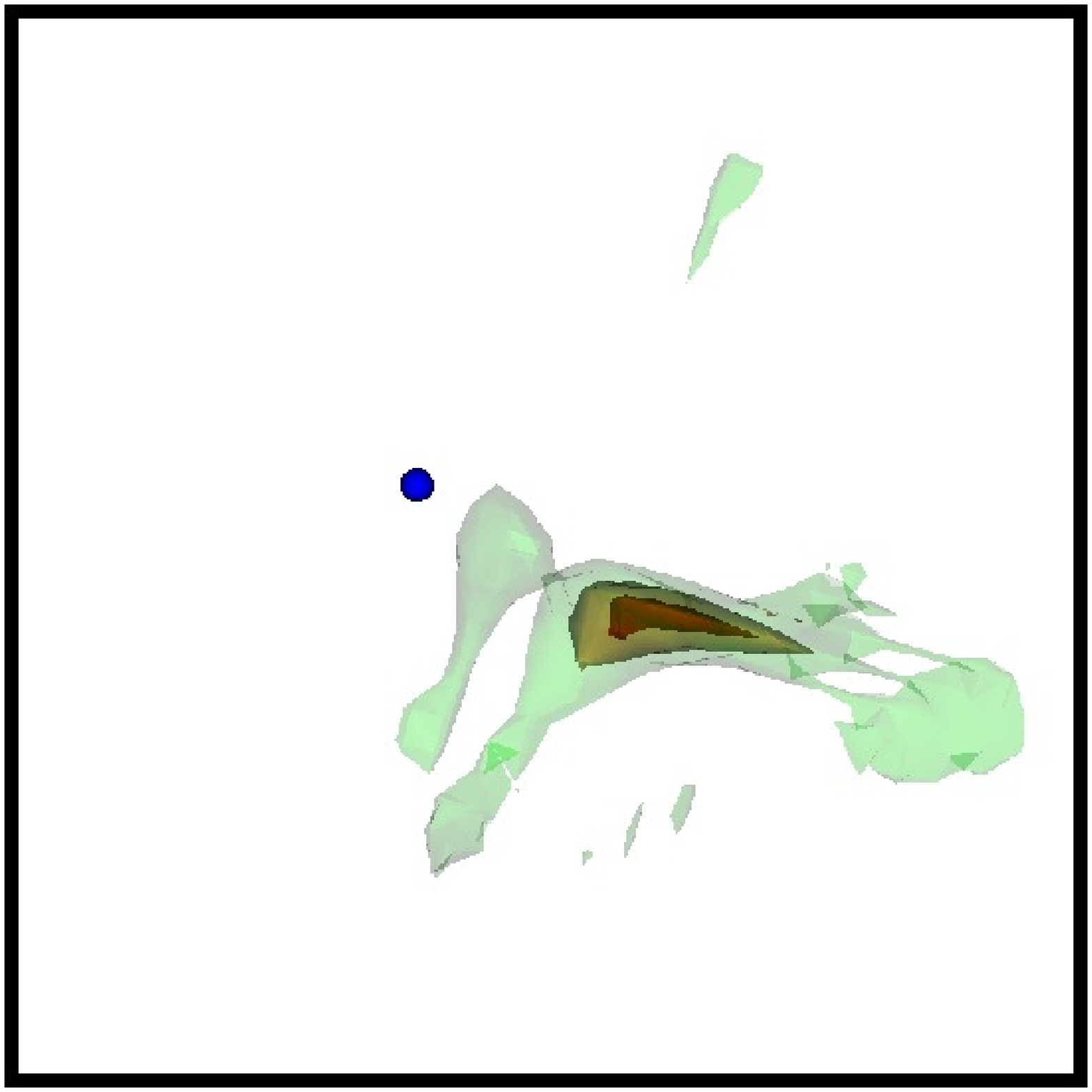}
\includegraphics[width=1.5in]{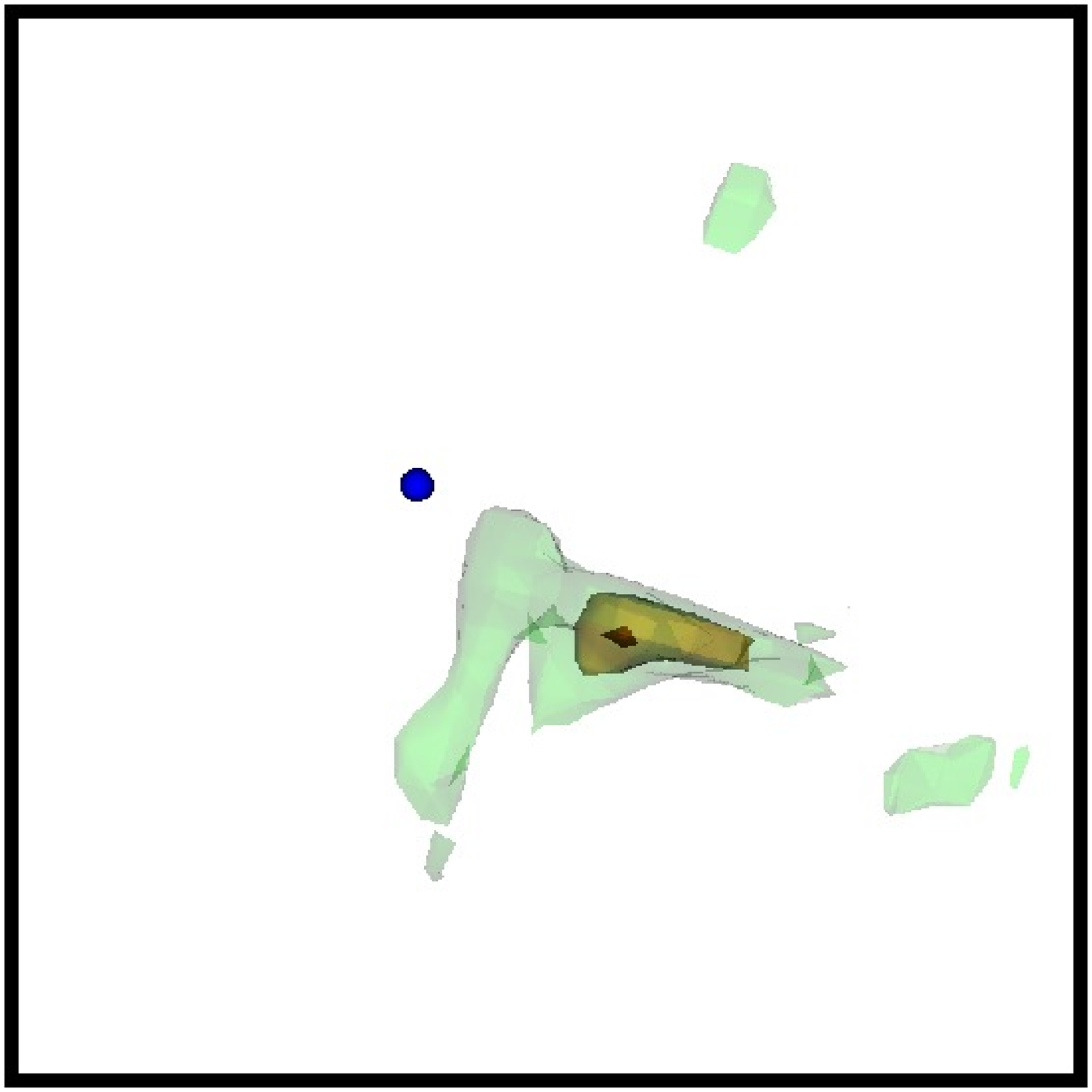}\\
\includegraphics[width=1.5in]{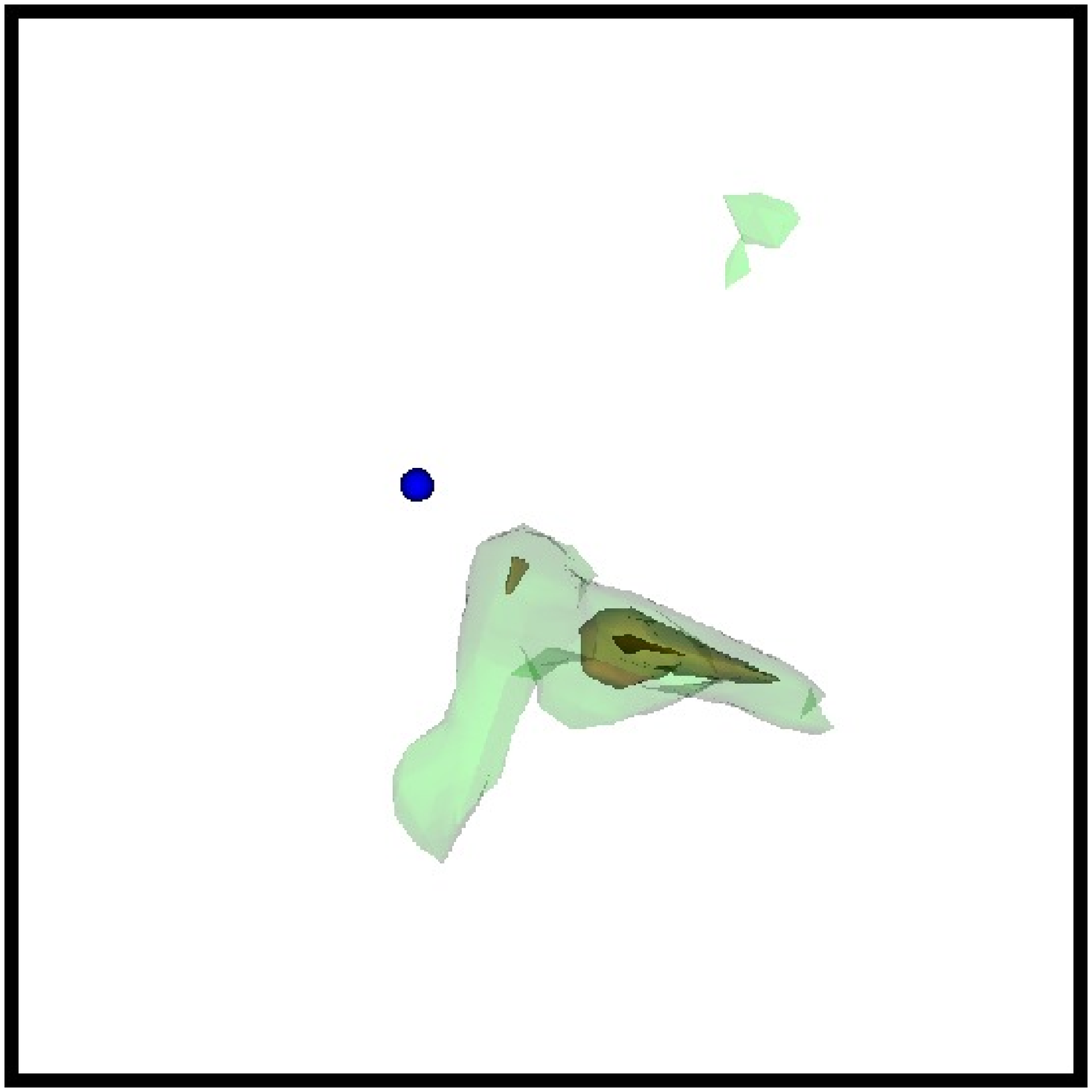}
\includegraphics[width=1.5in]{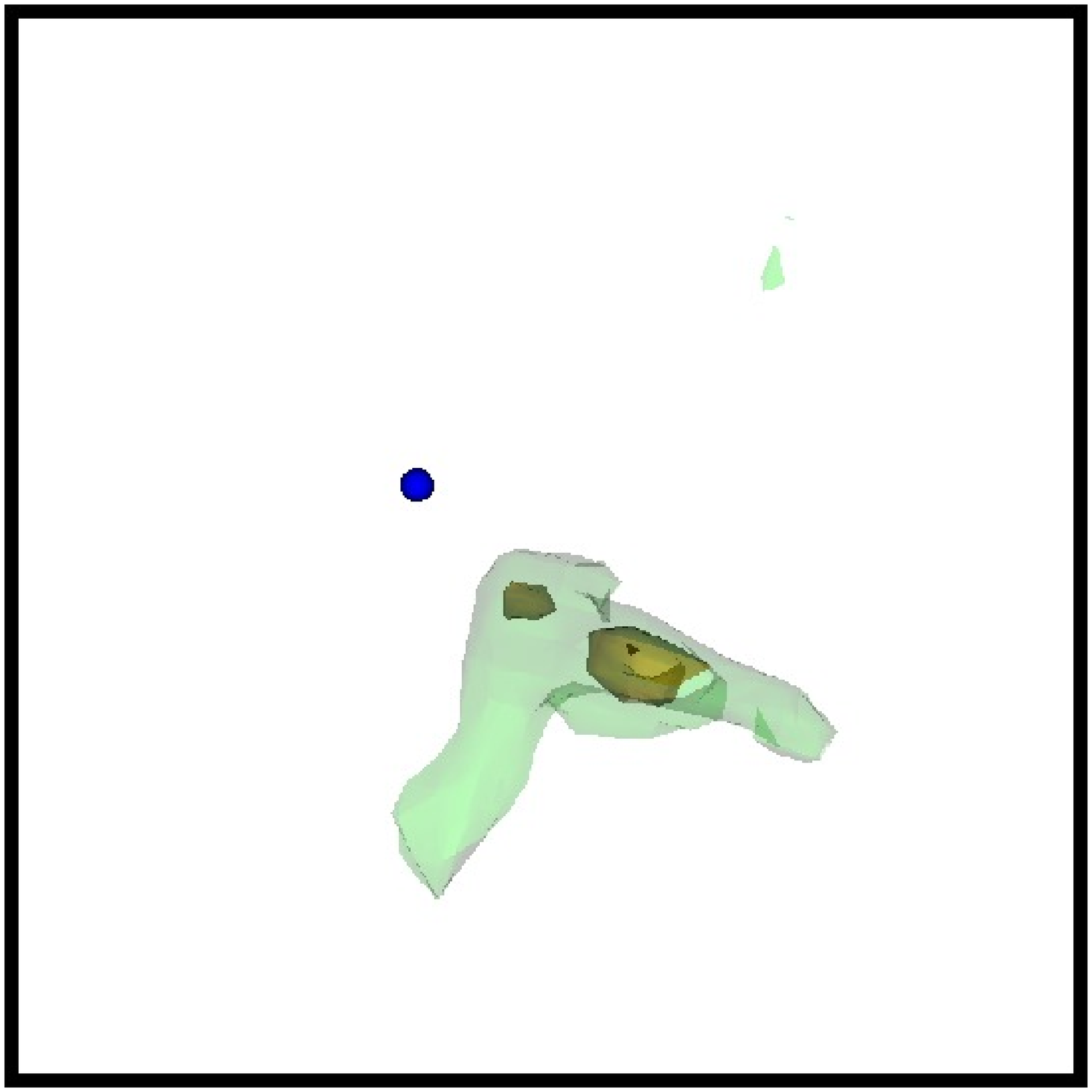}
\includegraphics[width=1.5in]{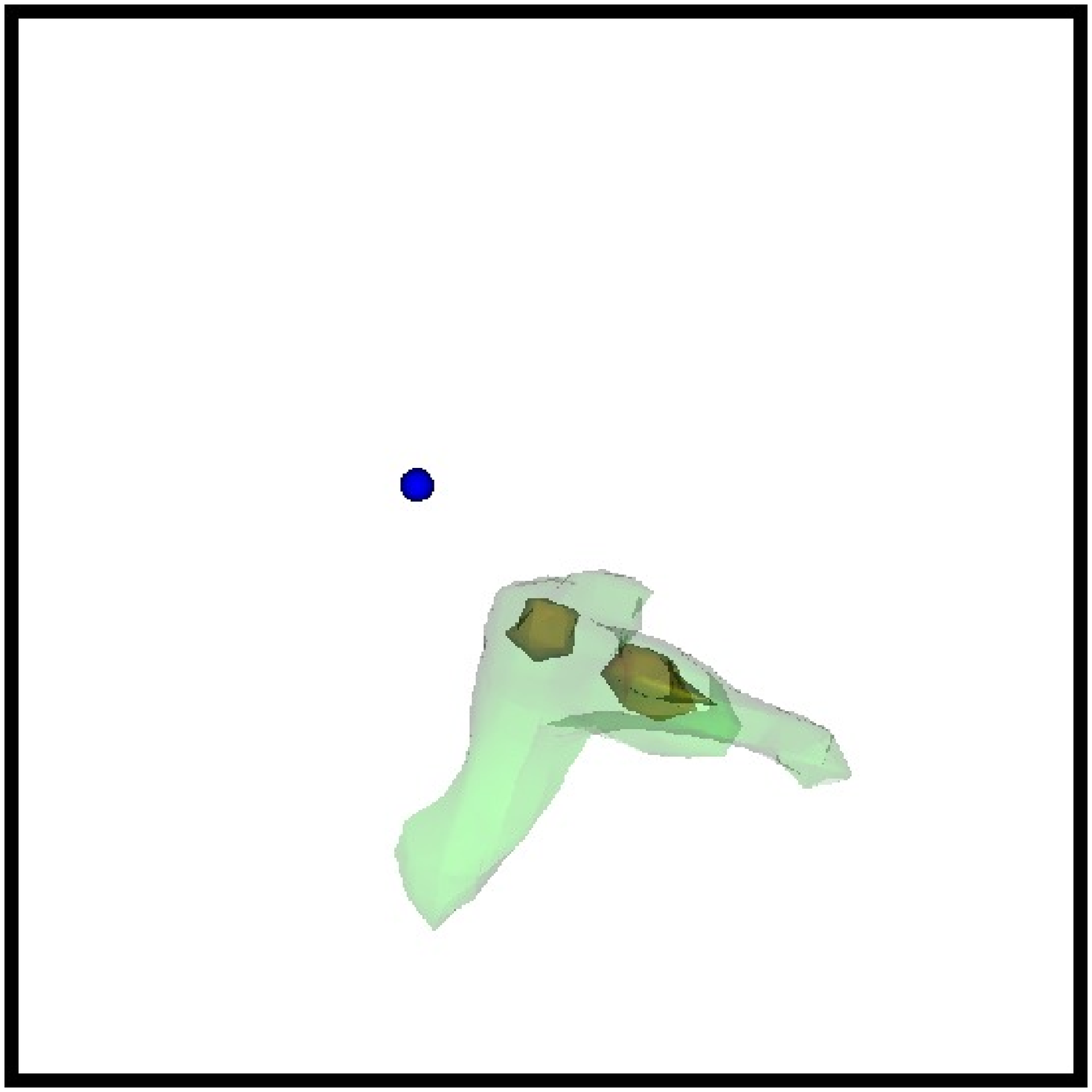}
\includegraphics[width=1.5in]{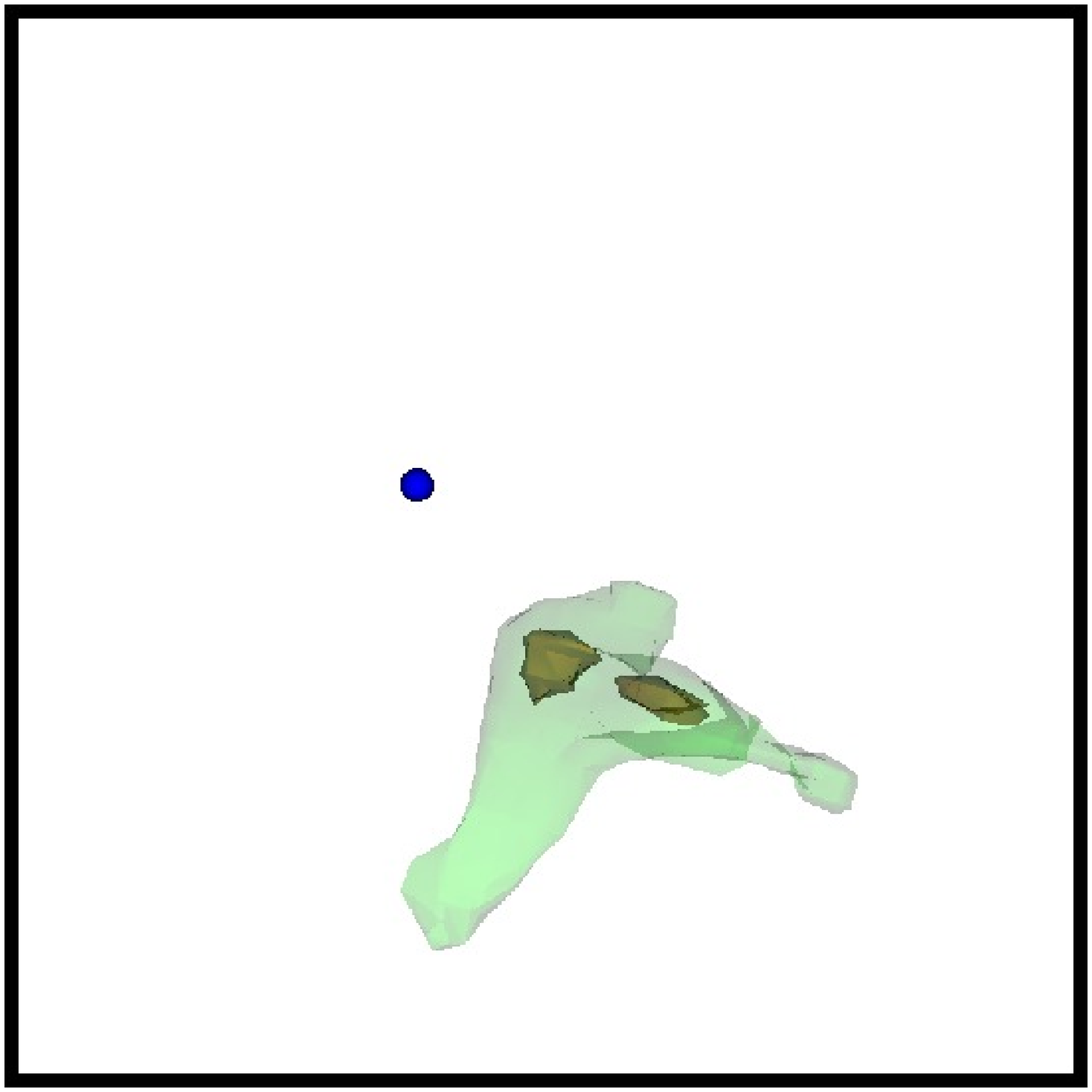}
\caption{\label{fig:hotspot_1} Temperature contours from $t=10404.0$~s to $t=10407.8$~s 
(corresponding to the green dotted time range in Figure \ref{fig:temp_1152}) spaced at 
$0.2$~s time intervals.  The contours are surfaces indicating where $T=7.15\times 10^8$~K 
(green), $T=7.2\times 10^8$~K (yellow), and $T=7.25\times 10^8$~K (orange).  The blue dot is
at the center of the star, and has a diameter of 4.34~km, which corresponds to the grid
cell width for this simulation.}
\end{center}
\end{figure}

\begin{figure}
\begin{center}
\includegraphics[width=1.5in]{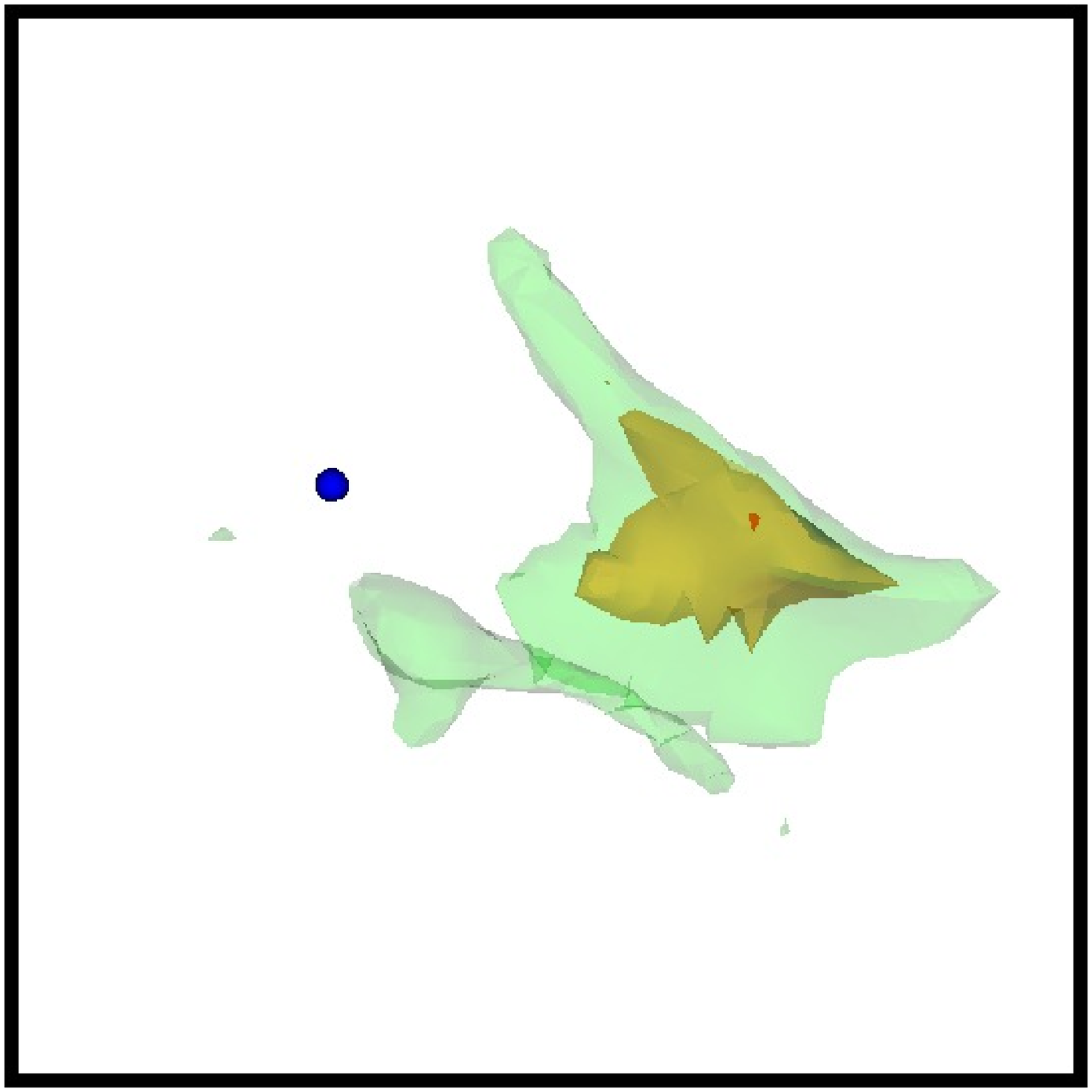}
\includegraphics[width=1.5in]{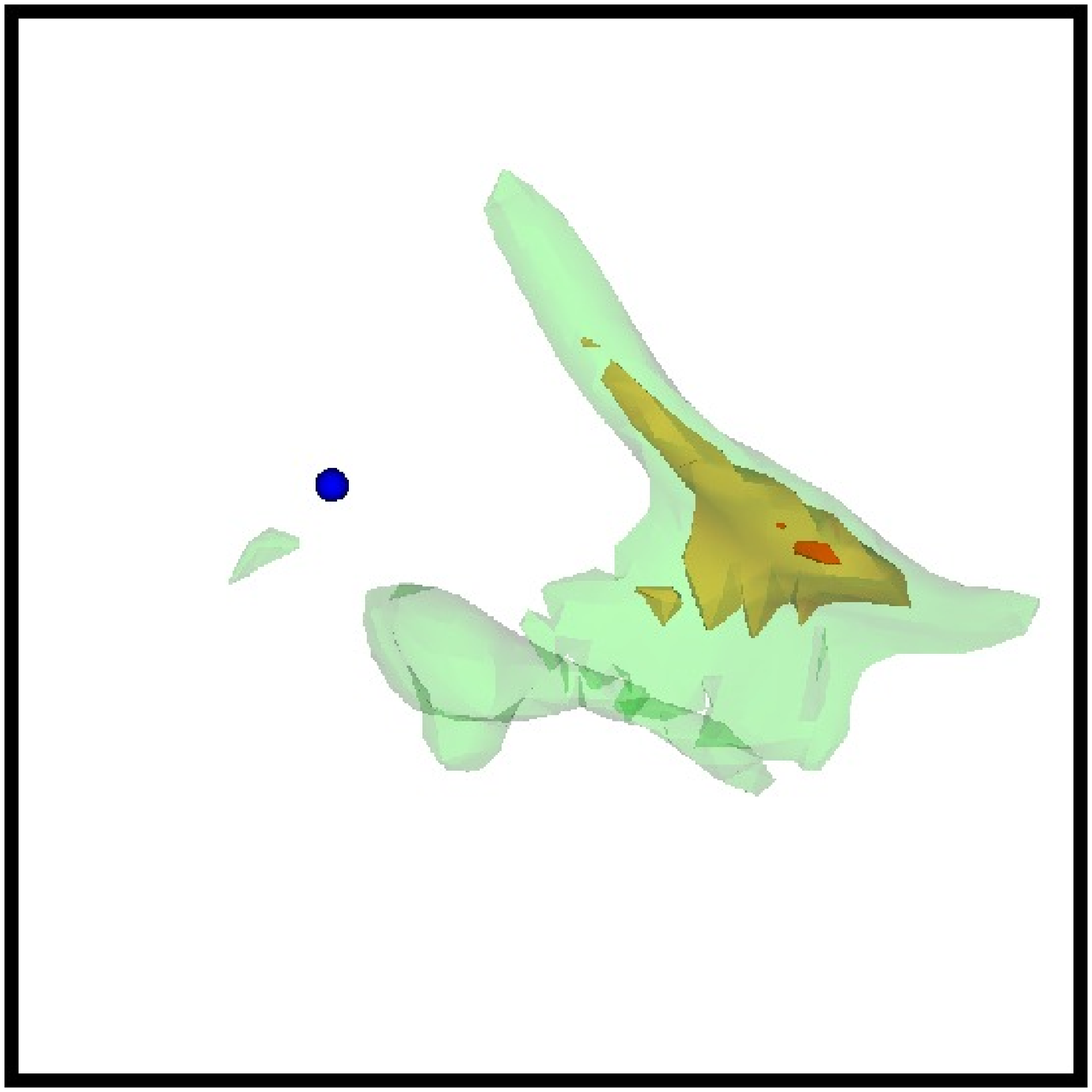}
\includegraphics[width=1.5in]{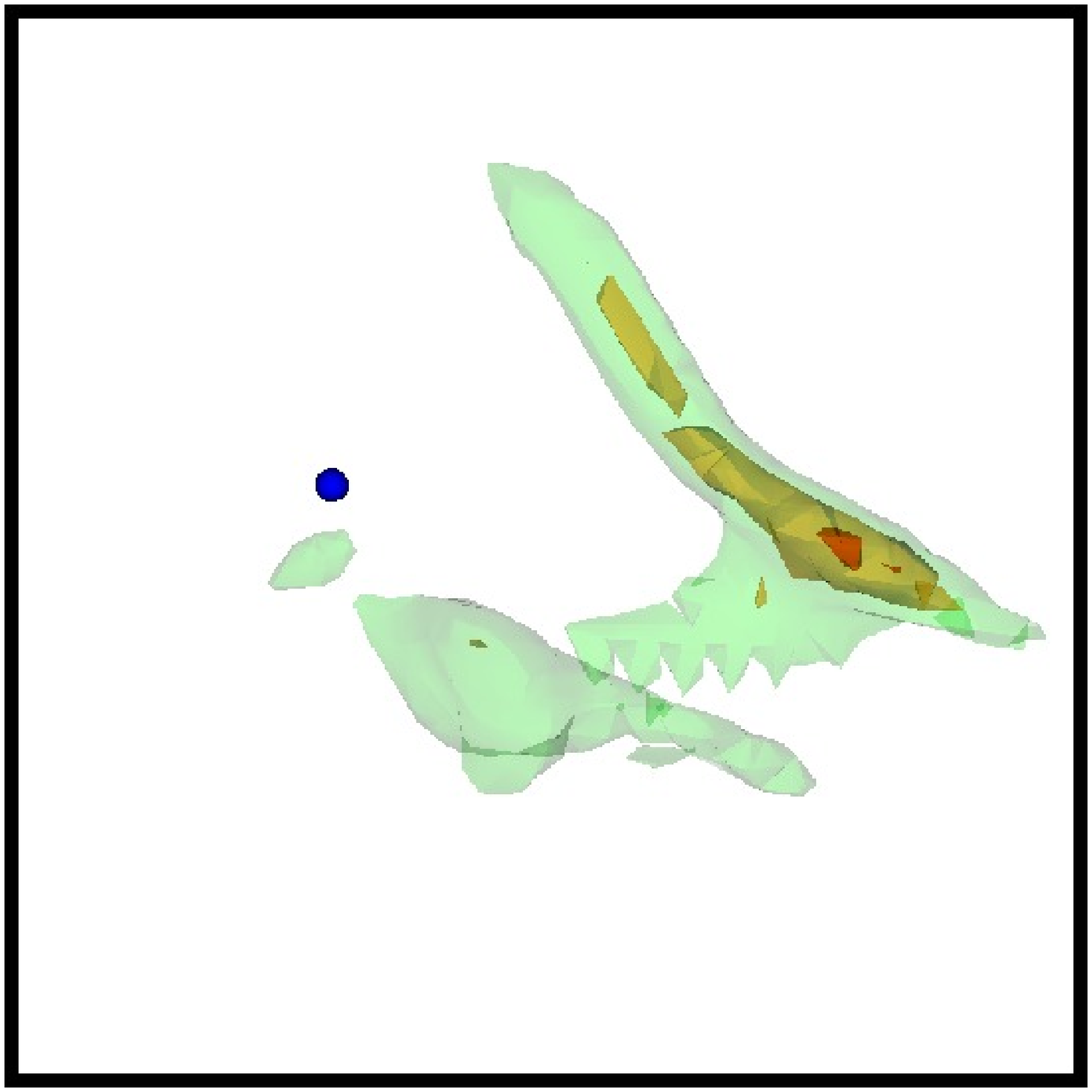}
\includegraphics[width=1.5in]{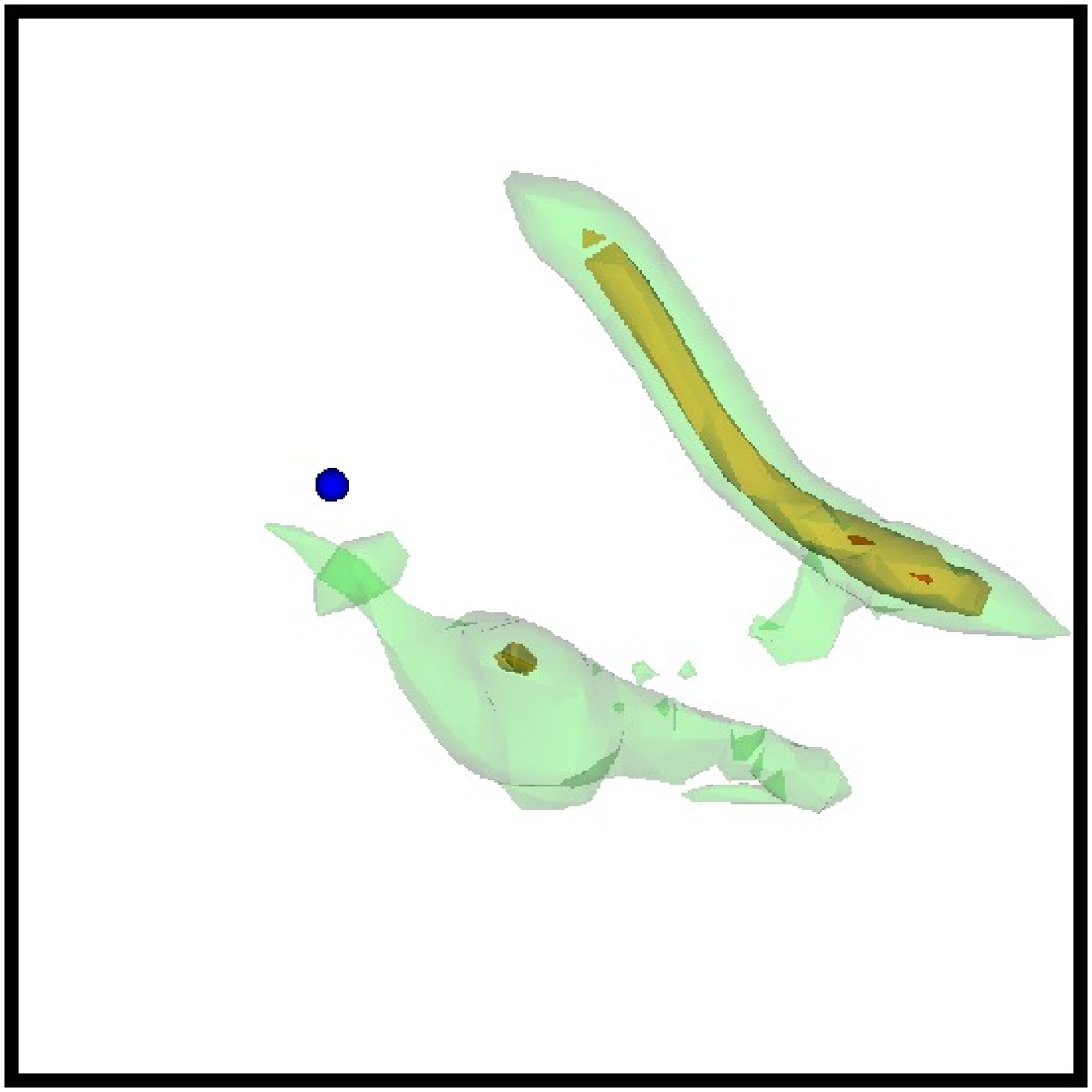}\\
\includegraphics[width=1.5in]{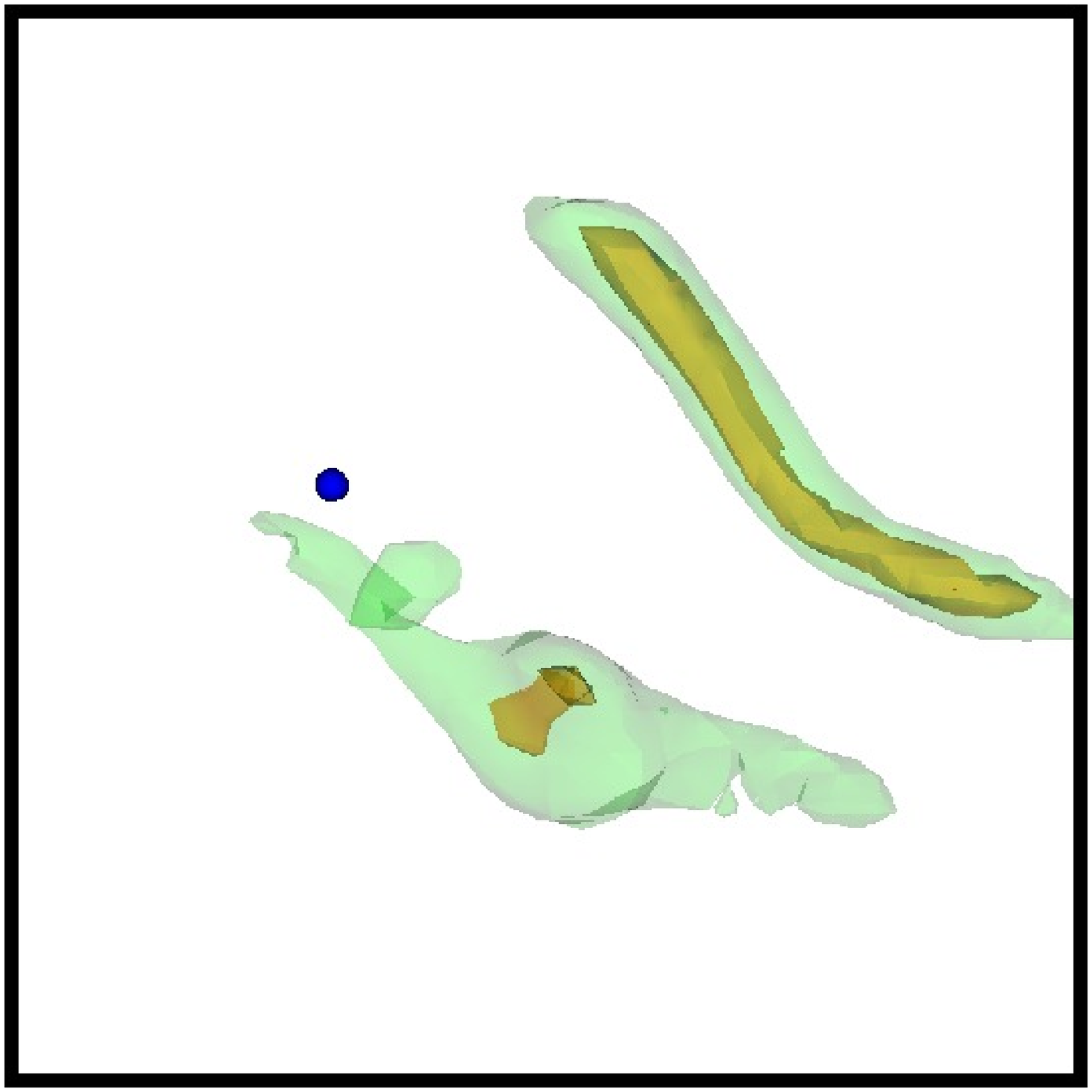}
\includegraphics[width=1.5in]{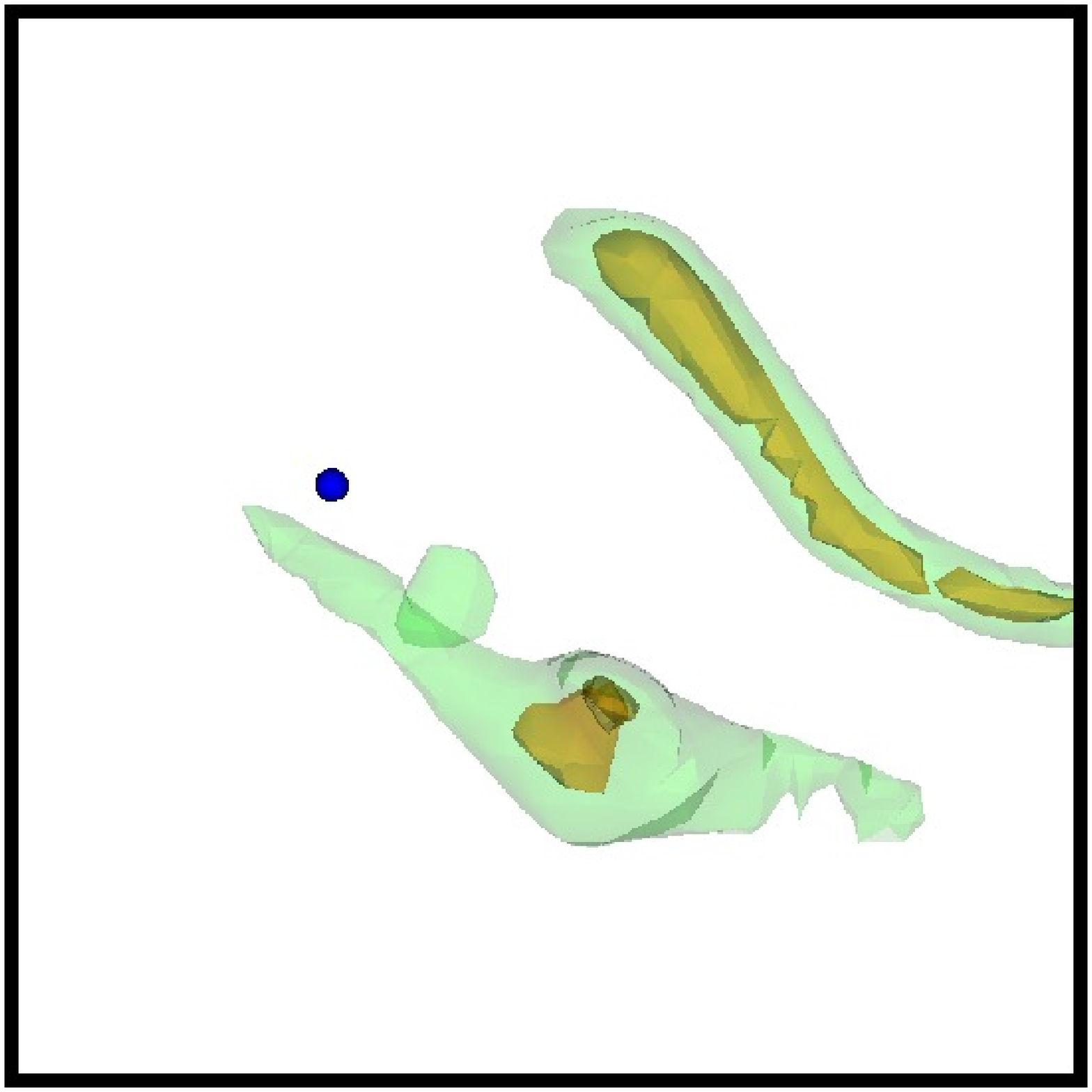}
\includegraphics[width=1.5in]{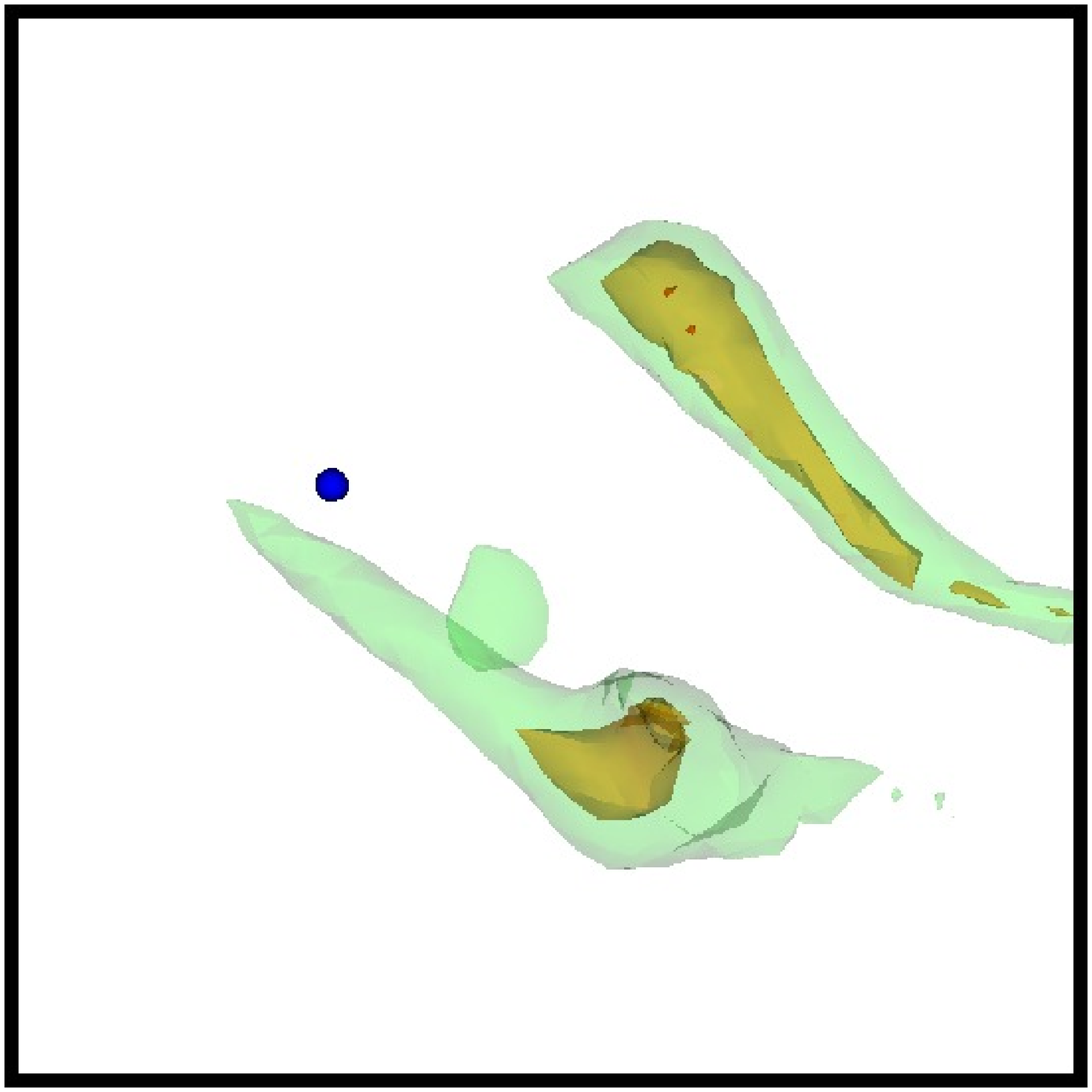}
\includegraphics[width=1.5in]{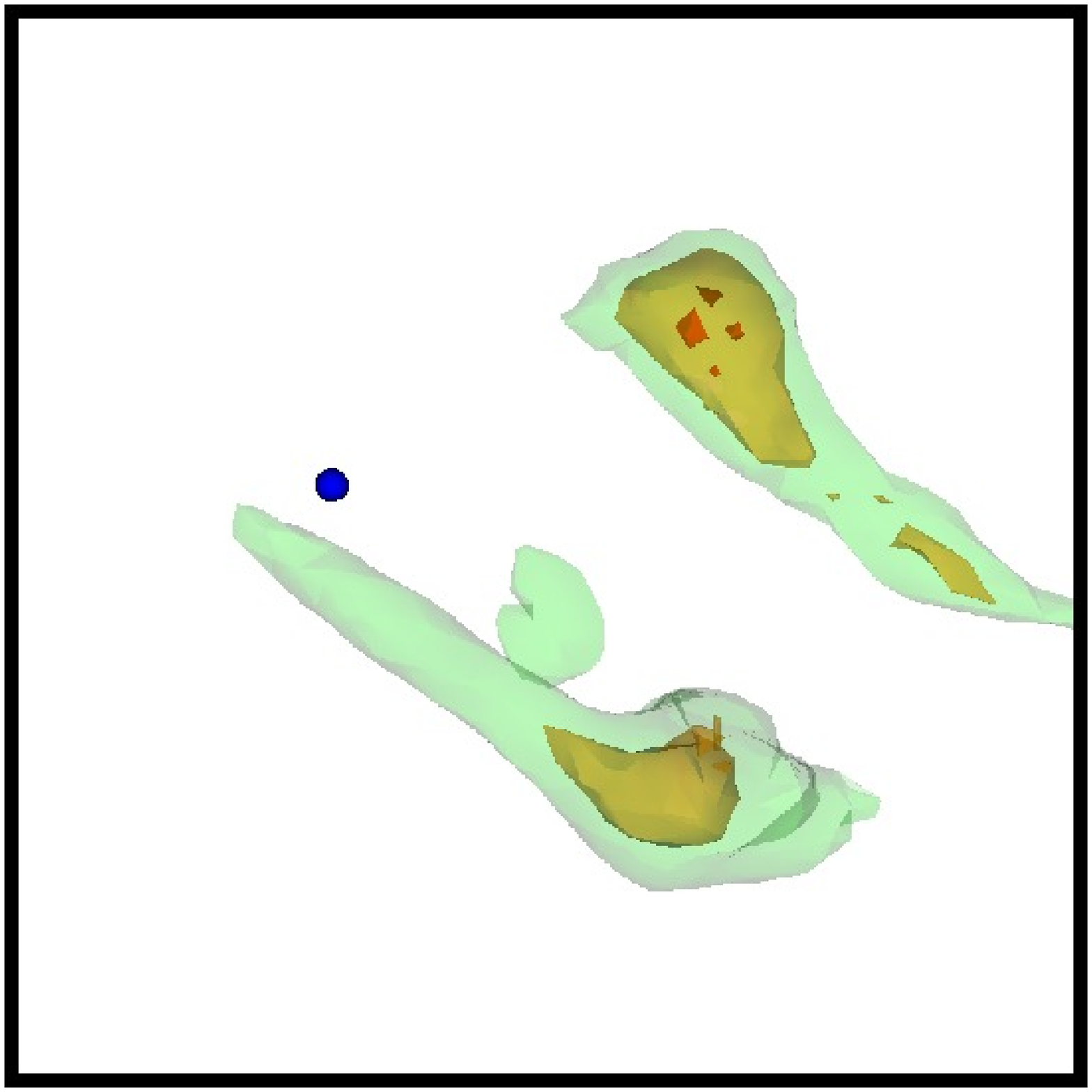}\\
\includegraphics[width=1.5in]{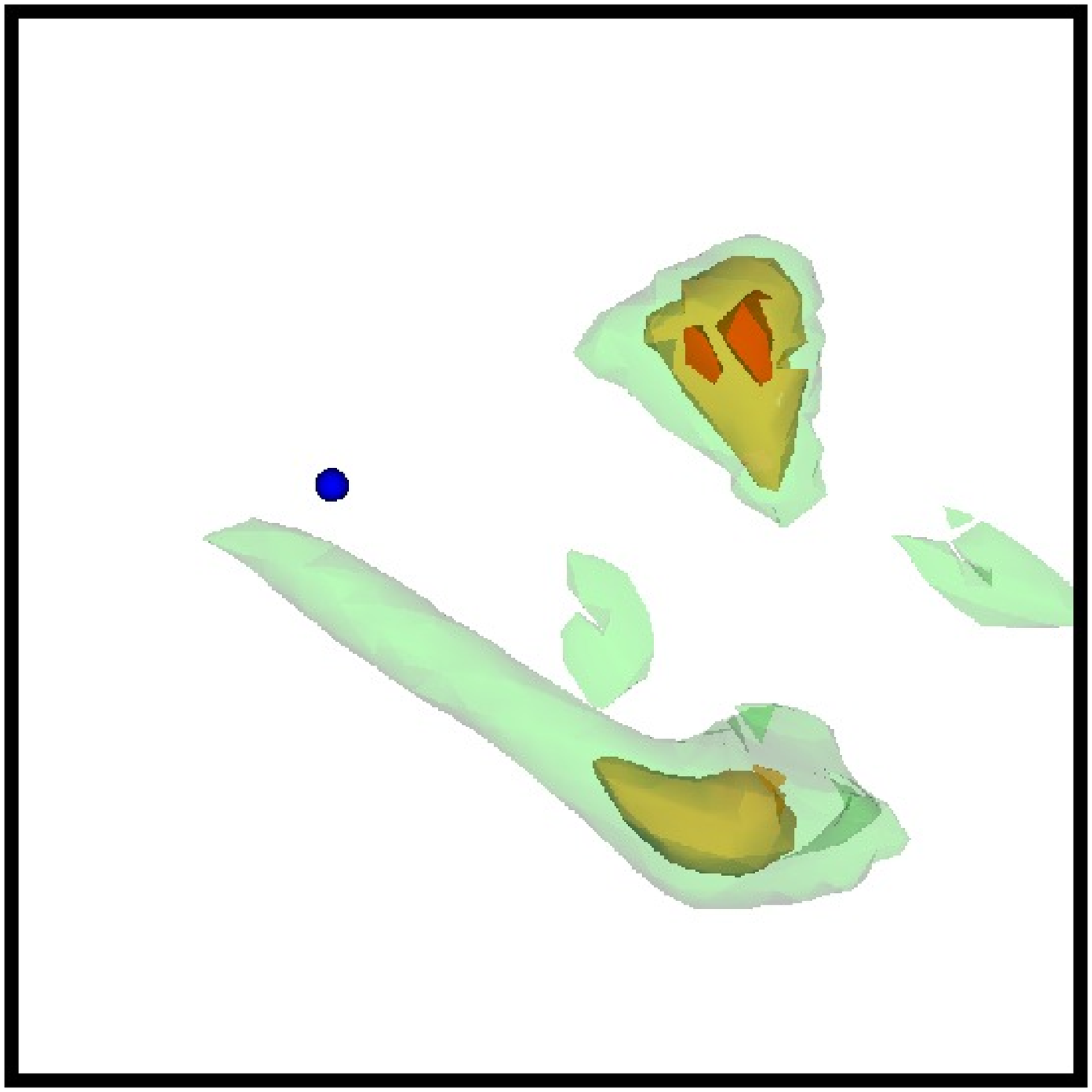}
\includegraphics[width=1.5in]{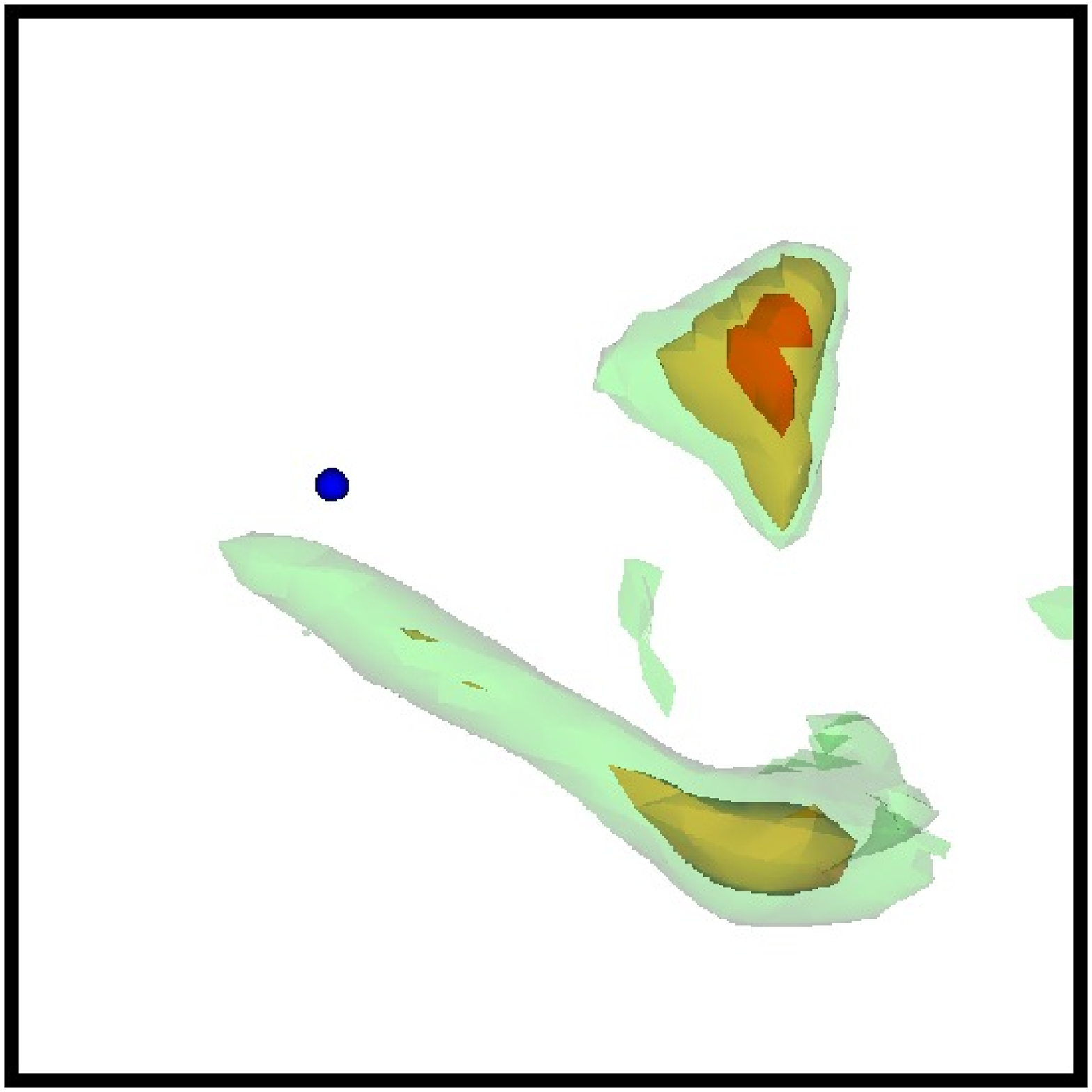}
\includegraphics[width=1.5in]{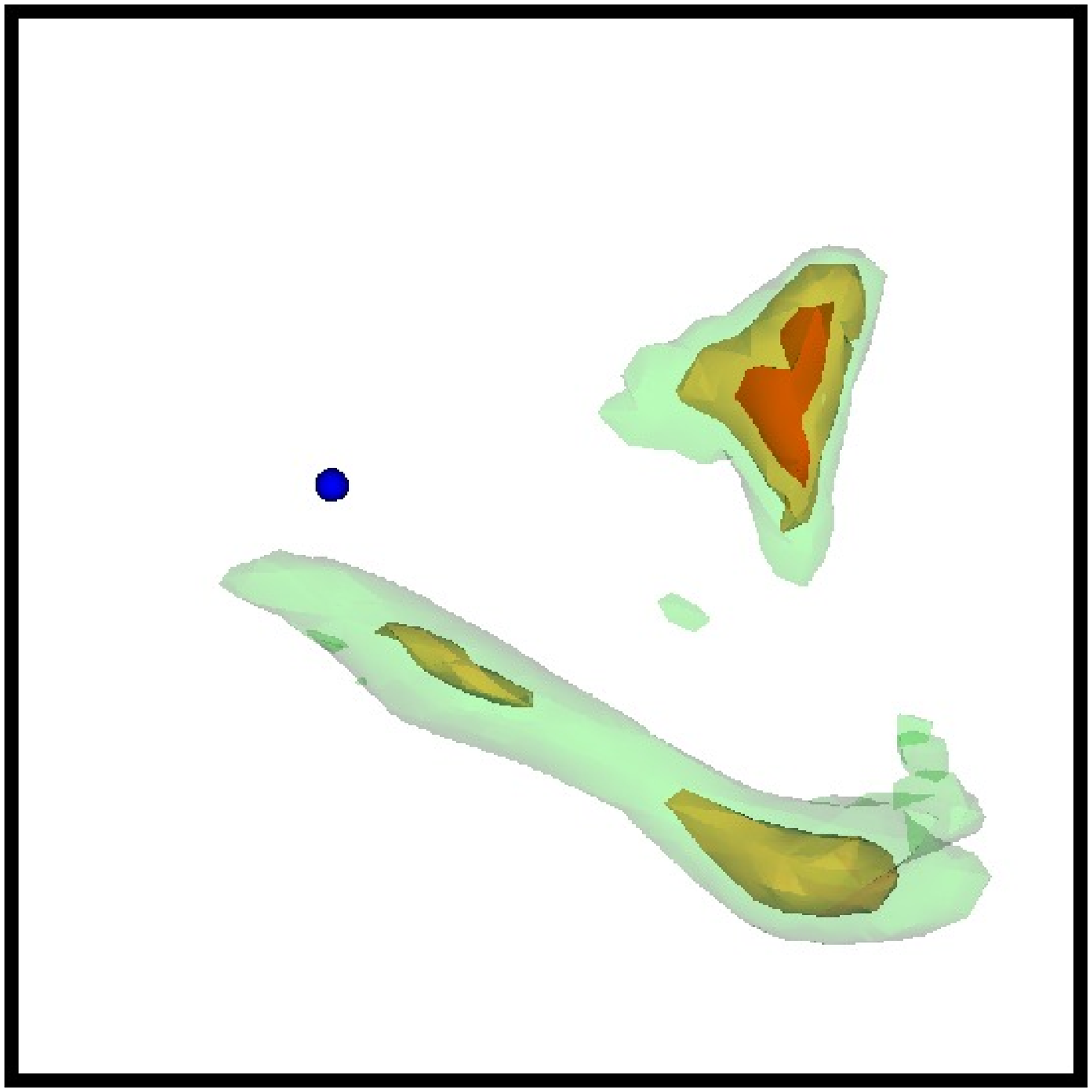}
\includegraphics[width=1.5in]{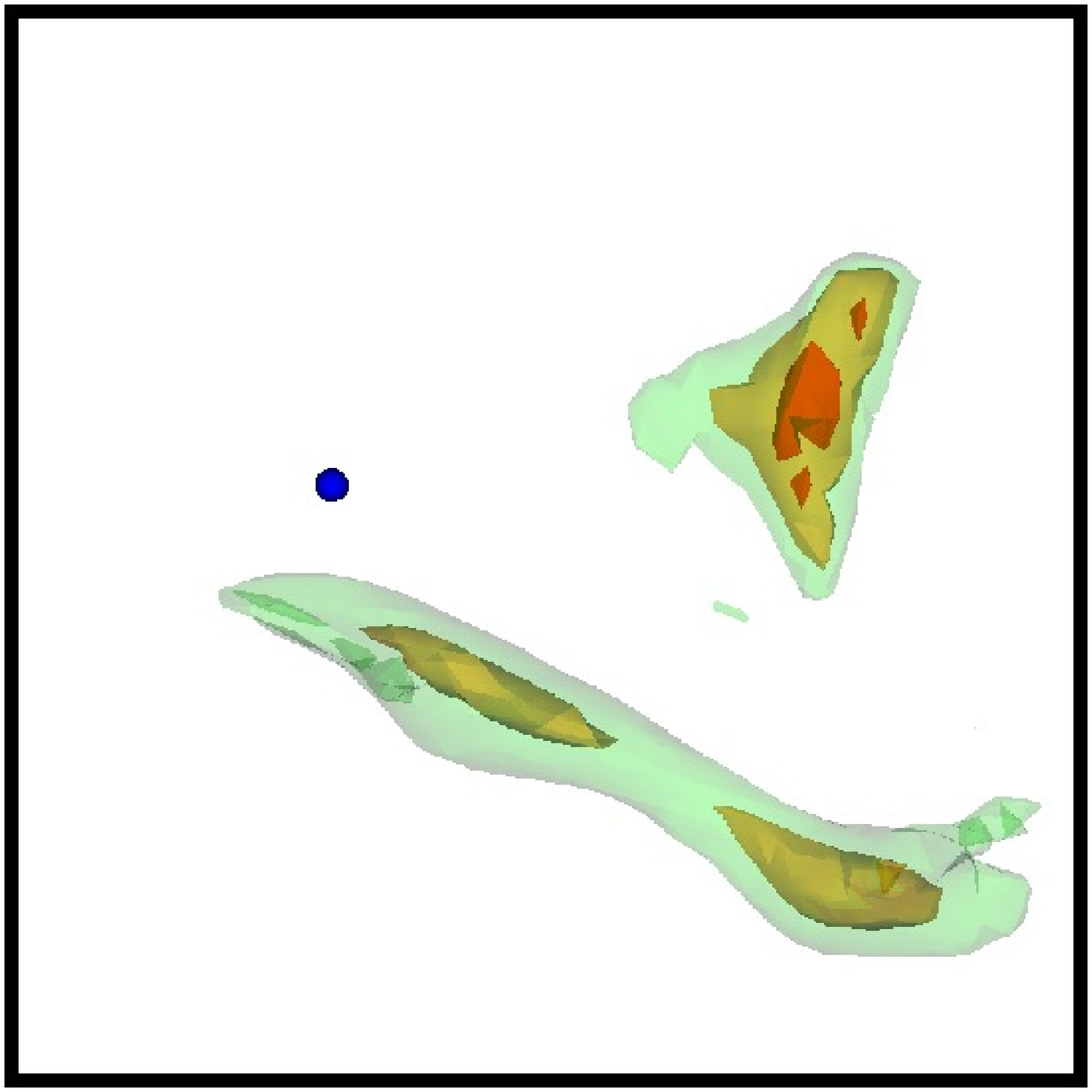}\\
\includegraphics[width=1.5in]{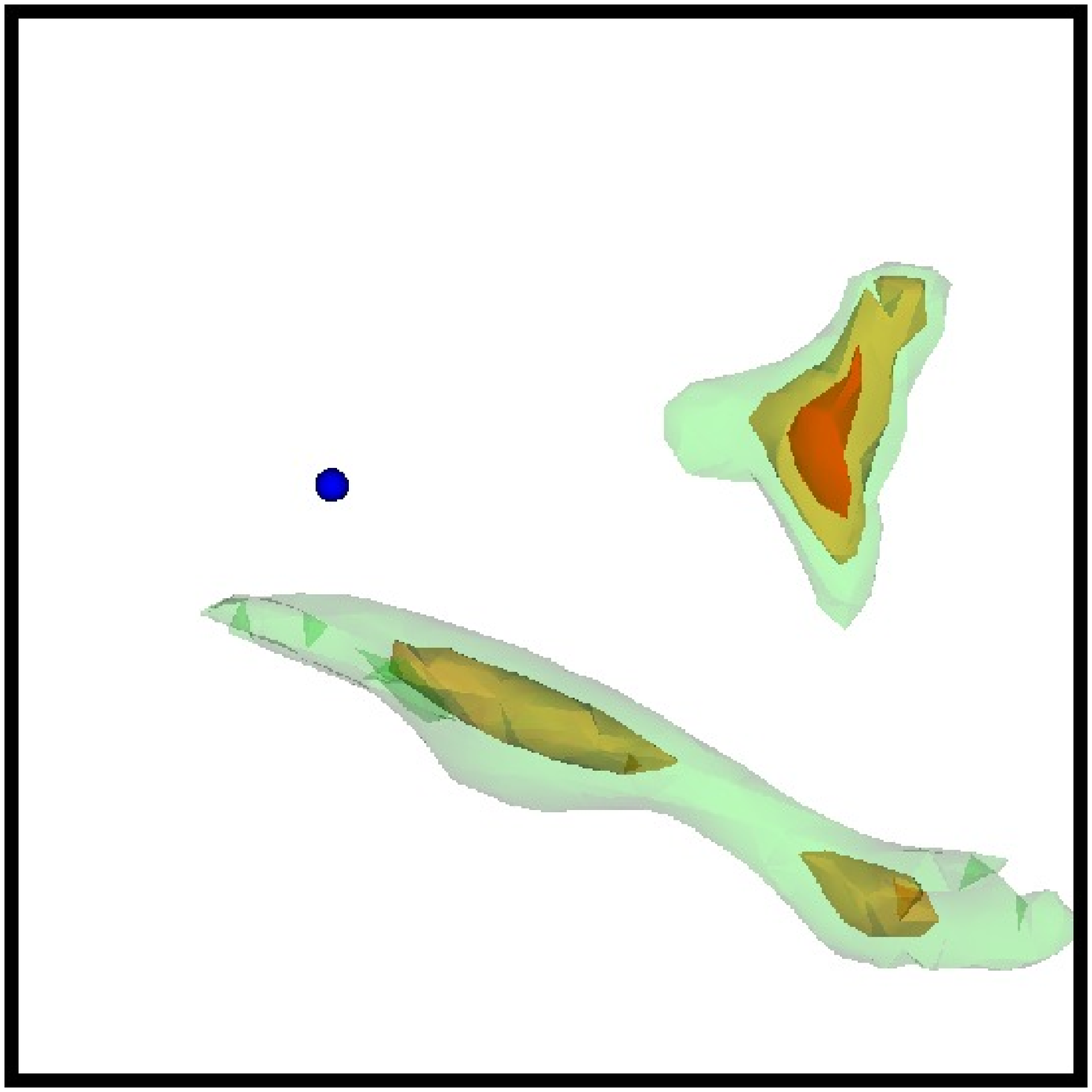}
\includegraphics[width=1.5in]{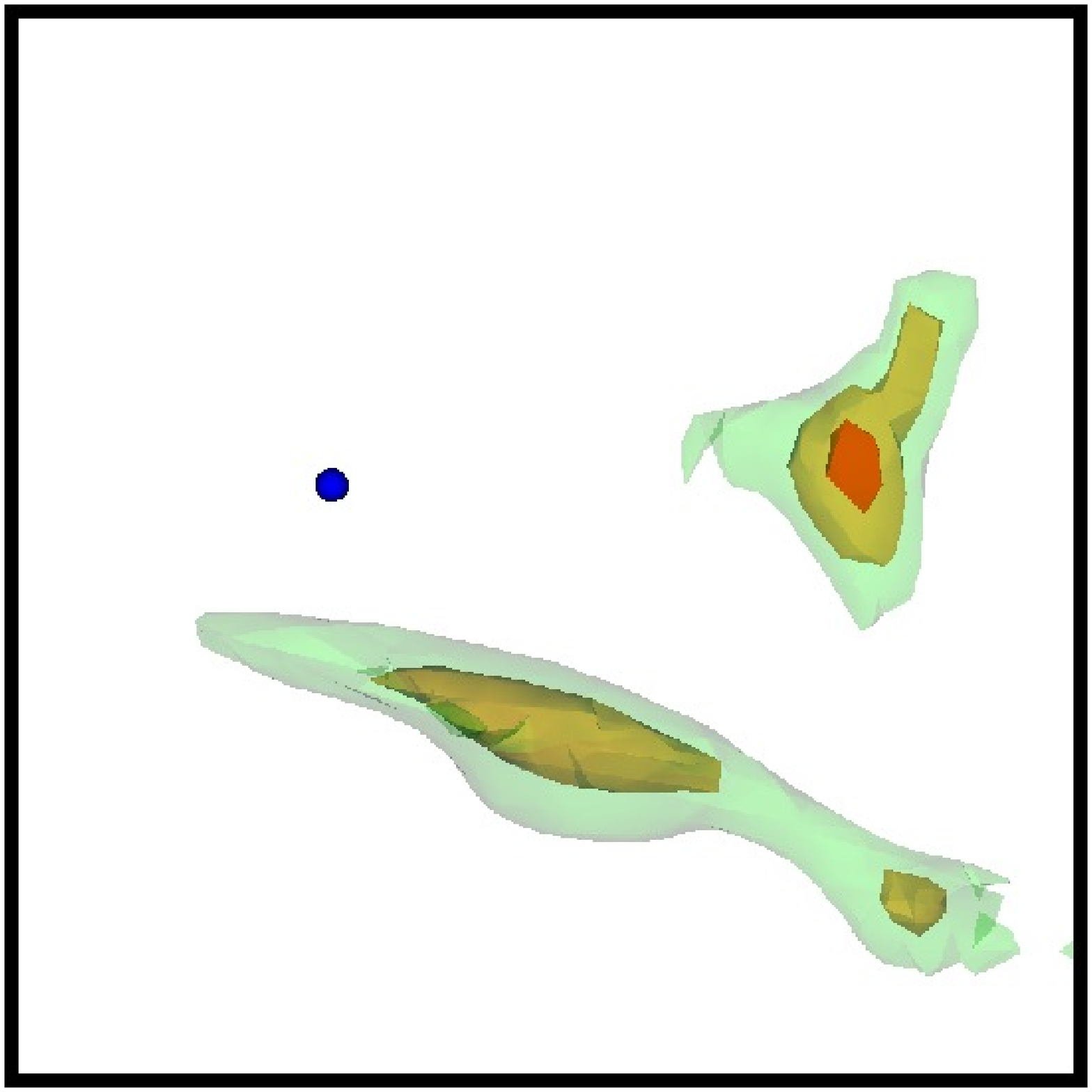}
\includegraphics[width=1.5in]{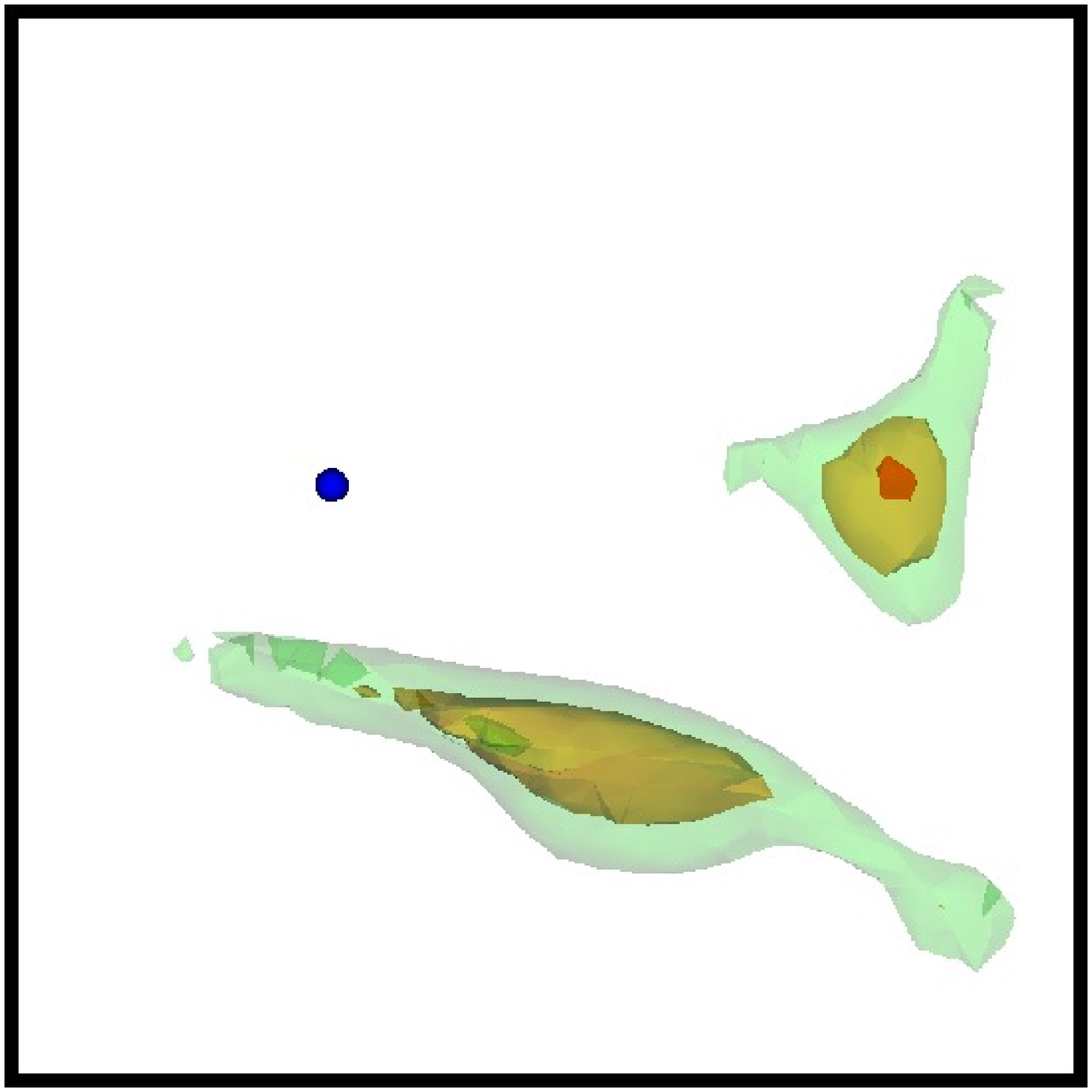}
\includegraphics[width=1.5in]{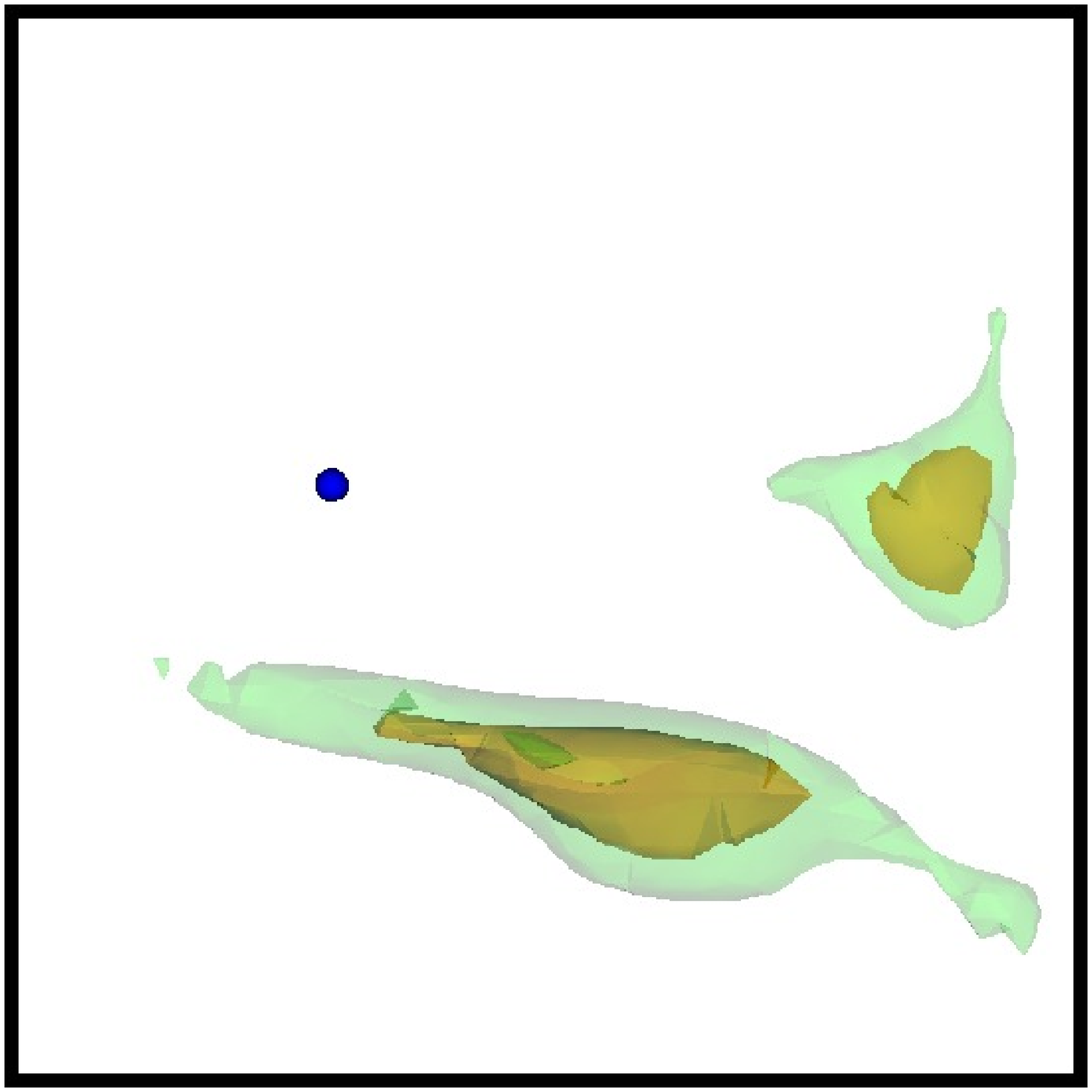}\\
\includegraphics[width=1.5in]{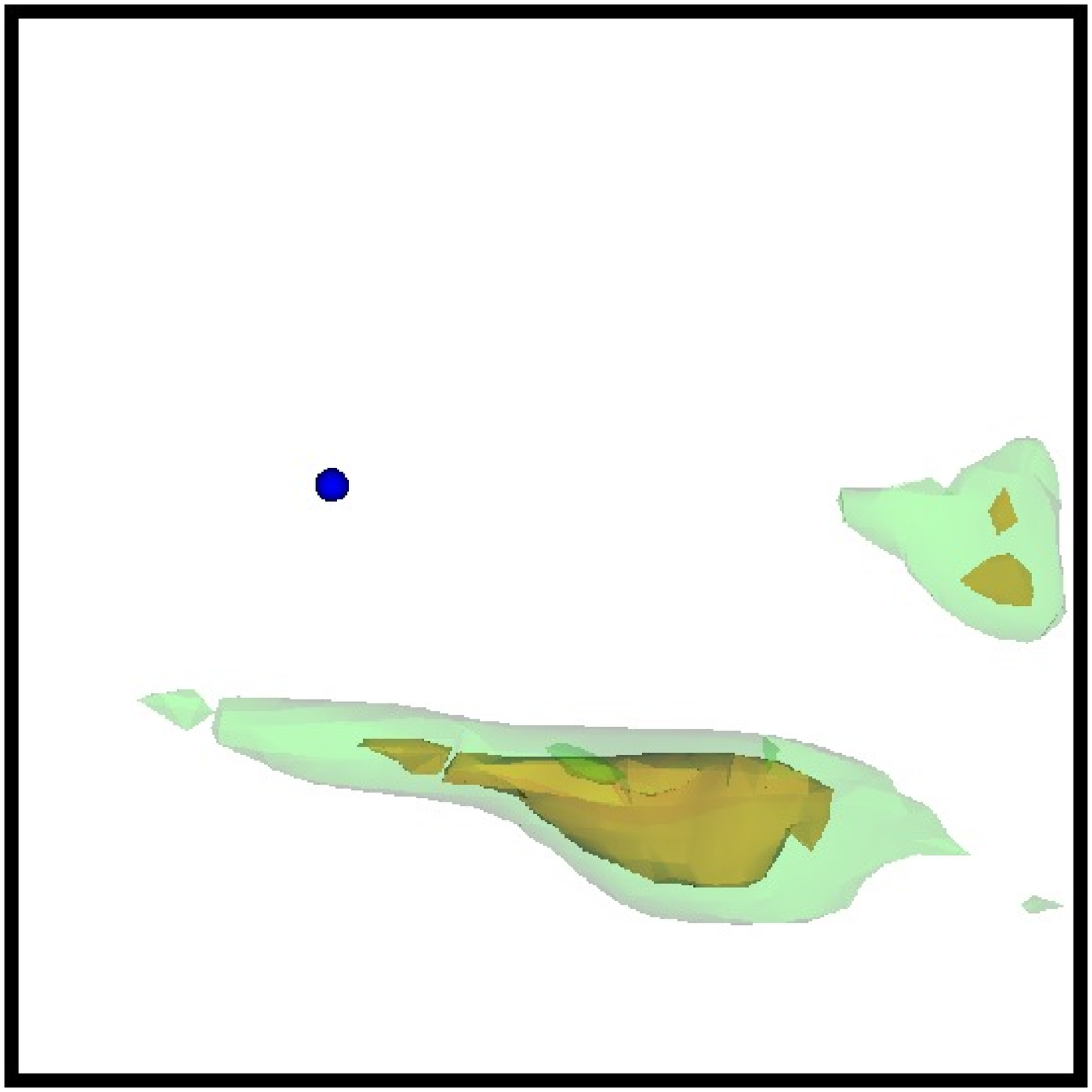}
\includegraphics[width=1.5in]{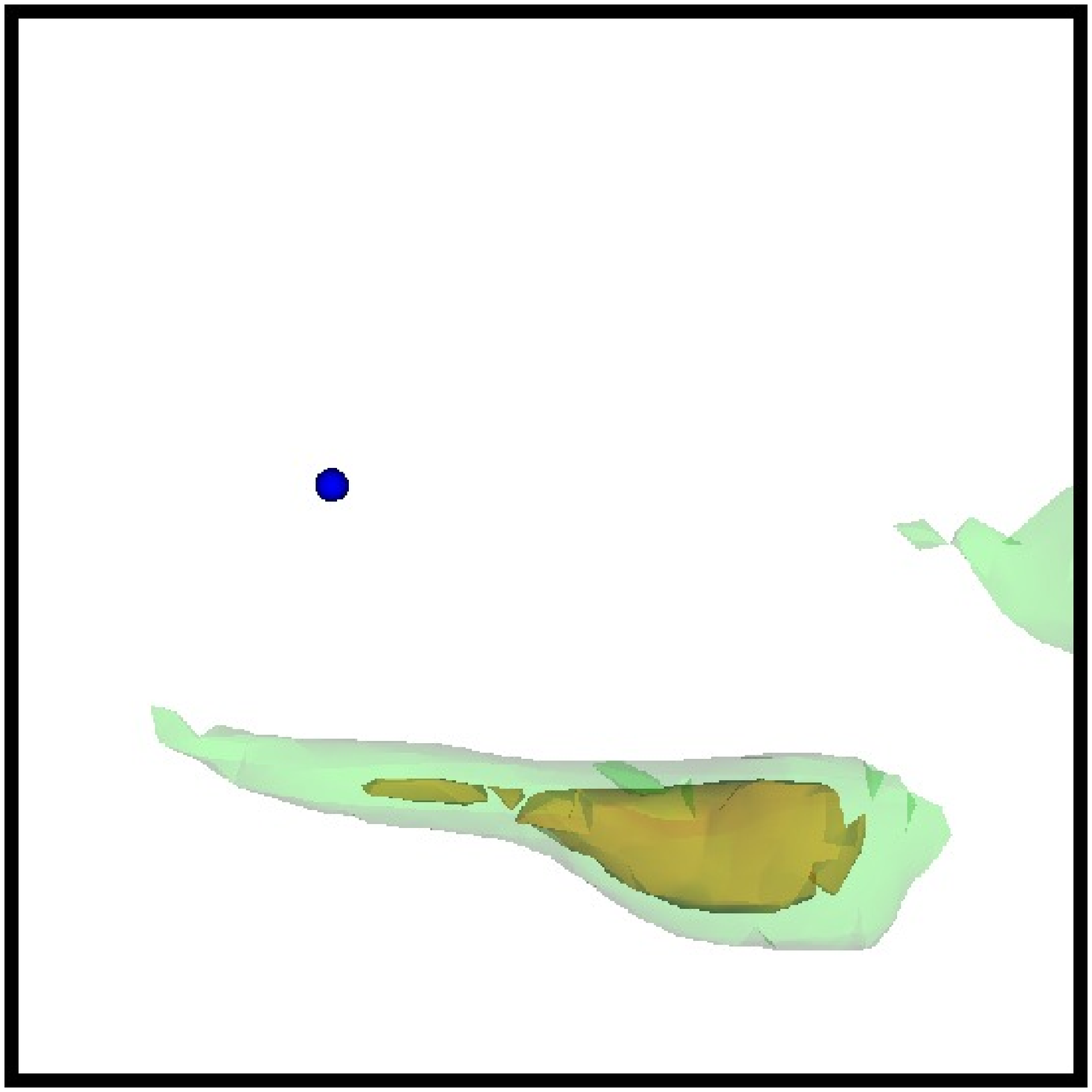}
\includegraphics[width=1.5in]{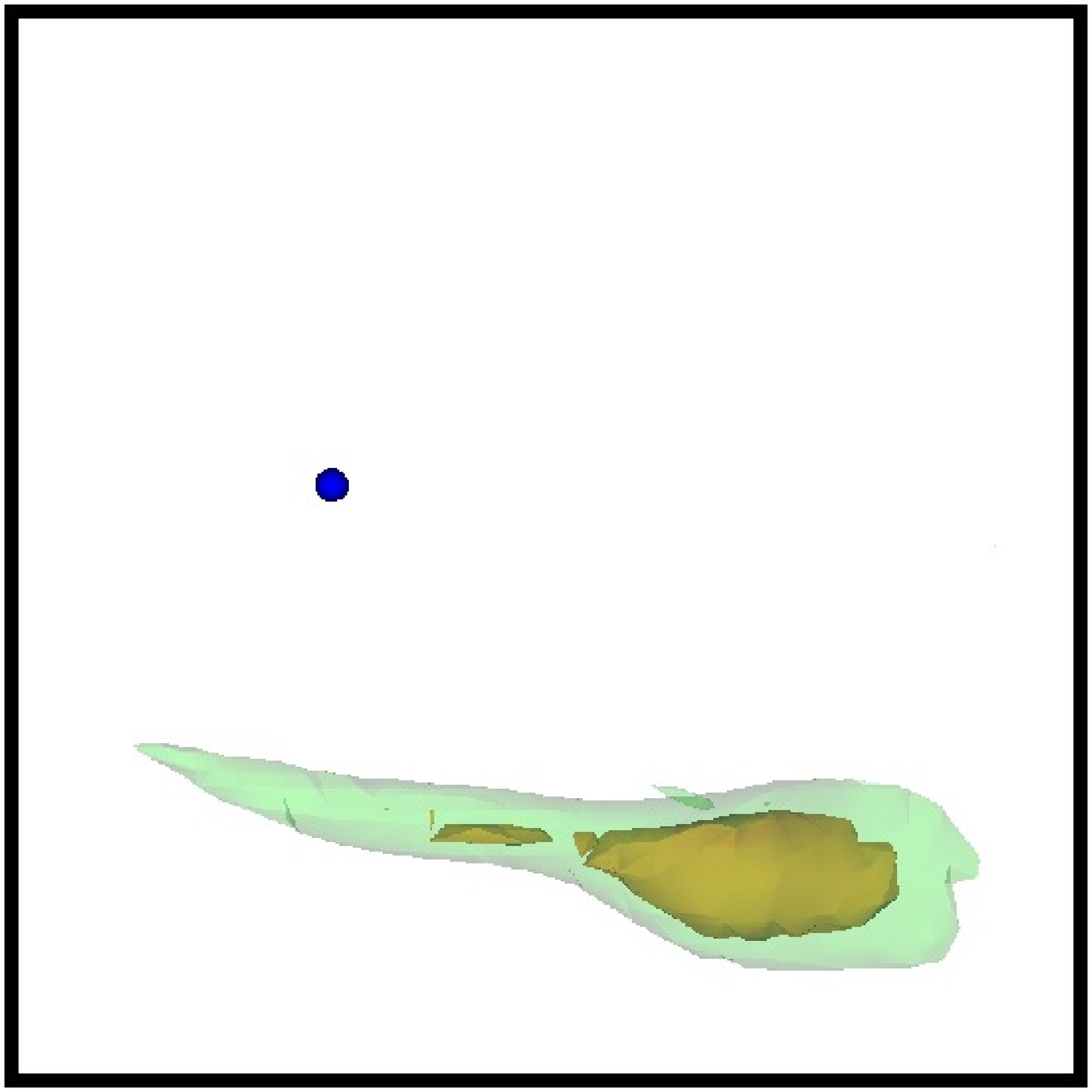}
\includegraphics[width=1.5in]{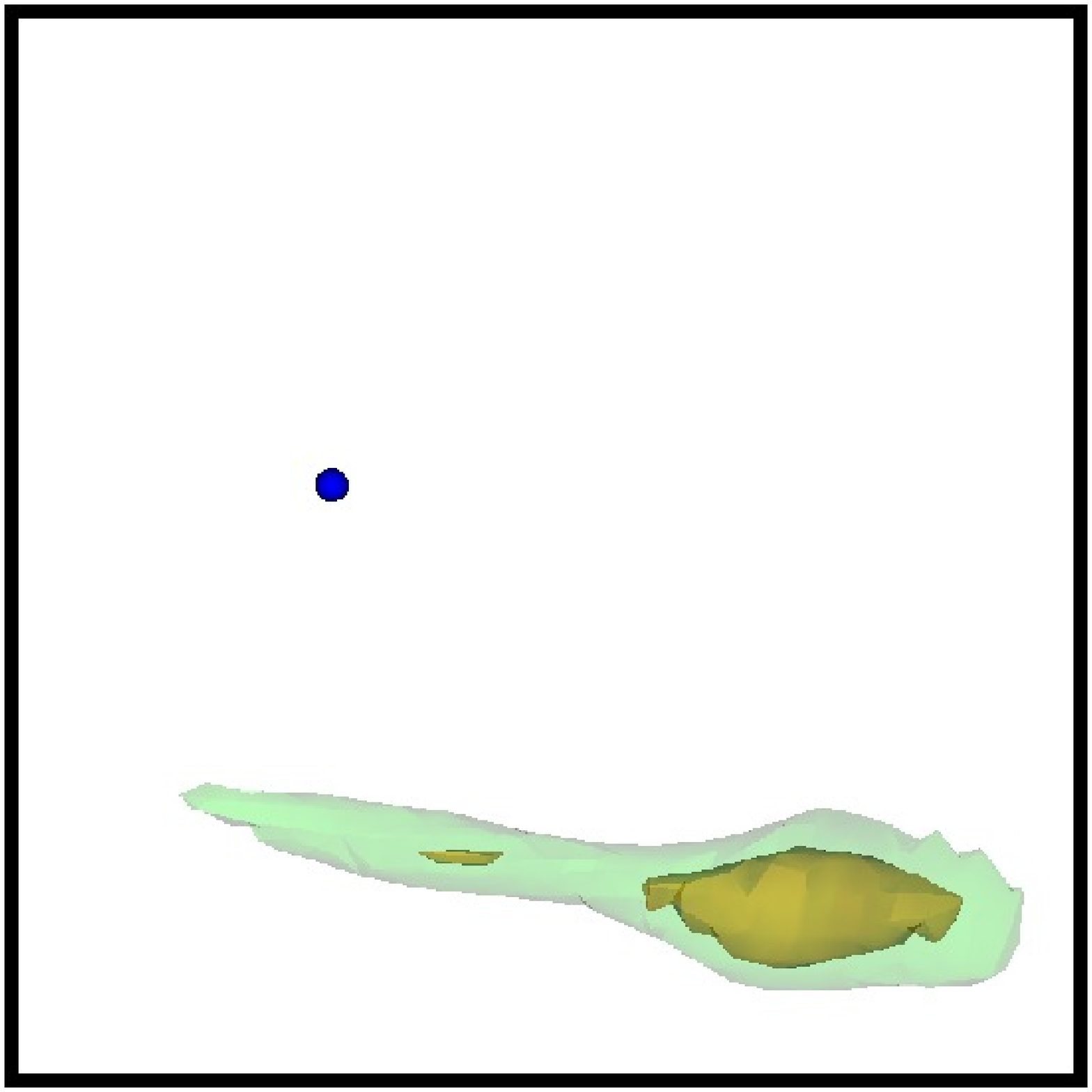}
\caption{\label{fig:hotspot_2} Temperature contours from $t=10454.6$~s to $t=10458.4$~s 
(corresponding to the blue dotted time range in Figure \ref{fig:temp_1152}) spaced at 
$0.2$~s time intervals.  The contours are surfaces indicating where $T=7.24\times 10^8$~K 
(green), $T=7.31\times 10^8$~K (yellow), and $T=7.38\times 10^8$~K (orange).  The blue dot is
at the center of the star, and has a diameter of 4.34~km, which corresponds to the grid
cell width for this simulation.}
\end{center}
\end{figure}

\begin{figure}
\begin{center}
\includegraphics[width=1.5in]{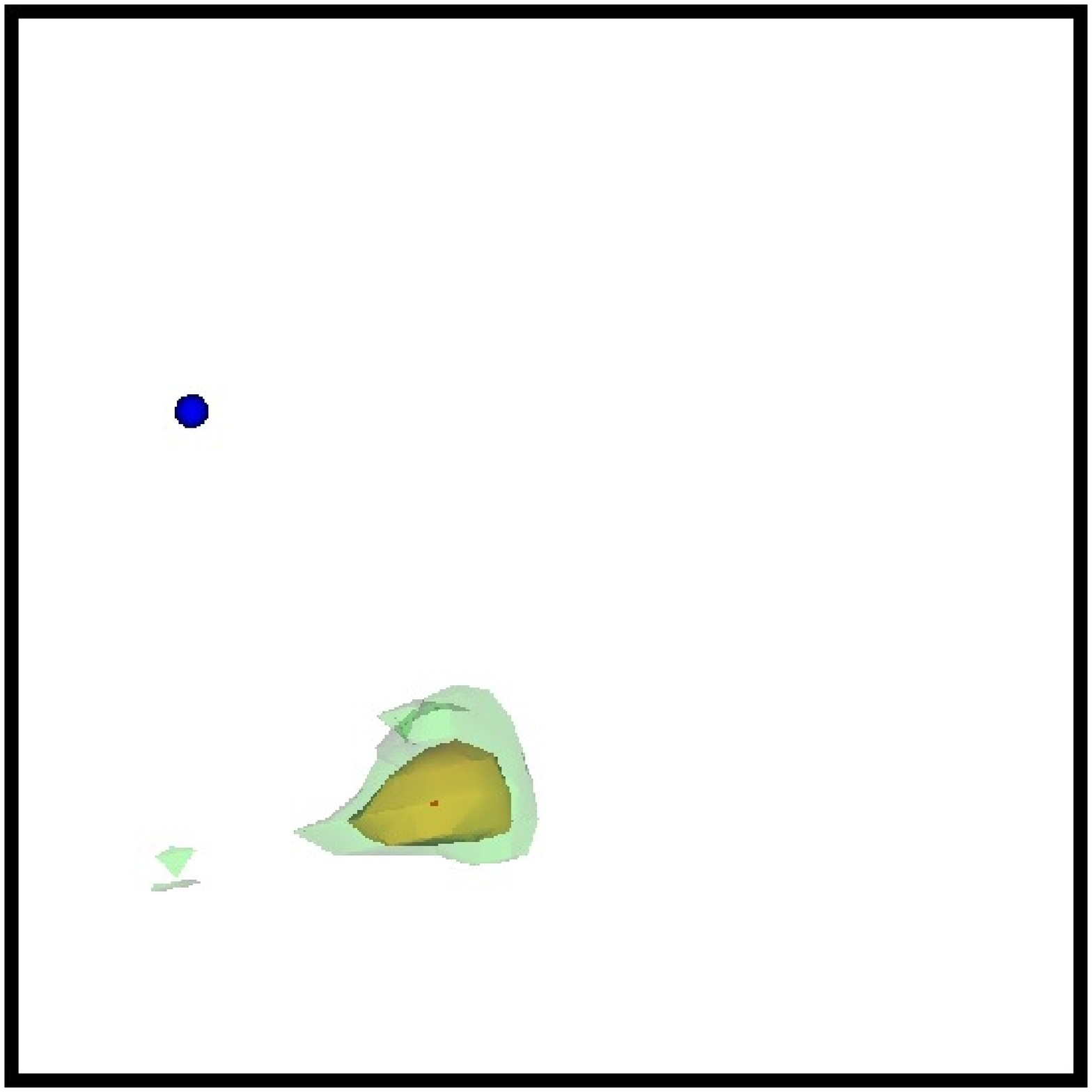}
\includegraphics[width=1.5in]{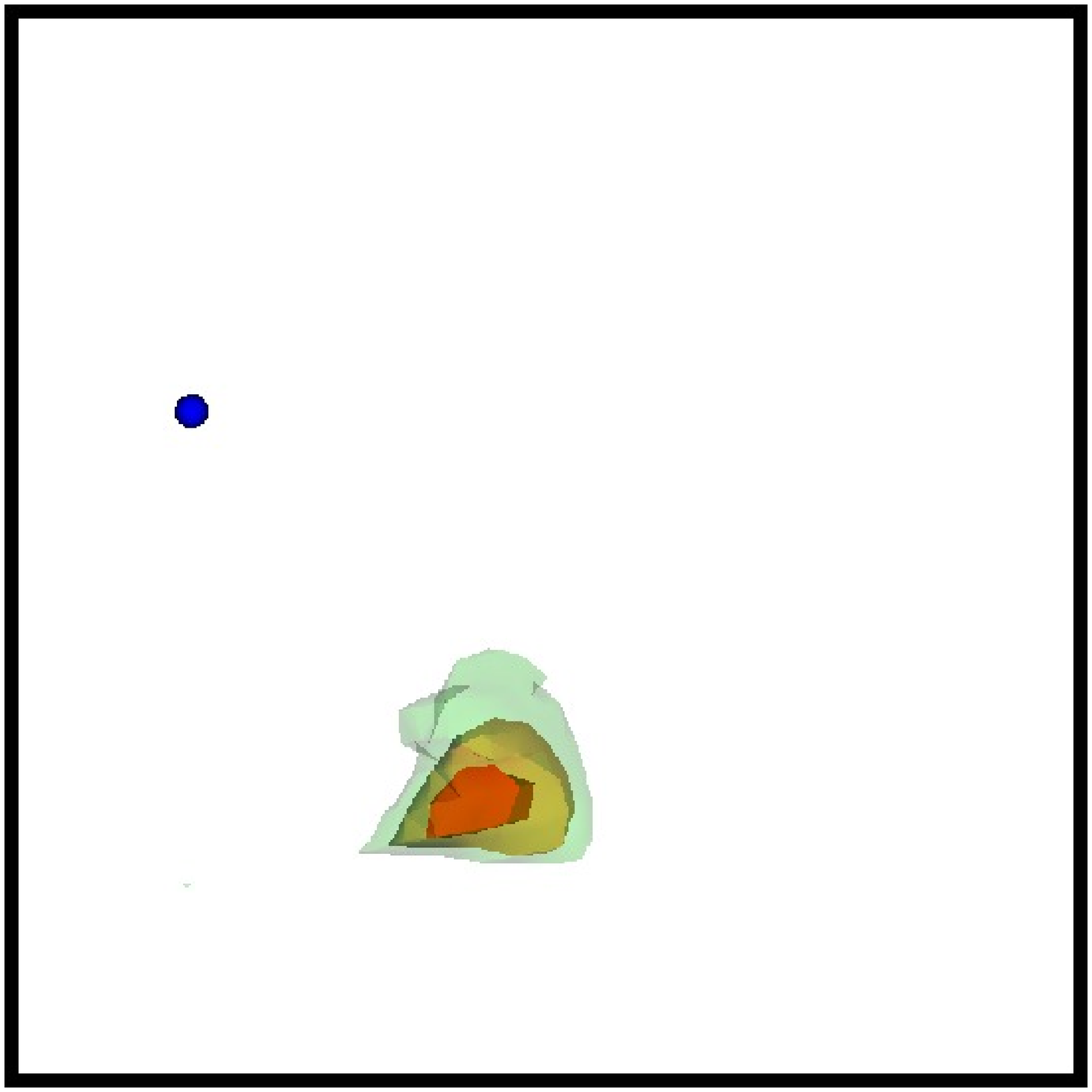}
\includegraphics[width=1.5in]{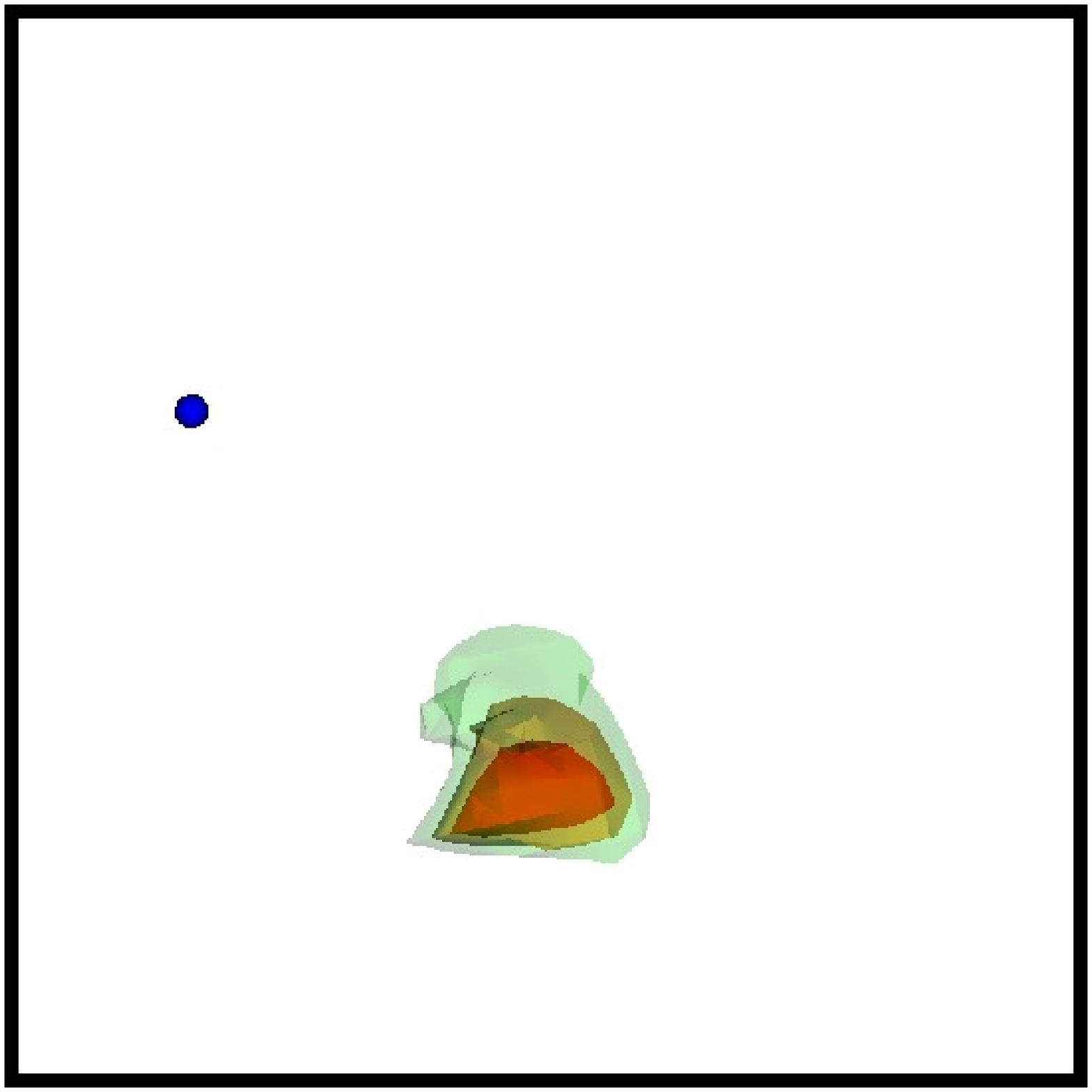}
\includegraphics[width=1.5in]{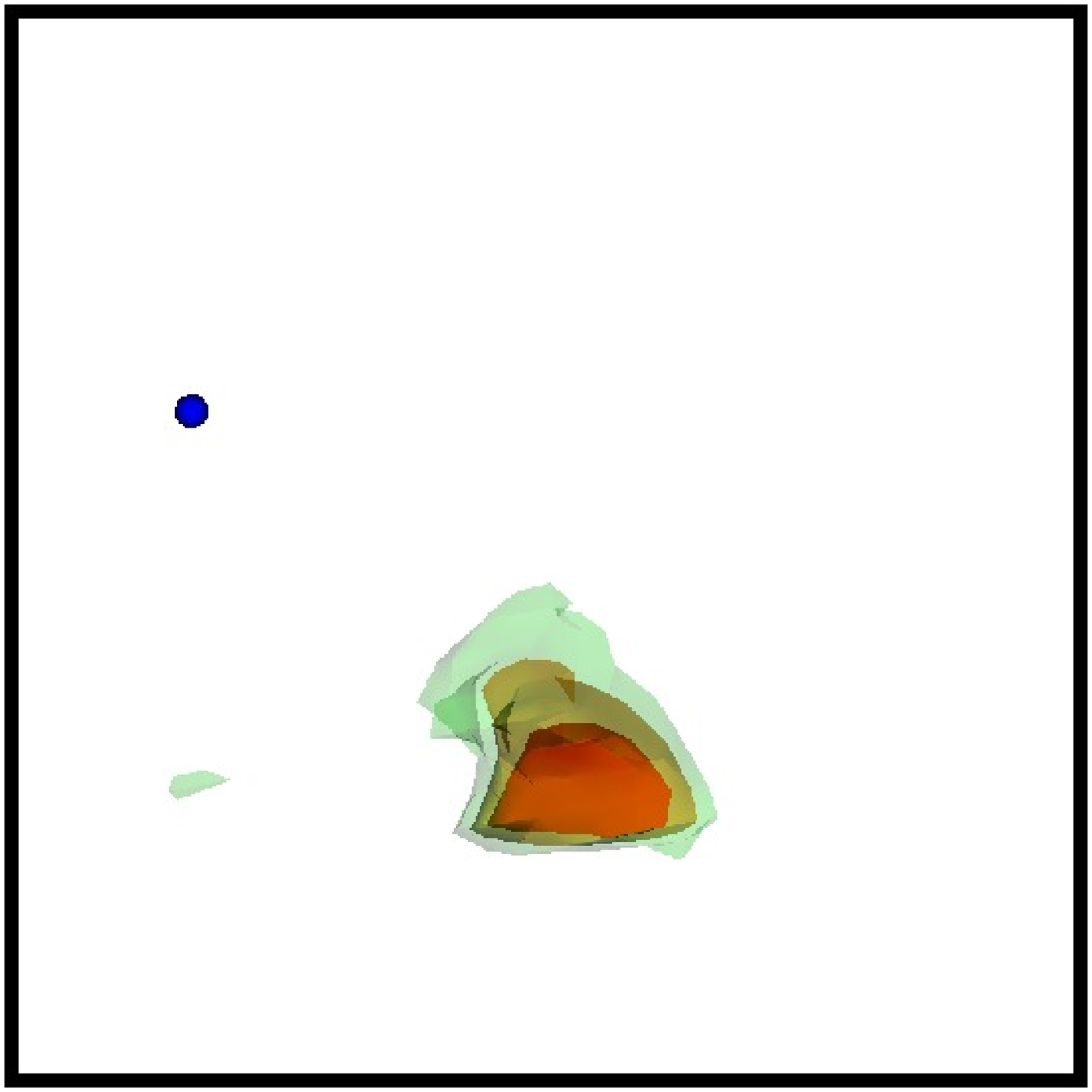}\\
\includegraphics[width=1.5in]{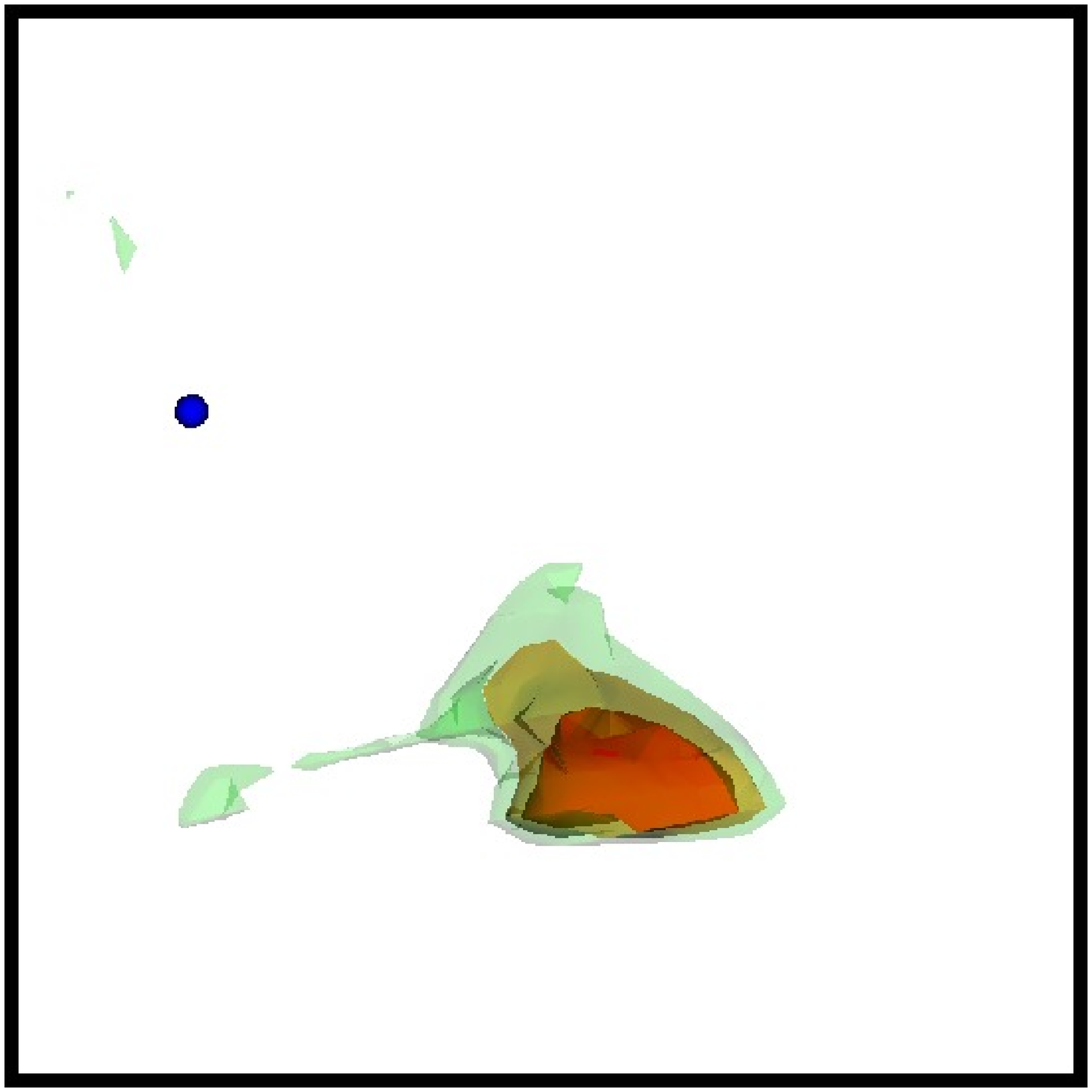}
\includegraphics[width=1.5in]{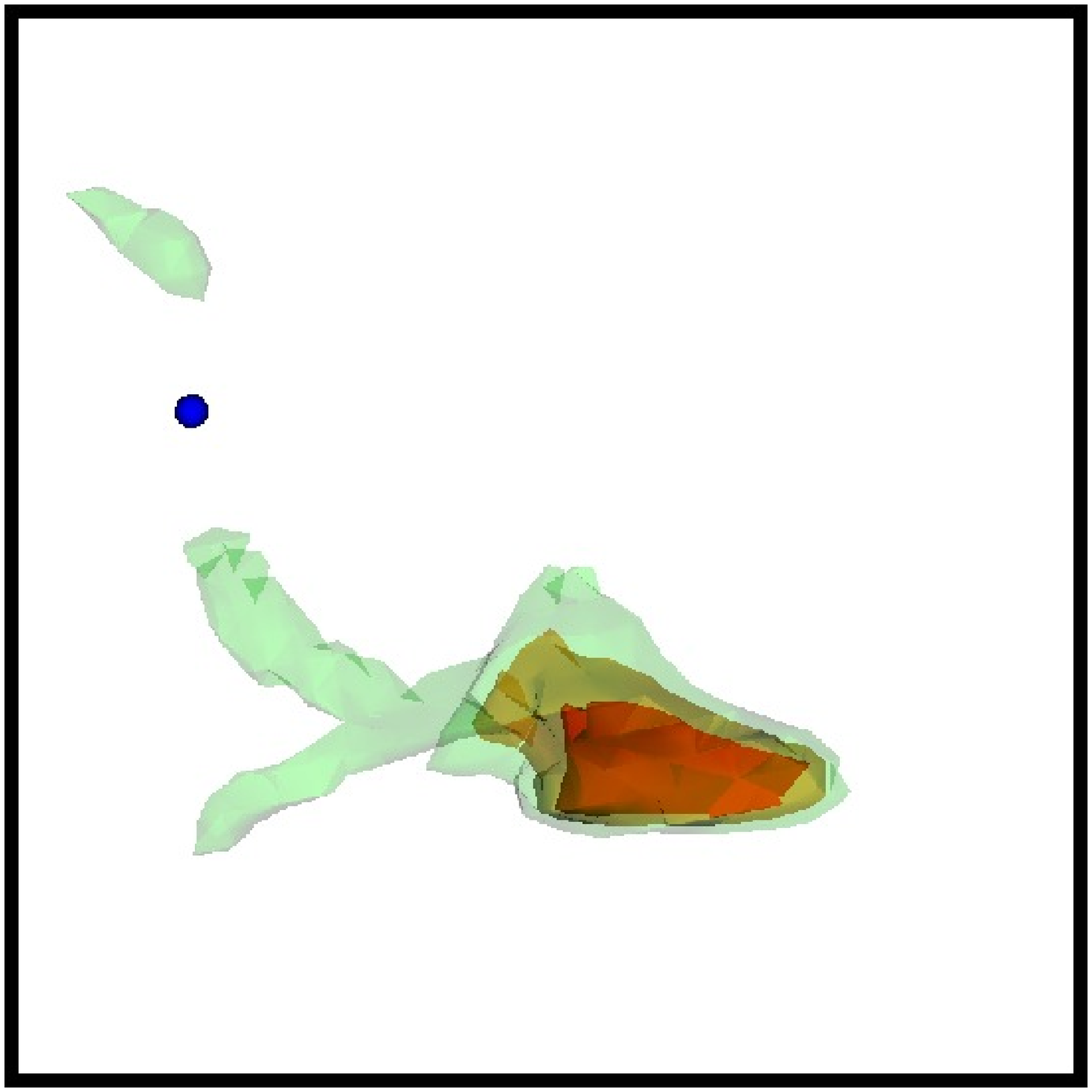}
\includegraphics[width=1.5in]{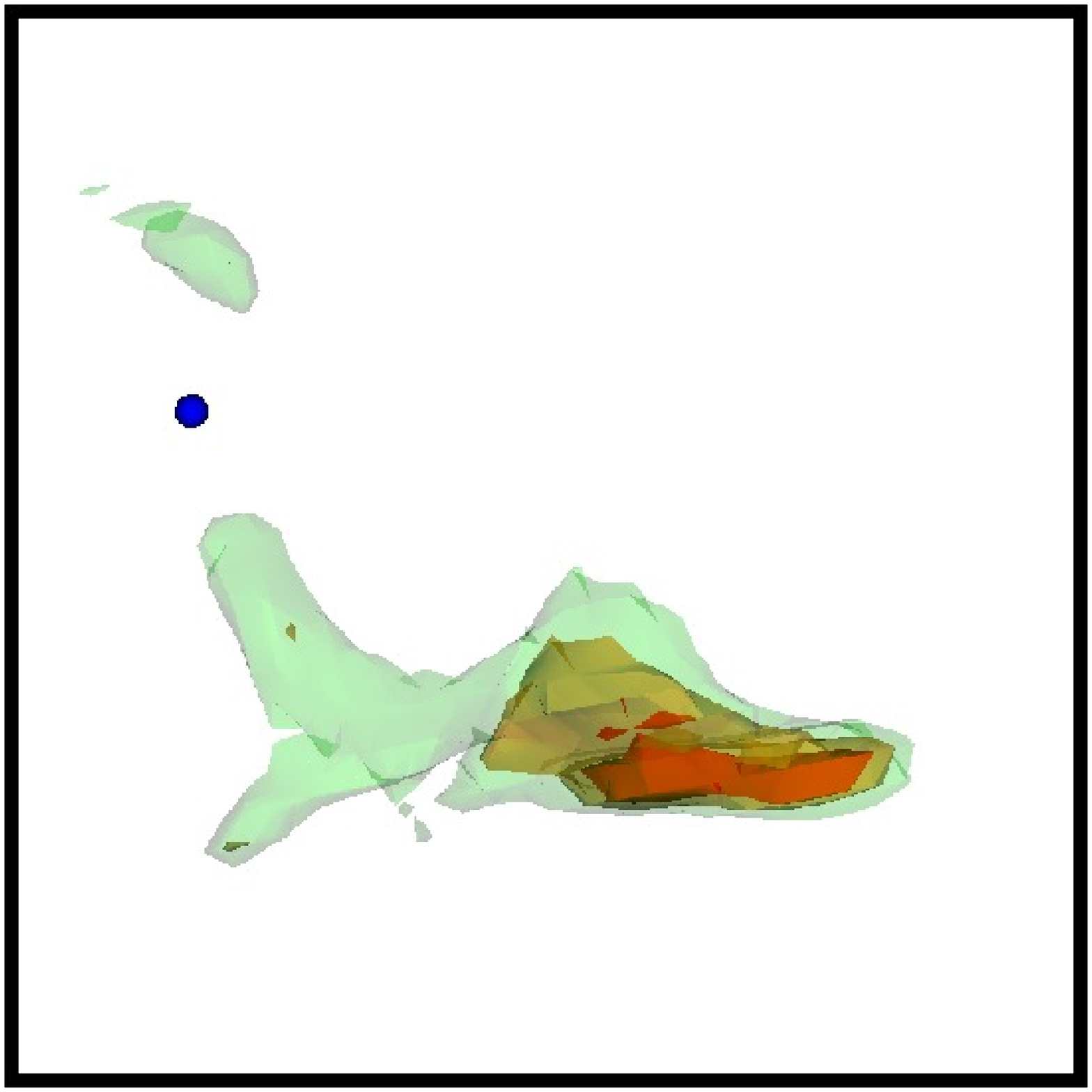}
\includegraphics[width=1.5in]{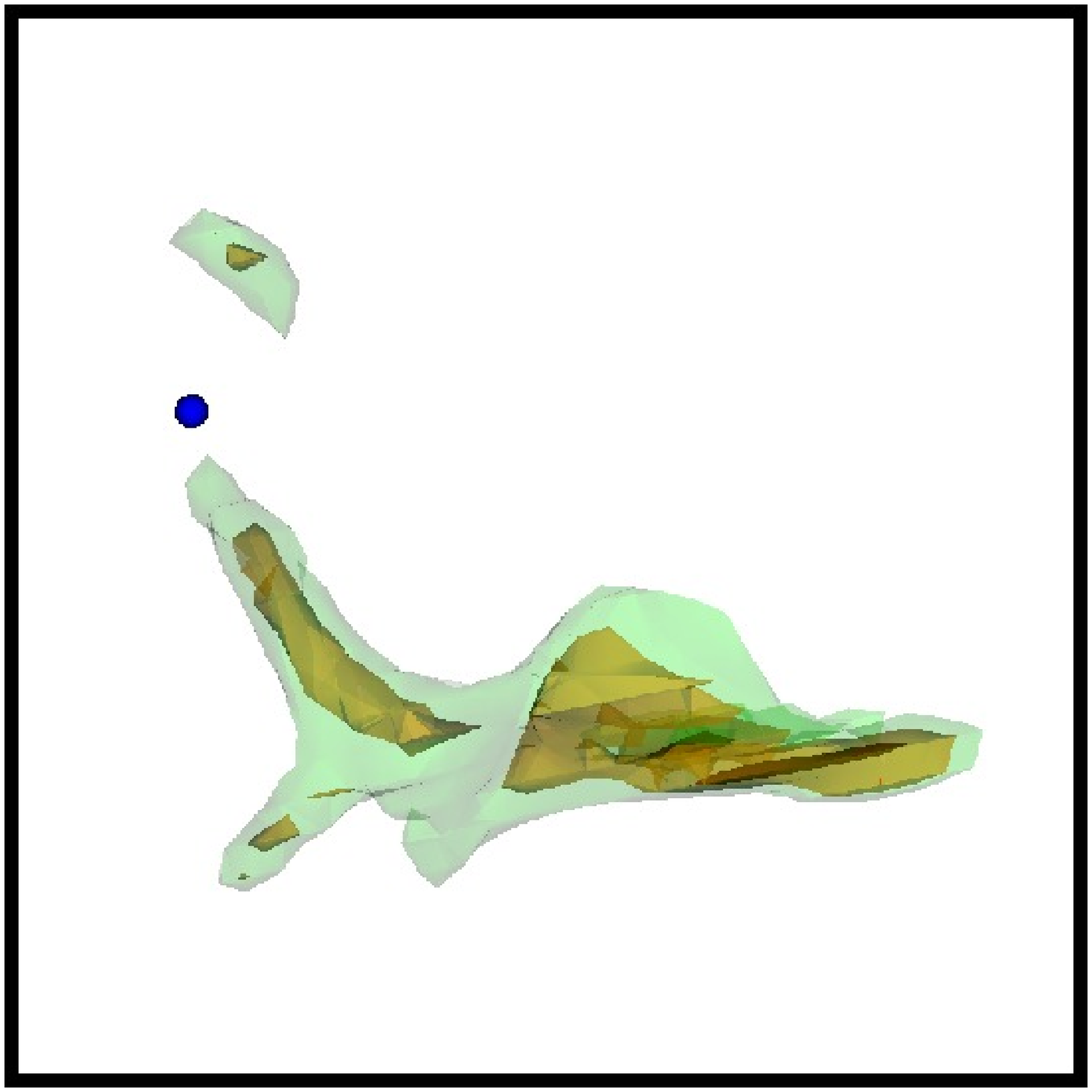}
\caption{\label{fig:hotspot_3} Temperature contours from $t=10559.2$~s to $t=10560.6$~s 
(corresponding to the black dotted time range in Figure \ref{fig:temp_1152}) spaced at 
$0.2$~s time intervals.  The contours are surfaces indicating where $T=7.48\times 10^8$~K 
(green), $T=7.54\times 10^8$~K (yellow), and $T=7.6\times 10^8$~K (orange).  The blue dot is
at the center of the star, and has a diameter of 4.34~km, which corresponds to the grid
cell width for this simulation.}
\end{center}
\end{figure}

Figure \ref{fig:hotspot_1} shows contours of temperature within the green dotted time
region from Figure \ref{fig:temp_1152}.  We note that for all subsequent temperature
visualizations, the blue dot represents the center of the star, and has a diameter
of 4.34~km, corresponding to the resolution of the simulation.  Also, all
visualization frames are spaced at 0.2~s time intervals.  The main observation is that in
the frames where an orange contour exists, indicative of a hot spot, that there are no
other regions in the star with comparable temperature.  This implies that if this
hot spot were to run away, there would be only one ignition point.
Figure \ref{fig:hotspot_2} shows contours of temperature within the blue dotted 
time region.  We do see that in the frames where an orange contour exists, there are 
other hot spots in different regions of the star.  But as the hottest spot floats away 
and cools off, the temperature of the other hot spot does not increase.  Again,
this implies only a single ignition point.
Figure \ref{fig:hotspot_3} shows contours of temperature within the black dotted 
time region.  This visualization is more like the green dotted time region, in that
there are no regions of the star with a peak temperature comparable to the main
hot spot, implying there would be only one ignition point if this hot spot were
to run away.

\begin{figure}
\begin{center}
\includegraphics[width=1.5in]{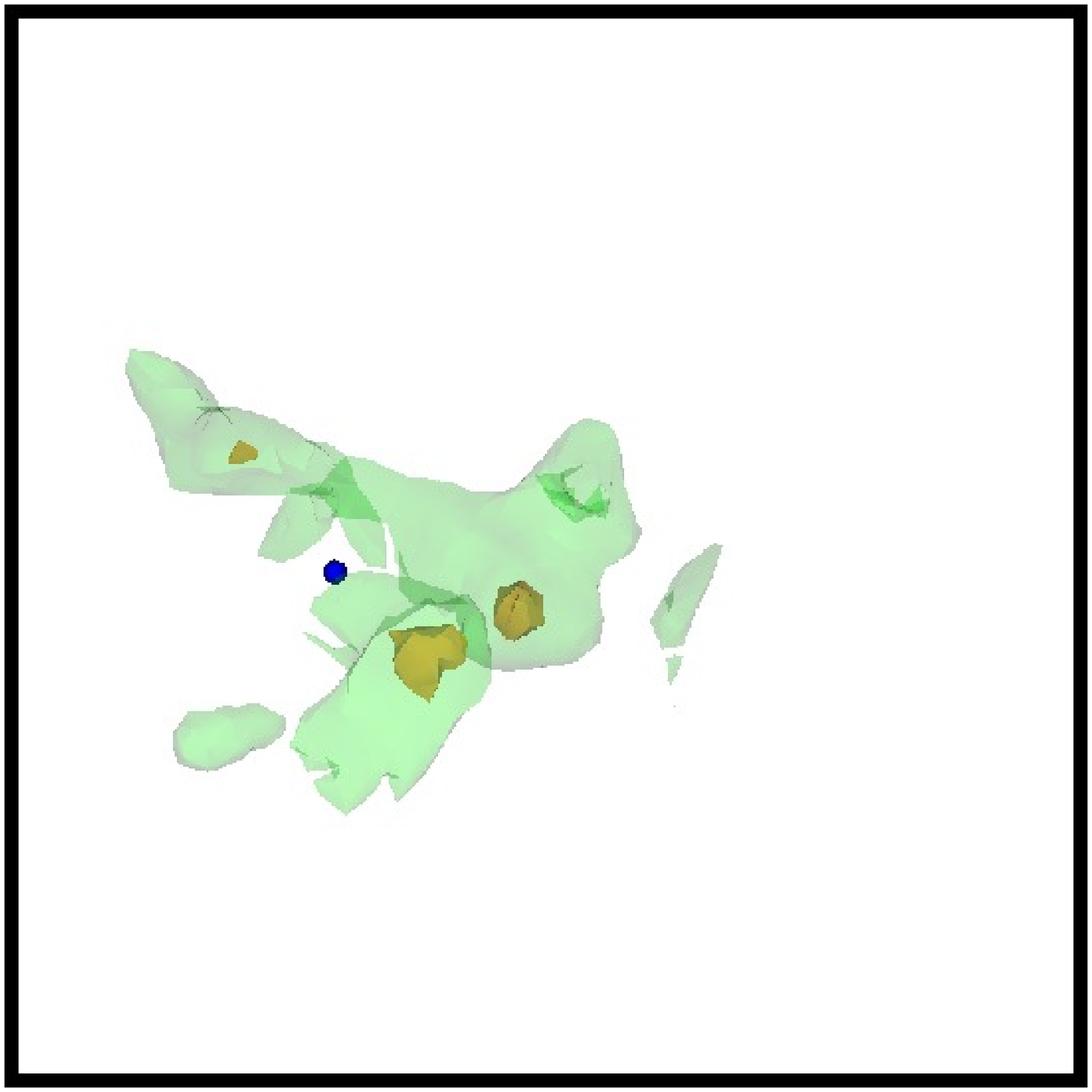}
\includegraphics[width=1.5in]{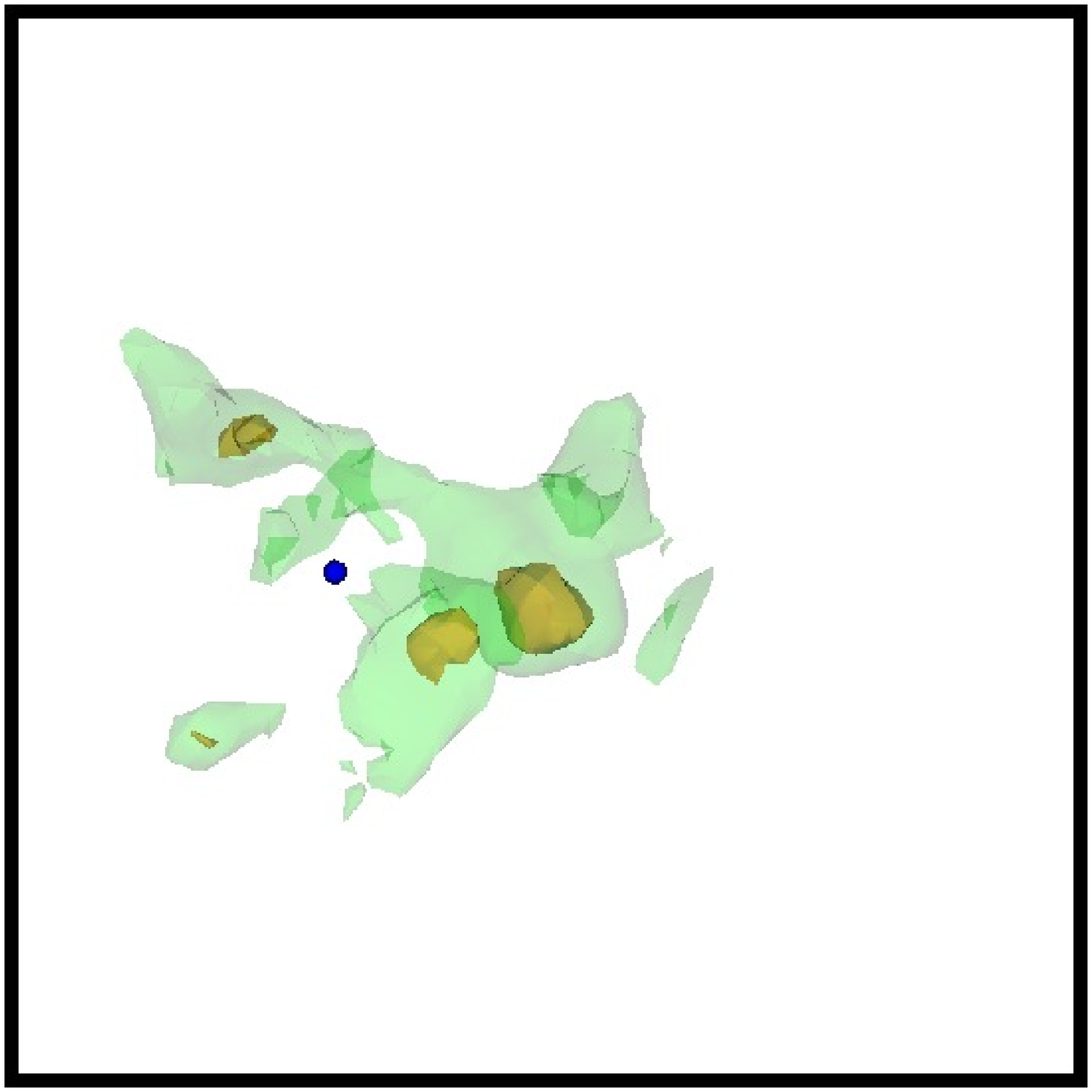}
\includegraphics[width=1.5in]{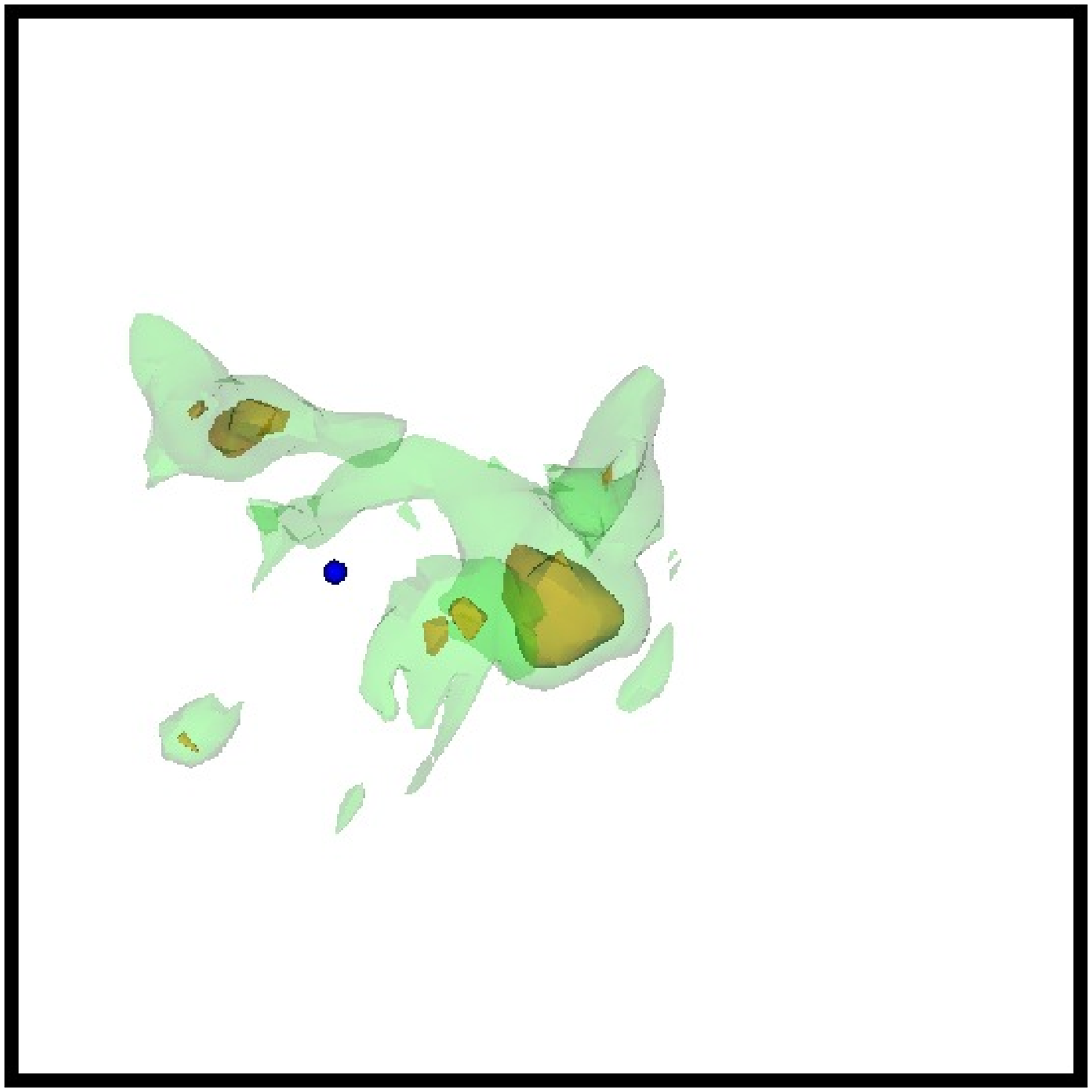}
\includegraphics[width=1.5in]{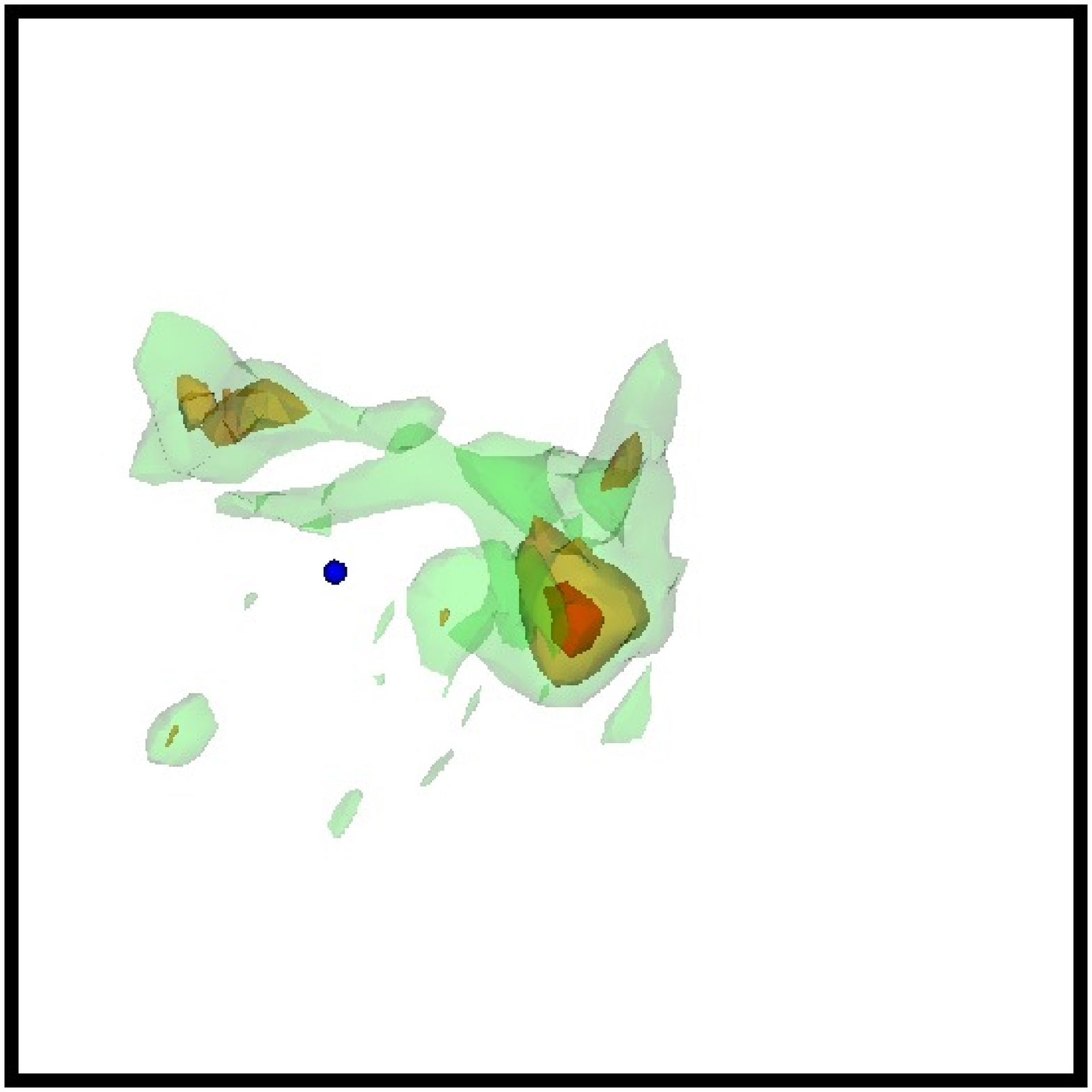}\\
\includegraphics[width=1.5in]{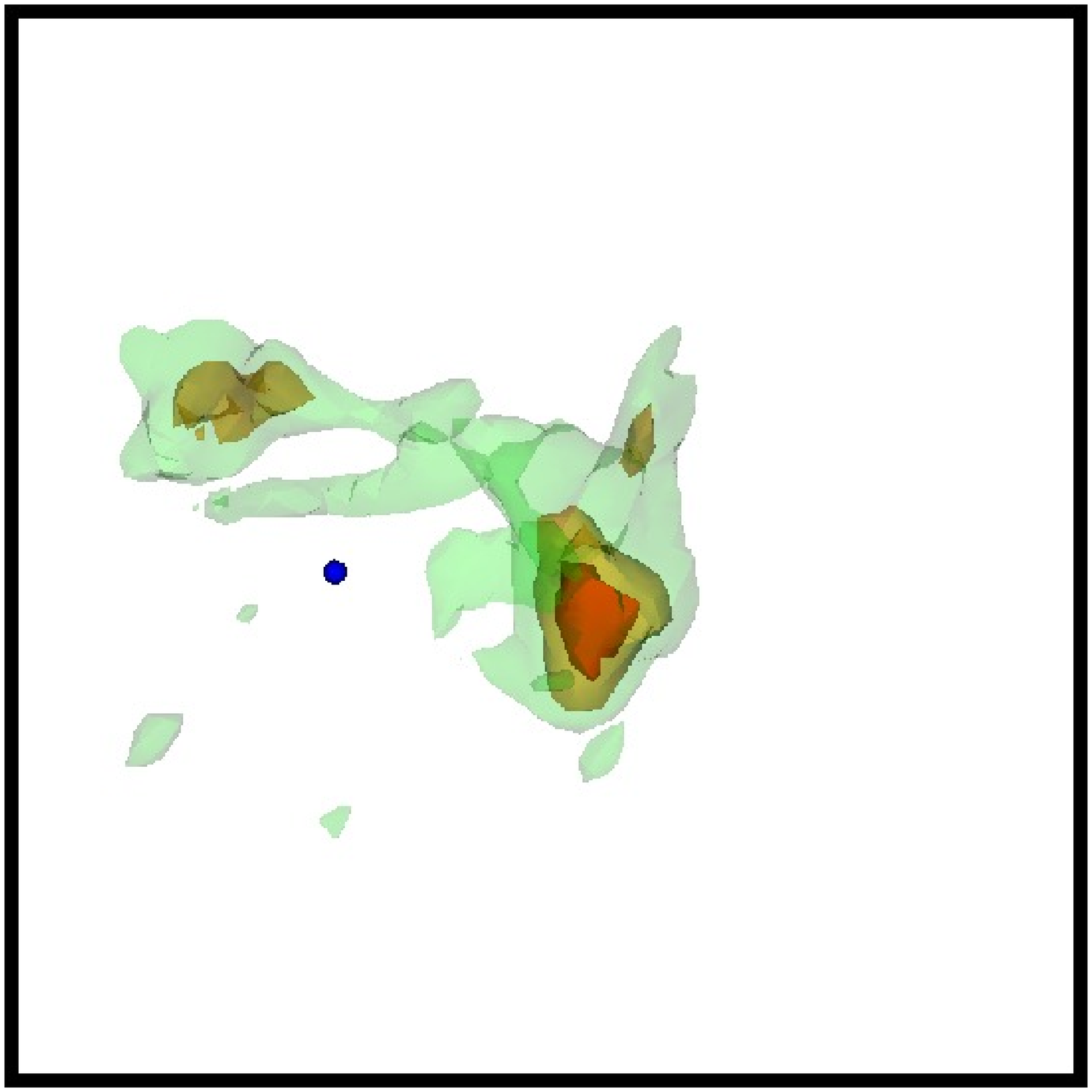}
\includegraphics[width=1.5in]{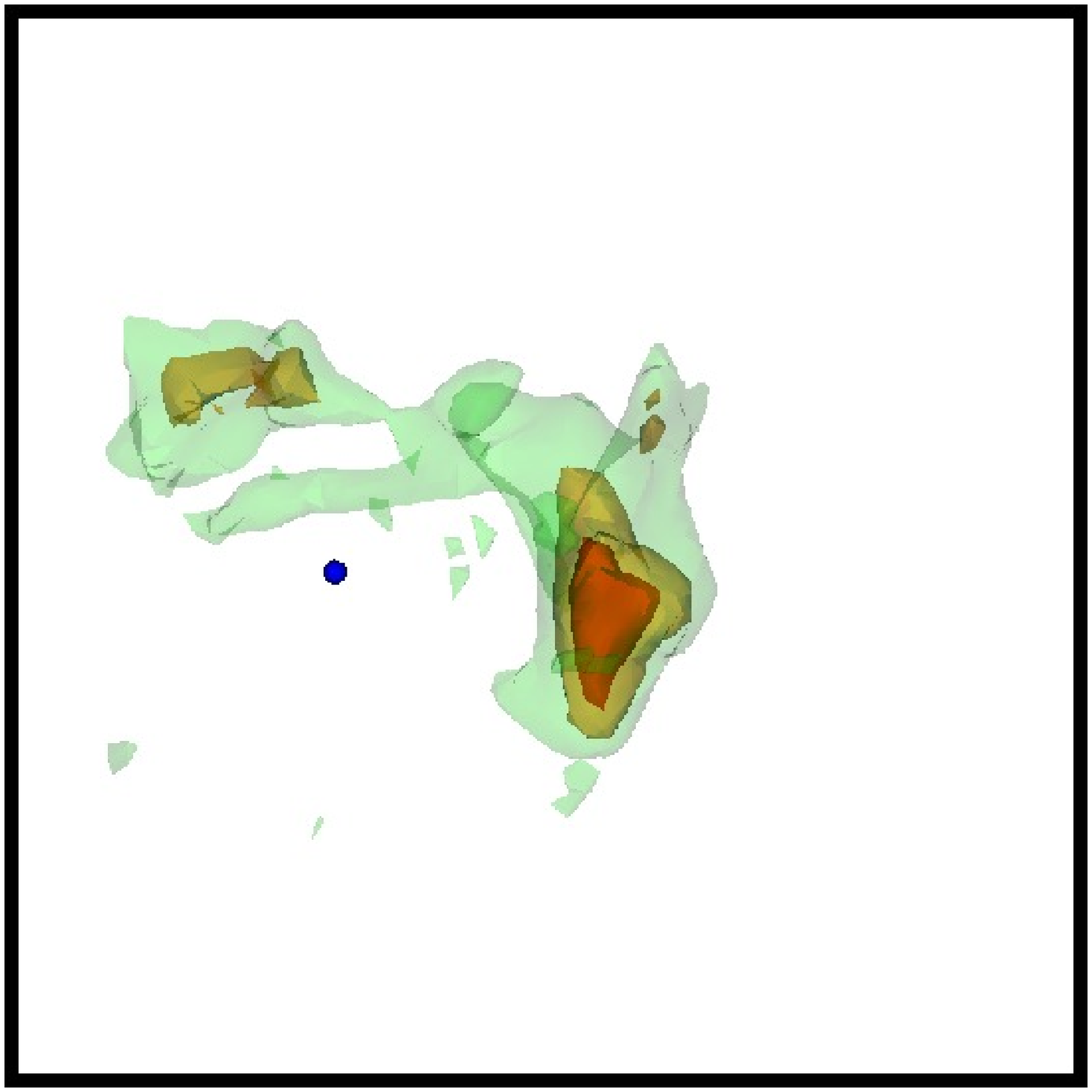}
\includegraphics[width=1.5in]{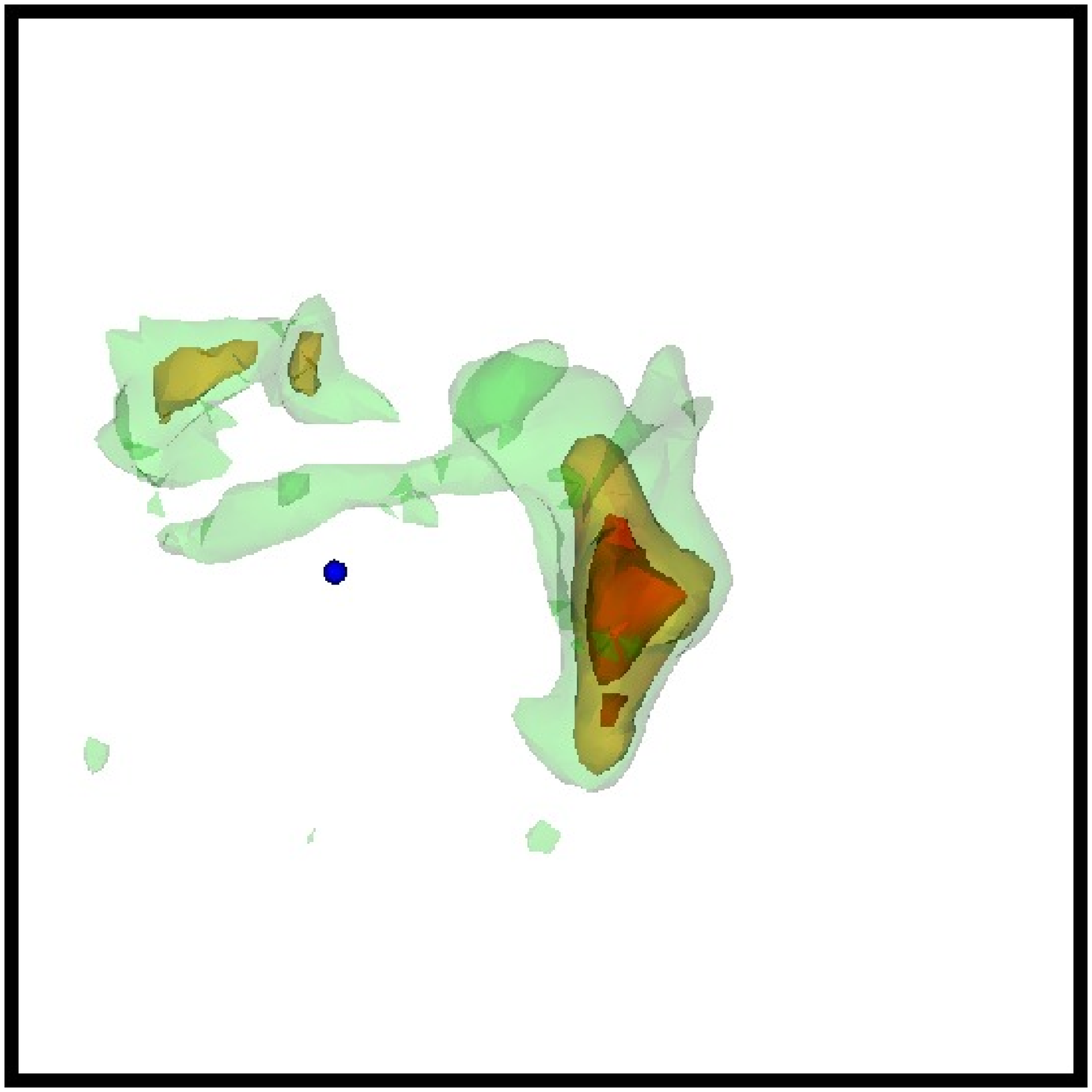}
\includegraphics[width=1.5in]{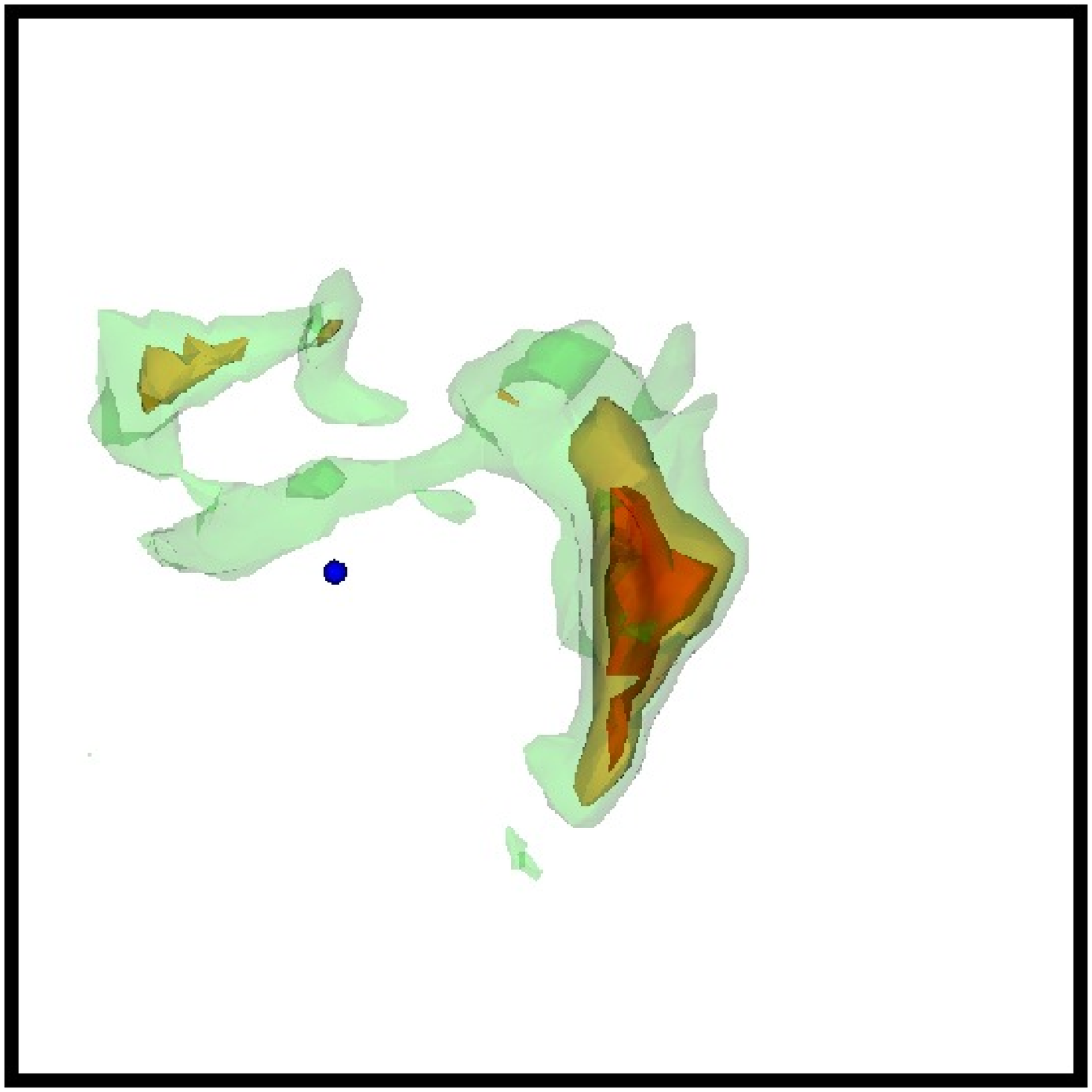}\\
\includegraphics[width=1.5in]{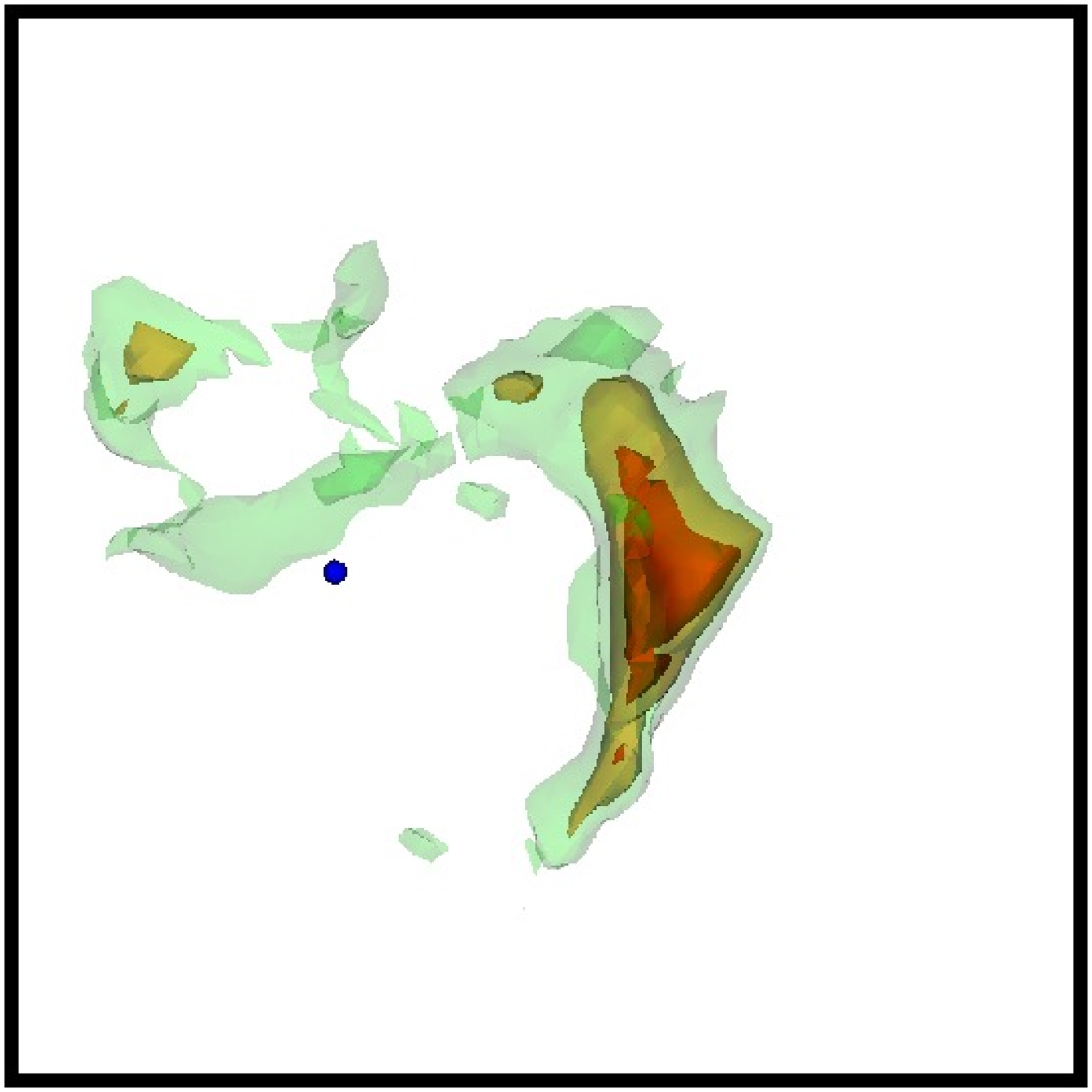}
\includegraphics[width=1.5in]{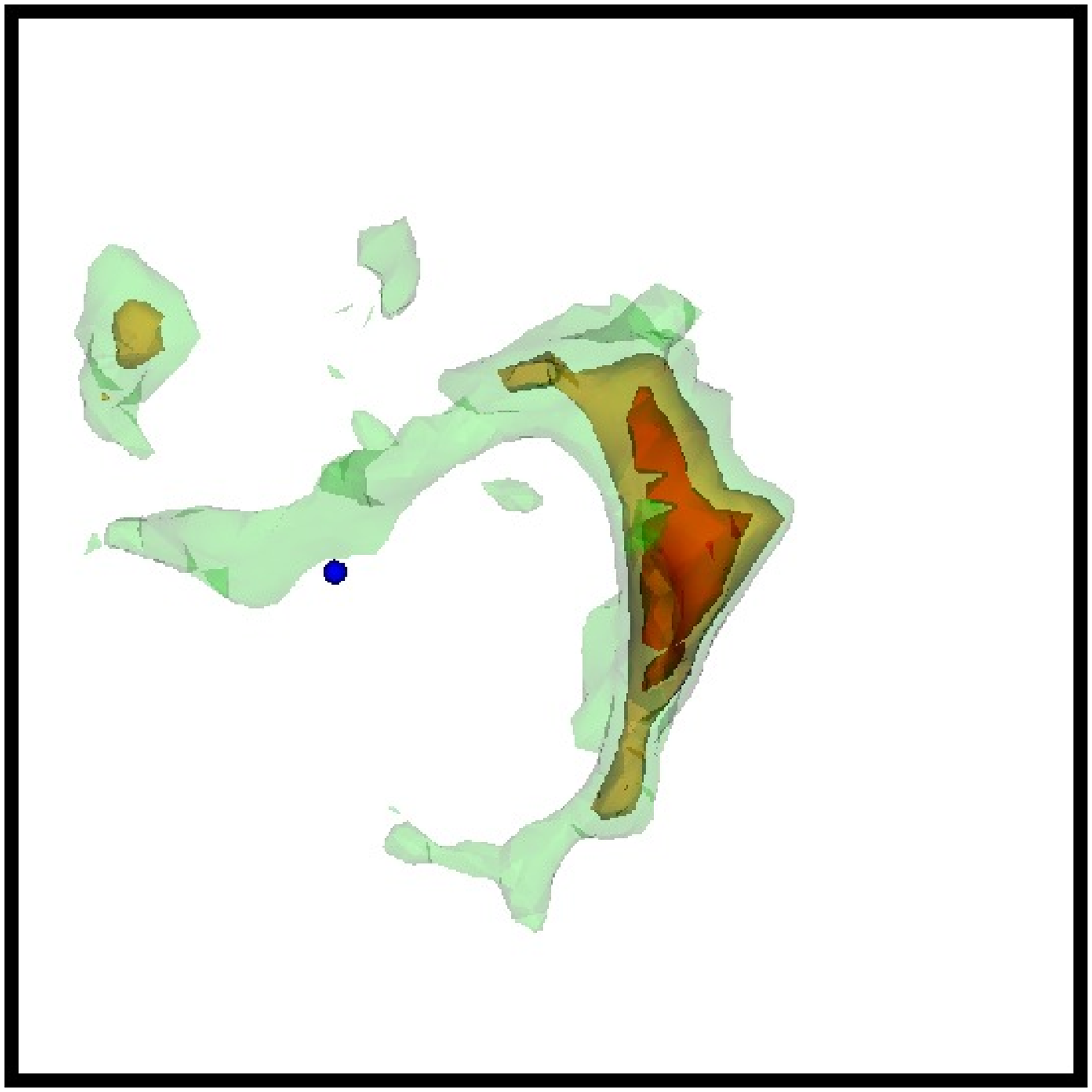}
\includegraphics[width=1.5in]{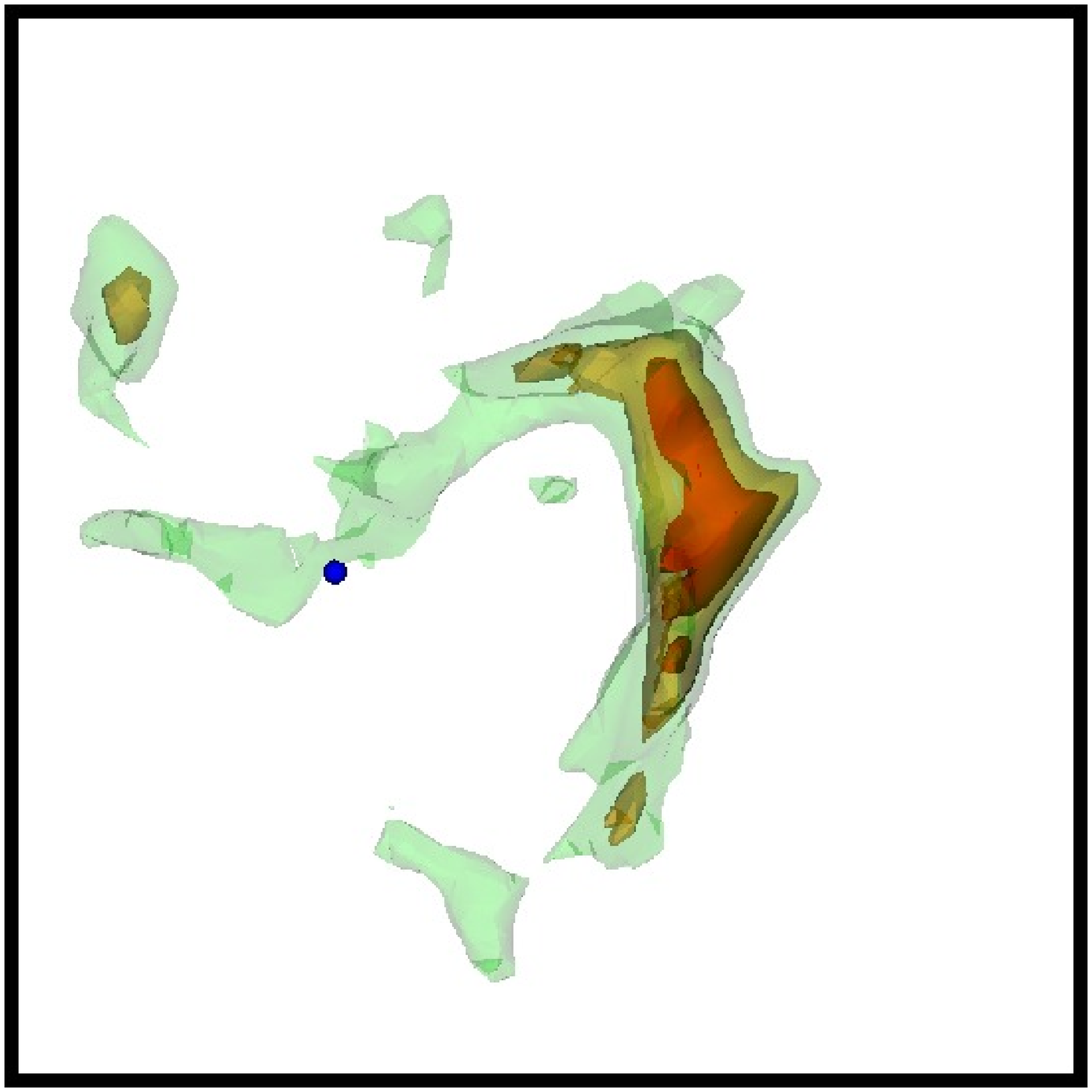}
\includegraphics[width=1.5in]{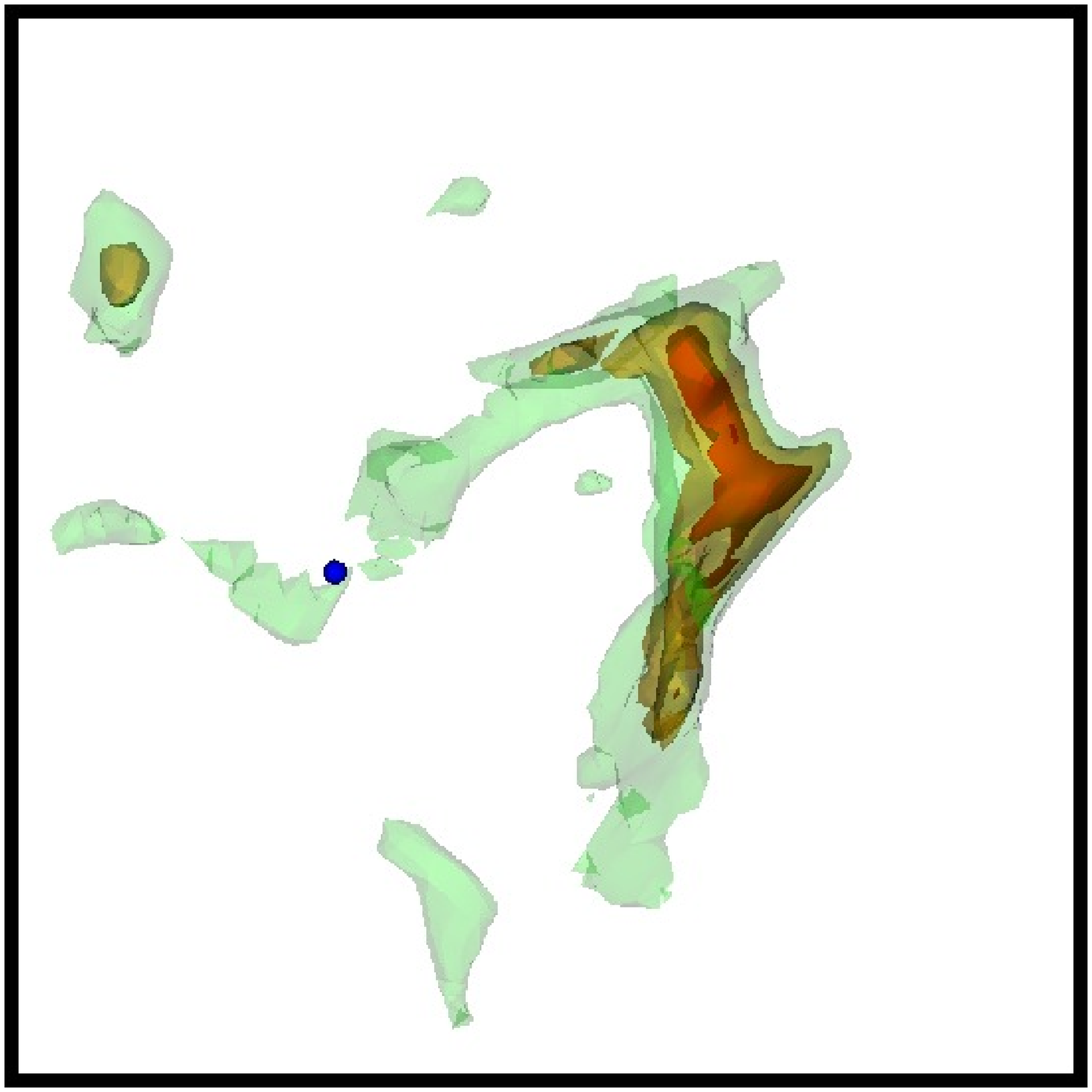}\\
\includegraphics[width=1.5in]{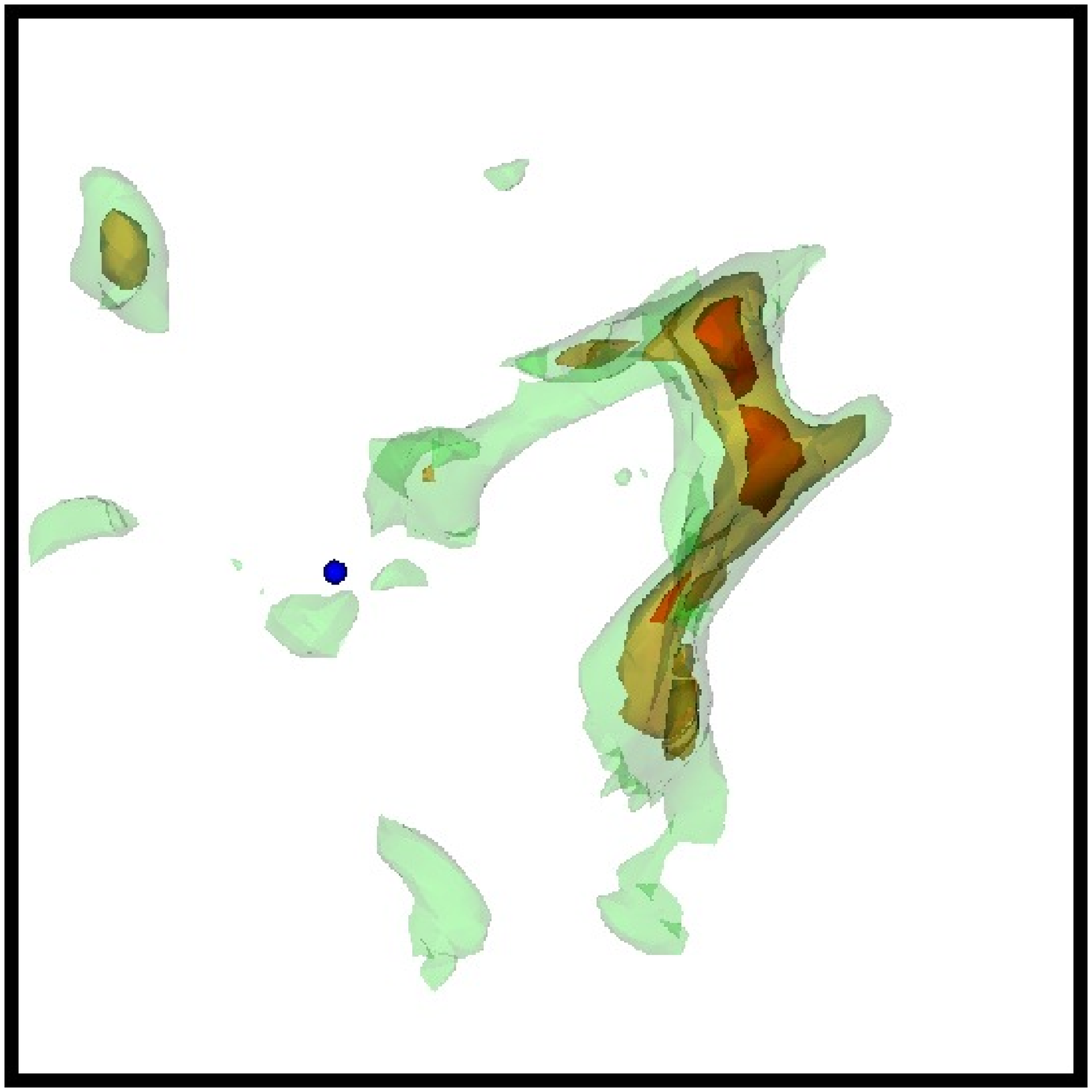}
\includegraphics[width=1.5in]{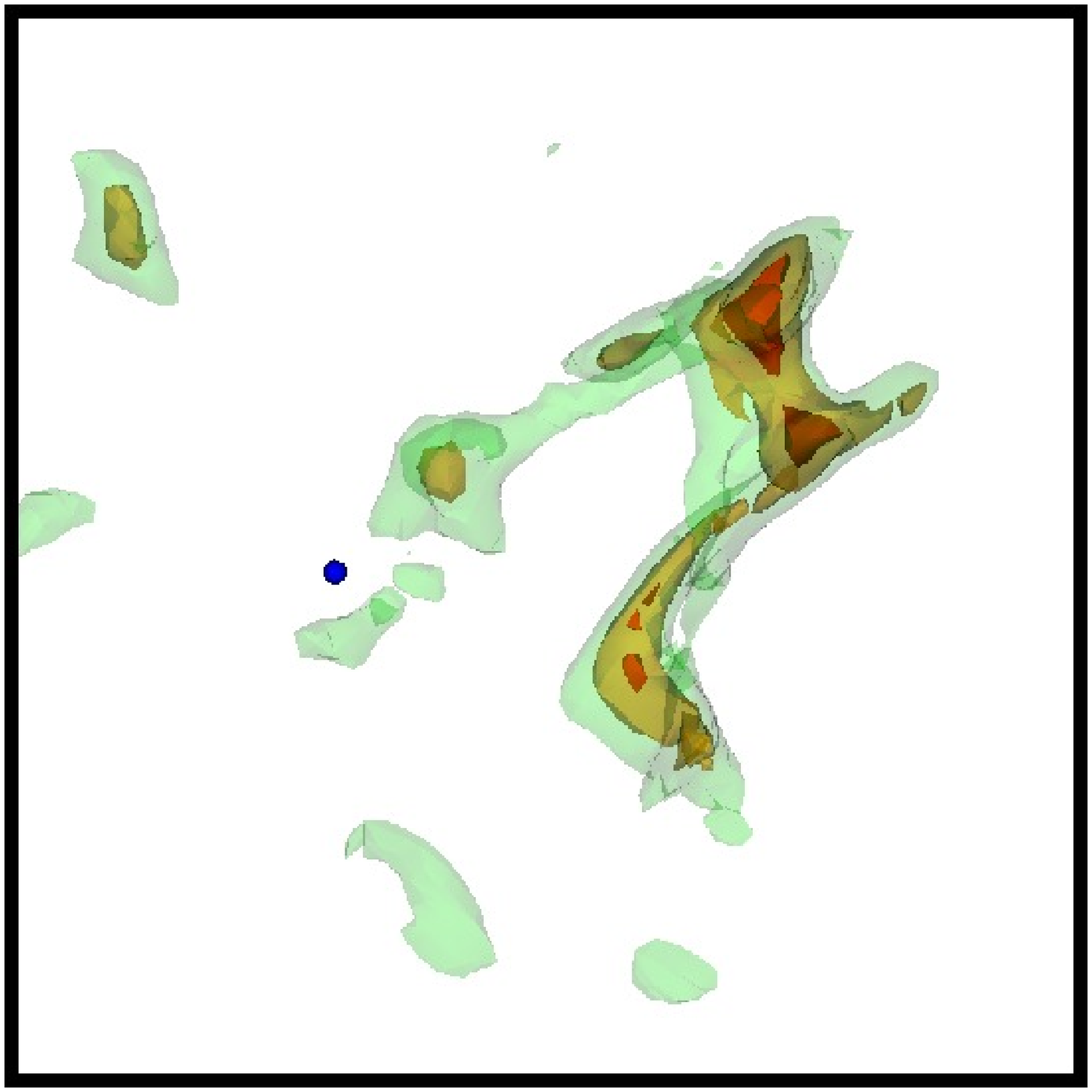}
\includegraphics[width=1.5in]{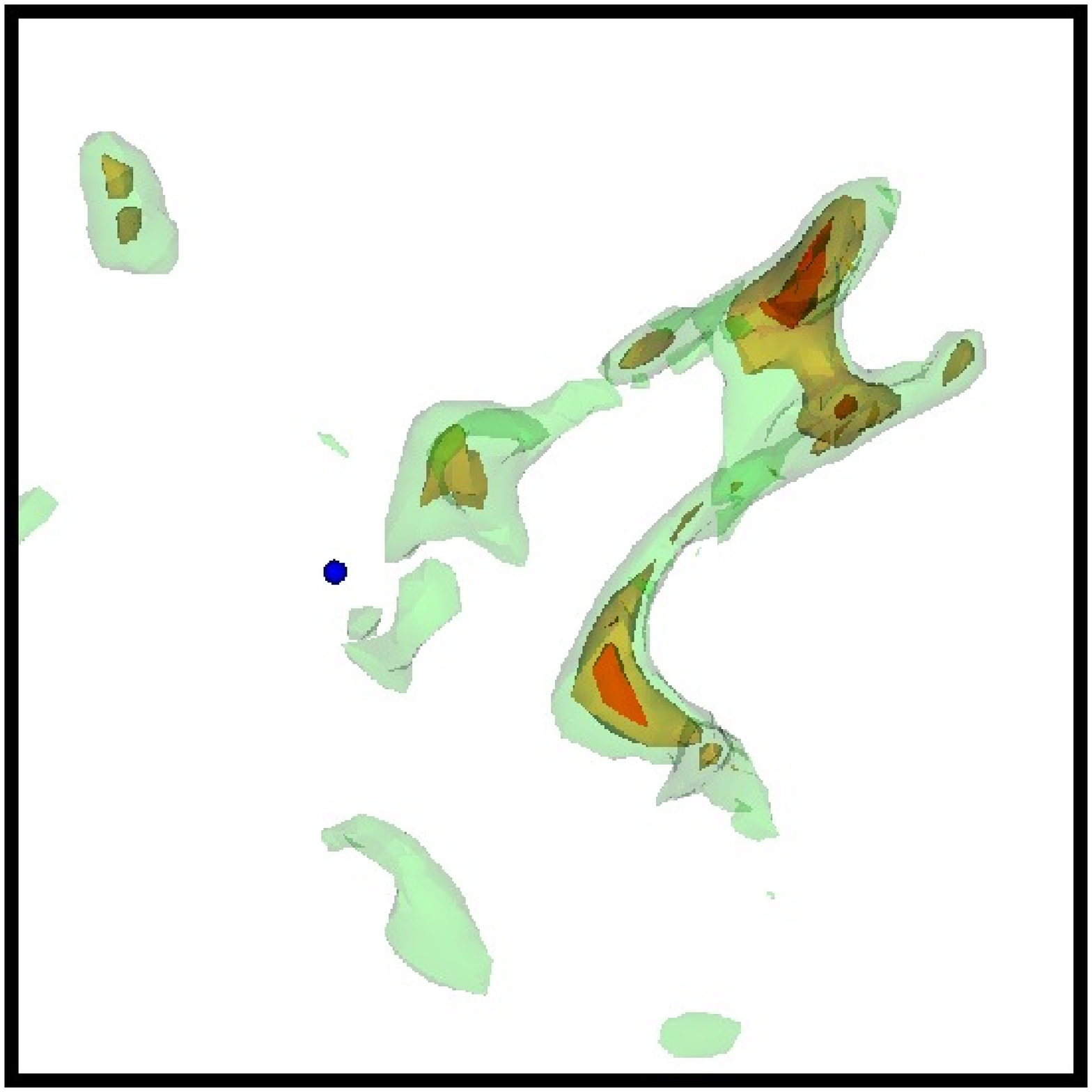}
\includegraphics[width=1.5in]{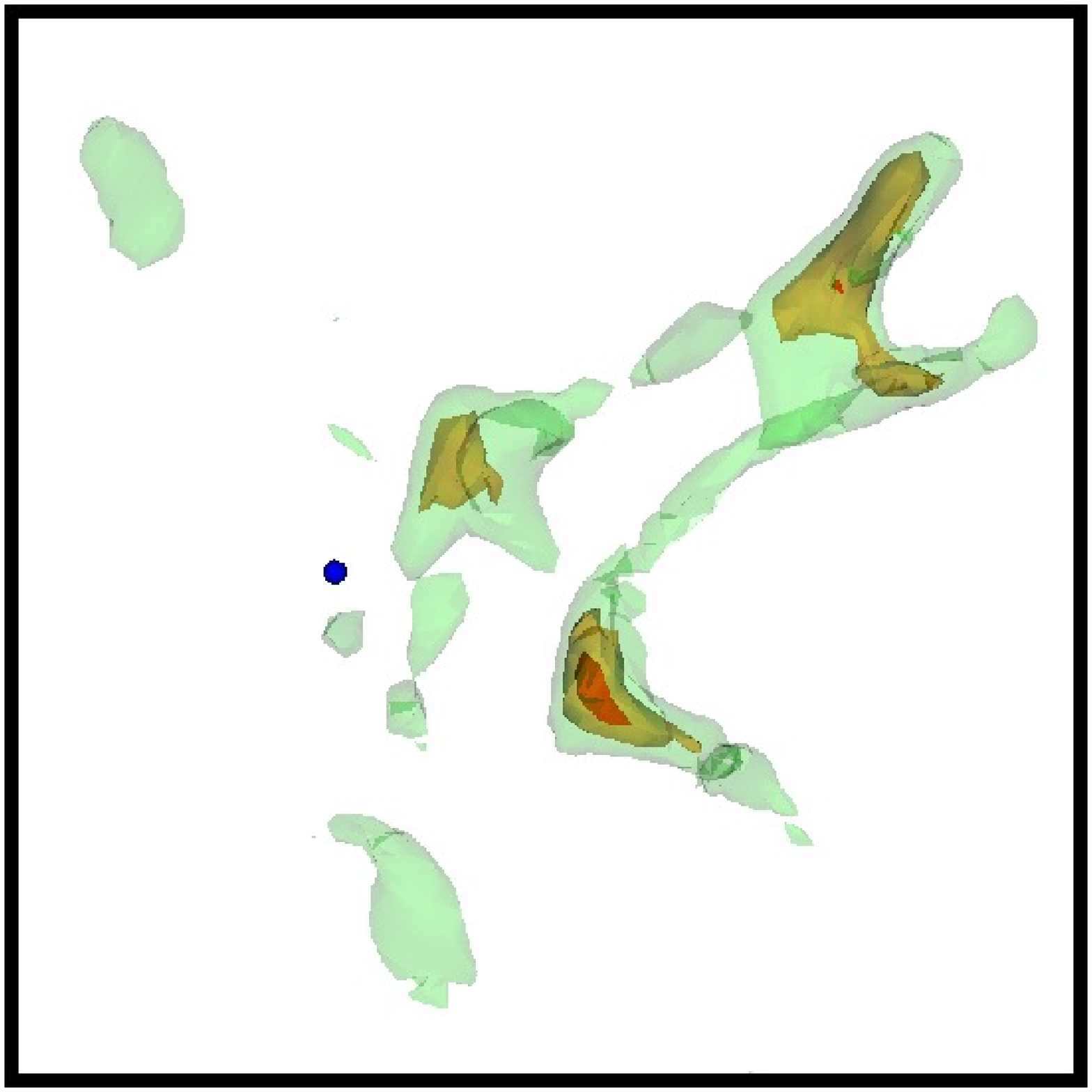}
\caption{\label{fig:hotspot_4} Temperature contours from $t=10562.0$~s to $t=10565.0$~s 
(corresponding to the pink dotted time range in Figure \ref{fig:temp_1152}) spaced at 
$0.2$~s time intervals.  The contours are surfaces indicating where $T=7.5\times 10^8$~K 
(green), $T=7.7\times 10^8$~K (yellow), and $T=7.9\times 10^8$~K (orange).  The blue dot is
at the center of the star, and has a diameter of 4.34~km, which corresponds to the grid
cell width for this simulation.}
\end{center}
\end{figure}

Next, we perform another simulation, beginning at the point of ignition, in which we have
disabled burning for all cells where $T > 8\times 10^8$~K.  This will give us a 
picture of the dynamics of nearby hot bubbles that did not initially ignite.  
The idea here is to let the initial ignition point float away and see if any other 
hot spots reach the ignition temperature soon afterwards.  The peak temperature
in this new simulation is given by the black solid line in Figure \ref{fig:temp_1152}.
The visualization of the temperature field within the pink dotted time
region from Figure \ref{fig:temp_1152} is shown in Figure \ref{fig:hotspot_4}.
We see that the hot bubble containing the ignition cell floats away from the center
of the star and cools off (because the burning is disabled) as it breaks up.  More importantly, none of the other 
hot bubbles not connected to the ignition cell increase in temperature to the point
of ignition.  Altogether, our analysis of the last few hot spots does not seem to
support multiple ignition points, implying that this scenario is much less likely
than a single ignition point.

A caveat to this analysis is that our resolution is still several
kilometers.  It is possible that if one could increase the resolution far
beyond what is possible today, even with AMR, that many smaller
hot spots could exist and the dynamics would be different.

\subsection{Convective Flow Pattern}
\begin{figure}
\begin{center}
\includegraphics[width=3in]{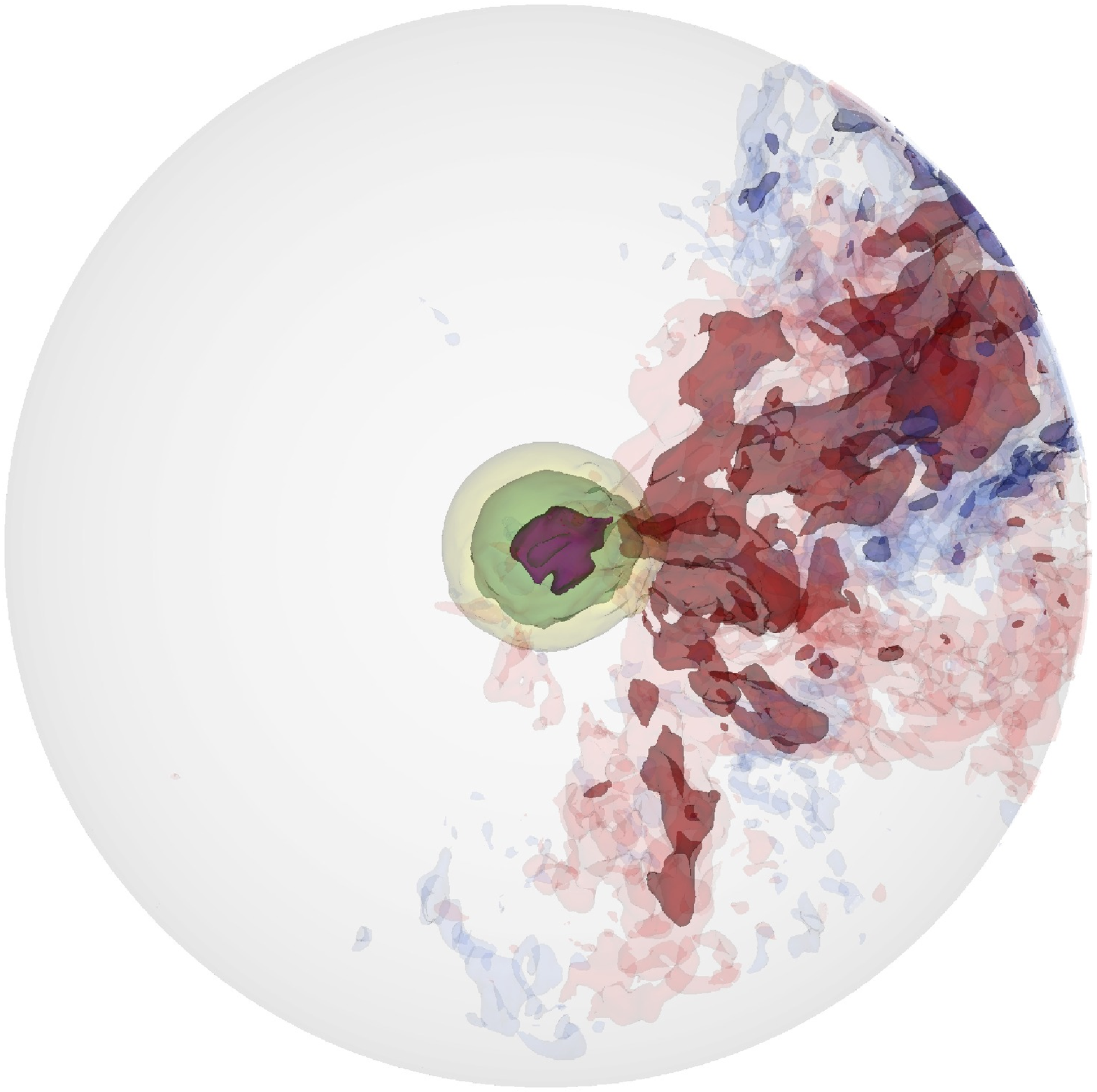}
\includegraphics[width=3in]{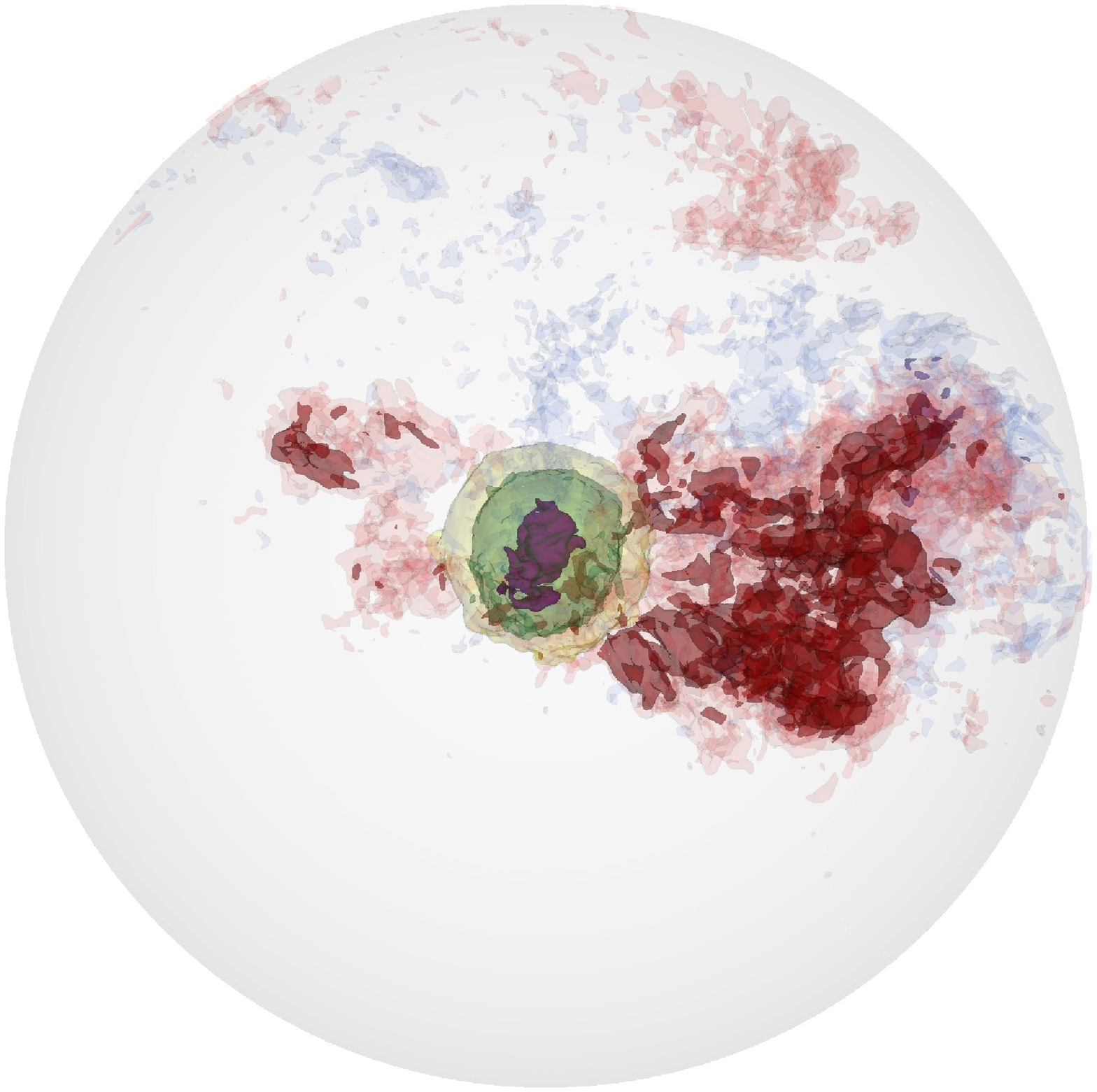}\\
\includegraphics[width=3in]{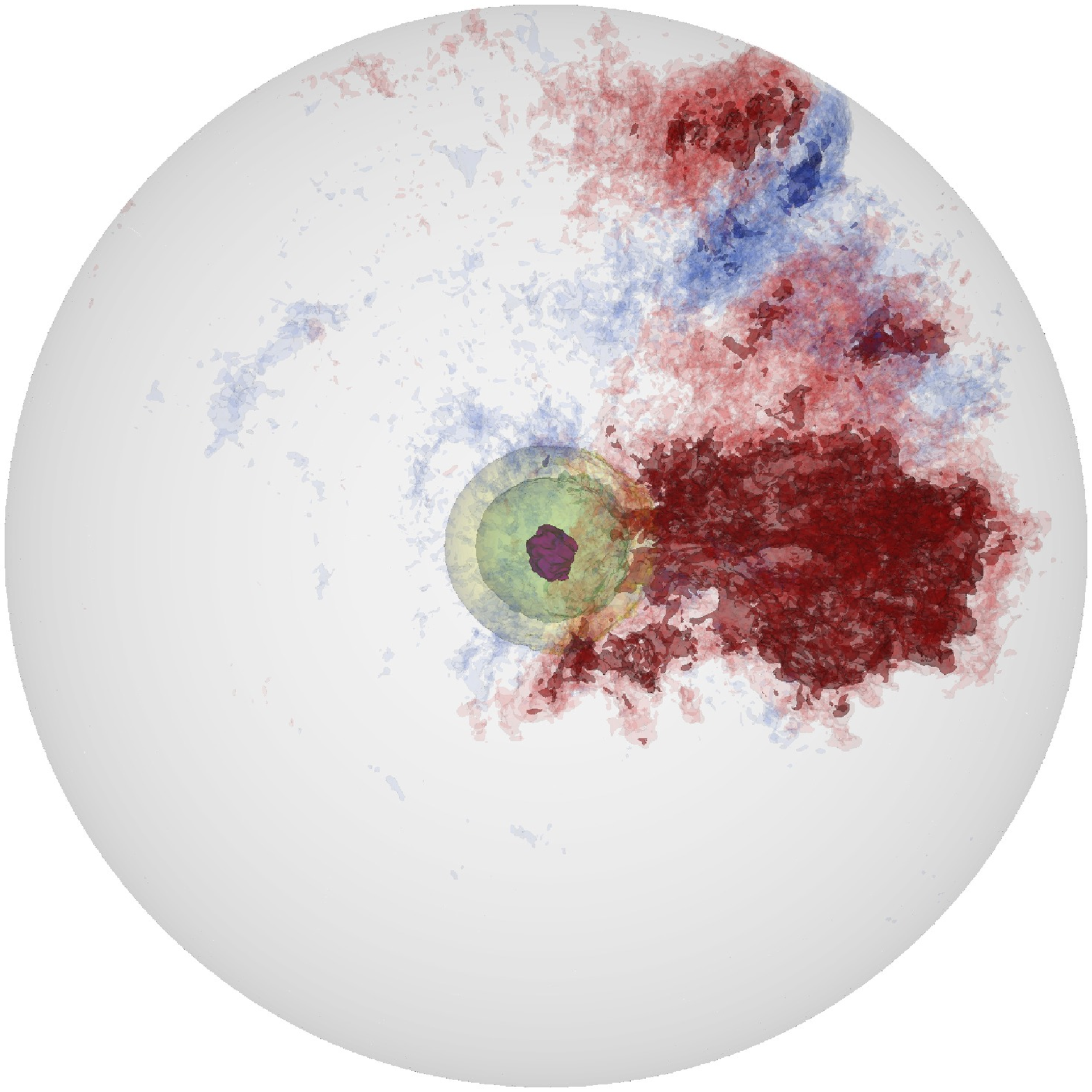}
\caption{\label{fig:wd_convect} Contours of nuclear energy
  generation rate (yellow to green to purple, corresponding to
  $4\times 10^{12}$, $1.27\times 10^{13}$, and 
  $4\times 10^{13}$~erg~g$^{-1}$~s$^{-1}$) and radial
  velocity (red is outflow, corresponding to $3\times 10^6$ and
  $6\times 10^6$~cm~s$^{-1}$; blue is inflow, corresponding to $-3\times
  10^6$ and $-6\times 10^6$~cm~s$^{-1}$) for the (clockwise, from top-left)
  8.68~km, 4.34~km, and 2.17~km simulations at
  t=10380~s.  Only the inner $r = 1000$~km are shown.}
\end{center}
\end{figure}

\begin{figure}
\begin{center}
\includegraphics[width=2.1in]{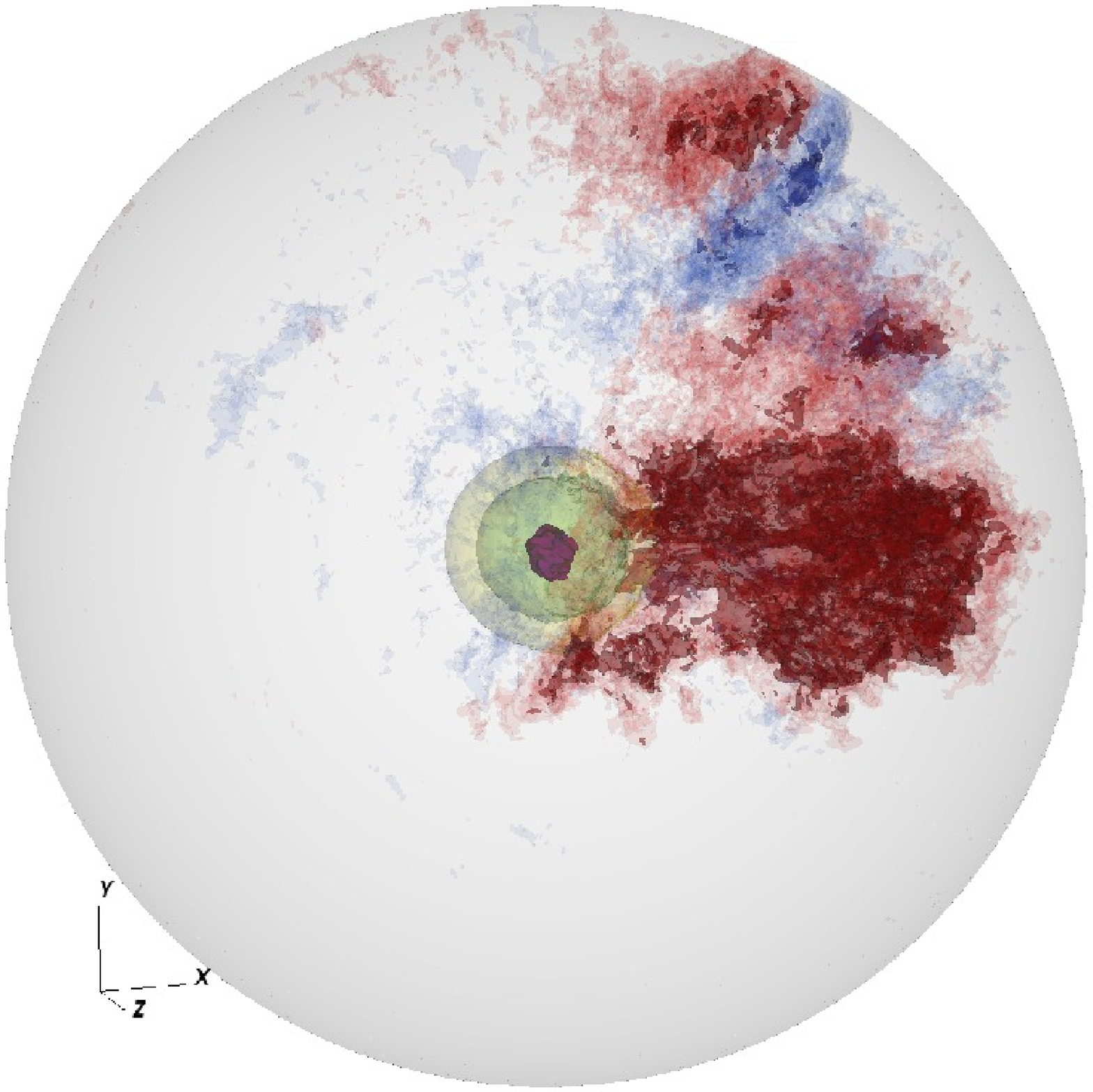}
\includegraphics[width=2.1in]{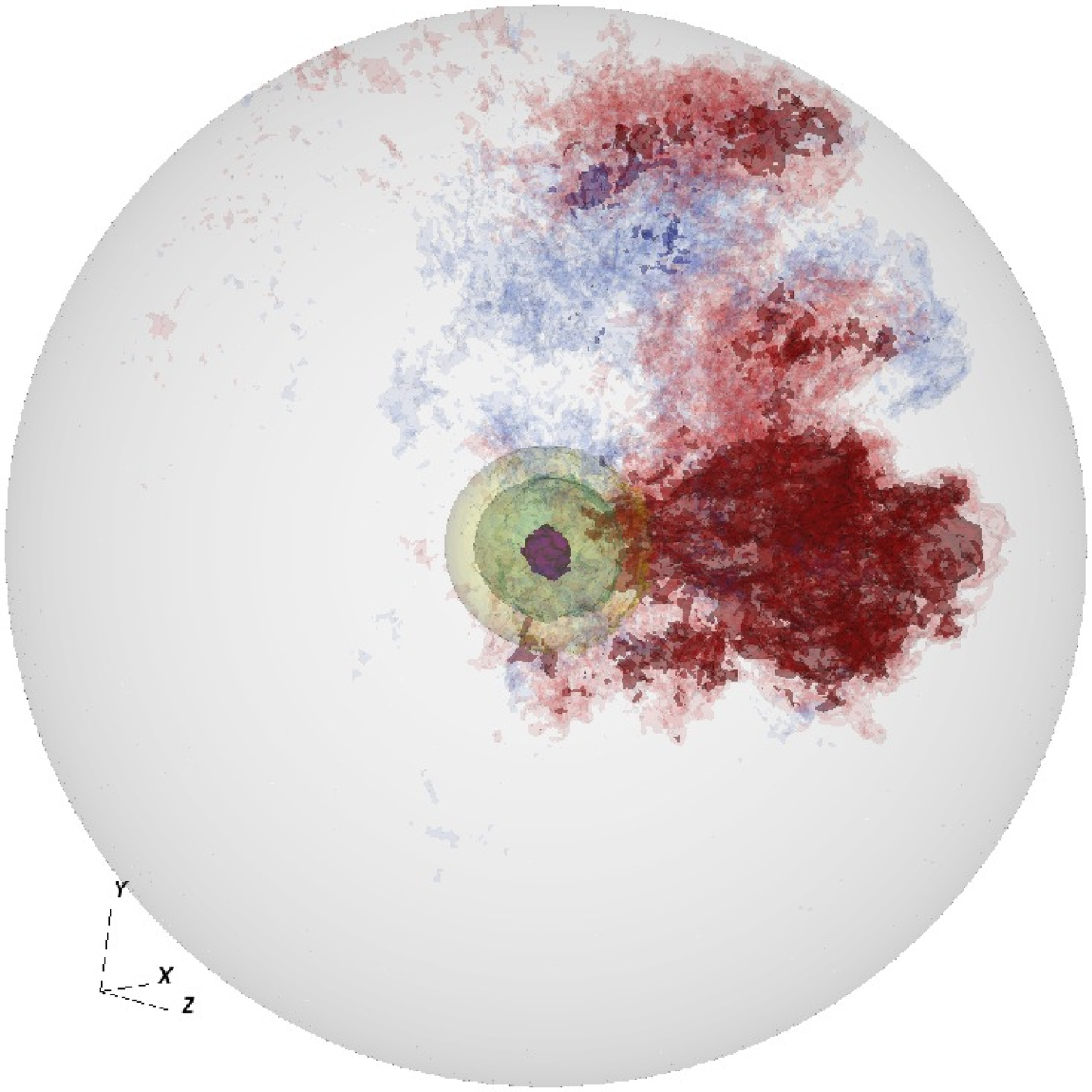}
\includegraphics[width=2.1in]{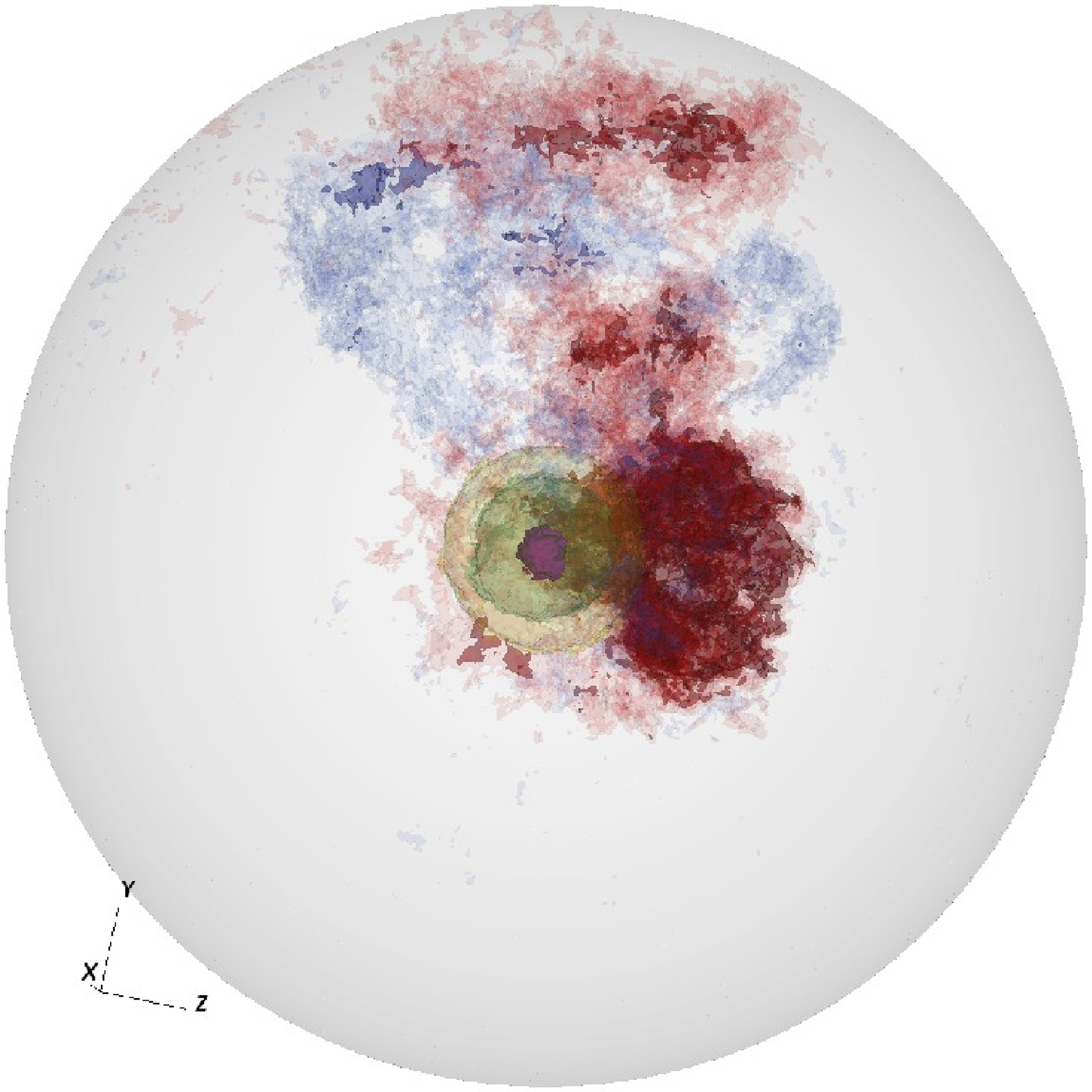}\\
\includegraphics[width=2.1in]{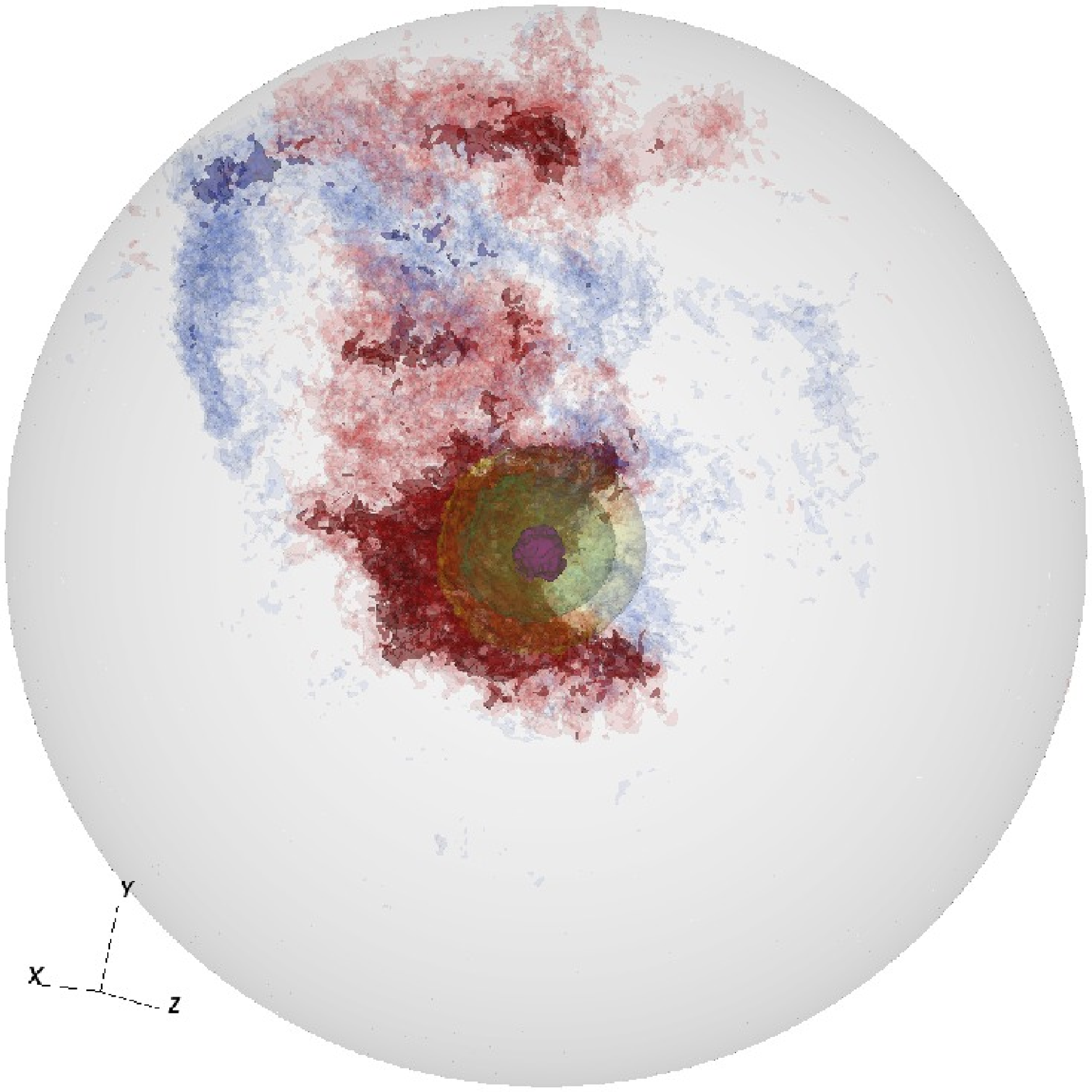}
\includegraphics[width=2.1in]{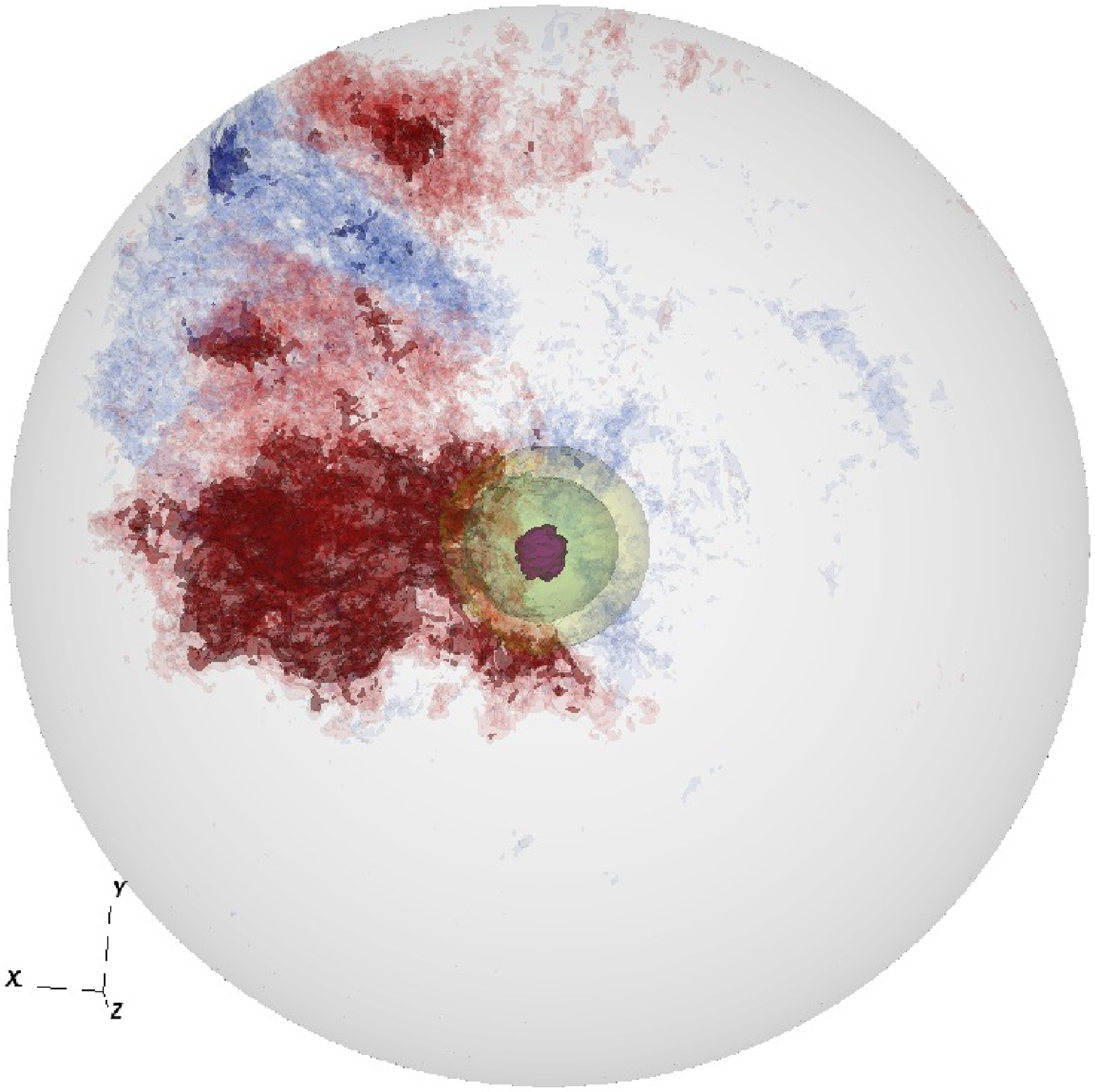}
\includegraphics[width=2.1in]{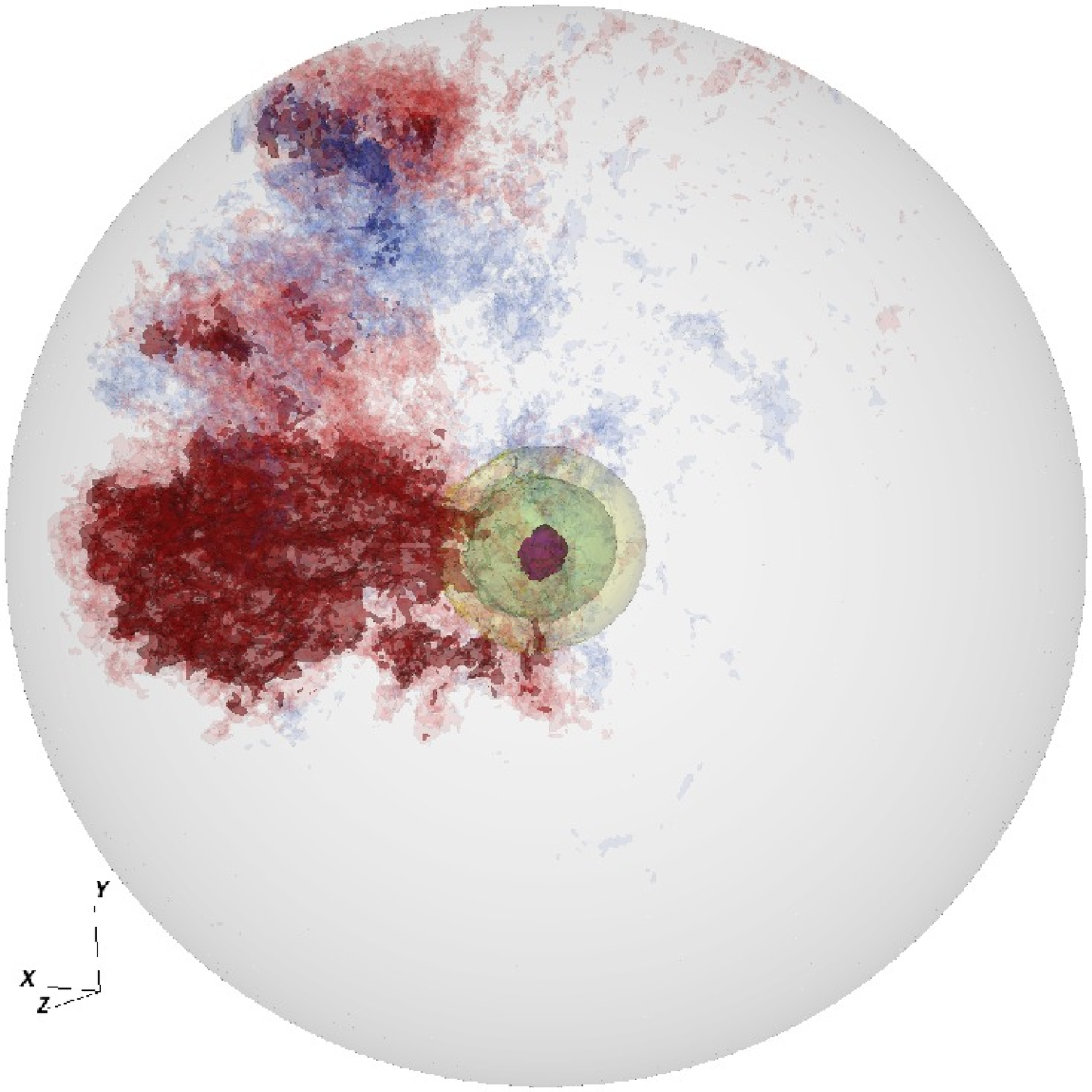}\\
\includegraphics[width=2.1in]{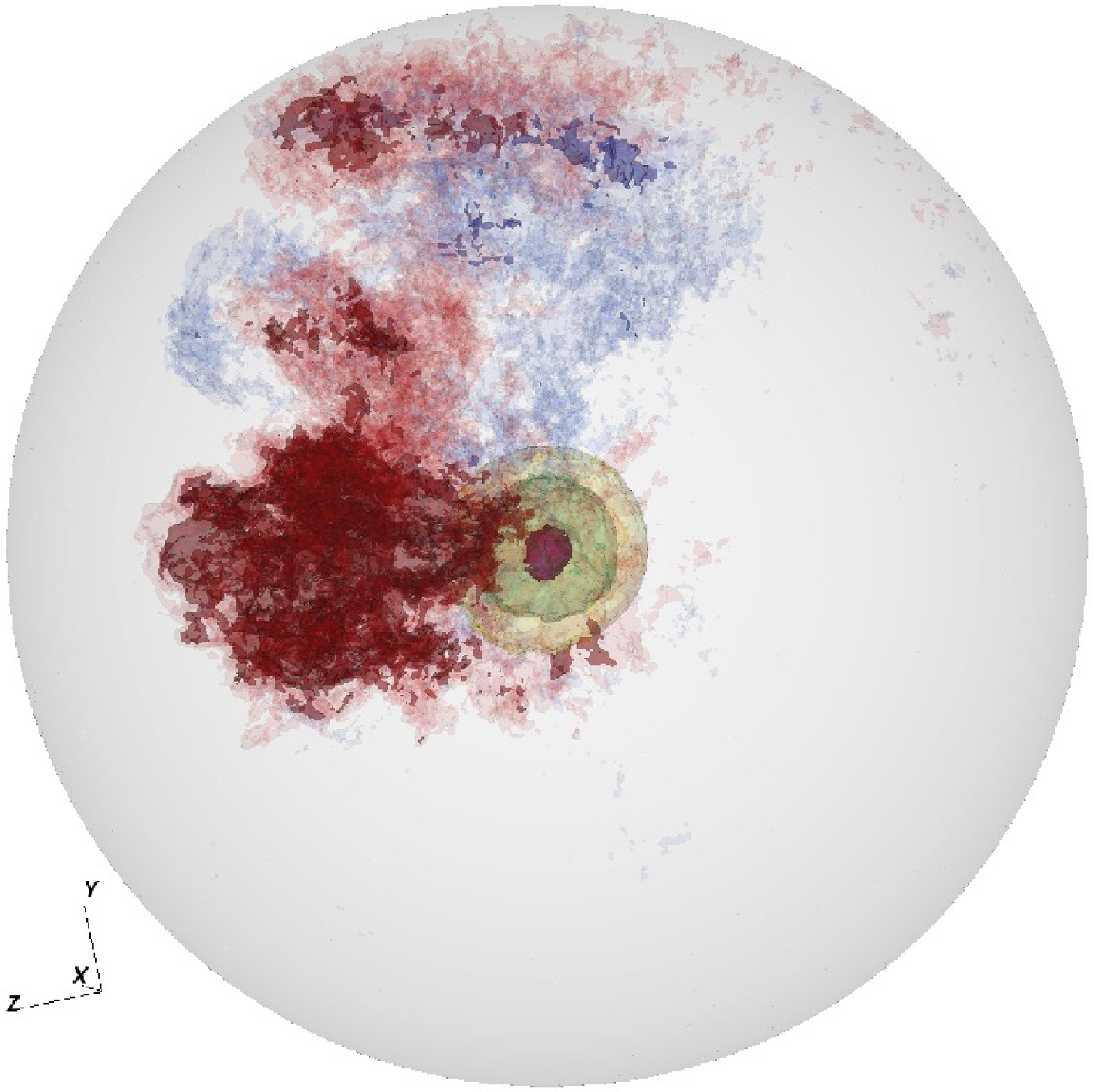}
\includegraphics[width=2.1in]{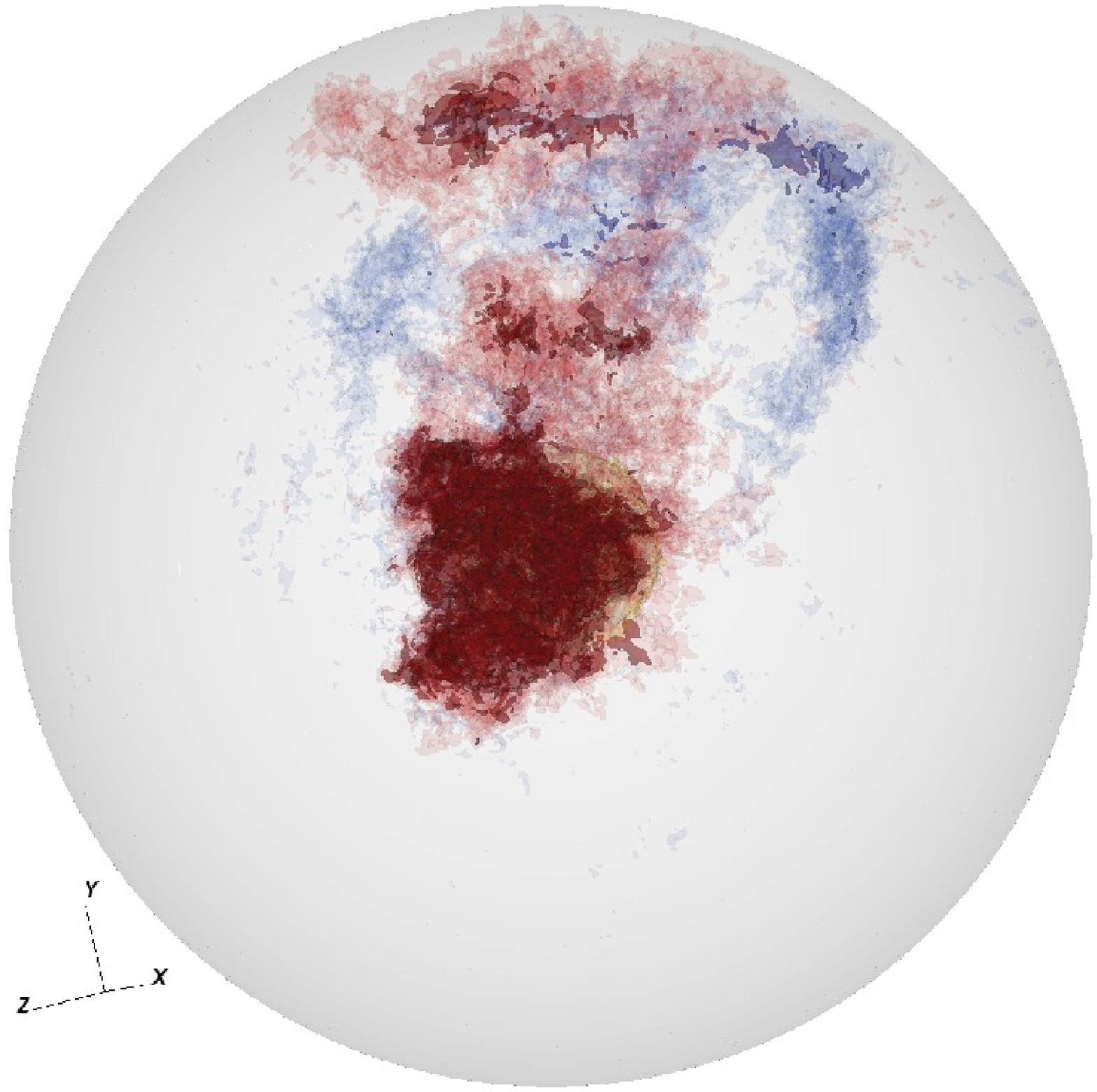}
\includegraphics[width=2.1in]{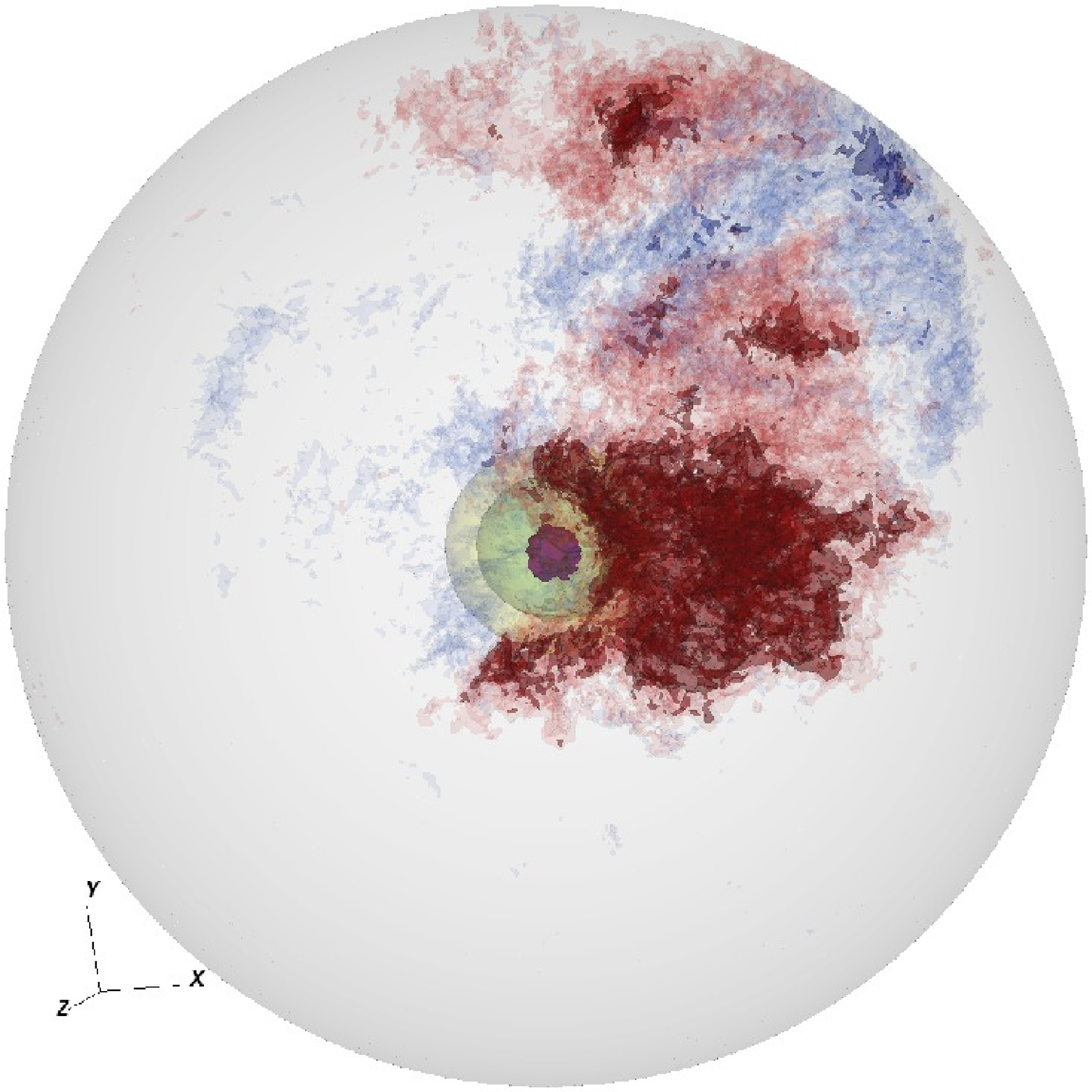}
\caption{\label{fig:wd_convect_rotate} Same data as Figure 
  \ref{fig:wd_convect}, but here we only show the 2.17~km
  grid cell simulation, and each image represents a view rotation of 40 
  degrees of the data from t=10380~s.}
\end{center}
\end{figure}

In Z09, Z11, we provided visualizations of the convective flow
field, noting the dipole feature observed in non-rotating white dwarfs.  We
recall that the convectively unstable region encompasses only an inner fraction
($r\ltaprx 1030$~km) of the star.  Outside of this, the flow is stable against convection
and dominated by large-scale structures with high circumferential velocities and 
a smaller radial component.
Figure \ref{fig:wd_convect} shows visualizations of the 8.68~km, 4.34~km, and 2.17~km
flow fields in the convective region at $t=10380$~s.  As before, we show 
contours of outward and inward radial velocity, as well as contours of increasing 
burning rates.  As expected, the burning is strongest near the center of the star.
Now, we see the effect that resolution has on visualization of the velocity contours.
Both the large-scale
nature of the flow as well as the smaller-scale eddies are much clearer
with increasing resolution.  This allows us to characterize the flow field as a plume,
with a small solid angle region and strong outward velocity, with a lower speed
recirculation.  This is in contrast to a dipole, where we would expect
the magnitude of the outflow and inflow to be more similar.
In Figure \ref{fig:wd_convect_rotate}, we highlight the plume
structure by showing the same 2.17~km flow field in more detail, where each 
frame represents a 40 degree rotation from the previous.

\begin{figure}
\begin{center}
\includegraphics[width=6in]{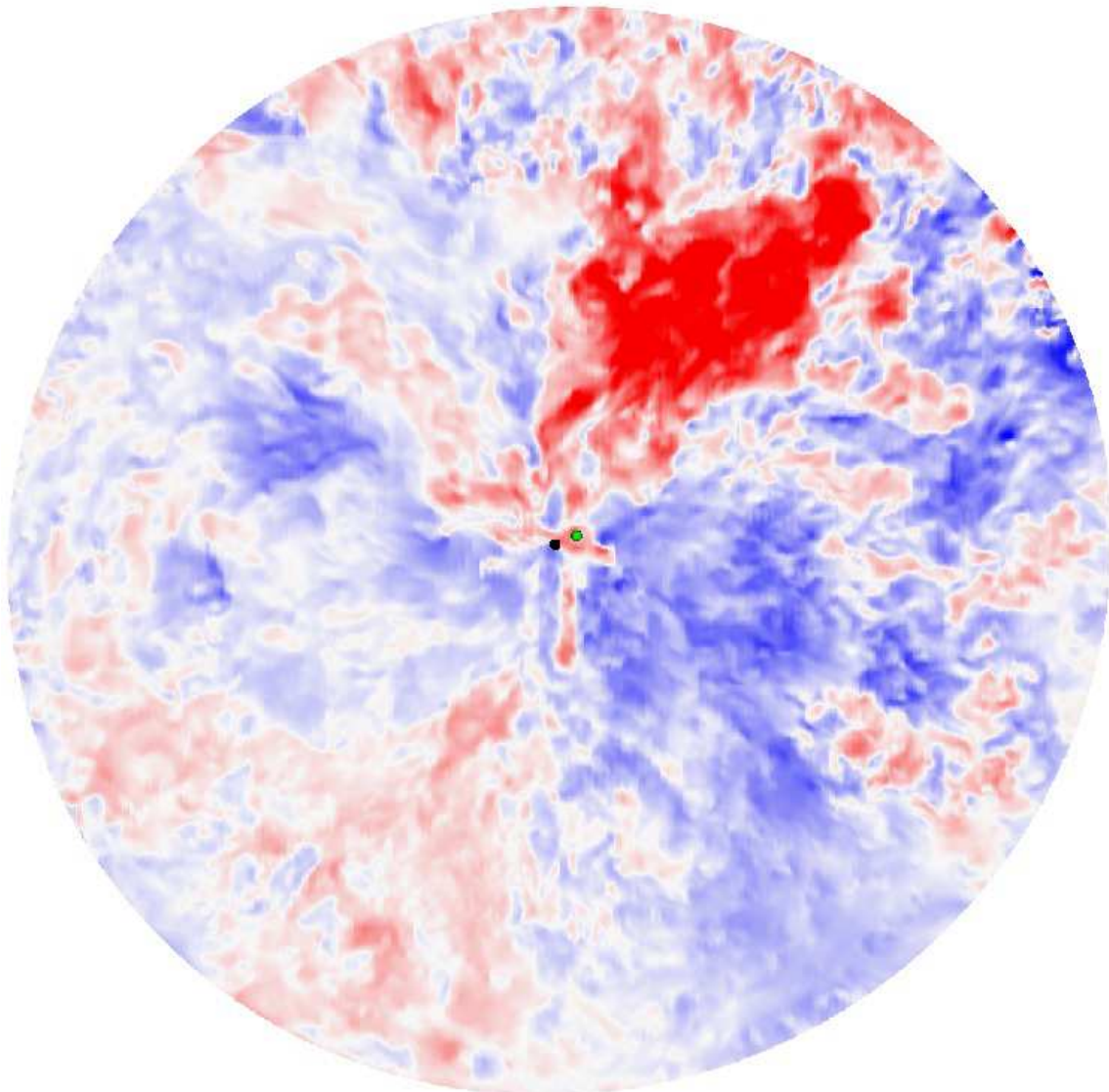}
\caption{\label{fig:hotspot_plane} Planar slice from the 4.34~km simulation at the
time of ignition, oriented so the center of the star (black dot), the
ignition location (green dot), and the center of the strongest outward plume lie in 
the plane.  The dots each have a radius of 20~km.
Red corresponds to $v_r > 60$~km~$s^{-1}$ and blue corresponds to $v_r < -60$~km~s$^{-1}$.
Only the inner $r = 1000$~km is shown.}
\end{center}
\end{figure}

In Figure \ref{fig:hotspot_plane} we observe that, for the 4.34~km simulation at the 
time of ignition, the ignition point 
lies in a region with positive $v_r$ (consistent with our
earlier report that $v_r = 9.5$~km~s$^{-1}$), and is almost aligned with
the strongest outward plume. We expect this hot ignition
point to accelerate radially outward to a significant fraction of the sound speed 
within a small fraction of a second; this does not give the parcel of fluid at
the ignition point enough time to change direction and align exactly with the strong 
outward plume.

\begin{figure}
\begin{center}
\includegraphics[width=3.5in]{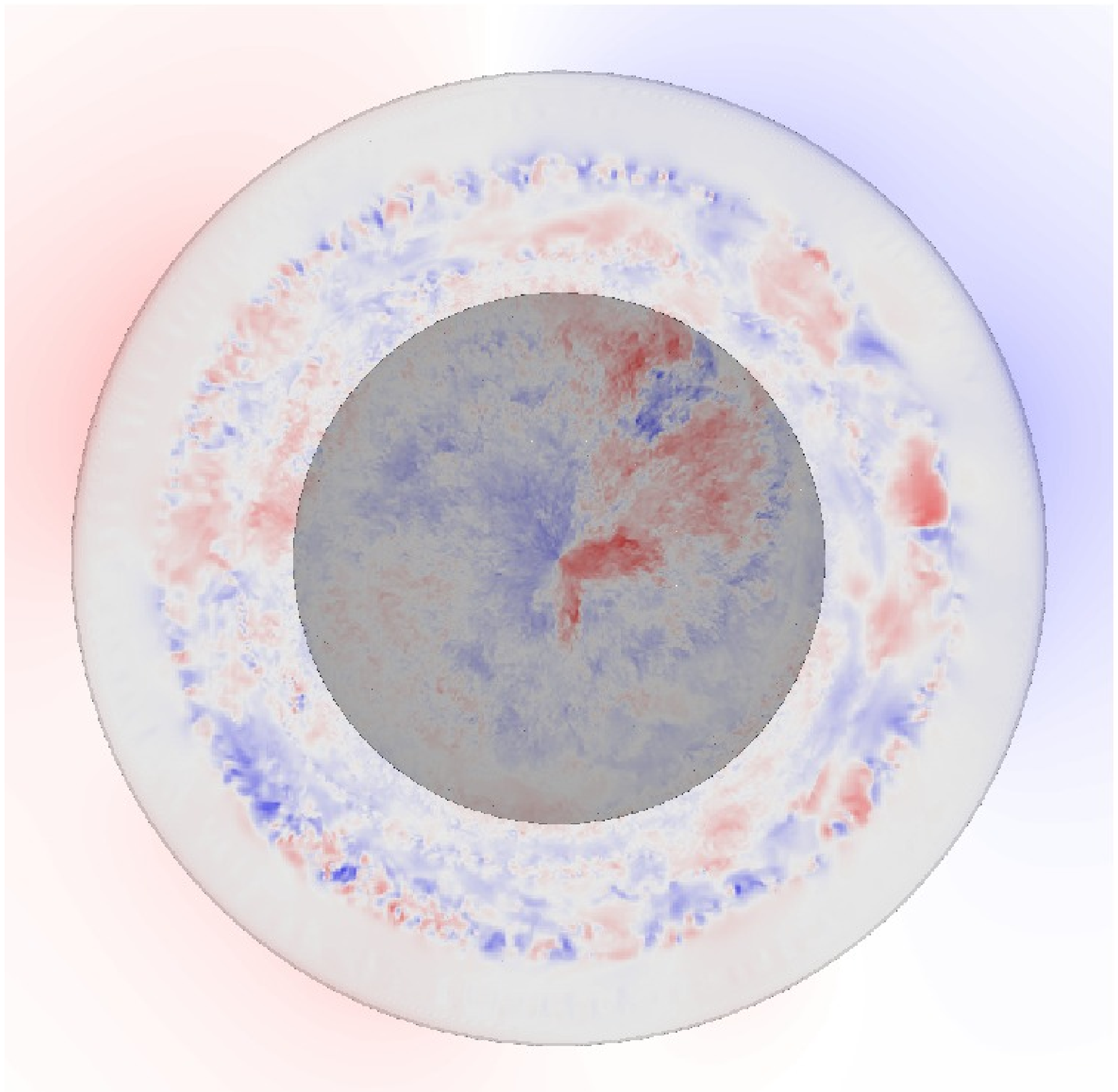}
\includegraphics[width=3.5in]{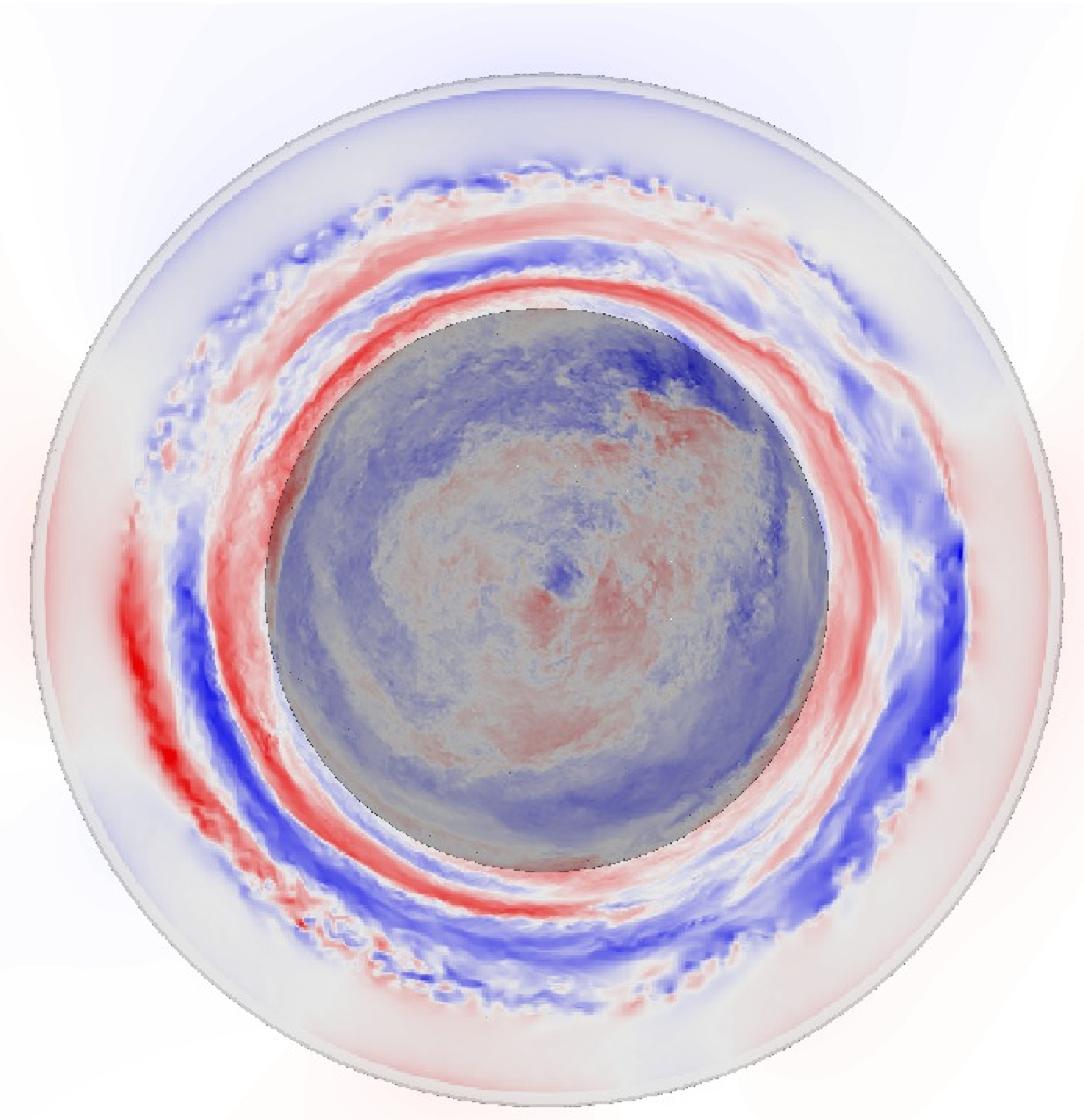}
\caption{\label{fig:wd_convect_urut} (Top) Plot of radial velocity ($\Ub\cdot\eb_r$) 
in the $x$-$y$ plane from the 2.17~km simulation at $t=10380$~s.  
(Bottom) Plot of $\Ub\cdot\eb_\theta$ in the $x$-$y$ plane from the same dataset.
In both plots, 
red = +100~km~s$^{-1}$ and blue = -100~km~s$^{-1}$.  The outer dark contour indicates the 
edge of the star, where $\rho_0\approx 1\times 10^5$~g~cm$^{-3}$ ($r\approx 1030$~km).  
The inner dark contour indicates the edge of the convective region, where 
$\rho_0\approx 1.26\times 10^8$~g~cm$^{-3}$ ($r\approx 1890$~km).}
\end{center}
\end{figure}

To get an idea of the structure of the flow outside of the convective
region, we visualize the flow field in the $x$-$y$ plane.
In Figure \ref{fig:wd_convect_urut}, we plot the radial velocity 
($\Ub\cdot\eb_r$), as well as one component of the circumferential velocity, 
$\Ub\cdot\eb_\theta$, where $\eb_\theta$ is the unit vector in the azimuthal direction
in the $x$-$y$ plane.  Both plots use the same scale for positive and negative
velocities, so we see that the circumferential velocities outside the convective 
region are generally larger than the radial velocities within the core.  These 
circumferential velocities in the stably stratified region may become important 
in explosion simulations in that they may deform hot bubbles or flames passing 
through that region.

\subsection{Turbulence Structure}
Predictive models for SNe Ia, in particular turbulent flame models,
depend critically on the structure of the turbulence in the star.  In
this section, we use the simulations to examine this structure and
extract estimates for the turbulent intensity and the integral length
scale.  This will help us understand the state of the turbulence that exists
at the start of the explosion phase.

The vast majority of literature on turbulence theory deals with 
flows that are assumed to have constant density.  In the present context, the 
significant variation in density due to stratification cannot be ignored, 
and has to be dealt with carefully.  Following \cite{vonWeizsacker51}, \cite{Fleck83,Fleck96} 
advocated casting the energy balance equation in terms of energy density
(energy per unit volume) as opposed to specific energy (energy per unit mass),
and we note that the difference is inconsequential for constant density turbulence.
Thus, the fundamental quantity relevant to the inertial range of a turbulent energy
cascade is the energy dissipation rate per unit volume $\varepsilon_V$, which 
should be expected to scale with $\varepsilon_V\sim\rho \check{u}^3/l$, 
where $\check{u}$ is the turbulent intensity (rms velocity fluctuation), 
and $l$ is the integral length scale.
Subsequently, \cite{Kritsuk:2007} used numerical simulation of compressible turbulence 
to demonstrate that an appropriately density-weighted velocity spectrum obeys a 
Kolmogorov-type five-thirds decay law.  Consequently, we consider a density-weighted 
velocity field,
\begin{equation}
\Vb_n(\xb) = \rho^n\Ub \quad [{\rm g}^n~{\rm cm}^{1-3n}~{\rm s}^{-1}],
\end{equation}
and its Fourier transform,
\begin{equation}
\widehat\Vb_n(\kappab) = \mathcal F[\Vb_n(\xb)]  \quad [{\rm g}^n~{\rm cm}^{4-3n}~{\rm s}^{-1}].
\end{equation}
We then define a generalized energy density spectrum as
\begin{equation}
E_n(\kappa) = \frac{1}{\Omega}\int_{\Sb(\kappa)}\frac{1}{2}\widehat\Vb_n(\kappab)\cdot\widehat\Vb_n^*(\kappab)\,\d\Sb  \qquad [{\rm g}^{2n}~{\rm cm}^{3-6n}~{\rm s}^{-2}].
\end{equation}
where $\Omega$ is the volume of the domain in physical space, the domain of the integral, $\Sb(\kappa)$, 
is the spherical surface defined by $|\kappab| = \kappa$, and $^*$ denotes 
the complex conjugate.
This generalized energy density spectrum can only be made dimensionless using $\varepsilon_V$
and $\kappa$ for $n=1/3$, resulting in the dimensionless group
$\varepsilon_V^{-2/3}\kappa^{5/3}E_{1/3}(\kappa)$.  Therefore, plotting $E_{1/3}(\kappa)$ should
present a five-thirds decay.  Henceforth, we only present energy density spectra appropriately
weighted and omit the 1/3 suffix.  We also note that only $n=1/2$ corresponds to a real
energy density.

In the diagnostics in this section, we consider the local velocity, $\Ubt$, 
rather than the total velocity, $\Ubt + w_0\eb_r$.  In Z11, we showed that the 
maximum magnitude of $w_0$ at ignition is $\sim$0.013~km~s$^{-1}$, so the effect 
of $w_0$ is insignificant on the scales we are interested in.

\begin{figure}
\begin{center}
\includegraphics[width=3in]{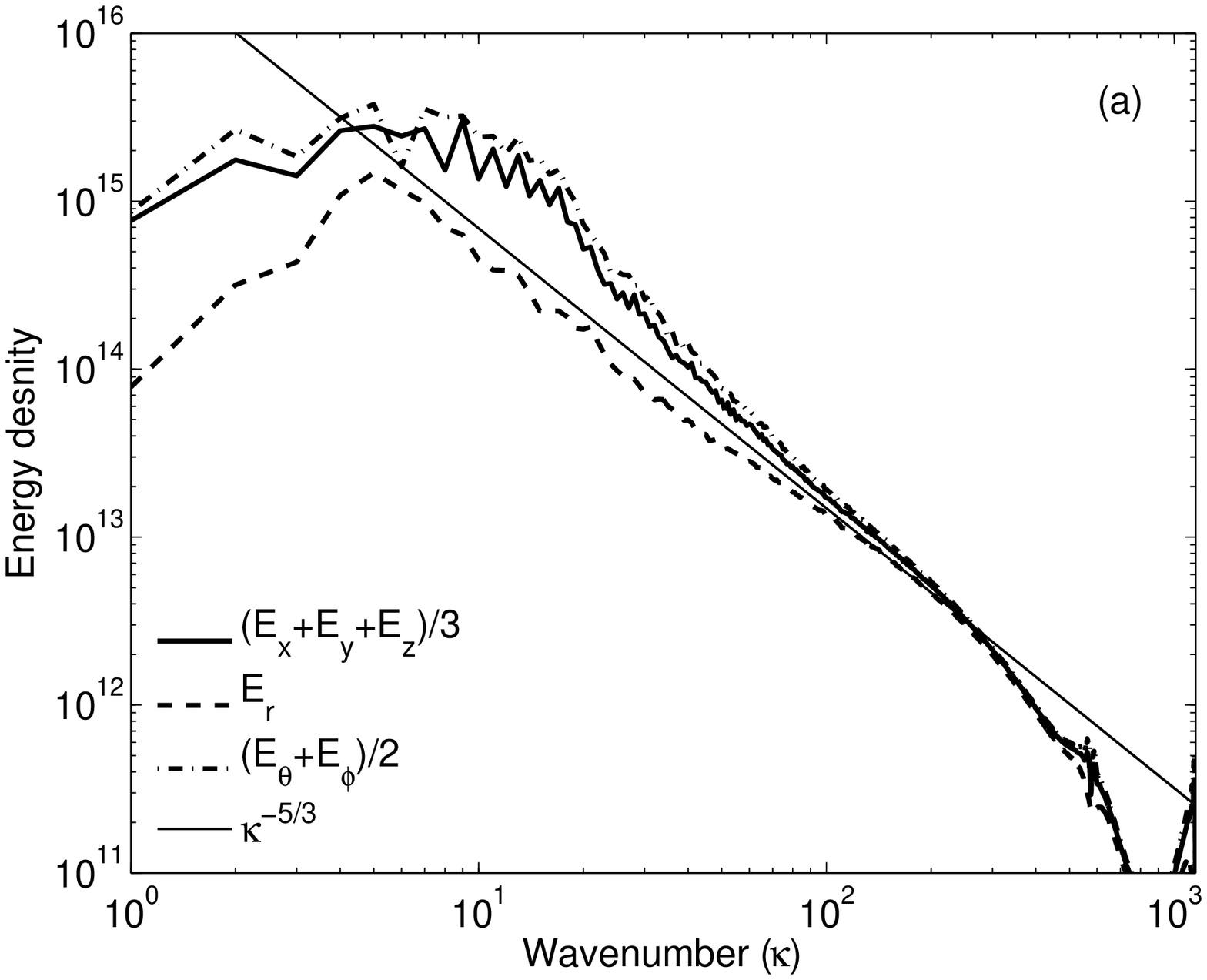}
\includegraphics[width=3in]{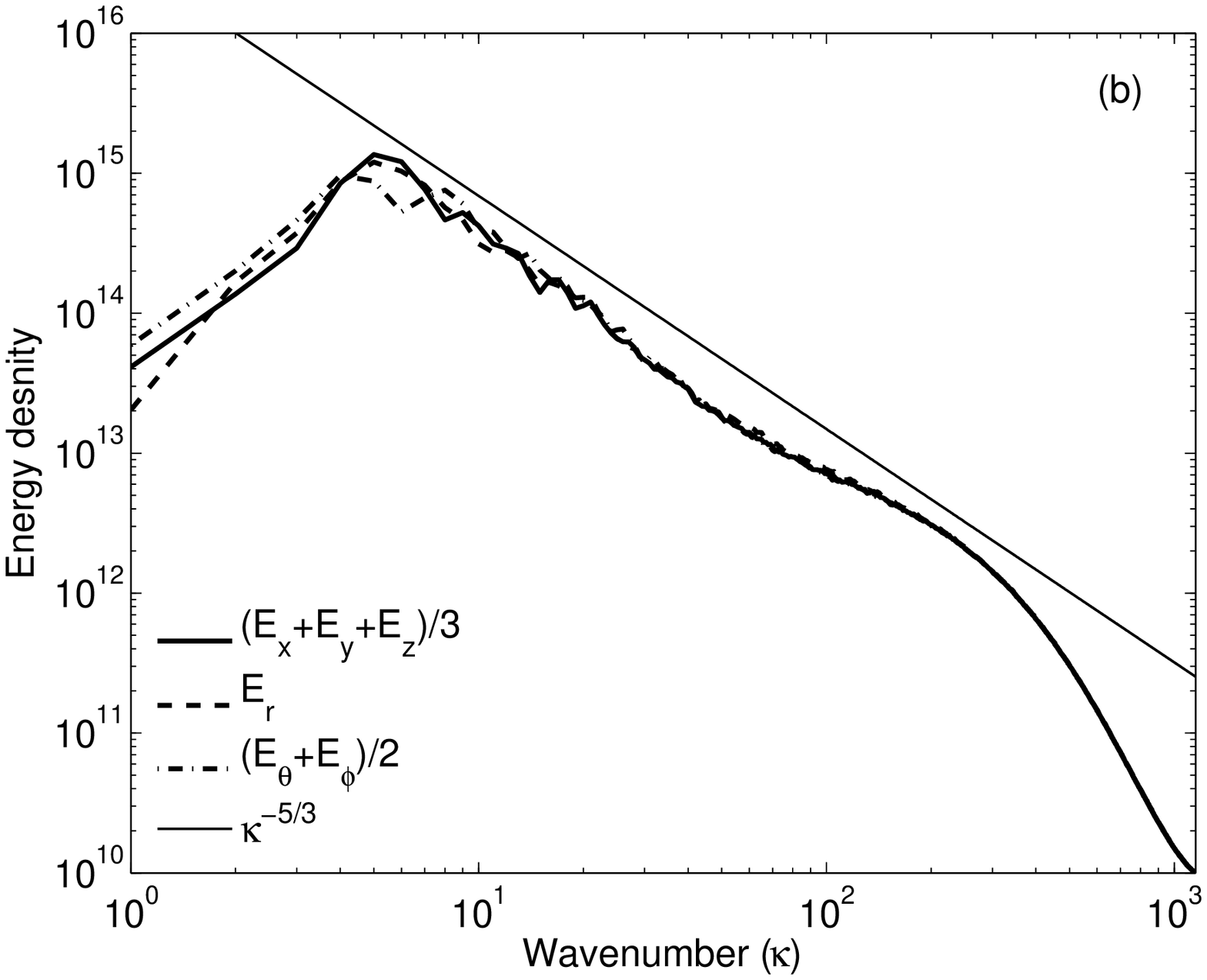}\\
\includegraphics[width=3in]{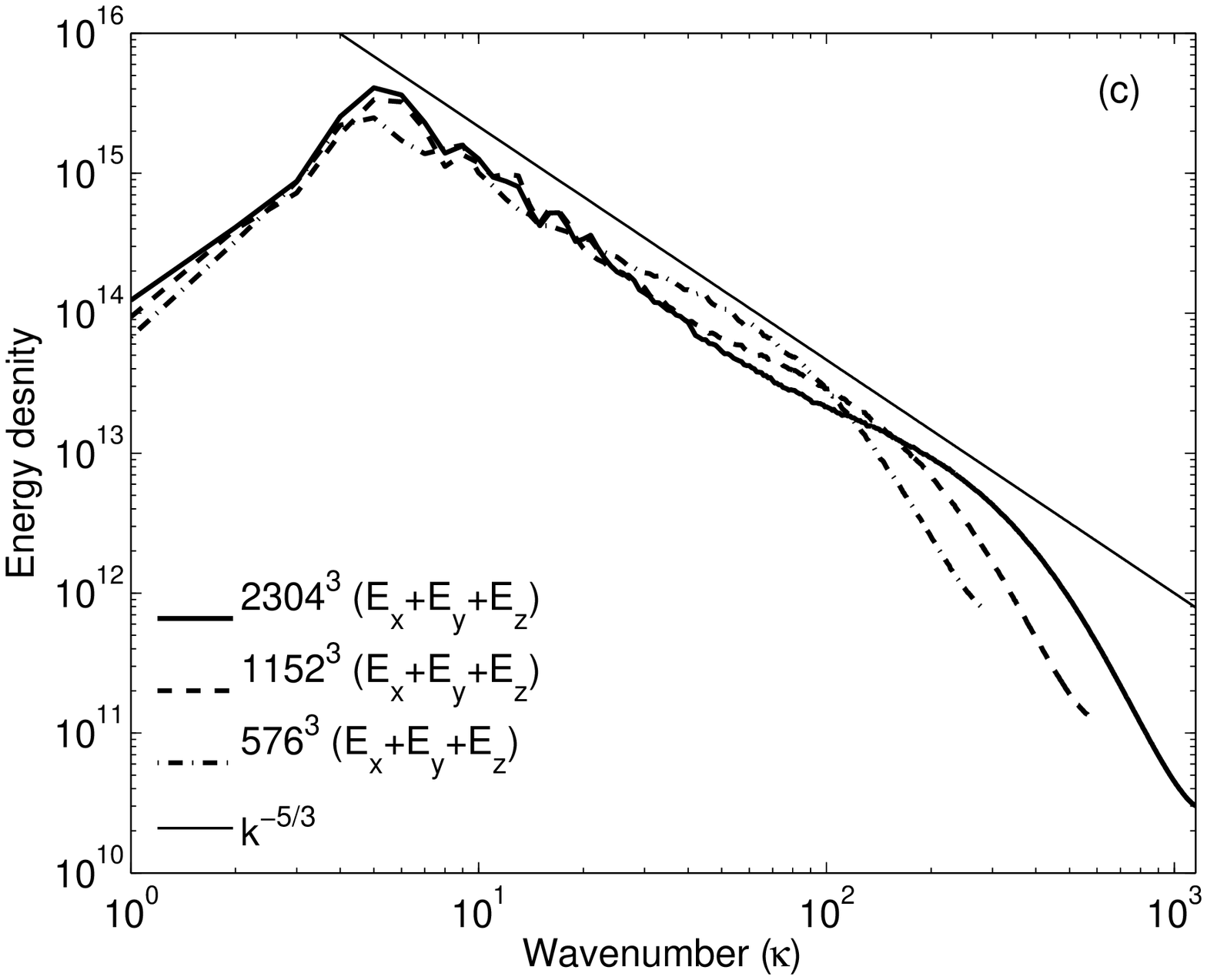}
\includegraphics[width=3in]{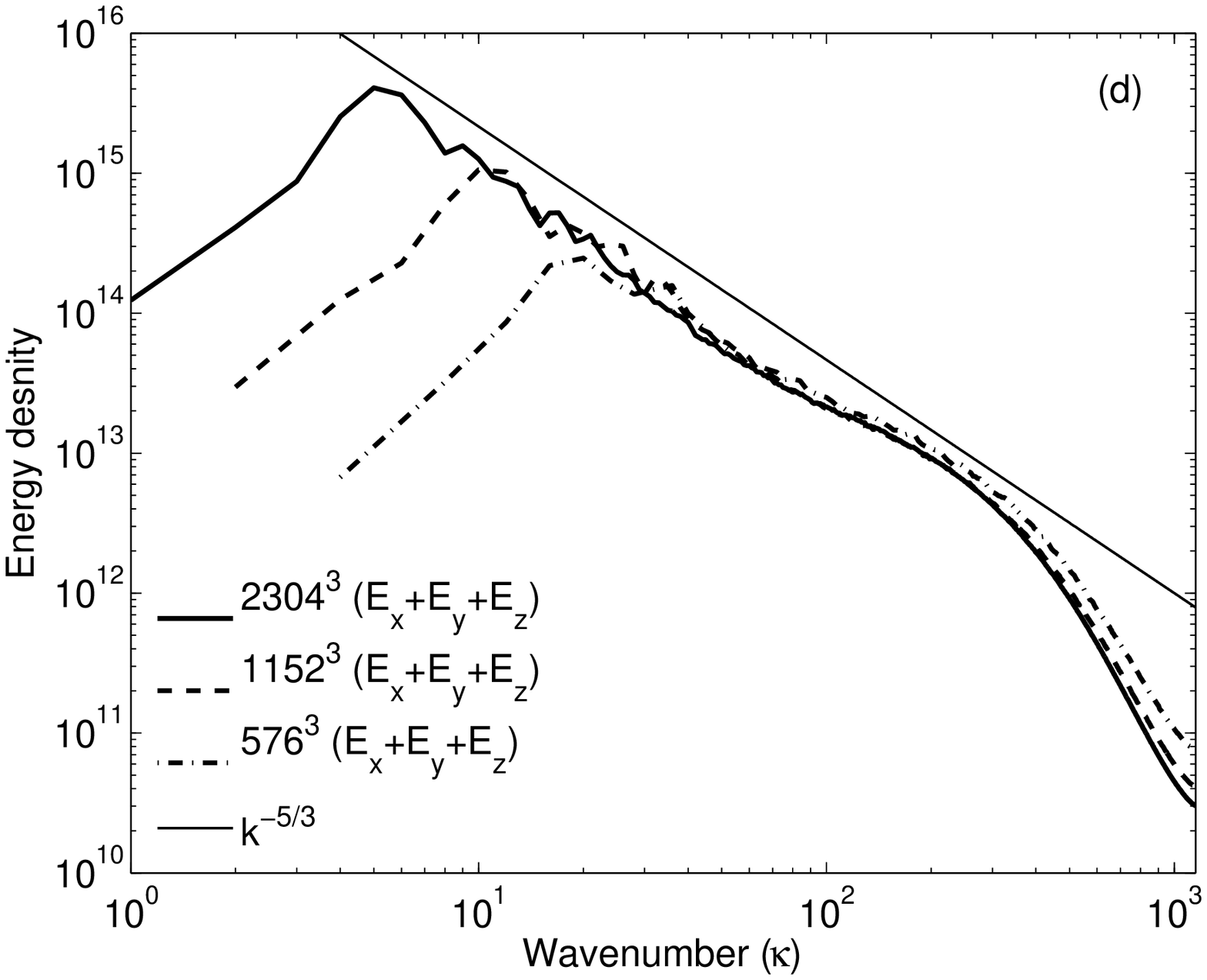}\\
\caption{\label{fig:spectra} (a) Energy density spectra of the entire
domain at 2.17~km resolution for $t=10380$.  Note how the radial
spectrum is much lower than the other components.  This is due to 
the large circumferential velocities outside the convective core.
(b) Energy density spectra for the convective core.  Note how the 
curves have collapsed to a single profile, especially for $\kappa\gtaprx20$, 
corresponding to about 250~km.
(c) Comparison of the energy density spectra at the three different
resolutions.  (d) As (c), but scaled to demonstrate that the effective
Kolmogorov length scale is proportional to the computational cell width.}
\end{center}
\end{figure}

Energy density spectra from the 2.17~km simulation at $t=10380$~s are shown in Figure \ref{fig:spectra}(a). 
The density-weighted velocity field has been decomposed into different components, 
specifically, the Cartesian components, $V_x$, $V_y$, and $V_z$,
and spherical polar components (using the convention that $\theta$ is the 
azimuthal angle with respect to the $x$-$y$ plane and $\varphi$ is the inclination angle 
measured from the $z$-axis),
\begin{equation}
V_r = \frac{x V_x + y V_y + z V_z}{r},\quad
V_\theta = \frac{-y V_x + x V_y}{R},\quad
V_\varphi = \frac{xzV_x + yzV_y - R^2V_z}{rR},
\end{equation}
where $r^2=x^2+y^2+z^2$ and $R^2=x^2+y^2$.  Three energy density spectra are 
plotted: first, the mean of the Cartesian components (individual 
components do not differ significantly from that shown); second, the 
radial component; and third, the circumferential component.  This decomposition demonstrates
that there is significantly less energy in the radial component than in the other components.
This is due to the large circumferential velocities in the layers outside the convection zone;
although the density is lower here, the volume is sufficiently large that the resulting
energy has a significant contribution to the spectrum.  It also appears that the radial component 
decays with an exponent close to five-thirds (if slightly smaller), and the other components have
a slightly higher exponent.

To circumvent the issue of large circumferential velocities in the stably stratified 
region, and to remove the signal from the coarse-fine 
interfaces at wavenumbers around 1152 and 576, the energy density spectra of a subdomain were constructed.  
This was achieved by applying a smoothing function to the velocity and density fields in such a way that 
the data outside the convection zone were set to zero.  
Specifically, each field was multiplied by the hyperbolic tangent function 
$(1-\tanh[(r-r_0)/\delta])/2$, where $r_0=875$~km and $\delta=30$~km.
All of the resulting non-trivial data were at the finest AMR level, and the resulting energy 
density spectra are shown in Figure \ref{fig:spectra}(b).  Now, each spectrum collapses to a single 
curve, especially for $\kappa \gtaprx 20$, which corresponds to a length of about 250~km.
The decay exponent of each spectrum is close to five-thirds and 
presents the characteristic ``bump'' between the inertial and dissipation ranges expected from
developed homogeneous isotropic turbulence (e.g.~\citealt{Saddoughi94,Porter94,Kaneda03,aspden-camcos}).

To explore the effect of resolution on the turbulence in the convective core, Figures \ref{fig:spectra}(c)
and (d) present the total kinetic energy density spectra for the three resolutions, first without scaling (c), 
and then scaled (d).  The spectra are scaled according to computational cell width and in keeping with a 
constant energy dissipation rate.  Specifically, the 4.34~km simulation spectrum 
is shifted to higher wavenumbers by a factor of 2, 
and to lower energy density by a factor of 2$^{-5/3}$, and the 8.68~km spectrum has been
shifted by factors of 4 and 4$^{-5/3}$, respectively.
The unscaled spectra demonstrate that the large scales are independent of resolution (as expected), in the sense
that increasing the resolution does not lead to an increased level of turbulent intensity.  This kind of
convective motion is not dominated by small-scale processes, and integral quantities are well-captured even
at moderate resolutions.  The 8.68~km simulation has a short inertial range, but this is more extensive at 
higher resolutions.  The scaled spectra demonstrate that the dissipation range depends on the computational
cell width as expected from an ILES-type simulation.  In particular, there is an effective Kolmogorov 
length scale that is a function of the cell width; the collapse is not exact, but is consistent with
previous work, see \cite{aspden-camcos}, for example, which also contains further discussion of the ILES 
approach and dependence on resolution.

The rms velocity in the convective core ($r<875$~km), $\check{u}$, was found by direct measurement to 
be approximately 14~km~s$^{-1}$ (the data ranged from 12~km~s$^{-1}$ to 18~km~s$^{-1}$ depending on component and resolution);
note that no density weighting was used.  Even though this estimate is smaller than previously suggested 
($\sim$100--500~km~s$^{-1}$),  we argue that it is actually an upper bound for the turbulence produced by 
convection because it includes large-scale plume-like flow, which artificially inflates the estimate.

To determine the integral length scale in the convective core, the longitudinal correlation
functions were evaluated for the Cartesian components of the velocity field, along with the
correlation functions of the radial velocity in each Cartesian direction, where the
(second-order) velocity correlation function (two-point, one-time) is defined as
\begin{equation}
Q_{ij}(\rb,t) = \frac{1}{\Omega} \int_\Omega U_i(\xb,t) U_j(\xb+\rb,t) \,\d\xb
\label{eq:corrFn}
\end{equation}
where $\rb$ denotes the separation vector.
The integral length scale in the $x$ direction, for example, is then defined as the integral 
of the longitudinal velocity correlation function
\begin{equation}
l_x = \frac{1}{\check{u}_x^2}\int Q_{xx}(r\eb_x) \,\d r.
\label{eq:intLen}
\end{equation}

\begin{figure}
\begin{center}
\includegraphics{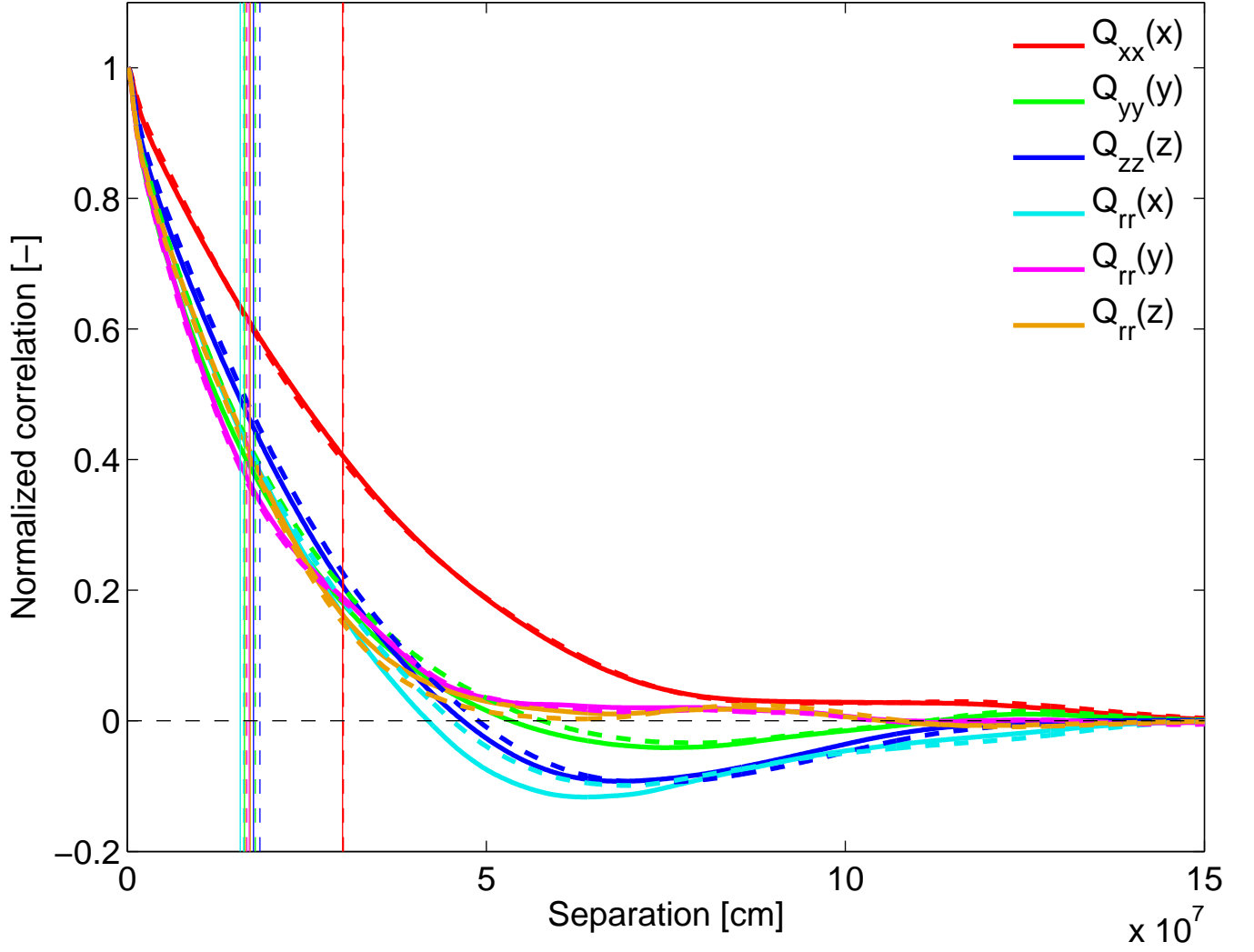}
\caption{\label{fig:correlation} Longitudinal correlation functions for 
the turbulence in the convective core at 2.17~km resolution for $t=10380$.
Density-weighted and non-weighted correlation functions are shown by solid 
and dashed lines, respectively.  The $x$ component presents a larger correlation
because there is a plume-like structure roughly aligned with the $x$-axis.  
The integral length scales are denoted by the vertical lines of the corresponding color.}
\end{center}
\end{figure}

The correlation functions were evaluated both for the density-weighted and non-weighted velocities 
(by replacing $U_i$ by $V_i$ in Equation (\ref{eq:corrFn}) and the appropriate normalization factor in 
Equation (\ref{eq:intLen}), and are compared in Figure \ref{fig:correlation} by solid and dashed lines, 
respectively.  The weighted and non-weighted correlation functions are in close agreement, suggesting 
that measuring the integral length scale is not affected by the variations in density.  By integrating
each correlation (ignoring the negative parts), integral length scales for each component
were evaluated and are shown by the vertical lines with the corresponding line style.
The $x$ component appears not be consistent with the other components, probably because
there is a large plume-like structure roughly aligned with the $x$-axis.  Taking this to
be an outlier, the mean integral length scale was found to be approximately 169~km 
(with a standard deviation of approximately 8.4~km).

Averages and standard deviations of integral length scale and rms velocity were evaluated using seven 
time points over 350~s at the 4.34~km resolution, and were found to be approximately 200$\pm$50~km and 
16$\pm$3~km~s$^{-1}$, respectively.

Taking the integral length scale to be 200~km and the turbulent intensity to be
16~km~s$^{-1}$, the specific energy dissipation rate $\varepsilon=\check{u}^3/l$ is approximately
$2\times10^{11}$~cm$^2$~s$^{-3}$.  The corresponding estimates that were suggested to be necessary for
a spontaneous detonation by \cite{Woosley:2011} were 10~km and 500~km~s$^{-1}$, respectively
(see also \citealt{lisewski:2000,roepke2007,timmeswoosley1992}).
This gives $\varepsilon\sim 10^{17}$~cm$^2$~s$^{-3}$, six orders of magnitude larger.  
The present simulations suggest that the turbulent intensity required for a spontaneous 
detonation can not be produced by convection within the core.

\section{Conclusions and Discussion}\label{Sec:Conclusions and Discussion}

Overall, our high-resolution simulations agreed with the findings of
Z11 regarding the ignition radius of 50~km with a likely range of
40~km to 75~km.  We do note that the outer limit of 100~km reported in Z11
is probably too large, as we do not see any hot bubbles at that radius that
are still increasing in temperature.
By looking closely at the dynamics of the last few hot spots, we conclude
that the multiple ignition scenario is unlikely.  With improved
resolution, we now describe the large-scale coherent structure in the
convective field as a plume, rather than a jet, and have a better
understanding of the turbulent nature of the flow.

These findings, together with those from Z11, indicate
that a single-point, off-center ignition is the most likely scenario
for SNe Ia.  At the radii we find ignition to be most likely, the
initial flame will float away faster than it can burn toward the
center (see e.g.~\citealt{plewa:2004,zingaledursi}), making for an asymmetric explosion.  This scenario has been
explored in explosion calculations, potentially giving rise to the
``gravitationally confined detonation'' \citep{plewa:2004,jordan:2008},
although other groups suggest that this mechanism may not be
robust~\citep{roepke-gcd}.  If a single off-center ignition fails to
blow up the star, then it is possible that we would need to wait for
the next ignition point, perhaps tens of seconds later, or cycle
through many widely spaced ignitions until we ignite closer to the
center (i.e.\ many successive false starts).  Alternately, some type of pulsational model may
ensue~\citep{ivanova1974,khokhlov1991b,bravogarciasenz:2006}.  With
these results, the challenge to the explosion modelers is to
demonstrate that the single-degenerate Chandrasekhar mass white dwarf
model can produce robust explosions resulting from single-point,
off-center ignition.  Observations may show support for asymmetric
models~\citep{maeda:2010}, but some radiative transfer calculations
seem to preclude extreme amounts of asymmetry~\citep{blondin:2011}.

\begin{figure}
\begin{center}
\includegraphics{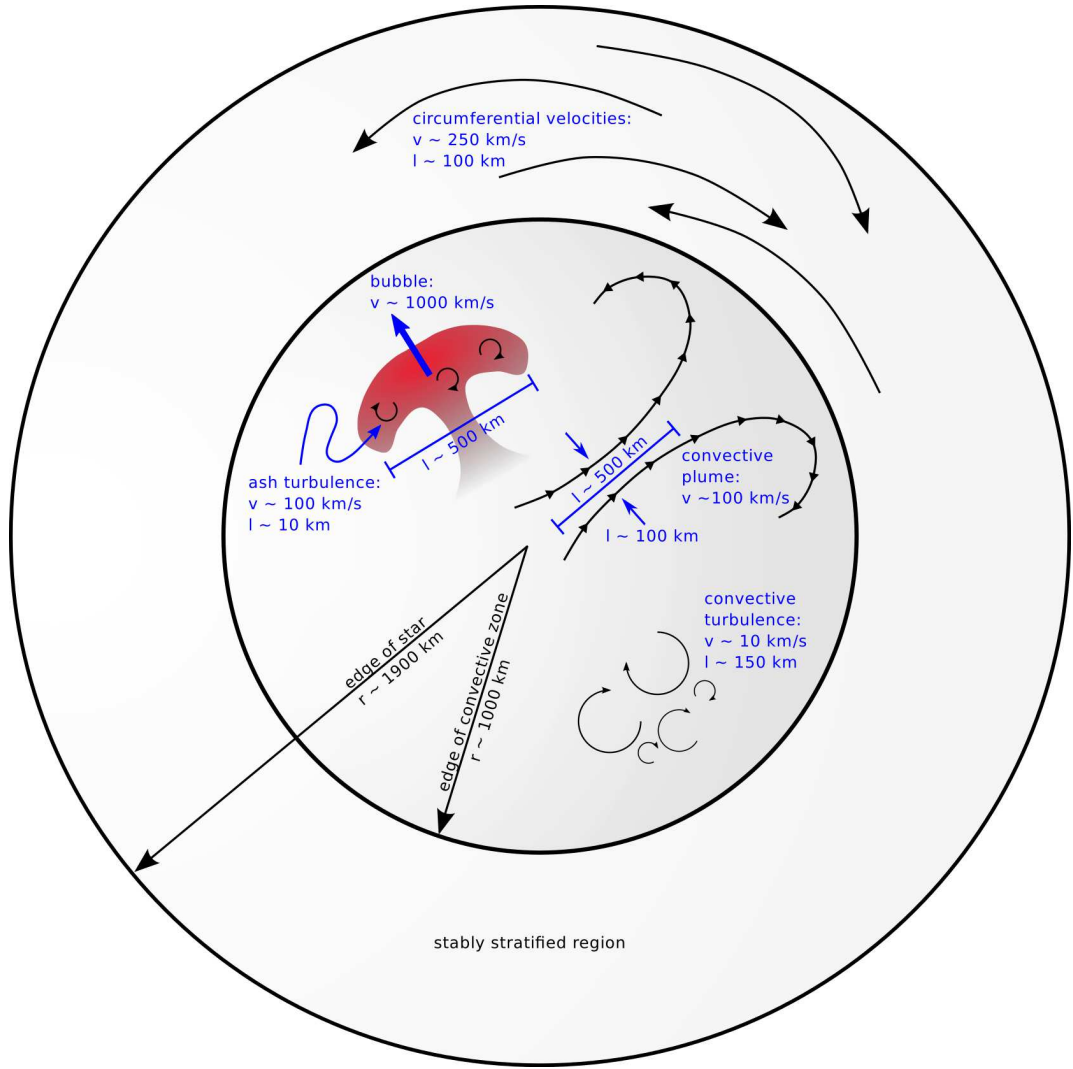}
\caption{\label{fig:schematic} Cartoon showing the various features
with associated velocities and length scales in the white dwarf at the end of 
convection/start of flame propagation.}
\end{center}
\end{figure}

We conclude by summarizing the various components of the convecting white dwarf
and give characteristic length and velocity scales for each; Figure 
\ref{fig:schematic} presents this information in schematic form.
Buoyancy drives a large-scale flow in the convective core, which extends to 
a radius on the order of 1000~km.  This large-scale flow is composed of 
plumes around 100~km wide and several hundred km long with
a bulk velocity around 100~km~s$^{-1}$.  These plumes
drive turbulence in the core with an rms velocity and integral length
scale that were estimated to be on the order of 16~km~s$^{-1}$ and 200~km,
respectively.  This level of turbulence is far below that required for
a spontaneous detonation to occur.  The stably stratified region outside the 
convective core, extending from $\sim$1000~km to $\sim$1900~km, is made up of 
circumferential shear layers, with a smaller radial velocity component.  These 
shear layers are on the order of 100~km deep, several hundred km long, with 
typical velocities on the order of 100~km~s$^{-1}$ and peak velocities that may be 
in excess of 250~km~s$^{-1}$.  The burning of a single off-center ignition 
would be dominated at early times by the laminar flame speed (on the order of 
50~km~s$^{-1}$), and the level of turbulence in the core is unlikely to 
deform the flame very much at all.  Furthermore, \cite{aspden-bubble} found 
that large-scale entrainment was the dominant process in the evolution of a 
burning bubble, and that the flame speed (turbulent or laminar) even up to 
100~km~s$^{-1}$ did not significantly affect the evolution.  Therefore, the turbulence
produced by convection in the core is unlikely to play a significant
role in the explosion.  As the bubble reaches the edge of the convective core,
it will be $\sim$500~km across moving with a rise speed on the order of 1000~km~s$^{-1}$.
The turbulence within the bubble itself is likely to have an rms velocity on the order 
of 100~km~s$^{-1}$ on an integral length scale of a few tens of km.
In the past, it has been suggested that the convective boundary lies at the density 
suggested for a deflagration-to-detonation transition~\citep{pirochang}.  
Although the velocities in the core are unlikely to affect the bubble as it rises,
the circumferential velocities in the stable region are much greater and may
interact strongly with the bubble as it passes through this region. We plan
to investigate this interaction in future work.

\acknowledgments

We thank Frank Timmes for
making his equation of state routines publicly available and for
helpful discussions on the thermodynamics.  The work at Stony Brook
was supported by a DOE/Office of Nuclear Physics grant
No.~DE-FG02-06ER41448 to Stony Brook.  The work at LBNL was supported
by the SciDAC Program of the DOE Office of High Energy Physics
and by the Applied Mathematics Program of the DOE Office of
Advance Scientific Computing Research
under U.S. Department of Energy under
contract No.~DE-AC02-05CH11231.  The work at UCSC was supported by
the DOE SciDAC program, under grant No.~DE-FC02-06ER41438.

Computer time for the calculations in this paper was provided
through a DOE INCITE award at the Oak Ridge Leadership Computational
Facility (OLCF) at Oak Ridge National Laboratory, which is supported
by the Office of Science of the U.S. Department of Energy under
Contract No.~DE-AC05-00OR22725.  Visualizations were performed using 
the VisIt package. We thank Gunther Weber and Hank Childs for their
assistance with VisIt.

\clearpage

\clearpage

\clearpage

\clearpage

\clearpage

\clearpage

\clearpage

\clearpage

\clearpage

\clearpage

\clearpage

\clearpage

\clearpage

\clearpage

\clearpage

\clearpage

\clearpage

\clearpage

\clearpage

\clearpage

\clearpage

\clearpage

\clearpage


\end{document}